\documentclass[fleqn,12pt]{report}
\usepackage{amssymb,amsfonts,latexsym,stmaryrd}
\newtheorem{theorem}{Theorem}[section]
\newtheorem{proposition}[theorem]{Proposition}
\newtheorem{lemma}[theorem]{Lemma}
\newtheorem{observation}[theorem]{Observation}
\newtheorem{definition}[theorem]{Definition}
\newtheorem{corollary}[theorem]{Corollary}

\newcommand{\proof}{\noindent {\sc Proof.}}
\newcommand{\qed}{\rule{2mm}{4mm}}
\newcommand{\consistent}{\bigtriangleup}
\newcommand{\true}{{\sl t}}
\newcommand{\false}{{\sl f}}	
\newcommand{\lsem}{[ \! [}
\newcommand{\rsem}{] \! ]}
\newcommand{\lsing}{\{ \! |}
\newcommand{\rsing}{ | \! \}}
\newcommand{\ltuple}{\mbox{${<}$}}
\newcommand{\rtuple}{\mbox{${>}$}}
\newcommand{\preord}{\mbox{${\lesssim}$}}
\newcommand{\notpreord}{\mbox{${\not\lesssim}$}}
\newcommand{\npreord}{\mbox{${\not\lesssim}$}}
\newcommand{\bisim}{\sim}
\newcommand{\Dom}{\mbox{${\cal D}$}}
\newcommand{\sierp}{\mbox{$\Bbb O$}}
\newcommand{\Nat}{\mbox{$\Bbb N$}}
\newcommand{\Oh}{\sierp}
\newcommand{\BLambda}{\mbox{${\bf \Lambda}$}}
\newcommand{\BOmega}{\mbox{${\bf \Omega}$}}
\newcommand{\labarrow}[3]{#1 \stackrel{#2}{\rightarrow} #3}
\newcommand{\nlabarrow}[3]{#1 \stackrel{#2}{\nrightarrow} #3}
\newcommand{\appl}{\mbox{$\cdot$}}
\newcommand{\Converges}{\mbox{${\Downarrow}$}}
\newcommand{\Diverges}{\mbox{${\Uparrow}$}}
\newcommand{\converges}{\mbox{${\downarrow}$}}
\newcommand{\diverges}{\mbox{${\uparrow}$}}

\newcommand{\Ell}{\mbox{${\cal L}$}}
\newcommand{\Ellomega}{\mbox{${\cal L}_{\omega}$}}
\newcommand{\Ellinfty}{\mbox{${\cal L}_{\infty}$}}
\newcommand{\mylet}[4]{{\sf let} \; #1 \; {\sf be} \; (#2,#3). \, #4}
\newcommand{\mycases}[5]{{\sf cases} \; #1 \; {\sf of} \; \imath (#2). \, #3 \;
{\sf else} \; \jmath (#4). \, #5}
\newcommand{\lift}[3]{{\sf lift} \; #1 \; {\sf to} \; {\sf up}(#2). \, #3}
\newcommand{\uextend}[3]{{\sf over} \; #1 \; {\sf extend} \; \lsing #2 \rsing_{u}. \, #3}
\newcommand{\lextend}[3]{{\sf over} \; #1 \; {\sf extend} \; \lsing #2 \rsing_{l}. \, #3}
\newcommand{\curry}[1]{\Lambda (#1)}
\newcommand{\stup}[2]{[#1, #2 ]}
\newcommand{\trans}[1]{(#1)^{\circ}}
\newcommand{\indic}{\bullet \;\;}
\newcommand{\DDL}{{\cal DDL}}
\newcommand{\foldalph}{\alpha}

\newenvironment{Eqarray}{\[ \begin{array}{rclr}}{\end{array} \]}
\begin{document}
\title{Domain Theory and the Logic of Observable Properties}
\author{Samson Abramsky\\
\\
\\
\\
Submitted for the degree of Doctor of Philosophy\\
Queen Mary College\\
University of London}
\date{October 31st 1987}
\bibliographystyle{alpha}
\maketitle
\chapter*{Abstract}
\addtocounter{page}{1}
The mathematical framework of Stone duality is used to synthesize a number of hitherto separate developments in Theoretical Computer Science:
\begin{itemize}
\item Domain Theory, the mathematical theory of computation introduced by Scott as a foundation for denotational semantics.
\item The theory of concurrency and systems behaviour developed by Milner, Hennessy {\it et al.} based on operational semantics.
\item Logics of programs.
\end{itemize}
Stone duality provides a junction between semantics (spaces of points = denotations of computational processes) and  logics (lattices of {\em properties} of 
processes).
Moreover, the underlying logic is {\em geometric}, which can be computationally
interpreted as the logic of {\em observable} properties---i.e. properties which
can be determined to hold of a process on the basis of a finite amount
of information about its execution.

These ideas lead to the following  programme:
\begin{enumerate}
\item A metalanguage is introduced, comprising
\begin{itemize}
\item types = universes of discourse for various computational situations.
\item terms = programs = syntactic intensions for models or points.
\end{itemize}
\item A standard denotational interpretation of the metalanguage is given, assigning domains to types and domain elements to terms.
\item The metalanguage is also given a {\em logical} interpretation, in which 
types are interpreted as  propositional theories and terms are interpreted 
{\it via} a program logic, which axiomatizes the properties they satisfy.
\item The two interpretations are related by showing that they are Stone duals of each other.
Hence, semantics and logic are guaranteed to be in harmony with each other, and in fact each determines the other up to isomorphism.
\item 
This opens the way to a whole range of applications.
Given a denotational description of a computational situation in our
meta-language, we can turn the handle to obtain a logic for that situation.
\end{enumerate}
\section*{Organization}
Chapter~1 is an introduction and overview.
Chapter~2 gives some background on domains and locales.
Chapters~3 and~4 are concerned with~1--4 above.
Chapters~5 and~6 each develop a major case study along the lines suggested by~5, in the areas of concurrency and $\lambda$-calculus respectively.
Finally, Chapter~7  discusses directions for further research.

\chapter*{Preface}
\section*{Acknowledgements}
My warmest thanks to the many people who have helped me along the way:
\begin{itemize}
\item To my colleagues at Queen Mary College (1978--83) for five
very happy and productive years.
\item To my supervisor, Richard Bornat, who gave me so much of his
time during my two years as a full-time Research Student, and also
gave me confidence in the worth of my ideas.
\item To Tom Maibaum for our regular meetings to work on semantics
in 1982--3; these were a life-line when my theoretical work had
previously been done in a vacuum.
\item To my colleagues in the Theory and Formal Methods Group
in the Department of Computing, Imperial College: 
Mark Dawson, Dov Gabbay, Chris Hankin, Yves Lafont, Tom Maibaum, Luke Ong, Iain Phillips, 
Martin Sadler, Mike Smyth,  Richard Sykes, 
Paul Taylor and Steve Vickers,
for creating such a stimulating and inspiring environment in which to
work.
\item To Axel Poign\'{e}, who has just returned to Germany to take
up a post at GMD, for being the most inspiring of colleagues,
whose interest in and encouragement of my work has meant
a great deal to me.
\item To Mark Dawson, for unfailingly finding elegant solutions to all
my computing problems.
\item To my hosts for two very enjoyable visits when much of the
work reported in Chapters 5 and 6 was done: the Programming
Methodology Group, Chalmers Technical University, G\"{o}teborg, Sweden,
March 1984; and Professor Raymond Boute and the Functional Languages 
and Architectures Group, University of Nijmegen, the Netherlands, 
March--April and
August, 1986.
\item To a number of colleagues for conversations, lectures and writings
which have provided inspiration and stimulus to this work: Henk Barendregt,
Peter Dybjer, Matthew Hennessy,  Per Martin-L\"{o}f, 
Robin Milner, 
Gordon Plotkin, Jan Smith, 
Mike Smyth,
Colin Stirling
and  Glynn Winskel.
Glynn's persistent enthusiasm for and encouragement of this work have
meant a great deal.
\end{itemize}
The ideas of  Mike Smyth, Gordon Plotkin and Per Martin-L\"{o}f have
been of particular importance to me in my work on this thesis.
Equally important has been the paradigm of how to do Computer Science
which I like many others have found in the work of Robin Milner and 
Gordon  Plotkin.
I thank them all for their inspiration and example.

I thank the Science and Engineering Research Council for supporting
my work, firstly with a Research Studentship and then with a number
of Research Grants.
Thanks also to the Alvey Programme for funding such ``long-term''
research, and in particular for providing the equipment on which
this document was produced (by me).

Finally, I thank my family for their love and support and, over the
past few months, their forbearance.
\section*{Chronology}
It may be worthwhile to make a few remarks about the chronology of the
work reported in this thesis, as a number of manuscripts describing
different versions of some of the material have been in circulation
over the past few years.
My first version of ``Domain Logic'' was worked out in October and
November of 1983, and presented to the Logic Programming Seminar
at Imperial (the invitation was never repeated), and again
at a seminar at Manchester arranged by Peter Aczel the following
February.
The slides of the talk, under the title ``Intuitionistic Logic of
Computable Functions'', were copied to a few researchers.
The main results of Chapter 6 were obtained, in the setting of
Martin-L\"{o}f's Domain Interpretation of his Type Theory,
during and shortly after a visit to Chalmers in March 1984.
A draft paper was begun in 1984 but never completed;
it formed the basis of a talk given at the CMU Seminar on Concurrency
in July 1984.
The outline of Chapter 5 was developed, with the benefit of
many discussions with Axel Poign\'{e}, in October and November 1984.
Thus the main ideas of the thesis had been formulated, admittedly in
rather inchoate form, by the end of 1984.
The following year was mainly taken up with other things;
but a manuscript on ``Domain Theory in Logical Form'',
essentially the skeleton of the present Chapter 4, minus the endogenous
logic, was written in December 1985, and circulated among a few researchers.
A manuscript on ``A Domain Equation for Bisimulation'' was written
during a visit to the University of Nijmegen in March--April 1986,
and another on ``Finitary Transition Systems'' soon afterwards.
A talk on ``The Lazy $\lambda$-Calculus'' was given at Nijmegen in
August 1986.
Chapters 3, 5 and 6 were written in September--December 1986, together
with a skeletal version of Chapter 4, which was presented at the
Second Symposium on Logic in Computer Science at Cornell, June 1987
\cite{Abr87a}.

\tableofcontents
\chapter{Introduction}
The main aim of this thesis is to synthesize a number of hitherto separate developments in Theoretical Computer Science and Logic:
\begin{itemize}
\item Domain Theory, the mathematical theory of computation introduced by Scott as a foundation for denotational semantics.
\item The theory of concurrency and systems behaviour developed by Milner, Hennessy {\it et al.} based on operational semantics.
\item Logics of programs.
\item Locale Theory.
\end{itemize}
The key to our synthesis is the mathematical theory of Stone duality, which provides a junction between semantics (topological spaces) and the {\em logic of observable properties} (locales).
As a worked example, we show how Domain Theory can be construed as a logic of observable properties;
and explore some applications to the study of programming languages.
\section{Background}
Domain Theory has been extensively studied since it was introduced by Scott 
\cite{Sco70}, both as regards the basic mathematical theory \cite{PloLN}, 
and the applications, particularly in denotational semantics 
\cite{MS76}, \cite{Sto77}, \cite{Gor79}, \cite{Sch86}, and more recently in static program analysis 
\cite{Myc81}, \cite{Nie84}, \cite{AH87}.
In the course of this development, a number of new perspectives have emerged.
\subsection*{Syntax vs. Semantics}
Domain theory was originally presented as a model theory for computation,
and this aspect was emphasised in \cite{Sco70,Sco80b}.
However, the effective character of domain constructions was immediately 
evident, and made fully explicit in \cite{EC76,Sco76,Smy77,Kan79}.
Moreover, in recent presentations of domains via neighbourhood systems and information systems \cite{Sco81,Sco82}, Scott has shown how the theory can be based on elementary, and finitary, set-theoretic representations, which in the case of information systems are deliberately suggestive of proof theory.

A further step towards explicitly syntactic presentations of domain theory was taken by Martin-L\"{o}f, in his Domain Interpretation of Intuitionistic Type Theory \cite{M-L83}.
His formulation also traces a line of descent from Kreisel's definition of the continuous functionals \cite{Kre59}, via \cite{M-L70,Ers72}.

The general tendency of these developments is to suggest that domains may as well be viewed in terms of {\em theories} as of {\em models}.
Our work should not only confirm this suggestion, but also show how it may be put to use.
\subsection*{Points vs. Properties}
An important recent development in mathematics has been the rise of {\it locale theory}, or ``topology without points'' \cite{Joh82},
in which the open-set lattices rather than the spaces of points become the primary objects of study.
That these mathematical developments have direct bearing on Computer Science was emphasised by Smyth in \cite{Smy83}.
If we think of the open sets as {\em properties} or {\em propositions},
we can think of spaces as logical {\em theories}; continuous maps act on these theories under inverse image as {\em predicate transformers} in the sense of Dijkstra \cite{Dij76}, or modal operators as studied in {\em dynamic logic} \cite{Pra79,Har79}.

There is also an important theme in Computer Science which emerges as confluent with these mathematical developments; namely, the use of notions of {\em observation} and {\em experiment} as a basis for the behavioural semantics of systems.
This plays a major role in the work of Milner, Hennessy {\it et al.} on concurrent systems \cite{Mil80,HM85,Win80}, and also in the theory of higher-order functional languages, e.g. \cite{Plo77,Mil77,BC85,BCL85}.
The leading idea here is to take some notion of {\em observable event} or {\em experiment} as an ``information quantum'', and to construct the meaning of a system out of its information quanta.
This corresponds to the leading idea of locale theory, that ``points'' are nothing but constructions out of properties.
By exploiting this correspondence, we may hope to obtain a {\it rapprochement} between domain theory and denotational semantics, on the one hand, and operationally formulated notions such as {\em observation equivalence} \cite{HM85} on the other.
\subsection*{Denotational vs. Axiomatic}
Another area in programming language theory which has received intensive 
development over the past 15 years has been {\em logics of programs}, 
e.g. Hoare logic \cite{Hoa69,deB80}, dynamic logic \cite{Pra79,Har79}, temporal logic \cite{Pnu77}, etc.
However, to date there has not been a satisfactory integration of this work with domain theory.
For example, dynamic logic deals with sets and relations, which from the perspective of domain theory corresponds only to an extremely naive and restricted fragment of programming language semantics.
One would like to see a dynamic logic of {\em domains} and {\em continuous functions}, which would encompass higher-order functions, quasi-infinite (or ``lazy'') data structures, self-application, non-determinism, and all the other computational phenomena for which domain theory provides a mathematical foundation.

The key mathematical idea which forms the basis of our attempt to draw all these diverse strands together is {\em Stone Duality}, which we now briefly review; a fuller discussion will be found in Chapter 2.
\section{Overview: Stone Duality}
The classic Stone Representation Theorem for Boolean algebras \cite{Sto36} 
is aimed at solving the following problem: 
\begin{quote}
show that every (abstract) Boolean algebra can be represented as a field of 
sets, in which the operations of meet, join and complement are represented by 
intersection, union and set complement.
\end{quote}

Stone's solution to the problem begins with observation that for any topological space $X$, the lattice ${\sf Clop} \; X$ of clopen subsets of $X$ forms a field of sets.
His radical step was to construct, from any Boolean lagebra $B$, a topological space ${\sf Spec} \; B$.
To understand the construction, think of $B$ as (the Lindenbaum algebra of) a classical propositional theory.
The elements of $B$ are thus to be thought of as (equivalence classes of) formulae, and the operations as logical conjunction, disjunction and negation.
Now a {\em model} of $B$ is an assignment of ``truth-values'' 0 or 1 to elements of $B$, in a manner consistent with the logical structure; e.g. so that $\neg b$ is assigned 1 if and only if $b$ is assigned 0.
In short, a model is a Boolean algebra homomorphism $f : B \rightarrow {\bf 2}$, where ${\bf 2} = \{ 0, 1 \}$ is the two-element lattice.
Identifying such an $f$ with $f^{-1}(1) \subseteq B$, which as is well-known is an {\em ultrafilter} over $B$ (see e.g. \cite{Joh82}), we can take ${\sf Spec} \; B$ as the set of ultrafilters over $B$, with the topology generated by
\[ U_{a} \equiv \{ x \in {\sf Spec} \; B : a \in x \} \;\;\; ( a \in B) . \]
The spaces arising as ${\sf Spec} \; B$ for Boolean algebras $B$ in this way were characterised by Stone as the totally disconnected compact Hausdorff spaces (subsequently named {\it Stone spaces} in his honour).
Moreover, we have the isomorphisms
\begin{equation}
B \cong {\sf Clop} \; {\sf Spec} \; B 
\end{equation}
\[ b \mapsto \{ x \in {\sf Spec} \; B : b \in x \} \]
\begin{equation}
S \cong {\sf Spec} \; {\sf Clop} \; S
\end{equation}
\[ s \mapsto \{ U \in {\sf Clop} \; S : s \in U \} . \]
The first of these isomorphisms solves the representation problem, and comprises Stone's Theorem in its classical form.
But we can go further; these correspondences also extend (contravariantly) to morphisms:
\[ \frac{S \stackrel{f}{\longrightarrow} T}{{\sf Clop} \; S 
\stackrel{f^{-1}}{\longleftarrow} {\sf Clop} \; T} \;\;\;\;\;\;
\frac{A \stackrel{h^{\star}}{\longleftarrow} B}{{\sf Spec} \; A \stackrel{h}{\longrightarrow} {\sf Spec} \; B} \]
where
\[ h : x \mapsto \{ b \in B : h^{\star} b \in x \} . \]
In modern terminology, this yields a {\em duality} (= contravariant equivalence of categories):
\[ {\bf Stone} \simeq {\bf Bool}^{\sf op} . \]
This is the prototype for a whole family of ``Stone-type duality theorems'', and leads to locale theory, as ``pointless topology'' or junior-grade (propositional) topos theory.
(An excellent reference for these topics is \cite{Joh82}).

But what has all this to do with Computer Science?
Two interpretations of Stone duality can be found in the existing literature from mathematics and logic:
\begin{itemize}
\item The topological view: Points vs. Open sets.
\item The logical view: Models vs. Formulas.
\end{itemize}
We wish to add a third interpretation:
\begin{itemize}
\item The Computer Science view: (Denotations of) computational processes vs. (extensions of) specifications.
\end{itemize}
The importance of Stone duality for Computer Science is that {\em it provides the right framework for understanding the relationship between denotational semantics and program logic}.
The fundamental logical relationship of program development is
\[ P \models \phi \]
to be read ``$P$ satisfies $\phi$'', where $P$ is a program (a syntactic description of a computational process), and $\phi$ is a formula (a syntactic description of a property of computations).
Thus $P$ is the ``how'' and $\phi$ the ``what'' in the dichotomy standardly used to explain the distinction between programs and specifications.
We can easily describe the main formal activities of the program development process in terms of this relation:
\begin{itemize}
\item {\em Program specification} is the task of defining (a list of) properties $\phi$ to be satisfied by the program.
\item {\em Program synthesis} is the task of finding $P$ given (a list of) $\phi$.
\item {\em Program verification} is the task of proving that $P \models \phi$.
\end{itemize}
The two sides of Stone duality---the spatial and the logical or localic---yield alternative but equivalent perspectives on this fundamental relationship:
\begin{itemize}
\item The spatial side of the duality, where points are taken as primary, 
properties are constructed as (open) sets of points, and the fundamental 
relationship is interpreted as $s \in U$ ($s$ a point, $U$ a property), 
corresponds to {\em denotational semantics}, where the data domains 
(i.e. the {\em types}) of a programming language are interpreted as spaces 
of points, and programs are given denotations as points in these spaces; 
this denotational perspective yields a topological interpretation of program logic.
\item The logical or localic side of the duality, where properties,  
as elements of an abstract (logical) lattice, are taken as primary, 
and points are constructed as sets (prime filters) of properties, 
with the fundamental relationship interpreted as $a \in x$ ($a$ a property, 
$x$ a point), corresponds to program logic, and yields a {\em logical interpretation of denotational semantics}.
The idea is that the structure of the open-set lattices and prime filters are presented {\em syntactically}, via axioms and inference rules, as a formal system.
\end{itemize}
We extract the following concrete research programme from these general perspectives on Stone duality:
\begin{enumerate}
\item A metalanguage is introduced, comprising
\begin{itemize}
\item types = data domains = universes of discourse for various computational situations.
\item terms = programs = syntactic intensions for models or points.
\end{itemize}
\item A standard denotational interpretation of the metalanguage, assigning domains to types and domain elements to terms, can be given using the spatial side of Stone duality.
\item The metalanguage is also given a {\em logical} interpretation, in which the localic side of the duality is presented as a formal system with axioms and inference rules.
Each type is interpreted as a propositional theory; and terms are interpreted by axiomatising the satisfaction relation $P \models \phi$.
This gives a program logic.
\item The denotational semantics from~2 and the program logic from~3 are related by showing that they are Stone duals of each other---a strengthened form of the logician's ``Soundness and Completeness''.
As a consequence of this, semantics and logic are guaranteed to be in harmony with each other, and in fact each determines the other up to isomorphism.
\item The framework developed in~1--4 is very {\em general}.
The metalanguage can be used to describe a wide variety of computational situations, following the ideas of ``classical'' denotational semantics.
Given such a description, we can turn the handle to obtain a logic for that situation.
This offers two exciting prospects: of replacing {\it ad hoc} ingenuity in the design of program logics to match a given semantics by the routine application of systematic general theory; and of bringing hitherto divergent fields of programming language theory (e.g. $\lambda$-calculus and concurrency) within the scope of a single unified framework.
\end{enumerate}
The main objective of this thesis is to elaborate the programme outlined in~1--5.
Chapter~2 is devoted to filling in some background on domains and locales.
Then Chapters~3 and~4 are concerned with~1--4 above.
Chapters~5 and~6 each develop a major case study along the lines suggested by~5, in the areas of concurrency and $\lambda$-calculus respectively.
Finally, Chapter~7  discusses directions for further research.

\chapter{Background: Domains and Locales}
The purpose of this Chapter is to summarise what we assume, to fix notation, and to review some basic definitions and results.
\section{Notation}
Most of the notation from elementary set theory and logic which we will use is standard and should cause no problems to the reader.
We shall use $\equiv$ for {\it definitional equality}; thus $M \equiv N$ 
means ``the expression $M$ is by definition equal to'' (or just: ``is defined to be'') ``$N$''.
We shall use $\omega$ to denote the natural numbers $\{ 0, 1, \ldots \}$ (thought of sometimes as an ordinal, and sometimes as just a set); and $\Bbb N$ to denote the set of {\em positive} integers $\{ 1, 2, \ldots \}$.
Given a set $X$, we write $\wp X$ for the powerset of $X$, $\wp_{\sf f} X$ for the set of {\em finite} subsets of $X$, and $\wp_{\sf fne} X$ for the {\em finite non-empty} subsets.
We write $X \subseteq_{\sf f} Y$ for ``$X$ is a finite subset of $Y$''.

We write substitution of $N$ for $x$ in $M$, where $M$, $N$ are expressions and $x$ is a variable, as $M[N/x]$.
We shall assume the usual notions of free and bound variables, as expounded e.g. in \cite{Bar}.
We shall always take expressions modulo $\alpha$-conversion, and treat substitution as a {\em total} operation in which variable capture is avoided by suitable renaming of bound variables.

Our notations for semantics will follow those standardly used in denotational semantics.
One operation we will frequently need is {\it updating} of environments.
Let ${\sf Env} = {\sf Var} \rightarrow {\cal V}$, where {\sf Var} is a set of variables, and $\cal V$ some value space.
Then for $\rho \in {\sf Env}$, $x \in {\sf Var}$, $v \in {\cal V}$, the expression $\rho [x \mapsto v]$ denotes the environment defined by
\[ ( \rho [ x \mapsto v]) y = \left\{ \begin{array}{ll}
v, & x = y \\
\rho y, & \mbox{otherwise.}
\end{array} \right. \]

Next, we recall some notions concerning posets (partially ordered sets).
Given a poset $P$ and $X \subseteq P$, we write
\[ \begin{array}{lcl}
\converges (X) & = & \{ y \in P : \exists x \in X. \, y \leq x \} \\
\diverges (X) & = & \{ y \in P : \exists x \in X. \, x \leq y \} \\
{\sf Con}(X) & = & \{ y \in P : \exists x, z \in X. \, x \leq y \leq z \}
\end{array} \]
We write $\converges (x)$, $\diverges (x)$ for $\converges (\{ x \} )$, $\diverges (\{ x \} )$.
A set $X$ is {\it left-closed} (or {\it lower-closed}) if $X = \converges (X)$,  {\it right-closed} (or {\it upper-closed}) if $X = \diverges (X)$, and {\it convex-closed} if $X = {\sf Con}(X)$.
When it is important to emphasise $P$ we write $\converges_{P}(X)$, $\diverges_{P}(X)$ etc.
We also have the lower, upper and Egli-Milner {\it preorders} (reflexive and transitive relations) on subsets of $P$:
\[ \begin{array}{lcl}
X \sqsubseteq_{l} Y & \equiv & \forall x \in X. \, \exists y \in Y. \, x \leq y   \\
X \sqsubseteq_{u} Y & \equiv & \forall y \in Y. \, \exists x \in X. \, x \leq y \\
X \sqsubseteq_{EM} Y & \equiv & X \sqsubseteq_{l} Y  \: \& \: X \sqsubseteq_{u} Y 
\end{array} \]
We write {\bf 2} for the two-element lattice $\{ 0, 1 \}$ with $0 < 1$, and $\Oh$ for {\em Sierpinski space}, which has the same carrier as {\bf 2}, and topology $\{ \varnothing, \{ 1 \} , \{ 0, 1 \} \}$.
As we shall see in the section on domains and locales, {\bf 2} and $\Oh$ are 
really two faces of the same structure (a ``schizophrenic object'' in the 
terminology of \cite[Chapter 6]{Joh82}), since $\Oh$ arises from the Scott topology on {\bf 2}, and {\bf 2} from the specialisation order on $\Oh$.
For other basic notions of the theory of partial orders and lattices, we refer to \cite{Compend,Joh82}.

Finally, we shall assume a modicum of familiarity with elementary category theory and general topology; suitable references are \cite{Mac71} and \cite{Dug66} respectively.
\section{Domains}
We shall assume some familiarity with \cite{PloLN}, and use it as our reference for Domain theory.
We shall not review such basic definitions as {\it cpo} (complete partial order---\cite[Chapter 1 p.\  7]{PloLN}), {\it continuous function} ({\it loc. cit.}) etc. here.

By a {\it category of domains} we shall mean a sub-category of {\bf CPO}, the category of complete partial orders and continuous functions ({\it loc. cit.}). ${\bf CPO}_{\bot}$ is the category of {\it strict} functions 
(\cite[Chapter 1 p.\  11]{PloLN}).

The properties of {\bf CPO} which make it a suitable mathematical universe for denotational semantics---a ``tool for making meanings'' in Plotkin's phrase---are:
\begin{enumerate}
\item It admits recursive definitions, both of elements of domains, and of domains themselves.
\item It supports a rich type structure.
\end{enumerate}
The mathematical content of (1) is given by the least fixed point theorem for 
continuous functions on cpo's (\cite[Chapter 1 Theorem 1]{PloLN}), 
and the initial fixed point theorem for continuous functors on {\bf CPO} 
(\cite[Chapter 5 Theorem 1]{PloLN}).
As for (2), the type constructions available over {\bf CPO} are extensively 
surveyed in \cite[Chapters 2 and 3]{PloLN}.
In order to fix notation, we shall catalogue the constructions of which mention will be made in this thesis, with references to the definitions in \cite{PloLN}:
\begin{center}
\begin{tabular}{|l|l|l|} \hline
$A \times B$ & product & Ch. 2 p.\  2 \\
$(A \rightarrow B)$ & function space & Ch. 2 p.\  9 \\
$A \oplus B$ & coalesced sum & Ch. 3 p.\  6 \\
$(A)_{\bot}$ & lifting & Ch. 3 p.\  9 \\
$(A \rightarrow_{\bot} B)$ & strict function space & Ch. 1 p.\  13 \\
$P_{l} A$ & lower (Hoare) powerdomain & Ch. 8 p.\  14 \\
$P_{u} A$ & upper (Smyth) powerdomain & Ch. 8 p.\  45 \\
$P_{p} A$ & convex (Plotkin) powerdomain & Ch. 8 p.\  28 \\ \hline
\end{tabular}
\end{center}
(Note that {\em separated sum} $A + B$ can be defined by: $A + B \equiv (A)_{\bot} \oplus (B)_{\bot}$.)

In this thesis, we shall mainly be concerned with {\em algebraic} domains, i.e. sub-categories of $\omega {\bf ALG}$, the category of $\omega$-algebraic cpo's 
\cite[Chapter 6 p.\  2]{PloLN}.
In particular, we shall be concerned with the following three full sub-categories of $\omega {\bf ALG}$:
\begin{enumerate}
\item {\bf AlgLat}: the category of $\omega$-algebraic lattices 
\cite[Chapter 6 p.\  13]{PloLN}.
\item {\bf SDom}: the category of {\em Scott domains}, i.e. the consistently complete $\omega$-algebraic cpo's ({\it loc. cit.}).
(The name comes from the fact that this is exactly the category presented in \cite{Sco81,Sco82}.)
\item {\bf SFP}: the category of {\em strongly algebraic} cpo's 
\cite[Chapter 6 p.\  17]{PloLN}.
The name is an acronym for ``Sequences of Finite Posets''---in more standard terminology, these are the $\omega$-profinite cpo's.
This category was introduced in \cite{Plo76}.
\end{enumerate}
Each of these categories is a full sub-category of the next.

The justification for studying these categories comes from the fact that 
{\bf SFP} is closed under all the type constructions listed above, while 
{\bf SDom} is closed under all but the Plotkin powerdomain.
In particular, both are cartesian closed; indeed, {\bf SFP} is the 
{\em largest} cartesian closed full sub-category of $\omega {\bf ALG}$ 
\cite{Smy83a}, while {\bf SDom} is the largest ``basis elementary'' such sub-category \cite{Gun86}.
Moreover, both categories admit initial solutions of domain equations built from these constructions (obviously excluding the Plotkin powerdomain in the case of {\bf SDom}).
Almost all the domains needed in denotational semantics to date can be defined from these constructions by composition and recursion (some exceptions of three different kinds: \cite{Abr83}, \cite{Ole85}, \cite{Plo82}).
The reason for including {\bf AlgLat} is that it is a usefully simpler special case, which will be applicable to our work in Chapter~6.

Given an algebraic domain $D$, we shall write ${\cal K}(D)$ for its {\it basis}, i.e. the sub-poset of finite elements.
Now algebraic domains are {\em freely constructed} from their bases, i.e.
\[ D \cong {\sf Idl}({\cal K}(D)) \]
where {\sf Idl} is the ideal completion described in \cite[Chapter 6 p.\  5]{PloLN}.
Thus we can in fact completely describe such categories as {\bf SDom} and {\bf SFP} in an elementary fashion in terms of the bases; various ways of doing this for {\bf SDom} are presented in \cite{Sco81,Sco82}.

An important part of this programme is to describe the type constructions listed above in terms of their effect on the bases.
We shall fix some concrete definitions of the constructions for use in later chapters.
\begin{itemize}
\item ${\cal K}(A \times B) = {\cal K}(A) \times {\cal K}(B)$; the ordering is component-wise.
\item ${\cal K}(A \oplus B) = {\cal K}(A) \oplus {\cal K}(B)$, i.e.
\[ \{ \bot \} \cup (\{ 0 \} \times ({\cal K}(A) - \{ \bot_{A} \} )) \cup (\{ 1 \} \times ({\cal K}(B) - \{ \bot_{B} \} )) \]
with the ordering defined by
\begin{eqnarray*}
x \sqsubseteq y & \equiv & x = \bot \\
& & \mbox{or} \; x = (0, a) \: \& \: y = (0, b) \: \& \: a \sqsubseteq_{A} b \\
& & \mbox{or} \; x = (1, c) \: \& \: y = (1, d) \: \& \: c \sqsubseteq_{B} d .
\end{eqnarray*}
\item ${\cal K}((A)_{\bot}) = \{ \bot \} \cup ( \{ 0 \} \times {\cal K}(A))$, with the ordering defined by
\begin{eqnarray*}
x \sqsubseteq y & \equiv & x = \bot \\
& & \mbox{or} \; x = (0, a) \: \& \: y = (0, b) \: \& \: a \sqsubseteq_{A} b.
\end{eqnarray*}
\item ${\cal K}(P_{l}(A)) = \{ \converges_{{\cal K}(A)}(X) : X \in \wp_{\sf fne}({\cal K}(A)) \}$, with the subset ordering.
\item ${\cal K}(P_{u}(A)) = \{ \diverges_{{\cal K}(A)}(X) : X \in \wp_{\sf fne}({\cal K}(A)) \}$, with the superset ordering.
\item ${\cal K}(P_{p}(A)) = \{ {\sf Con}_{{\cal K}(A)}(X) : X \in \wp_{\sf fne}({\cal K}(A)) \}$, with the Egli-Milner ordering (which {\em is} a partial order on the convex-closed sets).
\end{itemize}
All these definitions are valid for {\em any} algebraic cpo.
Since $\omega {\bf ALG}$ is not cartesian closed, we must obviously describe the function space construction for one of its cartesian closed sub-categories.
As the description for {\bf SFP} is rather complicated (see \cite{Gun85}),  we shall give the simpler description for {\bf SDom}.
\begin{definition}
{\rm (i) (\cite[Chapter 6 p.\  1]{PloLN}). Let $A$, $B$ be algebraic domains. For $a \in {\cal K}(A)$, $b \in {\cal K}(B)$,
\[ [a, b] : A \rightarrow B \]
is the {\em one-step} function defined by
\[ [a, b] d = \left\{ \begin{array}{ll}
b & \mbox{if $a \sqsubseteq d$} \\
\bot & \mbox{otherwise}
\end{array} \right. \]
(ii) (\cite[Chapter 6 p.\  13]{PloLN}). $X \subseteq A$ is {\em consistent}:
\[ \bigtriangleup (X) \equiv \exists d \in A. \, \forall x \in X. \, x \sqsubseteq d. \]
We write $x \bigtriangleup y$ for $\bigtriangleup \{ x, y \}$.}
\end{definition}
Note that Plotkin writes $(a \Rightarrow b)$ for $[a, b]$, and $\diverges X$ for $\bigtriangleup (X)$.
\begin{proposition}
\label{funcon}
(\cite[Chapter 6 pp.\  14--15]{PloLN}).
Let $A$, $B$ be Scott domains, and $\{ a_{i} \}_{i \in I} \subseteq {\cal K}(A)$, $\{ b_{i} \}_{i \in I} \subseteq {\cal K}(B)$ for some finite set $I$.

\noindent (i) $\bigtriangleup \{ [a_{i}, b_{i}] : i \in I \}$ if and only if
\[ \forall J \subseteq I. \, \bigtriangleup \{ a_{j} : j \in J \} \; \Rightarrow \; \bigtriangleup \{ b_{j} : j \in J \} \]

\noindent (ii)  $\bigtriangleup \{ [a_{i}, b_{i}] : i \in I \}$ implies that $\bigsqcup \{ [a_{i}, b_{i}] : i \in I \}$ exists and is defined by
\[ ( \bigsqcup \{ [a_{i}, b_{i}] : i \in I \}) d = \bigsqcup \{ b_{i} : a_{i} \sqsubseteq d \} . \]
\end{proposition}
Now we finally get our description of the function space:
\begin{itemize}
\item For Scott domains $A$, $B$:
\begin{eqnarray*}
{\cal K}(A \rightarrow B) & = & \{ \bigsqcup \{ [a_{i}, b_{i}] : i \in I \} : \mbox{$I$ finite}, \\
& & \{ a_{i} \}_{i \in I} \subseteq {\cal K}(A), \: \{ b_{i} \}_{i \in I} \subseteq {\cal K}(B), \\
& &  \bigtriangleup \{ [a_{i}, b_{i}] : i \in I \} \} .
\end{eqnarray*}
\end{itemize}

\section{Locales}
Our reference for locale theory and Stone duality will be \cite{Joh82}.
Since locale theory is not yet a staple of Computer Science, we shall briefly review some of the basic ideas.

Classically, the study of general topology is based on the category {\bf Top} of topological spaces and continuous maps.
However, in recent years mathematicicans influenced by categorical and constructive ideas have advocated that attention be shifted to the open-set lattices as the primary objects of study.
Given a space $X$, we write $\Omega (X)$ for the lattice of open subsets of $X$ ordered by inclusion.
Since $\Omega (X)$ is closed under arbitrary unions and finite intersections, it is a complete lattice satisfying the infinite distributive law
\[ a \wedge \bigvee S = \bigvee \{ a \wedge s : s \in S \} . \]
(By the Adjoint Functor Theorem, in any complete lattice this law is equivalent to the existence of a right adjoint to conjunction, i.e. to the fact that implication can be defined in a canonical way.)
Such a lattice is a {\em complete Heyting algebra}, i.e. the Lindenbaum algebra of an {\em intuitionistic} theory.
The continuous functions  between topological spaces preserve unions and intersections, and hence  all joins and finite meets of open sets, under inverse image; thus we get a functor
\[ \Omega : {\bf Top} \rightarrow {\bf Loc} \]
where {\bf Loc}, the category of {\em locales}, is the opposite of {\bf Frm}, the category of {\em frames}, which has complete Heyting algebras as objects, and maps preserving all joins and finite meets as morphisms.
Note that {\bf Frm} is a concrete category of structured sets and structure-preserving maps, and consequently convenient to deal with (for example, it is monadic over {\bf Set}).
Thus we study {\bf Loc} {\it via} {\bf Frm}; but it is {\bf Loc} which is the proposed alternative or replacement for {\bf Top}, and hence the ultimate object of study.

{\bf Notation.} Given a morphism $f : A \rightarrow B$ in {\bf Loc}, we write $f^{\star}$ for the corresponding morphism $B \rightarrow A$ in {\bf Frm}.

Now we can define a functor
\[ {\sf Pt} : {\bf Loc} \rightarrow {\bf Top} \]
as follows (for motivation, see our discussion of Stone's original construction in Chapter~1):
${\sf Pt} (A)$ is the set of all frame morphisms $f : A \rightarrow {\bf 2}$, where {\bf 2} is the two-point lattice.
Any such $f$ can be identified with the set $F = f^{-1}(1)$, which satisfies:
\[ 1 \in F \]
\[ a, b \in F \; \Rightarrow \; a \wedge b \in F \]
\[ a \in F, a \leq b \; \Rightarrow \; b \in F \]
\[ \bigvee_{i \in I} a_{i} \in F \; \Rightarrow \; \exists i \in I. \, a_{i} \in F . \]
Such a subset is called a {\em completely prime filter}.
Conversely, any completely prime filter $F$ determines a frame homomorphism 
$\chi_{F} : A \rightarrow {\bf 2}$.
Thus we can identify ${\sf Pt}(A)$ with the completely prime filters over $A$.
The topology on ${\sf Pt}(A)$ is given by the sets $U_{a}$ ($a \in A$):
\[ U_{a} \equiv \{ x \in {\sf Pt}(A) : a \in F \} . \]
Clearly, 
\[ {\sf Pt}(A) = U_{1}, \;\; U_{a} \cap U_{b} = U_{a \wedge b}, \;\; \bigcup_{i \in I} U_{a_{i}} = U_{\bigvee_{i \in I} a_{i}}, \]
so this is a topology.
{\sf Pt} is extended to morphisms by:
\[ \frac{A \stackrel{f^{\star}}{\longleftarrow} B}{{\sf Pt}(A) \stackrel{{\sf Pt}(f)}{\longrightarrow} {\sf Pt}(B)} \]
\[ {\sf Pt}(f) x = \{ b : f^{\star} b \in x \} . \]

We now define, for each $X$ in {\bf Top} and $A$ in {\bf Loc}:
\[ \eta_{X} : X \rightarrow {\sf Pt}( \Omega (X)) \]
\[ \eta_{X}(x) = \{ U : x \in U \} \]
\[ \epsilon_{A} : \Omega ({\sf Pt}(A)) \rightarrow A \]
\[ \epsilon_{A}^{\star}(a) = \{ x : a \in x \} . \]
Now we have
\begin{theorem}
\label{t.1}
(\cite[II.2.4]{Joh82}). $(\Omega , {\sf Pt}, \eta , \epsilon ) : {\bf Top} 
\rightharpoonup {\bf Loc}$ defines an adjunction between {\bf Top} and 
{\bf Loc}; moreover (\cite[II.2.7]{Joh82}), this cuts down to an equivalence between the full sub-categories {\bf Sob} of sober spaces and {\bf SLoc} of spatial locales.
\end{theorem}
The equivalence between {\bf Sob} and {\bf SLoc} (and therefore the {\em duality} or contravariant equivalence between {\bf Sob} and {\bf SFrm}) may be taken as the most general purely topological version of Stone duality.
For our purposes, some  dualities arising as restrictions of this one are of interest.
\begin{definition}
{\rm A space $X$ is {\em coherent} if the compact-open subsets of $X$ 
(notation: $K \Omega (X)$) form a basis closed under finite intersections, i.e. for which $K \Omega (X))$ is a distributive sub-lattice of $\Omega (X)$.
}
\end{definition}
\begin{theorem}
\label{t.2}
(i) (\cite[II.2.11]{Joh82}). The forgetful functor from {\bf Frm} to {\bf DLat}, the category of distributive lattices, has as left adjoint the functor {\sf Idl}, which takes a distributive lattice to its ideal completion.

\noindent (ii) (\cite[II.3.4]{Joh82}). Given a distributive lattice $A$, define ${\sf Spec} \; A$ as the set of prime filters over $A$ (i.e. sets of the form $f^{-1}(1)$ for lattice homomorphisms $f : A \rightarrow {\bf 2}$), with topology generated by
\[ U_{a} \equiv \{ x \in {\sf Spec} \; A : a \in x \} \;\;\; (a \in A). \]
Then ${\sf Spec} \; A \cong {\sf Pt}({\sf Idl}(A))$.

\noindent (iii) (\cite[II.3.3]{Joh82}). The duality of Theorem~\ref{t.1} cuts down to a duality
\[ {\bf CohSp} \simeq {\bf CohLoc} \simeq {\bf DLat}^{\sf op} \]
where {\bf CohSp} is the category of coherent $T_{0}$ spaces, and continuous maps which preserve compact-open subsets under inverse image; 
and ${\bf CohLoc}^{\sf op}$ is the image of {\bf DLat} under the functor {\sf Idl}.
\end{theorem}

The logical significance of the coherent case is that finitary syntax---specifically finite disjunctions---suffices.
The original Stone duality theorem discussed in Chapter~1 is obtained as the further  restriction of this duality to coherent Hausdorff spaces (which turns out to be another description of the Stone spaces) and Boolean algebras, i.e. complemented distributive lattices.
Note that under the compact Hausdorff condition, {\em all} continuous maps satisfy the special property in part (iii) of the Theorem.

As a further special case of Stone duality, we note:
\begin{theorem}
\label{msl}
(i) The forgetful functor from distributive lattices to the category {\bf MSL} of meet-semilattices has a left adjoint {\sf L}, where ${\sf L}(A) = \{ \converges(X) : X \in \wp_{\sf f}(A) \}$, ordered by inclusion. 
(Notice that this is the same construction as for the lower powerdomain; this fact {\em is} significant, but not in the scope of this thesis.)

\noindent (ii) For any meet-semilattice $A$, define ${\sf Filt}(A)$ as 
the set of all filters over $A$, with topology defined exactly as for ${\sf Spec}(A)$. Then
\[ {\sf Filt}(A) \cong {\sf Spec}({\sf L}(A)) \cong {\sf Pt}({\sf Idl}({\sf L}(A))). \]

\noindent (iii) The duality of Theorem~\ref{t.2} cuts down to a duality
\[ {\bf CohAlgLat} \simeq {\bf MSL}^{\sf op} \]
where {\bf CohAlgLat} is the full sub-category of {\bf CohSp} of algebraic 
lattices with the Scott topology (to be defined in the next section).
\end{theorem}

An extensive treatment of locale theory and Stone-type dualities can be found in \cite{Joh82}.
Our purpose in the remainder of this section is to give some conceptual perspectives on the theory.

Firstly, a {\em logical} perspective.
As already mentioned, locales are the Lindenbaum algebras of intuitionistic theories, more particularly of {\em propositional geometric theories}, i.e. the logic of finite conjunctions and infinite conjunctions.
The morphisms preserve this geometric structure, but are {\em not} required to preserve the additional ``logical'' structure of implication and negation (which can be defined in any complete Heyting algebra).
Thus from a logical point of view, locale theory is propositional geometric logic.
Moreover, Stone duality also has a logical interpretation.
The {\em points} of a space correspond to {\em models} in the logical sense; the {\em theory} of a model is the completely prime filter of opens it satisfies, where the satisfaction relation is just
\[ x \models a \equiv x \in a \]
in terms of spaces, (i.e. with $x \in X$ and $a \in \Omega (X)$), and
\[ x \models a \equiv a \in x \]
in terms of locales (i.e. with $x \in {\sf Pt}(A)$ and $a \in A$).
Spatiality of a class of locales is then a statement of {\em Completeness}: every consistent theory has a model.

Secondly, a {\em computational} perspective.
If we view the points of a space as the denotations of computational processes 
(programs, systems), then the elements of the corresponding locale can be 
seen as {\em properties} of computational processes.
More than this, these properties can in turn be thought of as computationally meaningful; we propose that they be interpreted as {\em observable properties}.
Intuitively, we say that a property is observable if we can tell whether or not it holds of a process on the basis of only a finite amount of information about that process\footnote{This is really only one facet of observability. Another
is {\em extensionality}, i.e. that we regard a process as a black box with
some specified interface to its environment, and only take what is observable
via this interface into account in determining the meaning of the process.
Extensionality in this sense is obviously {\em relative} to our choice of 
interface;
it is orthogonal to the notion being discussed in the main text.}.
Note that this is really {\em semi}-observability, since if the property is {\em not} satisfied, we do not expect that this is finitely observable.
This intuition of observability  motivates the asymmetry between conjunction and disjunction in geometric logic and topology.
Infinite disjunctions of observable properties are still observable---to see that $\bigvee_{i \in I} a_{i}$ holds of a process, we need only observe that {\em one} of the $a_{i}$ holds---while infinite conjunctions clearly do not preserve finite observability in general.
More precisely, consider Sierpinski space $\Oh$.
We can regard this space as representing the possible outcomes of an experiment to determine whether a property is satisfied; the topology is motivated by semi-observability, so an observable property on a space $X$ should be a {\em continuous} function to $\Oh$.
In fact, we have
\[ \Omega (X) \cong (X \rightarrow \Oh ) \]
where $(X \rightarrow \Oh )$ is the continuous function space, ordered pointwise (thinking of $\Oh$ as {\bf 2}).
Now for infinite $I$, $I$-ary disjunction, viewed as a function
\[ \Oh^{I} \rightarrow \Oh \]
is continuous, while $I$-ary conjunction is not.
Similarly, implication and negation, taken as functions
\[ \Rightarrow : \Oh^{2} \rightarrow \Oh , \;\;\; \neg : \Oh \rightarrow \Oh \]
are not continuous.
Thus from this perspective,
\begin{center}
geometric logic = observational logic.
\end{center}

These ideas follow those proposed by Smyth in his pioneering paper \cite{Smy83}, but with some differences.
In \cite{Smy83}, Smyth interprets ``open set'' as {\em semi-decidable} property; this represents an ultimate commitment to interpret our mathematics in some effective universe.
My preference is to do Theoretical Computer Science in as ontologically or foundationally {\em neutral} a manner as possible.
The distinction between semi-observability and semi-decidability is analogous 
to the distinction between the computational motivation for the basic axioms 
of domain theory in terms of ``physical feasibility'' given in 
\cite[Chapter 1]{PloLN}, without any appeal to notions of recursion theory; and a commitment to only considering computable elements and morphisms of effectively given domains, as advocated in \cite{Kan79}.
It should also be said that the link between observables and open sets in domain theory was clearly (though briefly!) stated in \cite[Chapter 8 p.\  16]{PloLN}, and used there to motivate the definition of the Plotkin powerdomain.

A final perpective is {\em algebraic}.
The category {\bf Frm} is algebraic over {\bf Set} (\cite[II.1.2]{Joh82}); thus working with locales, we can view topology as a species of (infinitary) algebra.
In particular,  constructions of universal objects of various kinds by ``generators and relations'' are possible.
Two highly relevant examples in the locale theory literature are \cite{Joh85} and \cite{Hyl81}. 
This provides a link with the information systems approach to domain theory as in \cite{Sco82,LW84}.
Some of our work in Chapters~3 and~4 can be seen as a systematization of these ideas in an explicitly syntactic framework.

\section{Domains and Locales}
We now turn to the connections between domains and locales.
Firstly, it is standard that domains can be viewed topologically.
\begin{definition}
{\rm (\cite[Chapter 1 p.\  16]{PloLN}). Given a poset $P$, the {\em Scott topology} on $P$ has as open sets those $U \subseteq P$ satisfying
\begin{enumerate}
\item $U$ is upper-closed, i.e. $U = \diverges (U)$.
\item $U$ is inaccessible by $\omega$-chains, i.e.
\[ \bigsqcup_{n \in \omega} x_{n} \in U \; \Rightarrow \;  \exists n. \, x_{n} \in U. \]
\end{enumerate}
We write $\sigma (D)$ for the Scott topology on a domain $D$.}
\end{definition}
\begin{proposition}
\label{cop}
(i) ({\it loc. cit.}) Let $D$, $E$ be cpo's; a function $f : D \rightarrow E$ 
is continuous in the cpo sense iff it is continuous with respect to the Scott topology.

\noindent (ii) (\cite[Chapter 6 p.\  3]{PloLN}). 
For algebraic domains $D$, the Scott topology has a particularly simple form: namely all sets of the form
\[ \bigcup_{i \in I} \diverges (b_{i}) \;\;\;\; (b_{i} \in {\cal K}(D), i \in I) \]
Moreover, the compact-open sets are just those of this form with $I$ {\em finite}.
\end{proposition}
Given a space $X$, we define the {\it specialisation order} on $X$ by
\[ x \leq_{\sf spec} y \equiv \forall U \in \Omega (X). \, x \in U \; \Rightarrow \; y \in U. \]
\begin{proposition}
(\cite[Chapter 1 p.\  16]{PloLN}). Let $D$ be a cpo. The specialisation order on the space $(D, \sigma (D))$ coincides with the original ordering on $D$.
\end{proposition}
Thus we may regard domains indifferently as posets or as spaces with the Scott topology, justifying some earlier abuses of notation.

We now relate domains to coherent spaces.
\begin{theorem}[The $2/3$ SFP Theorem]
(\cite[Chapter 8 p.\  41]{PloLN}). An algebraic cpo is coherent as a space iff it is ``$2/3$ SFP'' in the terminology of (\it loc. cit.).
Since coherent spaces are sober (\cite{Joh82} II.3.4), any such domain $D$ satisfies
\[ D \cong {\sf Spec}(K \Omega (D)). \]
\end{theorem}
We shall refer to such domains as {\em coherent algebraic}.
Thus {\bf SDom} and {\bf SFP} are categories of coherent spaces, and we need only consider the lattices of compact-open sets on the logical side of the duality.

We conclude with some observations which show how the finite elements in a coherent algebraic domain play an ambiguous role as both points and properties.
Firstly, we have
\[ D \cong {\sf Idl}({\cal K}(D)) \]
so the finite elements determine the structure of $D$ on the spatial side.
We can also recover the finite elements in purely lattice-theoretic terms from $A = K \Omega (D)$.
Say that $a \in A$ is {\it consistent} if $a \not= 0$, and {\it prime} if $a \leq {b \vee c}$ implies $a \leq b$ or $a \leq c$. (We should probably say coprime rather than prime, but as we will have no need for the dual concept, we will use the shorter term.)
Writing $cpr(A)$ for the set of consistent primes of $A$, we have
\begin{equation}
{\cal K}(D) = (cpr(A))^{\sf op}, \;\;\; A \cong {\sf L}(({\cal K}(D))^{\sf op}) . 
\end{equation}
(The fact that the latter construction produces a distributive lattice even 
though ${\cal K}(D)$ is not a meet-semilattice follows from the MUB axioms 
characterizing the coherent algebraic domains \cite[Chapter 8 p.\  41]{PloLN}.)

\begin{theorem}
Let $A$ be a distributive lattice. ${\sf Spec}(A)$ is coherent algebraic iff the following conditions are satisfied:
\[ \begin{array}{rl}
(1) & 1_{A} \in cpr(A) \\
(2) & \forall a \in A. \, \exists b_{1}, \ldots , b_{n} \in cpr(A). \, a = \bigvee_{i=1}^{n} b_{i}. 
\end{array} \]
\end{theorem}
Of these, (1) ensures the existence of a bottom point, and (2) says ``there are enough primes''.
This result will be proved as part of our work in the next Chapter.

\chapter{Domains and Theories}
\section{Introduction}
In this Chapter, we lay some of the foundations for the domain 
logic to be presented in Chapter 4. 
In section 2, a category of domain prelocales (coherent propositional theories) and approximable mappings is defined, 
and proved equivalent to {\bf SDom}. 
This is the category in which, implicitly, all the work of Chapter 4 is set. 
In section 3, following the ideas of a number of authors, particularly 
Larsen and Winskel in \cite{LW84}, 
a large cpo of domain prelocales is defined, and used to reduce the solution 
of domain equations to taking least fixpoints of continuous functions over this cpo. 
In section 4, a number of type constructions are defined as operations over domain prelocales. 
We prove in detail that these operations are naturally isomorphic to the corresponding constructions on domains. 
In section 5 a semantics for a language of recursive type expressions is given, in which each type is interpreted as a logical theory. 
This is related to a standard semantics in which types denote domains by 
showing that for each type its interpretation in the logical semantics is 
the Stone dual of its denotation in the standard semantics.

{\bf Important Notational Convention.}
Throughout this Chapter and the next, we shall use $I$, $J$, $K$, $L$
to range over {\em finite} index sets.
\section{A Category of Pre-Locales}
\begin{definition} 
{\rm A {\em coherent prelocale} is a structure
\[A = (|A|, \leq_{A}, =_{A}, 0_{A}, \vee_{A}, 1_{A}, \wedge_{A}) \]
where \begin{itemize}
\item $|A|$ is a set, the {\it carrier}
\item $\leq_{A}$, $=_{A}$ are binary relations over $|A|$
\item $0_{A}$, $1_{A}$ are constants, i.e. elements of $|A|$
\item $\vee_{A}$, $\wedge_{A}$ are binary operations over $|A|$
\end{itemize}
subject to the following axioms (subscripts omitted):}
\[(p1) \;\;\; a \leq a \;\;\; \frac{a \leq b \;\; b \leq c}{a \leq c} \;\;\;\;     \frac{a \leq b \;\; b \leq a}{a = b} \;\;\;\; \frac{a = b}{a \leq b \;\; b \leq a} \]
\[(p2) \;\;\; 0 \leq a \;\;\;\; \frac{a \leq c \;\; b \leq c}{a \vee b \leq c} \;\;\;\; a \leq a \vee b \;\;\;\; b \leq a \vee b \]
\[(p3) \;\;\; a \leq 1 \;\;\;\; \frac{a \leq b \;\; a \leq c}{a \leq b \wedge c} \;\;\;\; a \wedge b \leq a \;\;\;\; a \wedge b \leq b \]
\[(p4) \;\;\; a \wedge (b \vee c) \leq (a \wedge b) \vee (a \wedge c) \]
\end{definition}
Evidently, the quotient structure
\[ \tilde{A} = (|A|/{=_{A}}, {\leq} /{=_{A}}) \]
is a distributive lattice.
\begin{definition} 
{\rm Given a prelocale A, we define}
\[\begin{array}{rrcl}
(i) & pr(A) & \equiv & \{a \in |A| : \forall b, c \in |A|.\, a \leq b \vee c \Rightarrow a \leq b \; {\rm or} \; a \leq c \} \\
(ii) & con(A) & \equiv & \{a \in |A| : \neg (a =_{A} 0) \} \\
(iii) & cpr(A) & \equiv & con(A) \cap pr(A) \\
(iv) & t(A) & \equiv & \{a \in |A| : \neg (a =_{A} 1) \}
\end{array} \]
\end{definition}
\begin{definition} 
{\rm A {\it domain prelocale} is a coherent prelocale $A$ which satisfies 
the following additional axioms:}
\[ (d1) \;\;\; \forall a \in |A| . \, \exists b_{1}, \ldots b_{n} \in pr(A). \, a =_{A} \bigvee_{i=1}^{n}b_{i} \]
\[ (d2) \;\;\; 1_{A} \in cpr(A) \]
\[ (d3) \;\;\; a, b \in pr(A) \; \Rightarrow \; a \wedge b \in pr(A) \]
\end{definition}
We now introduce a notion of morphism for domain prelocales, based on Scott's {\it approximable mappings} \cite{Sco81,Sco82}.
\begin{definition} 
{\rm Let $A$, $B$, be domain prelocales. An {\it approximable mapping} $R : A \rightarrow B$ is a relation $R \subseteq |A| \times |B|$ satisfying}
\[ (r1) \;\;\; a R 1 \]
\[ (r2) \;\;\; a R b \: \& \: a R c \; \Rightarrow \; a R (b \wedge c) \]
\[ (r3) \;\;\; 0 R b\]
\[ (r4) \;\;\; a R c \: \& \: b R c \; \Rightarrow \; (a \vee b) R c \]
\[ (r5) \;\;\; a \leq a' R b' \leq c \; \Rightarrow \; a R b \]
\[ (r6) \;\;\; a R 0 \; \Rightarrow \; a =_{A} 0 \]
\[ (r7) \;\;\; a \in pr(A) \: \& \: a R (b \vee c) \; \Rightarrow \; a R b \: 
\; {\rm or} \; \: a R c. \]
\end{definition}

Approximable mappimgs are closed under relational composition. 
We verify the least trivial closure condition, $(r7)$. 
Suppose $R : A \rightarrow B$, $S : B \rightarrow C$, $a \in pr(A)$ and 
$a (R \circ S) b \vee c$. 
For some $d \in |B|$, $a R d$ and $d S b \vee c$. By $(d1)$,
\[ d =_{B} \bigvee_{i \in I}d_{i} \;\;(d_{i} \in pr(B), i \in I). \]
If $I = \varnothing$, $d =_{B} 0_{B}$, hence by $(r3)$ $d R b$, and so $a (R \circ S) b$. 
Otherwise, by $(r7)$, $a R d_{i}$ for some $i \in I$. Now
\[ d_{i} \leq \bigvee_{i \in I}d_{i} S (b \vee c) \]
\[ \Rightarrow \;\; d_{i} S (b \vee c) \;\;\; (r5) \]
\[ \Rightarrow \;\; d_{i} S b \; {\rm or} \; d_{i} S c \;\;\; (r7) \]
\[ \Rightarrow \;\; a (R \circ S) b \; {\rm or} \; a (R \circ S) c\]
as required.
Identities with respect to this composition are given by
\[ a \: {\rm id}_{A} \: b \; \equiv \; a \leq_{A} b. \]
Hence we can define a category {\bf DPL} of domain prelocales and approximable mappings.

\begin{definition} 
{\rm A {\em pre-isomorphism} $\varphi : A \simeq B$ of domain prelocales is a surjective function
\[ \varphi : |A| \rightarrow |B| \]
satisfying}
\[ \forall a, b \in |A|. \, a \leq_{A} b \; \Leftrightarrow \; \varphi (a) \leq_{B} \varphi (b). \]
\end{definition}
\begin{proposition}
If $\varphi : A \simeq B$ is a preisomorphism, the relation
\[ a R_{\varphi} b \; \equiv \; \varphi (a) \leq_{B} b \]
is an isomorphism in {\bf DPL}. \qed
\end{proposition}
\begin{theorem} \label{domtheq}
{\bf DPL} is equivalent to {\bf SDom}.
\end{theorem}

\proof\ We define functors
\[ F : {\bf SDom} \rightarrow {\bf DPL} \]
\[ G : {\bf DPL} \rightarrow {\bf SDom} \]
as follows:
\[ F(D) = (K \Omega (D), \subseteq , =, \varnothing , \cup , D, \cap ) \]
i.e. the distributive lattice of compact-open subsets of $D$;
\[ F(f) = R_{f}, \]
where
\[ a R_{f} b \; \equiv \; a \subseteq f^{- 1}(b). \]
The verification that F is well-defined is routine. Note that:
\[ \bullet \;\; pr(F(D)) = \{\diverges  u : u \in K(D) \} \cup \{\varnothing \} \]
\[ \bullet \;\; a \in con(F(D)) \; \Leftrightarrow \; a \not= \varnothing \]
\[ \bullet \;\; {\diverges  u} \cap {\diverges  v} \in con(F(D)) \; \Leftrightarrow \; u \consistent\ v \]
To verify $(r7)$ for $R_{f}$, note that, for $u \in K(D)$:
\begin{eqnarray*}
\diverges  u \subseteq f^{-1}(b \cup c)  & \Leftrightarrow &  u \in f^{-1}(b \cup c) \\
                     & \Leftrightarrow & f(u) \in b \cup c \\
                     & \Leftrightarrow & f(u) \in b \; {\rm or} \; f(u) \in c \\
                     & \Leftrightarrow & \diverges  u \subseteq  f^{-1}(b) 
\; {\rm or} \; \diverges  u \subseteq f^{-1}(c). 
\end{eqnarray*} 

\[ G(A) \equiv \hat{A}, \]
where $\hat{A}$ is the set of prime proper filters of $A$, i.e. sets $x \subseteq |A| - \{0_{A}\}$ closed under finite conjunction and entailment and satisfying
\[ a \vee b \in x \; \Rightarrow \; a \in x \; {\rm or} \; b \in x. \]
$\hat{A}$ is a partial order under set inclusion; or, equivalently, (via the specialisation order) a topological space with basic opens
\[ U_{a} \equiv \{x \in \hat{A} : a \in x \} \;\; (a \in |A|). \]
Note that, with either structure,
\[ \hat{A} \; \cong \; {\rm Spec} \: \tilde{A}. \]

\[ G(R) = f_{R}, \]
where
\[ f_{R}(x) = \{b \: | \: \exists a \in x. \, a R b \}. \]
We check that $G$ is well defined. By $(d2)$, the filter generated by 1
is prime, hence a least element for $\hat{A}$; while it is easy to see that $\hat{A}$ is closed under unions of directed families. 
Thus $\hat{A}$ is a cpo. 
Moreover, the principal filters $\diverges  (a)$ with $a \in cpr(A)$ are prime, and (using $(d1)$) form a basis of finite elements.
Finally, by $(d3)$ this basis is closed under consistent finite  joins. Thus $\hat{A}$ is a Scott domain.

Now we check that $f_{R}$ is well defined and continuous. Given $x \in \hat{A}$, it is easy to see that $f_{R}(x)$ is a filter. To check that it is prime, suppose $b \vee c \in f_{R}(x)$. 
Then for some $a \in x$, we must have $a R (b \vee c)$. 
By $(d1)$,
\[ a =_{A} \bigvee_{i \in I}a_{i}, \;\;\;(a_{i} \in cpr(A), i \in I). \]
Since $x$ is a proper filter, $a \not= 0$, hence $I \not= \varnothing$. 
Then since $x$ is prime, for some $i \in I$ $a_{i} \in x$. Now by $(r7)$,
\[ a_{i} R (b \vee c) \; \Rightarrow \; a_{i} R b \; {\rm or} \; a_{i} R c \]
and so $b \in f_{R}(x)$ or $c \in f_{R}(x)$.
Since directed joins in $\hat{A}$ are just unions, continuity of $f_{R}$ is trivial.

The remainder of the verification that G is a functor is routine.

We now define natural transformations
\[ \eta : I_{{\bf SDom}} \rightarrow GF \]
\[ \epsilon : I_{{\bf DPL}} \rightarrow FG \]

\[ \eta D(d) = \{U \in K\Omega (D) : d \in U\} \]
\[ \epsilon A = R_{\varphi A}, \]
where $\varphi A : A \simeq K\Omega (\hat{A})$ is the pre-isomorphism defined by
\[ \varphi A(a) = \{x \in \hat{A} : a \in x\}. \]
Note that $\eta$, $\varphi$ are the natural isomorphisms in the Stone duality for distributive lattices. This shows that the components of $\eta$, $\epsilon$ are isomorphisms, while naturality is easily checked to extend to our setting.

Altogether, we have shown that 
\[ (F, G, \eta , \epsilon ) : {\bf SDom} \simeq {\bf DPL} \]
is an equivalence of categories. \qed

\section{A Cpo of Pre-locales}
In this section, we follow the ideas of Larsen and Winskel \cite{LW84}, and 
define a (large) cpo of domain pre-locales, in such a way that type constructions can be represented as continuous functions over this cpo, and the process of solving recursive domain equations reduced to taking least fixed points of such functions.

\begin{definition} 
{\rm Let $A$, $B$ be domain prelocales. Then we define $A \Subset B$ iff
\begin{itemize}
\item $|A| \subseteq |B|$
\item $(|A|, 0_{A}, \vee_{A}, 1_{A}, \wedge_{A})$ is a subalgebra of $(|B|, 0_{B}, \vee_{B}, 1_{B}, \wedge_{B})$
\item $\leq_{A} \; \subseteq \; \leq_{B}$
\end{itemize}}
\end{definition}
Although this inclusion relation is simple, it is too weak, and has only been introduced for organisational purposes. What we need is
\begin{definition} 
{\rm $A \trianglelefteq B$ iff}
\[(s1) \;\;\; A \Subset B \]
\[(s2) \;\;\; \forall a, b \in |A|. \, a \leq_{B} b \; \Rightarrow a \leq_{A} b \]
\[(s3) \;\;\; pr(A) \subseteq pr(B) \]
\end{definition}
Note that apart from $(s3)$ this is just the usual notion of {\it submodel} 
(cf. e.g. \cite{CK73}).
\begin{proposition}
The class of domain prelocales under $\trianglelefteq$ is an $\omega$-chain complete partial order.
\end{proposition}

\proof\ The verification that $\trianglelefteq$ is a partial order is routine. Let $\{A_{n}\}$ be a $\trianglelefteq$-chain. Set
\[A_{\infty} \equiv (\bigcup_{n \in \omega}A_{n}, \bigcup_{n \in \omega}\leq_{A_{n}}, \ldots etc.). \]
We check that $A_{\infty}$ is a well-defined domain prelocale, for in that case it is clearly the least upper bound of the chain. We verify $(d1)$ for illustration.

Given $a \in |A_{\infty}|$, for some $n$, $a \in |A_{n}|$,  hence
\[a =_{A_{n}} \bigvee_{i \in I}a_{i}, \;\; (a_{i} \in pr(A_{n}), i \in I).\]
Clearly $a =_{A_{\infty}} \bigvee_{i \in I}a_{i}$; furthermore, $pr(A_{n}) \subseteq pr(A_{\infty})$. To see this, suppose $b \in pr(A_{n})$ and $b \leq_{A_{\infty}} c \vee d$. For some $m \geq n$, $\{a, b, c\} \subseteq |A_{m}|$, and so $b \leq_{A_{m}} c \vee d$. Since $A_{n} \trianglelefteq A_{m}$, $pr(A_{n}) \subseteq pr(A_{m})$, and so $b \leq_{A_{m}} c$ or $b \leq_{A_{m}} d$, which implies $b \leq_{A_{\infty}} c$ or $b \leq_{A_{\infty}} d$, as required. \qed

The class of domain prelocales is not a cpo under $\trianglelefteq$; it does not have a least element. However, we can easily remedy this deficiency.
\begin{definition} 
{\rm {\bf 1} is the domain prelocale defined as follows. 
The carrier $|{\bf 1}|$ is defined inductively by
\begin{itemize}
\item $\true , \false \in |{\bf 1}|$
\item $a, b \in |{\bf 1}| \; \Rightarrow \; a \wedge b, a \vee b \in |{\bf 1}|$
\end{itemize}
The operations are defined ``freely'' in the obvious way:
\[0_{{\bf 1}} \equiv \false , \;\; 1_{{\bf 1}} \equiv \true , \;\; a \vee_{{\bf 1}} b \equiv a \vee b , \;\; a \wedge_{{\bf 1}} b \equiv a \wedge b \]
Finally, $\leq_{{\bf 1}}$, $=_{{\bf 1}}$ are defined inductively as the least relations satisfying $(p1)$--$(p4)$.
It is easy to see that $\tilde{{\bf 1}}$ is the two-point lattice; hence {\bf 1} is a domain prelocale.}
\end{definition}
Now let {\bf DPL1} be the class of domain prelocales $A$ such that ${\bf 1} \trianglelefteq A$. Clearly {\bf DPL1} is still chain-complete. Thus we have
\begin{proposition}
{\bf DPL1} is a large cpo with least element {\bf 1}. \qed
\end{proposition}
{\bf DPL1} also determines a full subcategory of {\bf DPL}. To see that we are not losing anything in passing from {\bf DPL} to {\bf DPL1}, we note
\begin{proposition}
{\bf DPL1} is equivalent to {\bf DPL}. \qed
\end{proposition}

We now relate this partial order of prelocales to the category of domains and 
embeddings used in the standard category-theoretic treatment of the solution 
of domain equations \cite{SP82}. 
Recall that an {\it embedding-projection pair} between domains $D$, $E$ 
is a pair of continuous functions $e : D \rightarrow E$, $p : E \rightarrow D$ satisfying
\[p \circ e = {\sf id}_{D} \]
\[ e \circ p \sqsubseteq {\sf id}_{E}. \]
Each of these functions uniquely determines the other, since $e$ is left adjoint to $p$. We write $e^{R}$ for the projection determined by $e$.
\begin{proposition}
If $A \trianglelefteq B$, then $e : \hat{A} \rightarrow \hat{B}$ is an 
embedding, where 
\[ e : x \mapsto \diverges_{B}(x). \]
($\hat{A}$, $\hat{B}$ are defined as in the proof of Theorem~\ref{domtheq}).
\end{proposition}

\proof\ We define $p : \hat{B} \rightarrow \hat{A}$ by
\[ p(y) = y \cap |A|. \]
Since $A$ is a sublattice of $B$, $p$ is well defined and continuous (it is the surjection corresponding under Stone duality to the inclusion of $A$ in $B$). We check that $e$ is well defined, specifically that $e(x)$ is prime, $x \in \hat{A}$. Suppose $b \vee c \in e(x)$. Then for some $a \in x$, $a \leq_{B} b \vee c$. By $(d1)$,
\[a =_{A} \bigvee_{i \in I}a_{i}, \;\; (a_{i} \in pr(A), i \in I). \]
Since $x$ is a prime proper filter, $a_{i} \in x$ for some $i \in I$. Since $A \trianglelefteq B$, $a_{i} \in pr(B)$, and so
\begin{eqnarray*}
a_{i} \leq_{B} a \leq_{B} b \vee c & \Rightarrow & a_{i} \leq_{B} b \;
{\rm or} \; a_{i} \leq_{B} c \\
& \Rightarrow & b \in e(x) \; {\rm or} \; c \in e(x).
\end{eqnarray*} 
Moreover,
\[p \circ e(x) = \diverges_{B}(x) \cap |A| = x \]
\[e \circ p(y) = \diverges_{B}(y \cap |A|) \subseteq \diverges_{B}(y) = y. \]
Finally, $e$ preserves all joins since it is a left adjoint; in particular, it is continuous. \qed

Now given a (unary) type construction $T$, we will seek to represent it as a function
\[f_{T} : {\bf DPL1} \rightarrow {\bf DPL1} \]
which is $\trianglelefteq$-monotonic and chain continuous. We can then construct the initial solution of the domain equation
\[D = T(D) \]
as the least fixpoint of the function $f_{T}$, given in the usual way as
\[\bigsqcup_{n \in \omega}f_{T}^{(n)}({\bf 1}). \]

More generally, we can consider systems of domain equations by using powers of {\bf DPL1}; while $T$ can be built up by composition from various primitive operations. As long as each basic type construction is $\trianglelefteq$-monotonic and continuous, this approach will work.

The task of verifying continuity is eased by the following observation, 
adapted from \cite{LW84}.
\begin{proposition}
Suppose $f : {\bf DPL1} \rightarrow {\bf DPL1}$ is $\trianglelefteq$-monotonic and continuous on carriers, i.e. given a chain
$\{A_{n}\}_{n \in \omega}$,
\[ |f(\bigsqcup_{n \in \omega}A_{n}| = \bigcup_{n \in \omega}|f(A_{n})|,\]
then $f$ is continuous.
\end{proposition}

\proof\ Firstly, note that $A \trianglelefteq B$ and $|A| = |B|$ implies $A = B$. Now given a chain $\{A_{n}\}$, let
\[B \equiv \bigsqcup_{n}f(A_{n}), \]
\[C \equiv f(\bigsqcup_{n}A_{n}). \]
By monotonicity of $f$, $B \trianglelefteq C$, while by continuity on carriers, $|B| = |C|$. Hence $B = C$, and $f$ is continuous. \qed

\section{Constructions}
In this section, we fill in the programme outlined in the previous section by defining a number of type constructions as $\trianglelefteq$-monotonic and continuous functions over ${\bf DPL1}$. These definitions will follow a common pattern. We take a binary type construction $T(A,B)$ for illustration. Specific to each such construction will be a set of {\it generators} $G(T(A,B))$. Then the carrier $|T(A,B)|$ is defined inductively by
\[ \bullet \;\;\; G(T(A,B)) \subseteq |T(A,B)| \]
\[ \bullet \;\;\; \true , \false \in |T(A,B)| \]
\[ \bullet \;\;\; \frac{a, b \in |T(A,B)|}{a \wedge b, a \vee b \in |T(A,B)|} \]
The operations $0, 1, \wedge , \vee$ are then defined ``freely'' in the obvious way, i.e.
\[ 0_{T(A,B)} \equiv \false , \;\; a \vee_{T(A,B)} b \equiv a \vee b, \;\; 1_{T(A,B)} \equiv \true , \;\; a \wedge_{T(A,B)} b \equiv a \wedge b. \]
Finally, the relations $\leq_{T(A,B)}$, $=_{T(A,B)}$ are defined inductively as the least satisfying $(p1)$--$(p4)$ plus specific axioms on the generators. 
(Note that our definition of {\bf 1} in the previous section is the special case of this scheme where the set of generators is empty.)

As an essential part of the machinery for defining the type constructions, we shall introduce a number of meta-predicates over the carriers $|T(A,B)|$ of the constructed prelocales. These will be used as side-conditions on a number of axiom-schemes and rules. They will serve as ``syntactic'' analogues of the ``semantic'' predicates $con$, $pr$, $t$ introduced previously. 
The same predicates will be defined for each contruction:
\begin{itemize}
\item ${\sf PNF}$, prime normal form.
\item ${\sf CON}$, ${\sf T}$, defined over elements of the form $\bigwedge_{i \in I}a_{i}$, 
with each $a_{i}$ in ${\sf PNF}$. 
${\sf CON}$ is {\it consistency} (i.e. ${\sf CON}(a)$ means $a \not= 0$), 
and ${\sf T}$ is {\it termination} (i.e. ${\sf T}(a)$ means $a \not= 1$).
\item ${\sf CPNF}$, consistent prime normal forms, where ${\sf CPNF}(a)$ implies ${\sf PNF}(a)$ and ${\sf CON}(a)$.
\end{itemize}
Given these definitions, three further predicates are defined as follows:
\begin{itemize}
\item ${\sf CDNF}$, consistent disjunctive normal form:
\[ {\sf CDNF}(a) \; \equiv \; a = \bigvee_{i \in I} a_{i} \: \& \: \forall i \in I . \, {\sf CPNF}(a_{i}) \]
\end{itemize}
\[ \bullet \;\;\; a \converges \; \equiv \; a = \bigvee_{i \in I}a_{i} \;\& \; \forall i \in I . \, {\sf PNF}(a_{i}) \: \& \: {\sf T}(a_{i}) \]
\[ \bullet \;\;\; \#(a) \; \equiv \; a = \bigvee_{i \in I}a_{i} \; \& \; \forall i \in I . \, {\sf PNF}(a_{i}) \: \& \: \neg {\sf CON}(a_{i}). \]

It will follow from our general scheme of definition and the way that the generators are defined that the following points are immediate, for $A, A', B, B'$ in {\bf DPL1} with $A \trianglelefteq A'$ and $B \trianglelefteq B'$:
\begin{itemize}
\item $T(A,B)$ satisfies $(p1)$--$(p4)$
\item ${\bf 1} \trianglelefteq T(A,B)$
\item $T(A,B) \Subset T(A',B')$
\item $T$ is continuous on carriers.
\end{itemize}
We are left to focus our attention on proving that:
\begin{itemize}
\item $T(A,B)$ satisfies $(d1)$--$(d3)$
\item conditions $(s2)$ and $(s3)$ for $T(A,B) \trianglelefteq T(A',B')$ are satisfied.
\end{itemize}

Our method of establishing this for each $T$ is uniform, and goes via another essential verification, namely that $T$ does indeed correspond to the intended construction over domains. We define a semantic function
\[ \lsem \cdot \rsem_{T(A,B)} : |T(A,B)| \rightarrow K\Omega (F_{T}(\hat{A}, \hat{B})) \]
where $F_{T}$ is the functor over {\bf SDom} corresponding to $T$, and show 
that $\lsem \cdot \rsem_{T(A,B)}$ is a (pre)isomorphism; and moreover natural with respect to embeddings induced by $\trianglelefteq$. 
This allows us to read off the required ``proof-theoretic'' facts about $T$ from the known ``model-theoretic'' ones about $F_{T}$. 
Moreover, we can derive ``soundness and completeness'' theorems as byproducts.

For each type construction $T$, we prove the following sequence of results:

{\bf T1: Adequacy of Metapredicates.} {\it For each $a \in {\sf PNF}(T(A,B))$:}
\[\begin{array}{rl}
(i)   & \lsem a \rsem_{T(A,B)} \in pr(K\Omega (F_{T}(\hat{A},\hat{B}))) \\
(ii)  & {\sf CON}(a) \; \Longleftrightarrow \; \lsem a \rsem_{T(A,B)} \neq \varnothing \\
(iii) & {\sf T}(a) \; \Longleftrightarrow \; \bot_{F_{T}(\hat{A},\hat{B})} \not\in \lsem a \rsem_{T(A,B)}. 
\end{array} \]

{\bf T2: Normal Forms.} 
\[ \forall a \in |T(A,B)|. \, \exists b \in {\sf CDNF}(T(A,B)) . \, a =_{T(A,B)} b. \]

{\bf T3: Soundness.} {\it For all $a, b \in |T(A,B)|$:}
\[ a \leq_{T(A,B)} b \; \Rightarrow \; \lsem a \rsem_{T(A,B)} \subseteq \lsem b \rsem_{T(A,B)}. \]

{\bf T4: Prime Completeness.} {\it For all $a, b \in {\sf CPNF}(T(A,B))$:}
\[ \lsem a \rsem_{T(A,B)} \subseteq \lsem b \rsem_{T(A,B)} \; \Rightarrow \; a \leq_{T(A,B)} b. \]

{\bf T5: Definability.} 
\[ \forall u \in K(F_{T}(\hat{A},\hat{B})). \, \exists a \in {\sf CPNF}(T(A,B)). \, \lsem a \rsem_{T(A,B)} = \diverges (u). \]

{\bf T6: Naturality.} {\it Given $A \trianglelefteq A'$, $B \trianglelefteq B'$ in {\bf DPL1}, let $e_{1} : \hat{A} \rightarrow \hat{A'}$, $e_{2} : \hat{B} \rightarrow \hat{B'}$ be the corresponding embeddings. Given an embedding $e : D \rightarrow E$, let $e^{\dag} : K\Omega (D) \rightarrow K\Omega (E)$ be defined by
\[ e^{\dag}(\diverges X) = \diverges \{ e(x) : x \in X\} \]
which is well defined since embeddings map finite elements to finite elements. 
Let 
\[ \eta_{T(A,B)} : \hat{C} \rightarrow F_{T}(\hat{A},\hat{B}) \]
be the adjoint of $\lsem \cdot \rsem_{T(A, B)}$, where $C = T(A, B)$. Then:
\[ \begin{array}{lrcl}
(A) & (F_{T}(e_{1}, e_{2}))^{\dag} \circ \lsem \cdot \rsem_{T(A, B)} & = & 
\lsem \cdot \rsem_{T(A' , B' )} \\
(B) & F_{T}(e_{1}, e_{2}) \circ \eta_{T(A, B)} & = & \eta_{T(A' , B' )} 
\circ {\converges}_{T(A' , B' )} (\cdot )
\end{array} \]
(These equations make sense since $T(A, B) \Subset T(A' , B' )$ by assumption.)}

All the desired properties of our constructions can easily be derived from these results.

{\bf T7: Completeness.} {\it For $a, b \in |T(A,B)|$:}
\[ \lsem a \rsem_{T(A,B)} \subseteq \lsem b \rsem_{T(A,B)} \; \Rightarrow \; a \leq_{T(A,B)} b. \]

\proof\ By (T2),
\[ a =_{T(A,B)} \bigvee_{i \in I}a_{i}, \;\; b=_{T(A,B)} \bigvee_{j \in J}b_{j}, \]
with $a_{i}, b_{j} \in {\sf CPNF}(T(A,B))$ ($i \in I, j \in J$). By (T3),
\[ \lsem a \rsem_{T(A,B)} = \lsem \bigvee_{i \in I}a_{i} \rsem_{T(A,B)}, \;\;\; \lsem b \rsem_{T(A,B)} = \lsem \bigvee_{j \in J}b_{j} \rsem_{T(A,B)}. \]
By (T1),
\[ \lsem a_{i} \rsem_{T(A,B)}  =   \diverges (u_{i}),  
\lsem b_{j} 
\rsem_{T(A,B)} =  
\diverges (v_{j}) \]
\[ u_{i}, v_{j} \in K(F_{T}(\hat{A},\hat{B})) \;\;\; (i \in I, j \in J). \] 
Now,
\[\begin{array}{lll}
            & \lsem a \rsem_{T(A,B)} \subseteq \lsem b \rsem_{T(A,B)}   & \\
\Longrightarrow & \bigcup_{i \in I}\diverges (u_{i}) \subseteq \bigcup_{j \in J} \diverges (v_{j})  & \\
\Longrightarrow & \forall i \in I. \, \exists j \in J. \, \diverges (u_{i}) \subseteq \diverges (v_{j}) & \\
\Longrightarrow & \forall i \in I. \, \exists j \in J. \, a_{i} \leq_{T(A,B)} b_{j} &  \mbox{by (T4)} \\
\Longrightarrow & \bigvee_{i \in I}a_{i} \leq_{T(A,B)} \bigvee_{j \in J}b_{j} & \mbox{by (p2)} \\
\Longrightarrow & a \leq_{T(A,B)} b & \mbox{by (p1).} \; \qed 
\end{array} \]

{\bf (T8): Stone Duality.} {\it $T(A,B)$ is the Stone dual of $F_{T}(\hat{A},
\hat{B})$, i.e.}
\[\begin{array}{rl}
(i)   & F_{T}(\hat{A},\hat{B}) \; \cong \;  \hat{C} \;\;\; (C = T(A,B)) \\
(ii)  & \lsem \cdot \rsem : |T(A,B)| \rightarrow K\Omega (F_{T}(\hat{A},\hat{B})) 
\; \mbox{is a pre-isomorphism.}
\end{array} \]

\proof\ $(i)$ and $(ii)$ are equivalent since Scott domains are coherent.
$(ii)$ is an immediate consequence of (T3), (T5) and (T7). \qed

{\bf (T9).} {\it $T$ is a well defined, $\trianglelefteq$-monotonic and continuous operation on {\bf DPL1}.}

\proof\ T(A,B) is a domain prelocale by (T8), 
since $K\Omega (F_{T}(\hat{A},\hat{B}))$ is. 
Given $A \trianglelefteq A'$, $B \trianglelefteq B'$, 
$T(A,B) \trianglelefteq T(A',B')$ follows from (T6)(A) and the following 
general properties of $e^{\dag}$ for embeddings $e : D \rightarrow E$:
\begin{enumerate}
\item $e^{\dag}$ is an order-mono, i.e. for $U, V \in K\Omega (D)$:
\[ U \subseteq V \; \Longleftrightarrow \; e^{\dag}(U) \subseteq e^{\dag}(V) \]
\item $e^{\dag}$ preserves primes.
\end{enumerate}
To prove (1), we take $U = \diverges  X$, $V = \diverges  Y$, and calculate:
\begin{eqnarray*}
\diverges  X \subseteq \diverges  Y & \Longleftrightarrow & X \sqsubseteq_{u} Y \\
& \Longleftrightarrow & e(X) \sqsubseteq_{u} e(Y) \;\;\; \mbox{{\mit e} 
is an order-mono} \\
& \Longleftrightarrow & \diverges  e(X) \subseteq \diverges  e(Y) \\
& \Longleftrightarrow & e^{\dag}(U) \subseteq e^{\dag}(V).
\end{eqnarray*}
For (2), we recall that $U \in pr(K\Omega (D))$ implies $U = \varnothing$ or $U = \diverges (u)$ for some $u \in K(D)$. But $e^{\dag}(\varnothing) = \varnothing$, $e^{\dag}(\diverges (u)) = \diverges (e(u))$.

By the remarks at the beginning of the section, the proof is now complete. \qed

{\bf Notation.} Given a domain prelocale $A$, we write
\[ \lsem \cdot \rsem_{A} : |A| \rightarrow K\Omega (\hat{A}) \]
for the pre-isomorphism $\varphi A$ defined in the proof of Theorem~\ref{domtheq}.

We note a further trivial but useful fact about direct images of embeddings for future use.
\begin{proposition}
\label{embim}
If $A \trianglelefteq B$, and $e : \hat{A} \rightarrow \hat{B}$ is the induced embedding, then
\[ e^{\dag} \circ \lsem \cdot \rsem_{A} = \lsem \cdot \rsem_{B}. \; \qed \]
\end{proposition}

\begin{definition} 
{\rm The {\em function space} construction $A \rightarrow B$.

\noindent (i) The generators:
\[ G(A \rightarrow B) \; \equiv \; \{ (a \rightarrow b) : a \in |A|, b \in |B| \}. \]
This fixes $|A \rightarrow B|$ according to the general scheme described above.

\noindent (ii) The metapredicates:
\begin{eqnarray*}
{\sf PNF}(A \rightarrow B) & \equiv & \{\bigwedge_{i \in I}(a_{i} \rightarrow b_{i}) : a_{i} \in pr(A), b_{i} \in pr(B), i \in I \} \\
{\sf CON}(\bigwedge_{i \in I}(a_{i} \rightarrow b_{i})) & \equiv & \forall J \subseteq I. \\ 
& & \bigwedge_{j \in J}a_{j} \in con(A) \; \Longrightarrow \bigwedge_{j \in J}b_{j} \in con(B) \\
{\sf T}(\bigwedge_{i \in I}(a_{i} \rightarrow b_{i})) & \equiv & \exists i \in I. \, a_{i} \in con(A) \& b_{i} \in t(B) \\
{\sf CPNF}(\bigwedge_{i \in I}(a_{i} \rightarrow b_{i})) & \equiv & {\sf CON}(\bigwedge_{i \in I}(a_{i} \rightarrow b_{i})) \\ 
& & \& \; \forall i \in I. \, a_{i} \in con(A) \:
\& \: b_{i} \in con(B)
\end{eqnarray*}
The predicates ${\sf CDNF}$, $\#(.)$, $\underline{\ }\converges$ are then defined according to our general scheme.

\noindent (iii) The relations $\leq_{A \rightarrow B}$, $=_{A \rightarrow B}$ are then defined inductively by the following axioms and rules in addition to $(p1)$--$(p4)$ (subscripts omitted).
\[ (\rightarrow - \leq ) \;\;\; \frac{a' \leq a, \; b \leq b'}{(a \rightarrow b) \leq (a' \rightarrow b' )} \]
\[ (\rightarrow - \wedge) \;\;\; (a \rightarrow \bigwedge_{i \in I}b_{i}) = \bigwedge_{\ \in I}(a \rightarrow b_{i}) \]
\[ (\rightarrow - \vee - L) \;\;\; (\bigvee_{i \in I}a_{i} \rightarrow b) = \bigwedge_{i \in I}(a_{i} \rightarrow b) \]
\[ (\rightarrow - \vee - R) \;\;\;  (a \rightarrow \bigvee_{i \in I}b_{i}) = \bigvee_{i \in I}(a \rightarrow b_{i}) \;\;\; (a \in cpr(A)) \]
\[ (\#) \;\;\; a \leq 0 \;\;\; (\#(a)) \] 

\noindent (iv) The semantic function
\[ \lsem \cdot \rsem_{A \rightarrow B} : |A \rightarrow B| \longrightarrow K\Omega ([\hat{A} \rightarrow \hat{B}]) \]
is defined by
\[ \lsem (a \rightarrow b) \rsem_{A \rightarrow B} = (\lsem a \rsem_{A}, \lsem b \rsem_{B}) \]
where for spaces $X$, $Y$ and subsets $U \in K\Omega (X)$, $V \in K\Omega (Y)$,
\[ (U,V) \; \equiv \; \{ f : X \rightarrow Y \; | \; f \; {\rm continuous,} \; f(U) \subseteq V \} \]
is a sub-basic open set in the compact-open topology. The further clauses
\[ \lsem \bigwedge_{i \in I}a_{i} \rsem = \bigcap_{i \in I} \lsem a_{i} \rsem \]
\[ \lsem \bigvee_{i \in I}a_{i} \rsem = \bigcup_{i \in I} \lsem a_{i} \rsem \]
will apply to all type constructions.}
\end{definition}

We will now establish that the function space construction satisfies (T1)--(T6) in a sequence of propositions.

\begin{proposition}[T1]
\label{funT1}
For all $a \in {\sf PNF}(A \rightarrow B)$:
\[ \begin{array}{rl}
(i)   & \lsem a \rsem_{A \rightarrow B} \in pr(K \Omega ([\hat{A} \rightarrow \hat{B}])) \\
(ii)  & {\sf CON}(a) \; \Longleftrightarrow \; \lsem a \rsem_{A \rightarrow B} \neq \varnothing \\
(iii) & {\sf T}(a) \; \Longleftrightarrow \; \bot \not\in \lsem a \rsem_{A \rightarrow B}.
\end{array} \]
\end{proposition}

\proof\ (i) Let $a \in pr(A)$, $b \in pr(B)$. If $a \not\in con(A)$,
\[ \lsem (a \rightarrow b) \rsem_{A \rightarrow B} = [\hat{A} \rightarrow \hat{B}] = 1_{K\Omega ([\hat{A} \rightarrow \hat{B}])}; \]
while if $a \in con(A)$, $b \not\in con(B)$, 
\[ \lsem (a \rightarrow b) \rsem_{A \rightarrow B} = \varnothing . \]
Otherwise, $a \in con(A)$ and $b \in con(B)$. 
Let $u = \diverges (a)$, $v = \diverges (b)$. 
Then $u \in K(\hat{A})$, $v \in K(\hat{B})$, and so
\begin{eqnarray*}
\lsem (a \rightarrow b) \rsem_{A \rightarrow B} & = & (\lsem a \rsem_{A}, \lsem b \rsem_{B}) \\
& = & (\diverges u, \diverges v) \\
& = & \diverges [u, v],
\end{eqnarray*} 
where $[u, v]$ is the step function in $[\hat{A} \rightarrow \hat{B}]$. Similarly, for $a_{i} \in cpr(A)$, $b_{i} \in cpr(B)$:
\begin{eqnarray*}
\lsem \bigwedge_{i \in I}(a_{i} \rightarrow b_{i}) \rsem_{A \rightarrow B} & = & \bigcap_{i \in I} \diverges [u_{i}, v_{i}] \\
& = & \left\{ \begin{array}{ll}
               \diverges ( \bigsqcup_{i \in I}[u_{i}, v_{i}]) & \mbox{if 
               $\consistent \{[u_{i}, v_{i}] : i \in I \}$} \\
               \varnothing & \mbox{otherwise.}
               \end{array}
      \right.
\end{eqnarray*}
(ii) Let $a = \bigwedge_{i \in I}(a_{i} \rightarrow b_{i})$. We use the notation of (i). Suppose ${\sf CON}(a)$. Then for $i \in I$,
\[ b_{i} \not\in con(B) \; \Longrightarrow \; a_{i} \not\in con(A) \; \Longrightarrow \; \lsem (a_{i} \rightarrow b_{i}) \rsem_{A \rightarrow B} = 1_{K\Omega ([\hat{A} \rightarrow \hat{B}])}, \]
and so
\begin{eqnarray*}
\lsem a \rsem_{A \rightarrow B} & = & \lsem \bigwedge 
\{ (a_{j} \rightarrow b_{j}) : a_{j} \in cpr(A), b_{j} \in cpr(B) \} \rsem_{A \rightarrow B} \\
& = & \diverges (\bigsqcup \{ [u_{j}, v_{j}] : a_{j} \in cpr(A), b_{j} \in cpr(B) \} ),
\end{eqnarray*} 
which is well-defined by \ref{funcon}.
For the converse, suppose $\neg {\sf CON}(a)$. Then for some $J \subseteq I$,
$\bigwedge_{j \in J}a_{j} \in con(A)$ and $\bigwedge_{j \in J}b_{j} \not\in con(B)$. But then we have
\[ \lsem a \rsem_{A \rightarrow B} \subseteq \lsem 
( \bigwedge_{j \in J}a_{j} \rightarrow \bigwedge_{j \in J}b_{j} ) 
\rsem_{A \rightarrow B} = \varnothing . \]
(iii) With notation as in (ii),
\[ \bot \not\in \lsem a \rsem_{A \rightarrow B} \; \Longleftrightarrow \; \exists i \in I. \, \bot \not\in \lsem (a_{i} \rightarrow b_{i}) \rsem_{A \rightarrow B}. \]
Now if $a_{i} \not\in con(A)$, 
\[ \bot \in 1_{K\Omega ([\hat{A} \rightarrow \hat{B}])} = \lsem (a_{i} \rightarrow b_{i})\rsem_{A \rightarrow B}; \]
while if $a_{i} \in con(A)$, $b_{i} \not\in con(B)$, then
\[ \bot \not\in \varnothing = \lsem (a_{i} \rightarrow b_{i})\rsem_{A \rightarrow B}. \]
Finally, if $a_{i} \in con(A)$ and $b_{i} \in con(B)$, then $\lsem (a_{i} \rightarrow b_{i})\rsem_{A \rightarrow B} = \diverges [u_{i}, v_{i}]$, and
\[ \bot \not\in \lsem (a_{i} \rightarrow b_{i})\rsem_{A \rightarrow B} \;
\Longleftrightarrow \; v_{i} \neq \bot \; \Longleftrightarrow \; b_{i} \in t(B). \]
\[ \mbox{Thus } \bot \not\in \lsem (a_{i} \rightarrow b_{i})\rsem_{A \rightarrow B} \; \Longleftrightarrow \; a_{i} \in con(A) \: \& \: b_{i} \in t(B).  \;\;\; \qed \]

As corollaries we have:
\[ \begin{array}{rl}
\mbox{(iv)} & {\sf CPNF}(\bigwedge_{i \in I}(a_{i} \rightarrow b_{i})) \; \Longrightarrow \; \lsem \bigwedge_{i \in I}(a_{i} \rightarrow b_{i})\rsem_{A \rightarrow B} = \diverges (\bigsqcup_{i \in I}[u_{i}, v_{i}]), \\
&  \mbox{where } \diverges u_{i} = \lsem a_{i} \rsem_{A}, \diverges v_{i} = \lsem b_{i} \rsem_{B}, i \in I. \\
\mbox{(v)} & \#(a) \; \Longleftrightarrow \; \lsem a \rsem_{A \rightarrow B} = \varnothing . \\
\mbox{(vi)} & a \converges \; \Longleftrightarrow \; \bot \not\in \lsem a \rsem_{A \rightarrow B}.
\end{array} \]

\begin{proposition}[T2]
$\forall a \in |A \rightarrow B|. \, \exists b \in {\sf CDNF}(A \rightarrow B). \, a =_{A \rightarrow B} b.$ 
\end{proposition}

\proof\ Using the distributive lattice laws, $a$ can be put in the form
\[ \bigvee_{i \in I} \bigwedge_{j \in J_{i}} (a_{ij} \rightarrow b_{ij}). \]
By $(d1)$, each $a_{ij}$ is equal to
\[ \bigvee_{k \in K_{ij}}c_{k}, \;\;\; (c_{k} \in pr(A), k \in K_{ij}), \]
and each $b_{ij}$ is equal to 
\[ \bigvee_{l \in L_{ij}}d_{l}, \;\;\; (d_{l} \in pr(B), l \in L_{ij}). \]
Moreover, we may assume that $c_{k} \in con(A)$ for all $k \in K_{ij}$, since otherwise
\[ \bigvee_{k \in K_{ij}}c_{k} =_{A} \bigvee_{k' \in K_{ij} - \{k\}}c_{k'}, \]
and so any inconsistent disjuncts can be deleted; and similarly for the $d_{l}$. Now
\begin{eqnarray*}
(\bigvee_{k \in K_{ij}}c_{k} \rightarrow \bigvee_{l \in L_{ij}}d_{l}) & =_{A \rightarrow B} &
\bigwedge_{k \in K_{ij}}(c_{k} \rightarrow \bigvee_{l \in L_{ij}}d_{l}) \;\;\; {\rm by } \; (\rightarrow - \vee -L) \\
& =_{A \rightarrow B} & \bigwedge_{k \in K_{ij}} \bigvee_{l \in L_{ij}} (c_{k} \rightarrow d_{l}) \;\;\; {\rm by } \; (\rightarrow - \vee - R).
\end{eqnarray*}
Using the distributive lattice laws again, we obtain the required normal form. \qed
\begin{proposition}[T3]
$\forall a, b \in |A \rightarrow B|. \, a \leq_{A \rightarrow B} \; \Rightarrow \; \lsem a \rsem_{A \rightarrow B} \subseteq \lsem b \rsem_{A \rightarrow B}.$ 
\end{proposition}

\proof\ $\lsem \rsem_{A \rightarrow B}$ preserves meets and joins by definition, and $(p1)$--$(p4)$ are valid in any distributive lattice. Moreover, given any spaces $X$, $Y$ and subsets $U \subseteq X$, $V \subseteq Y$,
\[ U' \subseteq U, V \subseteq V' \; \Longleftrightarrow \; (U, V) \subseteq (U', V') \]
\[ (U, \bigcap_{i \in I}V_{i}) = \bigcap_{i \in I}(U, V_{i}) \]
\[ (\bigcup_{i \in I}U_{i}, V) = \bigcap_{i \in I}(U_{i}, V) \]
are simple set-theoretic calculations. The soundness of ($\rightarrow$-$\#$) 
follows from Corollary (v) to Proposition~\ref{funT1}. 
Finally, suppose $a \in cpr(A)$. 
Then $\lsem a \rsem_{A} = \diverges  u$ with $u \in K(\hat{A})$, and
\begin{eqnarray*}
\lsem (a \rightarrow \bigvee_{i \in I}b_{i}) \rsem_{A \rightarrow B} & = & (\diverges u, \bigcup_{i \in I}\lsem b_{i} \rsem_{B}) \\
& = & \{ f : f(u) \in \bigcup_{i \in I}\lsem b_{i} \rsem_{B} \} \;\;\; \mbox{by monotonicity}\\
& = & \bigcup_{i \in I}\{ f : f(u) \in \lsem b_{i} \rsem_{B} \} \\
& = & \bigcup_{i \in I}(\diverges u, \lsem b_{i} \rsem_{B}) \\
& = & \lsem \bigvee_{i \in I}(a \rightarrow b_{i}) \rsem_{A \rightarrow B}
\end{eqnarray*} 
and so $(\rightarrow - \vee - R)$ is sound. \qed

\begin{proposition}[T4]
For $\bigwedge_{i \in I}(a_{i} \rightarrow b_{i})$, $\bigwedge_{j \in J}(a_{j} \rightarrow b_{j})$ in ${\sf CPNF}(A \rightarrow B)$:
\[ \lsem \bigwedge_{i \in I}(a_{i} \rightarrow b_{i}) \rsem_{A \rightarrow B} \subseteq \lsem \bigwedge_{j \in J}(a_{j} \rightarrow b_{j})\rsem_{A \rightarrow B} \]
implies
\[ \bigwedge_{i \in I}(a_{i} \rightarrow b_{i}) \leq_{A \rightarrow B} \bigwedge_{j \in J}(a_{j} \rightarrow b_{j}). \]
\end{proposition}

\proof\ By Corollary (iv) to Proposition~\ref{funT1},
\[  \lsem \bigwedge_{i \in I}(a_{i} \rightarrow b_{i}) \rsem_{A \rightarrow B} = \diverges \bigsqcup_{i \in I}[u_{i}, v_{i}], \]
\[  \lsem \bigwedge_{j \in J}(a_{j} \rightarrow b_{j}) \rsem_{A \rightarrow B} = \diverges \bigsqcup_{j \in J}[u_{j}, v_{j}], \]
where
\[ \diverges u_{i} = \lsem a_{i} \rsem_{A}, \ldots \; etc. \]
Now,
\[ \lsem \bigwedge_{i \in I}(a_{i} \rightarrow b_{i}) \rsem_{A \rightarrow B} \subseteq  \lsem \bigwedge_{j \in J}(a_{j} \rightarrow b_{j}) \rsem_{A \rightarrow B} \]
\[ \Longleftrightarrow \;\; \bigsqcup_{j \in J}[u_{j},v_{j}] \sqsubseteq 
\bigsqcup_{i \in I}[u_{i}, v_{i}] \]
\[ \Longleftrightarrow \;\; \forall j \in J. \, v_{j} \sqsubseteq \bigsqcup \{ v_{i} : u_{i} \sqsubseteq u_{j} \} \]
\[ \Longleftrightarrow \;\; \forall j \in J. \, \lsem \bigwedge \{ b_{i} : \lsem a_{j} \rsem_{A} \subseteq \lsem a_{i} \rsem_{A} \} \rsem_{B} \subseteq \lsem b_{j} \rsem_{B} \]
\[ \Longleftrightarrow \;\; \forall j \in J. \, \bigwedge \{b_{i} : a_{j} \leq_{A} a_{i} \} \leq_{B} b_{j} \;\;\; (*). \]
Thus, for all $j \in J$:
\begin{Eqarray}
\bigwedge_{i \in I}(a_{i} \rightarrow b_{i}) & \leq_{A \rightarrow B} & \bigwedge \{ (a_{i} \rightarrow b_{i}) : a_{j} \leq_{A} a_{i} \} & \mbox{by (p3)} \\
& \leq_{A \rightarrow B} & \bigwedge \{ (a_{j} \rightarrow b_{i}) : a_{j} \leq_{A} a_{i} \} & \mbox{by $(\rightarrow - \leq )$} \\
& =_{A \rightarrow B} & (a_{j} \rightarrow \bigwedge \{ b_{i} : a_{j} \leq_{A} a_{i} \}) & \mbox{by $(\rightarrow - \wedge )$} \\
& \leq_{A \rightarrow B} & (a_{j} \rightarrow b_{j}) & \mbox{by (*)}
\end{Eqarray}
and so by $(p2)$
\[ \bigwedge_{i \in I}(a_{i} \rightarrow b_{i}) \leq_{A \rightarrow B} \bigwedge_{j \in J}(a_{j} \rightarrow b_{j}). \;\;\; \qed \]

\begin{proposition}[T5] 
$\forall U \in K\Omega ([\hat{A} \rightarrow \hat{B}]). \, \exists a \in |A \rightarrow B|. \, \lsem a \rsem_{A \rightarrow B} = U.$ 
\end{proposition}

\proof\ Directly from Propositions~\ref{cop} and~\ref{funT1}. \qed

\begin{proposition}[T6]
Given $A \trianglelefteq A'$, $B \trianglelefteq B'$, let $e_{1} : \hat{A} \rightarrow \hat{A'}$, $e_{2} : \hat{B} \rightarrow \hat{B'}$ be the corresponding embeddings. Then
\[ (A) \;\;\; (e_{1} \rightarrow e_{2})^{\dag} \circ \lsem \cdot \rsem_{A \rightarrow B} = \lsem \cdot \rsem_{A' \rightarrow B'} \]
\[ (B) \;\;\; (e_{1} \rightarrow e_{2}) \circ \eta_{A \rightarrow B} =  \eta_{A' \rightarrow B'} \circ \converges (\cdot ). \]
\end{proposition}

\proof\ Firstly, we recall the definition of $e_{1} \rightarrow e_{2}$:
\[ (e_{1} \rightarrow e_{2})(f) = e_{2} \circ f \circ e_{1}^{R}, \]
where $e_{1}^{R}$ is the right adjoint of $e_{1}$, i.e. the corresponding projection. Now in fact we can eliminate the use of the projection in describing $(e_{1} \rightarrow e_{2})^{\dag}$, since we have
\[ (e_{1} \rightarrow e_{2})(\bigsqcup_{i \in I}[u_{i}, v_{i}]) = \bigsqcup_{i \in I}[e_{1}(u_{i}), e_{2}(v_{i})]. \]
Indeed,
\[ \begin{array}{cl}
  & (e_{1} \rightarrow e_{2})(\bigsqcup_{i \in I}[u_{i}, v_{i}])(d) \\ 
= & e_{2} \circ \bigsqcup_{i \in I}[u_{i}, v_{i}] \circ e_{1}^{R} (d) \\
= & e_{2}(\bigsqcup_{i \in I}\{ v_{i} : u_{i} \sqsubseteq e_{1}^{R}(d) \}) \\
= & e_{2}(\bigsqcup_{i \in I}\{ v_{i} : e_{1}(u_{i}) \sqsubseteq d \} ) \\
= & \bigsqcup_{i \in I}\{ e_{2}(v_{i}) : e_{1}(u_{i}) \sqsubseteq d \} \\
  & \mbox{($e_{2}$ preserves joins since it is a left adjoint)} \\
= & (\bigsqcup_{i \in I}[e_{1}(u_{i}), e_{2}(v_{i})])(d).
\end{array} \] 
Now for (A), given
\[ a =_{A \rightarrow B} \bigvee_{i \in I} \bigwedge_{j \in J_{i}}(a_{ij} \rightarrow b_{ij}) \in {\sf CDNF}(A \rightarrow B), \]
we calculate
\begin{eqnarray*}
(e_{1} \rightarrow e_{2})^{\dag} \lsem a \rsem_{A \rightarrow B} & = & \bigcup_{i \in I} \bigcap_{j \in J_{i}} (e_{1}^{\dag} \lsem a_{ij} \rsem_{A}, e_{2}^{\dag}\lsem b_{ij} \rsem_{B}) \\
& = & \bigcup_{i \in I} \bigcap_{j \in J_{i}} (\lsem a_{ij} \rsem_{A'}, \lsem b_{ij} \rsem_{B'}) \;\;\; \mbox{by~\ref{embim}} \\
& = & \lsem a \rsem_{A' \rightarrow B'}.
\end{eqnarray*} 
Similarly for (B) we have:
\[ \begin{array}{cl}
  & (e_{1} \rightarrow e_{2}) \circ \eta_{A \rightarrow B}(x) \\ 
= & \bigsqcup \{ [u, v] : \exists (a \rightarrow b) \in x. \, \diverges u = \lsem a \rsem_{A} \: \& \: \diverges v = \lsem b \rsem_{B} \} \\
= & \bigsqcup \{ [u, v] : \exists (a \rightarrow b) \in x. \, \diverges u = \lsem a \rsem_{A'} \: \& \: \diverges v = \lsem b \rsem_{B'} \} \\
= & \eta_{A' \rightarrow B'}( \converges (x)). \;\;\; \qed
\end{array} \] 

To illustrate the uniformity in our treatment of all the type constructions, we shall deal with two more: the upper or Smyth powerdomain, and the coalesced sum.

\begin{definition} 
{\rm The {\it upper powerdomain} $P_{u}(A)$.

\noindent (i) The generators:
\[ G(P_{u}(A)) \; \equiv \; \{ \Box a | a \in |A| \]

\noindent (ii) Metapredicates:
\begin{eqnarray*}
{\sf PNF}(P_{u}(A)) & \equiv & \{ \Box \bigvee_{i \in I}a_{i} : a_{i} \in pr(A), i \in I \} \\
{\sf CON}(t) & & \\
{\sf CON}(\bigwedge_{i \in I} \Box \bigvee_{j \in J_{i}} a_{ij}) & \equiv & \exists 
f \in \prod_{i \in I}J_{i}. \, \bigwedge_{i \in I}a_{i, f(i)} \in con(A) \\
{\sf T}(\bigwedge_{i \in I} \Box \bigvee_{j \in J_{i}} a_{ij}) & \equiv & \exists i \in I. \, \forall j \in J_{i}. \, a_{ij} \in t(A) \\
{\sf CPNF}( \Box \bigvee_{i \in I}a_{i}) & \equiv & {\sf CON}(\Box \bigvee_{i \in I}a_{i}) 
\: \& \: I \neq \varnothing \\
& & \& \: \forall i \in I. \, a_{i} \in con(A)
\end{eqnarray*} 

\noindent (iii) Axioms in addition to $(p1)$ -- $(p4)$:
\[ (\Box - \leq) \;\;\; \frac{a \leq b}{\Box a \leq \Box b} \]
\[ (\Box - \wedge) \;\;\; \Box \bigwedge_{i \in I}a_{i} = \bigwedge_{i \in I} \Box a_{i} \]
\[ (\Box - 0) \;\;\; \Box 0 = 0 \]

\noindent (iv) The semantic function:
\[ \lsem \cdot \rsem_{P_{u}(A)} : | P_{u}(A) | \longrightarrow K\Omega (P_{u}(\hat{A})) \]
\[ \lsem \Box a \rsem_{P_{u}(A)} = \{ S \in P_{u}(\hat{A}) : S \subseteq \lsem a \rsem_{A} \} \]
(The further clauses are the standard ones described in the definition of function space.)}
\end{definition}
\begin{proposition}[T1]
\label{pdomT1}
For all $a, \{ a_{i}\}_{i \in I} \in {\sf PNF}(P_{u}(A))$:
\[ \begin{array}{rl}
(i)   & \lsem a \rsem_{P_{u}(A)} \in pr(K\Omega (P_{u}(A))) \\
(ii)  & {\sf CON}(\bigwedge_{i \in I}a_{i}) \; \Longleftrightarrow \; \lsem \bigwedge_{i \in I}a_{i} \rsem_{P_{u}(A)} \neq \varnothing \\
(iii) & {\sf T}(\bigwedge_{i \in I}a_{i}) \; \Longleftrightarrow \; \bot \not\in \lsem \bigwedge_{i \in I}a_{i} \rsem_{P_{u}(A)}  \\
\end{array} \]
\end{proposition}

\proof\ $(i)$. Let $\Box \bigvee_{i \in I}a_{i} \in {\sf PNF}(P_{u}(A))$. Then either $\bigvee_{i \in I}a_{i} \not\in con(A)$, and
\[ \lsem \Box \bigvee_{i \in I}a_{i} \rsem_{P_{u}(A)} = \varnothing \in  pr(K\Omega (P_{u}(A))); \]
or for some $X \subseteq_{\sf f} {\cal K}(\hat{A})$, $X \neq \varnothing$ and
\[ \lsem \bigvee_{i \in I}a_{i} \rsem_{A} = \diverges_{\hat{A}} X. \]
In the latter case,
\begin{eqnarray*}
\lsem \Box \bigvee_{i \in I}a_{i} \rsem_{P_{u}(A)} & = & \{ S \in P_{u}(\hat{A}) : S \subseteq \lsem \bigvee_{i \in I}a_{i} \rsem_{A} \} \\
& = & \{ S \in P_{u}(\hat{A}) : \diverges_{\hat{A}} X \sqsubseteq_{u} S \} \\
& = & \diverges_{P_{u}(\hat{A})}(\lsem \bigvee_{i \in I}a_{i} \rsem_{A}).
\end{eqnarray*} 
(ii) Firstly,
\[ \lsem \bigwedge_{i \in I} \Box \bigvee_{j \in J_{i}} a_{ij}  \rsem_{P_{u}(A)} = \lsem \Box \bigvee_{f \in \prod_{i \in I}J_{i}} \bigwedge_{i \in I}a_{i, f(i)} \rsem_{P_{u}(A)}, \]
by $(\Box - \wedge )$ (see the proof of (T3)) and distributivity. Now by (i),
\[ \lsem \Box \bigvee_{f \in \prod_{i \in I}J_{i}} \bigwedge_{i \in I}a_{i, f(i)} \rsem_{P_{u}(A)} \neq \varnothing \]
\[ \Longleftrightarrow \;\; \lsem \bigvee_{f \in \prod_{i \in I}J_{i}} \bigwedge_{i \in I}a_{i, f(i)} \rsem_{A} \neq \varnothing \]
\[ \Longleftrightarrow \;\; \exists f \in \prod_{i \in I}J_{i}. \, \bigwedge_{i \in I}a_{i, f(i)} \in con(A). \]
(iii) This follows from the fact that
\[ \bot \not\in \lsem \Box a \rsem_{P_{u}(A)} \; \Longleftrightarrow \; \bot \not\in \lsem a \rsem_{A}. \;\;\; \qed \]

\begin{proposition}[T2] 
\label{pdomT2}
$\forall a \in |P_{u}(A)|. \, \exists b \in {\sf CDNF}(P_{u}(A)). \, a =_{P_{u}(A)} b.$
\end{proposition}

\proof\ We can use the distributive lattice laws to put $a$ in the form
\[ \bigvee_{i \in I} \bigwedge_{j \in J_{i}} \Box a_{ij}. \]
By $(d1)$, each $a_{ij}$ can be written as
\[ \bigvee_{k \in K_{ij}}b_{k}, \]
where each $b_{k} \in cpr(A)$. We can now use $(\Box - \wedge )$ and the distributive laws to obtain an expression of the form
\[ \bigvee_{i' \in I'} \Box \bigvee_{l \in L_{i'}} c_{l}, \]
where each $c_{l} \in cpr(A)$. Moreover disjuncts with $L_{i'} = \varnothing$ can be deleted using $(\Box - 0)$. This yields the required normal form. \qed

\begin{proposition}[T3]
For all $a, b \in |P_{u}(A)|$:
\[ a \leq_{P_{u}(A)} b \; \Longrightarrow \; \lsem a \rsem_{P_{u}(A)} \subseteq \lsem b \rsem_{P_{u}(A)}. \]
\end{proposition}

\proof\ Given $U \in K\Omega (\hat{A}))$, define
\[ \Box U \; \equiv \; \{ S \in P_{u}(\hat{A}) : S \subseteq U \}. \]
Then
\[ U \subseteq V \; \Longrightarrow \Box U \subseteq \Box V, \]
\[ \Box \bigcap_{i \in I}U_{i} = \bigcap_{i \in I} \Box U_{i} \]
are simple set calculations, which validate $(\Box - \leq )$ and 
$(\Box - \wedge )$. $(\Box - 0)$ is valid because the empty set is 
excluded from $P_{u}(\hat{A})$. 
(In fact, dropping $(\Box - 0)$ exactly corresponds to retaining the empty set). \qed

\begin{proposition}[T4]
For all $\Box a, \Box b \in {\sf CPNF}(P_{u}(A))$:
\[ \lsem \Box a \rsem_{P_{u}(A)} \subseteq \lsem \Box b \rsem_{P_{u}(A)} \; \Longrightarrow \; \Box a \leq_{P_{u}(A)} \Box b. \]
\end{proposition}

\proof\ Using the description of $\lsem \Box a \rsem_{P_{u}(A)}$, $\lsem \Box b \rsem_{P_{u}(A)}$ from the proof of Proposition~\ref{pdomT1}(i),
\[ \lsem \Box a \rsem_{P_{u}(A)} \subseteq \lsem \Box b \rsem_{P_{u}(A)} \]
\[ \Longrightarrow \;\; \lsem a \rsem_{A} \subseteq \lsem b \rsem_{A} \]
\[ \Longrightarrow \;\; a \leq_{A} b \]
\[ \Longrightarrow \;\; \Box a \leq_{P_{u}(A)} \Box b \;\;\; (\Box - \leq). \;\;\; \qed \]

\begin{proposition}[T6(A)]
Let $A \trianglelefteq B$, with $e : \hat{A} \rightarrow \hat{B}$ the corresponding projection. Then
\[ (P_{u}(e))^{\dag} \circ \lsem \cdot \rsem_{P_{u}(A)} = \lsem \cdot \rsem_{P_{u}(B)}. \]
\end{proposition}

\proof\ From the proof of Proposition~\ref{pdomT1}(i), for $a \in con(A)$:
\[ (*) \;\;\;\; \lsem \Box a \rsem_{P_{u}(A)} = \diverges_{P_{u}(A)} \lsem a \rsem_{P_{u}(A)}, \]
while for $a \in con(A)$ we have, directly from the definitions,
\[ (**) \;\;\; P_{u}(e)(\lsem a \rsem_{A}) = e^{\dag}(\lsem a \rsem_{A}). \]
Now given $a \in |P_{u}(A)|$, by \ref{pdomT2}
\[ a =_{P_{u}(A)} \bigvee_{i \in I} \Box a_{i}, \;\;\; (a_{i} \in con(A), i \in I), \]
and we can calculate:
\begin{Eqarray}
P_{u}(e)^{\dag}(\lsem a \rsem_{P_{u}(A)}) & = & \bigcup_{i \in I}P_{u}(e)^{\dag}( \lsem \Box a_{i} \rsem_{P_{u}(A)}) & \\
& = & \bigcup_{i \in I}P_{u}(e)^{\dag}( \diverges_{P_{u}(\hat{A})} \lsem a_{i} \rsem_{A}) & (*) \\
& = & \bigcup_{i \in I} \diverges_{P_{u}(\hat{B})} (P_{u}(e) \lsem a_{i} \rsem_{A}) & \\
& = & \bigcup_{i \in I} \diverges_{P_{u}(\hat{B})} (e^{\dag} \lsem a_{i} \rsem_{A}) & (**) \\
& = & \bigcup_{i \in I} \diverges_{P_{u}(\hat{B})} (\lsem a_{i} \rsem_{B}) & \ref{embim} \\
& = & \bigcup_{i \in I} \lsem \Box a_{i} \rsem_{P_{u}(B)} & (*) \\
& = & \lsem a \rsem_{P_{u}(B)}. & \qed 
\end{Eqarray}

\begin{definition} 
{\rm The {\it coalesced sum.}

\noindent (i) The generators:
\[ G(A \oplus B) \; \equiv \; \{ (a \oplus \false ) : a \in |A| \} \cup \{ (\false \oplus b) : b \in |B| \}. \]

\noindent (ii) Metapredicates:
\[ {\sf PNF}(A \oplus B) \; \equiv \; \{ (a \oplus \false ) : a \in pr(A) \} \cup \{ (\false \oplus b) : b \in pr(B) \} \cup \{ \true \} \]
\[ {\sf CON}( \true ) \]
\begin{eqnarray*}
{\sf CON}( \bigwedge_{i \in I}(a_{i} \oplus \false ) \wedge \bigwedge_{j \in J}(\false \oplus b_{j})) & \equiv & 
\neg (\bigwedge_{i \in I}a_{i} \in t(A) 
\: \& \: \bigwedge_{j \in J}b_{j} \in t(B)) \\
& & \& \: \bigwedge_{i \in I}a_{i} \in con(A) \\ 
& & \& \: \bigwedge_{j \in J}b_{j} \in con(B)  
\end{eqnarray*}
\[ {\sf T}( \bigwedge_{i \in I}(a_{i} \oplus \false ) \wedge \bigwedge_{j \in J}(\false \oplus b_{j})) \; \equiv \; \exists i \in I. \, a_{i} \in t(A) \; {\rm or} \; \exists j \in J. \, b_{j} \in t(B) \]
\[ {\sf CPNF}(a) \; \equiv \; {\sf CON}(a) \]

\noindent (iii) Axioms:
\[ (\oplus - \leq ) \;\;\; \frac{a \leq b}{(a \oplus \false ) \leq (b \oplus \false )} \;\;\;\;\; \frac{a \leq b}{(\false \oplus a) \leq (\false \oplus b)} \]
\[ (\oplus - \wedge ) \;\;\; \bigwedge_{i \in I}(a_{i} \oplus \false ) = (\bigwedge_{i \in I}a_{i} \oplus \false ) \;\;\;\;\; \bigwedge_{i \in I}(\false \oplus a_{i} ) = (\false \oplus \bigwedge_{i \in I}a_{i}) \]
\[ (\oplus - \vee ) \;\;\; \bigvee_{i \in I}(a_{i} \oplus \false ) = (\bigvee_{i \in I}a_{i} \oplus \false ) \;\;\;\;\; \bigvee_{i \in I}(\false \oplus a_{i} ) = (\false \oplus \bigvee_{i \in I}a_{i}) \]
\[ (\oplus - \#) \;\;\; a \leq \false \;\;\;\;\; (\#(a)) \]

\noindent (iv) Semantic function:}
\[ \lsem \cdot \rsem_{A \oplus B} : |A \oplus B| \longrightarrow K\Omega (\hat{A} \oplus \hat{B}) \]
\begin{eqnarray*}
\lsem (a \oplus \false ) \rsem_{A \oplus B} & = &  \{ <0, d> : d \in \lsem a \rsem_{A}, d \neq \bot \} \\
& & \mbox{} \cup  \{ x \in \hat{A} \oplus \hat{B} : \bot \in \lsem a \rsem_{A} \}
\end{eqnarray*} 
\begin{eqnarray*}
\lsem (\false \oplus b) \rsem_{A \oplus B} & = & \{ <1, d> : d \in \lsem b \rsem_{B}, d \neq \bot \} \\
& & \mbox{} \cup  \{ x \in \hat{A} \oplus \hat{B} : \bot \in \lsem b \rsem_{B} \}
\end{eqnarray*} 
\end{definition}

\begin{proposition}[T1]
For all $c, \{ c_{i} \}_{i \in I} \in {\sf PNF}(A \oplus B)$:
\[ \begin{array}{rl}
(i)   & \lsem c \rsem_{A \oplus B} \in pr(K\Omega (\hat{A} \oplus \hat{B})) \\
(ii)  & {\sf CON}(\bigwedge_{i \in I}c_{i}) \; \Longleftrightarrow \; \lsem \bigwedge_{i \in I}c_{i} \rsem_{A \oplus B} \neq \varnothing \\
(iii) & {\sf T}(\bigwedge_{i \in I}c_{i}) \; \Longleftrightarrow \; \bot \not\in \lsem \bigwedge_{i \in I}c_{i} \rsem_{A \oplus B}.
\end{array} \]
\end{proposition}

\proof\ (i) If $c = (a \oplus \false )$, $a \in pr(A)$, we can distinguish three cases: \\
(1): $a \not\in con(A)$. In this case, 
\[ \lsem c \rsem_{A \oplus B} = \varnothing . \]
(2): $\lsem a \rsem_{A} = 1_{K\Omega (\hat{A})} = \diverges (\bot )$. In this case,
\[ \lsem c \rsem_{A \oplus B} = \diverges (\bot ) \in pr(K\Omega (\hat{A} \oplus \hat{B})). \]
(3): $a \in con(A)$, $\bot \not\in \lsem a \rsem_{A}$. In this case, for some $u \in K(\hat{A})$, $u \neq \bot$, $\lsem a \rsem_{A} = \diverges u$. Then
\begin{eqnarray*}
\lsem c \rsem_{A \oplus B} & = & \{ <0, d> : u \sqsubseteq d \} \\
& = & \diverges_{\hat{A} \oplus \hat{B}}(<0, u>).
\end{eqnarray*} 
The case for $c = (\false \oplus b)$ is similar.

(ii), (iii). Straightforward. \qed

\begin{proposition}[T2]
$\forall a \in |A \oplus B|. \, \exists b \in {\sf CDNF}(A \oplus B). \, a =_{A \oplus B} b.$
\end{proposition}

\proof\ We can use the distributive lattice laws to put $a$ in the form
\[ \bigvee_{i \in I}(\bigwedge_{j \in J_{i}}(a_{ij} \oplus \false ) 
\wedge \bigwedge_{k \in K_{i}}(\false \oplus b_{ik})). \]
Moreover, we can write each $a_{ij}$ as $\bigvee_{l \in L_{ij}}c_{l}$, $b_{ik}$ as $\bigvee_{m \in M_{ik}}d_{m}$, with $c_{l} \in cpr(A)$, $d_{m} \in cpr(B)$. Using $(\oplus - \vee )$, we obtain
\[ \bigvee_{i \in I'}(\bigwedge_{j \in {J_{i}}'}(a_{ij} \oplus \false ) \wedge \bigwedge_{k \in {K_{i}}'}(\false \oplus b_{ik})) \]
with $a_{ij} \in cpr(A)$, $b_{ik} \in cpr(B)$. Now using $(\oplus - \wedge )$, we obtain
\[ \bigvee_{i \in I'}((\bigwedge_{j \in {J_{i}}'}a_{ij} \oplus \false ) \wedge (\false \oplus \bigwedge_{k \in {K_{i}}'} b_{ik})). \]
For each $i \in I'$, if both 
\[ \bigwedge_{j \in {J_{i}}'}a_{ij} \in t(A) \] 
and 
\[ \bigwedge_{k \in {K_{i}}'}b_{ik} \in t(B), \] 
we may delete the $i$'th disjunct by $(\oplus - \# )$. If either 
\[ \bigwedge_{j \in {J_{i}}'}a_{ij} \not\in con(A) \]
or 
\[ \bigwedge_{k \in {K_{i}}'}b_{ik} \not\in con(B), \] 
we can delete the $i$'th disjunct by $(\oplus - \vee )$. Otherwise, either 
\[ \bigwedge_{j \in {J_{i}}'}a_{ij} =_{A} 1_{A} \]
or 
\[ \bigwedge_{k \in {K_{i}}'}b_{ik} =_{B} 1_{B}, \] 
and we can delete one of these conjuncts by $(\oplus - \wedge )$. In this way we obtain an expression of the form
\[ \bigvee \{ (a \oplus \false ) \} \vee \bigvee \{ (\false \oplus b) \} , \]
with each $a \in cpr(A)$, $b \in cpr(B)$, as required. \qed

\begin{proposition}[T4]
For all $c, d \in {\sf CPNF}(A \oplus B)$:
\[ \lsem c \rsem_{A \oplus B} \subseteq \lsem d \rsem_{A \oplus B} \; \Longrightarrow \; c \leq_{A \oplus B} d. \]
\end{proposition}

\proof\ Take $c = (a \oplus \false )$. We consider two subcases. \\
(1): $d = (b \oplus \false )$.
\begin{eqnarray*}
\lsem c \rsem_{A \oplus B} \subseteq \lsem d \rsem_{A \oplus B} & \Longrightarrow  & \lsem a \rsem_{A} \subseteq \lsem b \rsem_{A} \\
& \Longrightarrow & a \leq_{A} b \\
& \Longrightarrow & (a \oplus \false ) \leq_{A \oplus B} (b \oplus \false ) \;\;\; {\rm by} \; (\oplus - \leq ). 
\end{eqnarray*} 
(2): $d = (\false \oplus b)$.
\begin{eqnarray*}
\lsem c \rsem_{A \oplus B} \subseteq \lsem d \rsem_{A \oplus B} & \Longrightarrow  & \bot \in \lsem b \rsem_{B} \\
& \Longrightarrow & \true \leq_{B} b \\
& \Longrightarrow & c \leq_{A \oplus B} \true \\
& &  =_{A \oplus B} (\false \oplus \true ) \;\;\; (\oplus - \wedge) \\
& &  \leq_{A \oplus B} (\false \oplus b) \;\;\; (\oplus - \leq ).
\end{eqnarray*} 
The case for $c = (\false \oplus a)$ is similar. \qed

\section{Logical Semantics of Types}
We now build on the work of the previous sections to give a {\it logical semantics} for a language of type expressions, in which each type is interpreted as a propositional theory (domain prelocale).
\subsection*{Syntax of Type Expressions}
We define a set of type expressions {\sf TExp} by
\[ \sigma \;\; ::= \;\; \mbox{OP}(\sigma_{1}, \ldots \sigma_{n}) \; (\mbox{OP} \in \Sigma_{n}) \; | \; t \; | \; {\sf rec} \: t. \sigma \]
where $t$ ranges over a set of type variables {\sf TVar}, $\sigma$ over type expressions, and 
$\Sigma = \{\Sigma_{n}\}_{n \in \omega}$ is a ranked alphabet of type constructors. 
For each such constructor $\mbox{OP} \in \Sigma_{n}$, we assume we have an operation ${\rm op}^{{\cal L}} : {\bf DPL1}^{n} \rightarrow {\bf DPL1}$ which satisfies properties (T1) -- (T6) from the previous section with respect to a functor ${\rm op}^{{\cal D}} : {\bf SDom}^{n} \rightarrow {\bf SDom}$.
\subsection*{Logical Semantics of Type Expressions}
We define a semantic function
\[ {\cal L} : {\sf TExp} \longrightarrow {\sf LEnv} \longrightarrow {\bf DPL1} \]
where ${\sf LEnv}$ is the set of type environments
\[ {\sf TVar} \longrightarrow {\bf DPL1} \]
as follows:
\begin{eqnarray*}
{\cal L} \lsem \mbox{OP}(\sigma_{1}, \ldots ,\sigma_{n}) \rsem \rho & = & {\rm op}^{{\cal L}}({\cal L}\lsem \sigma_{1} \rsem \rho , \ldots ,{\cal L}\lsem \sigma_{n} \rsem \rho ) \\
{\cal L} \lsem t \rsem \rho & = &  \rho t \\
{\cal L} \lsem {\sf rec} \: t. \sigma \rsem \rho & = & {\sf fix}(F) = \bigsqcup_{k \in \omega}F^{k}({\bf 1}),
\end{eqnarray*} 
where $F : {\bf DPL1} \rightarrow {\bf DPL1}$ is defined by
\[ F(A) = {\cal L} \lsem \sigma \rsem \rho [t \mapsto A]. \]
We write ${\cal LA}(\sigma ) \rho$ for $\tilde{A}$, where $A = {\cal L} \lsem
\sigma \rsem \rho$.
\subsection*{Denotational Semantics of Type Expressions}
Similarly to the logical semantics, we define
\[ {\cal D} : {\sf TExp} \longrightarrow {\sf DEnv} \longrightarrow {\bf SDom} \]
where ${\sf DEnv} = {\sf TVar} \longrightarrow {\bf SDom}$. In this semantics, each $\mbox{OP} \in \Sigma_{n}$ is interpreted by the corresponding functor
\[ \mbox{op}^{{\cal D}} : ({\bf SDom^{E}})^{n} \longrightarrow {\bf SDom^{E}} \]
and ${\sf rec} \: t. \sigma$ as the inititial fixed point of the endofunctor 
${\bf SDom^{E}} \longrightarrow {\bf SDom^{E}}$ induced from  
$t \mapsto \sigma (t)$. 
See \cite[Chapter 5]{PloLN} and \cite{SP82,Nie84}.

\begin{theorem}[Stone Duality]
Let $\rho_{L} \in {\sf LEnv}$, $\rho_{D} \in {\sf DEnv}$ satisfy:
\[ \forall t \in {\sf TVar}. \, K \Omega (\rho_{D} t) \; \cong \; \rho_{L} t. \]
Then for any type expression $\sigma$, ${\cal LA} \lsem \sigma \rsem \rho_{L}$ is the Stone dual of ${\cal D} \lsem \sigma \rsem \rho_{D}$, i.e.
\[ \begin{array}{rl}
(i) & {\cal D} \lsem \sigma \rsem \rho_{D} \; \cong \; {\sf Spec} \; {\cal LA} \lsem \sigma \rsem \rho_{L} \\
(ii) & K \Omega ({\cal D} \lsem \sigma \rsem \rho_{D}) \; \cong \; 
{\cal LA} \lsem \sigma \rsem \rho_{L}. 
\end{array} \]
\end{theorem}

\proof\ Firstly, note that the two conclusions of the Theorem are equivalent, since Scott domains are coherent spaces. Thus it suffices to prove $(i)$.

It will be convenient to consider systems of simultaneous domain equations
\begin{equation}
\left. \begin{array}{rcl} 
\xi_{1} & = & \sigma_{1}(\xi_{1}, \ldots , \xi_{n})  \\
         & \vdots &  \label{systeq} \\
\xi_{n} & = & \sigma_{n}(\xi_{1}, \ldots , \xi_{n})
\end{array} \right\} 
\end{equation}
where each $\sigma_{i}$ is a type expression not containing any occurrences 
of ${\sf rec}$. 
It is standard that any $\sigma \in {\sf TExp}$ is equivalent to a system of 
equations of this form, in the sense that the denotation of $\sigma$ 
is isomorphic to a component of the solution of such a system. 
Thus what we shall show is that $\hat{A} \cong D$, where $A$ is the solution 
of~\ref{systeq} in {\bf DPL1} and $D$ is the solution in {\bf SDom}. 
To make this more precise, we need some definitions.

Firstly, we define a diagram $\Delta^{D}$ in $({\bf SDom}^{E})^{n}$ as follows:
\[ \Delta^{D} = (D_{n}, f_{n})_{n \in \omega} \]
where
\begin{eqnarray*}
D_{0} & = & ({\bf 1}^{\cal D}, \ldots , {\bf 1}^{\cal D}) \\
D_{k+1} & = & ({\cal D} \lsem \sigma_{1} \rsem \rho^{D} [\vec{\xi} \mapsto D_{k}], \ldots , {\cal D} \lsem \sigma_{n} \rsem \rho^{D} [\vec{\xi} \mapsto D_{k}])
\end{eqnarray*}
and $f_{k} : D_{k} \rightarrow D_{k+1}$ is defined as follows: $f_{0}$ is the unique morphism given by initiality of $D_{0}$ in $({\bf SDom}^{E})^{n}$;
\[ f_{k+1} =  ({\cal D}_{m} \lsem \sigma_{1} \rsem \rho_{m}^{D} [\vec{\xi} \mapsto f_{n}], \ldots , {\cal D}_{m} \lsem \sigma_{n} \rsem \rho_{m}^{D} [\vec{\xi} \mapsto f_{n}]) \]
where ${\cal D}_{m}$ gives the morphism part of the functor corresponding to $\sigma$, 
and $\rho_{m}^{D} t = {\sf id}_{\rho^{D} t}$. 
Now it is standard that the solution of \ref{systeq} in {\bf SDom} is given by 
\[ \lim_{\rightarrow} \Delta^{D}. \]

Similarly, we define a $\unlhd$--chain $\{ A_{n} \}$ in ${\bf DPL1}^{n}$ by 
\begin{eqnarray*}
A_{0} & = &  ({\bf 1}^{\cal L}, \ldots , {\bf 1}^{\cal L}) \\
A_{k+1} & = & ({\cal L} \lsem \sigma_{1} \rsem \rho^{L} [\vec{\xi} \mapsto A_{k}], \ldots , {\cal L} \lsem \sigma_{n} \rsem \rho^{L} [\vec{\xi} \mapsto A_{k}])
\end{eqnarray*}
and we let $\Delta^{L}$ be the diagram $(\hat{A}_{k}, e_{k})$ in $({\bf SDom}^{E})^{n}$, where $e_{k} : \hat{A}_{k} \rightarrow \hat{A}_{k+1}$ is the tuple of embeddings 
\[ e_{k, i} : \hat{A}_{k, i} \rightarrow \hat{A}_{k+1, i} \;\;\; (1 \leq i \leq n) \] 
induced by $A_{k, i} \unlhd A_{k+1, i}$. 
Now the solution of \ref{systeq} in {\bf DPL1} is given by 
\[ A_{\infty} = \bigsqcup_{k} A_{k} = (\bigsqcup_{k} A_{k, 1}, \ldots , \bigsqcup_{k} A_{k, n}). \]
It is easily verified that the cone 
$\mu : \Delta^{L} \rightarrow \hat{A}_{\infty}$ with $\mu_{k}$ the embedding induced by $A_{k} \unlhd A_{\infty}$ is colimiting in $({\bf SDom}^{E})^{n}$. 
Thus our task reduces to proving
\[ \lim_{\rightarrow} \Delta^{L} \; \cong \; \lim_{\rightarrow} \Delta^{D}, \]
for which it suffices to construct a natural isomorphism $\nu : \Delta^{L} \; \cong \; \Delta^{D}$.

We fix $\vec{\sigma} = (\sigma_{1}, \ldots , \sigma_{n})$ as the system of equations under consideration. For each $\vec{\tau} = (\tau_{1}, \ldots , \tau_{n})$ where each $\tau_{i}$ contains no occurrences of ${\sf rec}$, and $k \in \omega$, we shall define:
\begin{itemize}
\item objects $D_{\vec{\tau}, k}$ and morphisms 
\[ f_{\vec{\tau}, k} : D_{\vec{\tau}, k} \rightarrow D_{\vec{\tau}, k+1} \]
in $({\bf SDom}^{E})^{n}$;
\item objects $A_{\vec{\tau}, k}$ in ${\bf DPL1}^{n}$ and morphisms
\[ e_{\vec{\tau}, k} : \hat{A}_{\vec{\tau}, k} \rightarrow \hat{A}_{\vec{\tau}, k+1} \]
\item morphisms $\nu_{\vec{\tau}, k} : \hat{A}_{\vec{\tau}, k} \rightarrow D_{\vec{\tau}, k}$.
\end{itemize}
\[ D_{\vec{\tau}, 0} = ({\bf 1}^{\cal D}, \ldots , {\bf 1}^{\cal D}); \;\;\;  A_{\vec{\tau}, 0} = ({\bf 1}^{\cal L}, \ldots , {\bf 1}^{\cal L}) \]
\[ D_{\vec{\tau}, k+1} = ({\cal D} \lsem \tau_{1} \rsem \rho^{D} [\vec{\xi} \mapsto D_{\vec{\sigma}, k}], \ldots , {\cal D} \lsem \tau_{n} \rsem \rho^{D} [\vec{\xi} \mapsto D_{\vec{\sigma}, k}]) \]
\[ A_{\vec{\tau}, k+1} = ({\cal L} \lsem \tau_{1} \rsem \rho^{L} [\vec{\xi} \mapsto A_{\vec{\sigma}, k}], \ldots , {\cal L} \lsem \tau_{n} \rsem \rho^{L} [\vec{\xi} \mapsto A_{\vec{\sigma}, k}]) \] 
$f_{\vec{\tau}, 0}$ is the unique morphism given by initiality.
\[ f_{\vec{\tau}, k+1} = ({\cal D}_{m} \lsem \tau_{1} \rsem \rho^{D} [\vec{\xi} \mapsto f_{\vec{\sigma}, k}], \ldots , {\cal D}_{m} \lsem \tau_{n} \rsem \rho^{D} [\vec{\xi} \mapsto f_{\vec{\sigma}, k}]) \]
$e_{\vec{\tau}, k+1}$ is the embedding induced by
\[ A_{\vec{\tau}, k} \trianglelefteq A_{\vec{\tau}, k+1} \]
which holds since $A_{\vec{\sigma}, k} \trianglelefteq A_{\vec{\sigma}, k+1}$ by the usual argument.
$\nu_{\vec{\tau}, 0}$ is the unique isomorphism arising from $\hat{{\bf 1}}^{\cal L} \cong {\bf 1}^{\cal D}$.
\[ \nu_{\vec{\tau}, k+1} = (\nu_{\tau_{1}, k+1}, \ldots , \nu_{\tau_{n}, k+1}), \]
where $\nu_{\tau , k+1}$ is defined by induction on $\tau$:
\[ \nu_{\xi_{i}, k+1} = \nu_{\sigma_{i}, k} \]
\[ \nu_{t, k+1} = {\hat{\rho}}^{L} t \cong {\rho}^{D} t, \]
the isomorphism given in the hypothesis of the theorem.
For $\tau = \mbox{OP}(\theta_{1}, \ldots , \theta_{m})$,
\[ \nu_{\tau , k+1} = \mbox{op}^{\cal D}(\nu_{\theta_{1}, k+1}, \ldots , 
\nu_{\theta_{m}, k+1}) \circ \eta_{\tau , k+1}, \]
where $\eta_{\tau , k+1} : \hat{A}_{\tau , k+1} \cong \mbox{op}^{\cal D}(\hat{A}_{\theta_{1}, k+1}, \ldots , \hat{A}_{\theta_{m}, k+1})$ is the isomorphism given by property (T6)(B) for $\mbox{OP}$.

Note that
\[ {\Delta}^{D} = (D_{\vec{\sigma}, k}, f_{\vec{\sigma}, k})_{k \in \omega}, \]
\[ {\Delta}^{L} = (\hat{A}_{\vec{\sigma}, k}, e_{\vec{\sigma}, k})_{k \in \omega}, \]
and so, defining $\nu : \Delta^{L} \rightarrow \Delta^{D}$ by $\nu_{k} \equiv \nu_{\vec{\sigma}, k}$, it remains to verify that for all $k$:
\begin{itemize}
\item $\nu_{k}$ is an isomorphism
\item $\nu_{k+1} \circ e_{k} = f_{k} \circ \nu_{k}$.
\end{itemize}
We argue by induction on $k$. 
The basis follows from the fact that $\hat{{\bf 1}}^{\cal L} \cong {\bf 1}^{\cal D}$, and the initiality of $({\bf 1}^{\cal D}, \ldots , {\bf 1}^{\cal D})$ in $({\bf SDom}^{E})^{n}$. For the inductive step, we assume:
\[ (i) \;\; \nu_{k} = \nu_{\vec{\sigma}, k} \; \mbox{is an isomorphism} \]
\[ (ii) \;\; \nu_{k+1} \circ e_{k} = \nu_{\vec{\sigma}, k+1} \circ e_{\vec{\sigma}, k} = f_{\vec{\sigma}, k} \circ \nu_{\vec{\sigma}, k} = f_{k} \circ \nu_{k} \]
and prove that for all $\tau$ with no occurrences of ${\sf rec}$,
\[ (iii) \;\; \nu_{\tau , k+1} \; \mbox{is an isomorphism} \]
\[ (iv) \;\; \nu_{\tau , k+2} \circ e_{\tau , k+1} = f_{\tau , k+1} \circ \nu_{\tau , k+1} \]
(where $(e_{\tau , k+1}, \ldots , e_{\tau , k+1}) = e_{(\tau , \ldots , \tau ), k+1}$,
and similarly for $f_{\tau , k+1}$).
Taking $\tau = \sigma_{i}$, $1 \leq i \leq n$ in $(iii)$ and $(iv)$ then yields
\[ (v) \;\; \nu_{k+1} = \nu_{\vec{\sigma}, k+1} \; \mbox{is an isomorphism} \]
and
\begin{eqnarray*}
 (vi) \;\; \nu_{k+2} \circ e_{k+1} = \nu_{\vec{\sigma}, k+2} \circ e_{\vec{\sigma}, k+1} & = & f_{\vec{\sigma}, k+1} \circ \nu_{\vec{\sigma}, k+1}\\
& = & f_{k+1} \circ \nu_{k+1}, 
\end{eqnarray*}
as required. We prove $(iii)$ and $(iv)$ by induction on $\tau$.

\noindent Case 1: $\tau = \xi_{i}$. In this case, $(iii)$ just says that $\nu_{\sigma_{i}, k}$ is an isomorphism, and $(iv)$ that
\[ \nu_{\sigma_{i}, k+1} \circ e_{\sigma_{i}, k} = f_{\sigma_{i}, k} \circ \nu_{\sigma_{i}, k}, \]
and we can use our outer induction hypothesis on $k$.

\noindent Case 2: $\tau = t$. In this case, $\tau$ denotes a constant functor, and
\[ f_{\tau , k+1} = {\sl id}_{D_{\tau , k+1}}, \]
\[ e_{\tau , k+1} = {\sl id}_{\hat{A}_{\tau , k+1}}, \]
\[ \nu_{\tau , k+1} = \nu_{\tau , k+2} = 
(\hat{\rho}^{L} t \cong \rho^{D} t), \]
so $(iii)$ and $(iv)$ hold trivially.

\noindent Case 3: $\tau = \mbox{OP}(\theta_{1}, \ldots , \theta_{m})$. Applying our inner induction hypothesis to each $\theta_{i}$, we have
\[ (vii) \;\; \nu_{\theta_{i}, k+1} \; \mbox{is an isomorphism} \]
\[ (viii) \;\; \nu_{\theta{i}, k+2} \circ e_{\theta{i}, k+1} = f_{\theta_{i}, k+1} \circ \nu_{\theta_{i}, k+1}. \]
By definition,
\[ \nu_{\tau , k+1} = \mbox{op}^{\cal D}(\nu_{\theta_{1}, k+1}, \ldots , \nu_{\theta_{m}, k+1}) \circ \eta_{\tau , k+1}. \]
Since $\mbox{op}^{\cal D}$ is a functor, by $(vii)$ $\mbox{op}^{\cal D}(\nu_{\theta_{1}, k+1}, \ldots , \nu_{\theta_{m}, k+1})$ is an isomorphism; while $\eta_{\tau , k+1}$ is given as an isomorphism by (T6)(B). 
This proves~$(iii)$. Finally,
\[ \begin{array}{clr}
& \nu_{\tau , k+2} \circ e_{\tau , k+1} & \\
= & \mbox{op}^{\cal D}(\nu_{\theta_{1}, k+2}, \ldots , \nu_{\theta_{m}, k+2}) \circ \eta_{\tau , k+2} \circ e_{\tau , k+1} & \\
= & \mbox{op}^{\cal D}(\nu_{\theta_{1}, k+2}, \ldots , \nu_{\theta_{m}, k+2}) \circ \mbox{op}^{\cal D}(e_{\theta_{1}, k+1}, \ldots , e_{\theta_{m}, k+1}) \circ \eta_{\tau , k+1} \\
  & \mbox{by (T6)(B)} \\
= & \mbox{op}^{\cal D}(\nu_{\theta_{1}, k+2} \circ e_{\theta_{1}, k+1}, \ldots , \nu_{\theta_{m}, k+2} \circ e_{\theta_{m}, k+1}) \circ \eta_{\tau , k+1} & \\ 
= & \mbox{op}^{\cal D}(f_{\theta_{1}, k+2} \circ \nu_{\theta_{1}, k+1}, \ldots , f_{\theta_{m}, k+2} \circ \nu_{\theta_{m}, k+1}) \circ \eta_{\tau , k+1} \\
  & \mbox{by $(viii)$}  \\ 
= & \mbox{op}^{\cal D}(f_{\theta_{1}, k+2}, \ldots , f_{\theta_{m}, k+2}) \circ \mbox{op}^{\cal D}(\nu_{\theta_{1}, k+1}, \ldots , \nu_{\theta_{m}, k+1}) \circ \eta_{\tau , k+1} \\
= & f_{\tau , k+2} \circ \nu_{\tau , k+1}, &
\end{array} \] 
which proves $(iv)$. \qed

We finish with an observation that will be useful in the next Chapter. 
In our definitions of the constructions $A \rightarrow B$ etc. in section 4, 
we used the ``semantic'' predicates $pr$, $con$, $t$ at the argument types $A$, $B$. 
Now suppose we are forming a theory as the denotation of a type expression, 
e.g. ${\cal L} \lsem \sigma \rightarrow \tau \rsem \rho$; 
the arguments are $A = \lsem \sigma \rsem \rho$, $B = \lsem \tau \rsem \rho$. 
Then it makes sense to use the {\it syntactic} predicates ${\sf PNF}(A)$, 
${\sf CON}(A)$, 
${\sf T}(A)$ etc. in our definition of 
\[ A \rightarrow B = {\cal L} \lsem \sigma \rightarrow \tau \rsem \rho. \] 
Using properties (T1), (T2) and (T8) for each type construction, 
it is straightforward to prove the

\begin{observation}
\label{obs}
For all $\sigma$, $\rho$ the same theory is obtained as 
${\cal L} \lsem \sigma \rsem \rho$ whether syntactic or semantic predicates are used in each application of a type construction. \qed
\end{observation}

\chapter{Domain Theory In Logical Form}
\section{Introduction}
In this Chapter we shall complete the core of our 
research programme, as set out in Chapter~1.
We shall introduce a meta-language for denotational semantics,
give it a logical interpretation {\it via} the localic side of Stone duality,
and relate this logical interpretation to the standard denotational one
by showing that they are Stone duals of each other.

Denotational semantics is always based, more or less explicitly,
on a typed functional meta-language.
The types are interpreted as topological spaces
(usually domains in the sense of Scott \cite{Sco81,Sco82}, but
sometimes metric spaces, as in \cite{deBZ82,Niv81}), while the
terms denote elements of or functions between these spaces.
A {\em program logic} comprises an assertion language of formulas
for expressing properties of programs, and an interface between these
properties and the programs themselves.
Two main types of interface can be identified \cite{Pnu77}:
\begin{description}
\item[Endogenous logic] In this style, formulas describe properties
pertaining to the ``world'' of a single program.
Notation: \[ P \models \phi \]
where $P$ is a program and $\phi$ is a formula. Examples:
temporal logic as used e.g. in \cite{Pnu77}; Hennessy-Milner logic \cite{HM85};
type inference \cite{DM82}.
\item[Exogenous logic] Here, programs are embedded in formulas as
{\em modal operators}. Notation: \[ [P]\phi \]
where $P$ is now a program denoting a function or relation.
Examples: dynamic logic \cite{Har79,Pra79}, including as special cases
Hoare logic \cite{Hoa69}, since ``Hoare triples''
$\{ \phi \} P \{ \psi \}$ can be represented by
\[ \phi \rightarrow [P] \psi , \]
and Dijkstra's wlp-calculus \cite{Dij76}, since $wlp( P, \psi )$
can be represented as $[P] \psi$. (Total correctness assertions can also
be catered for; see \cite{Har79}.)
\end{description}

Extensionally, formulas denote sets of points in our denotational
domains, i.e. $\phi$ is a syntactic description of
$\{ x : x \: {\rm satisfies} \: \phi \}$.
Then $P \models \phi$ can be interpreted as $x \in U$, where
$x$ is the point denoted by $P$, and $U$ is the set denoted by 
$\phi$.
Similarly, $[M] \phi$ can be interpreted as $f^{-1}(U)$, where $f$
is the function denoted by $M$ (and elaborations of this when
$M$ denotes a relation or multifunction).
In this way, we can give a topological interpretation of program
logic.

But this is not all: duality cuts both ways.
We can also use it to give a {\em logical interpretation of
denotational semantics}.
Rather than starting with the denotational domains as spaces of points,
and then interpreting formulas as sets of points, we can give an
axiomatic presentation of the topologies on our spaces,
viewed as abstract lattices (logical theories), and then reconstruct
the points from the properties they satisfy.
In other words, we can present denotational semantics in axiomatic
form, as a logic of programs.
This has a number of attractions:
\begin{itemize}
\item It unifies semantics and program logic in a general and systematic
setting.
\item It extends the scope of program logic to the entire range of
denotational semantics -- higher-order functions, recursive types,
powerdomains etc.
\item The syntactic presentation of recursive types, powerdomains etc.
makes these constructions more ``visible'' and easier to calculate with. 
\item The construction of ``points'', i.e. denotations of computational
processes, from the properties they satisfy is very compatible
with work currently being done in a mainly operational setting in
concurrency \cite {HM85,Win80} and elsewhere \cite{BC85}, and offers
a promising approach to unification of this work with denotational semantics.
\end{itemize}

The setting we shall take for our work in this Chapter is {\bf SDom}, the category of Scott domains.
The significance of this as far as the meta-language is concerned is that 
we omit
the Plotkin powerdomain construction.
However, this construction will be treated, in the context of a particular domain equation,
in Chapter 5.
Our reason for not including the Plotkin powerdomain, and extending the duality
to {\bf SFP}, is that this creates some additional technical complications,
though certainly not insuperable ones; lack of time and energy supervened.
For further discussion, see Chapter 7.

The remainder of the Chapter is organised as follows.
In section~2, we interpret the types of our denotational meta-language
as propositional theories.
We can then apply the results of Chapter~3 to show that each such theory is the Stone dual of the domain obtained as the denotation of the type in the standard interpretation.
In section~3, we extend the meta-language to include typed terms, i.e. {\em functional programs}.
We extend our logic to an axiomatisation of the satisfaction relation $P \models \phi$ ($P$ a term, $\phi$ a formula of the logic introduced in section 2),
and prove that this axiomatisation is sound and complete with respect to the
spatial interpretation $x \in U$, where $x$ is the point denoted by $P$, and $U$ the open set denoted by $\phi$.
In section~4, we consider an alternative formulation of the meta-language, 
in which terms are formed at the morphism level rather than the element level;
the comparison between these formulations extends the standard one between
$\lambda$-calculus (element level) and cartesian closed categories (morphism level).
We find a pleasing correspondence between the two known, but hitherto quite unrelated, dichotomies:
\begin{center}
\begin{tabular}{ccc}
cartesian closed categories & & exogenous logic \\
{\it vs.} & ${\Large \sim}$ & {\it vs.} \\
$\lambda$-calculus & & endogenous logic.
\end{tabular}
\end{center}
Our axiomatisation of the morphism-level language comprises an extended and 
generalised {\em dynamic logic} \cite{Pra79,Har79}.
We prove a restricted Completeness Theorem for this axiomatisation,
and show that the general validity problem for this logic is undecidable.
Finally, in section~5 we indicate how the results of this Chapter pave the way
for a whole class of applications, and set the scene for the two case studies
to be described in Chapters~5 and~6.

\section{Domains as Propositional Theories}

We begin by introducing the first part of a meta-language for
denotational semantics, the {\em type expressions}, with syntax
\[ \sigma \;\; ::= \;\; {\bf 1} \; | \; \sigma \times \tau \; | \;
\sigma \rightarrow \tau \; | \; 
\sigma \oplus \tau \; | \; (\sigma )_{\bot} \; | \;
P_{u} \sigma \; | \; 
P_{l} \sigma \; | \; t  \; | \; 
{\sf rec} \, t.\sigma \]
where $t$ ranges over type variables, and $\sigma , \tau$ over
type expressions.

The standard way of interpreting these expressions is as objects
of  {\bf SDom} (more generally as cpo's, but {\bf SDom} is closed
under all the above constructions as a subcategory of  {\bf CPO}).
Thus for each type expression $\sigma$ we define a domain
${\cal D} ( \sigma ) = (D(\sigma ), \sqsubseteq_{\sigma} )$ in
{\bf SDom}; $\sigma \times \tau$ is interpreted as product,
$\sigma \rightarrow \tau$ as function space, $\sigma \oplus \tau$
as coalesced sum, $(\sigma )_{\bot}$ as lifting, $P_{u} \sigma$
and $P_{l} \sigma$ as the upper and lower (or Smyth and Hoare)
powerdomains, and ${\rm rec} \, t. \sigma$ as the solution of the
domain equation
\[t = \sigma (t),\]
i.e. as the initial fixpoint of an endofunctor over {\bf SDom}.
Other constructions (e.g. strict function space, smash product)
can be added to the list.

So far, all this is standard (\cite {PloLN,SP82}). 
Now we begin our alternative approach.
For each type expression $\sigma$, we shall define a propositional
theory ${\cal L}(\sigma ) = (L(\sigma ), \: \leq_{\sigma}, \: =_{\sigma})$,
where:
\begin{itemize}
\item $L(\sigma )$ is a set of formulae
\item $\leq_{\sigma}$, $=_{\sigma}$ are the relations of logical
{\em entailment} and {\em equivalence} between formulae.
\end{itemize}

${\cal L}(\sigma )$ is defined inductively via formation rules,
axioms and inference rules in the usual way.
\subsection*{Formation Rules}
\[ \bullet \;\; {\sl t, f} \in L(\sigma ) \;\;\;\;\;\;\;\;
\bullet \;\; \frac{\phi , \psi \in L(\sigma )}{\phi \wedge \psi , \phi \vee \psi
\in L(\sigma )}\]
\[\bullet \;\; \frac{\phi \in L(\sigma ), \; \psi \in L(\tau )}
{(\phi \times \psi ) \in L(\sigma \times \tau ), \;
(\phi \rightarrow \psi ) \in L(\sigma \rightarrow \tau ) } \]
\[ \bullet \;\; \frac{\phi \in L(\sigma ), \; \psi \in L(\tau )}
{(\phi \oplus \false ), \; (\false \oplus \psi ) \in L(\sigma \oplus \tau )} \;\;\;\;\;\;\;\; 
\bullet \;\; \frac{\phi \in L(\sigma )}{(\phi )_{\bot} \in L((\sigma )_{\bot})} \]
\[ \bullet \;\; \frac{\phi \in L(\sigma )}
{\Box \phi \in L(P_{u} \sigma ), \; 
\Diamond \phi \in L(P_{l} \sigma ) } \;\;\;\;\;\;\;\;
\bullet \;\; \frac{\phi \in L(\sigma [{\sf rec} \, t.\sigma / t])}
{\phi \in L({\sf rec} \, t.\sigma )} \]

We should think of $(\phi \rightarrow \psi )$, $\Box \phi$ etc.
as ``constructors'' or ``generators'', which build basic formulae
at complex types from arbitrary formulae at simpler types.
Note that no constructors are introduced for recursive types;
we are taking advantage of the observation, familiar from work on information
systems \cite{LW84}, that if we work with preorders it is easy to solve
domain equations up to {\em identity}.
\subsection*{Examples}
We define separated sum as a derived operation:
\[ \sigma + \tau \equiv (\sigma )_{\bot} \oplus (\tau )_{\bot} \]
Also, we define the Sierpinski space (two-point domain):
\[ \Oh \equiv ({\bf 1})_{\bot} \]
Now we construct a number of familiar semantic domains:
\begin{center}
\begin{tabular}{|l|c|l|} \hline
name &  expression &  description \\ \hline
{\sf B} & ${\bf 1} + {\bf 1}$ & flat domain of booleans \\
{\sf N} & ${\sf rec} \: t. \, \Oh \oplus t$ & flat domain of natural numbers \\
{\sf LN} & ${\sf rec} \: t. \, {\bf 1} + t$ & lazy natural numbers \\
{\sf List(N)} & ${\sf rec} \: t. \, {\bf 1} + ({\sf N} \times t)$ & lazy lists of eager numbers \\
{\sf CBN} & ${\sf rec} \: t. \, {\sf N} + (t \rightarrow t)$ & call-by-name untyped $\lambda$-calculus \\ \hline
\end{tabular}
\end{center}
Now we define some formulas in these types, to suggest how the expected
structure emerges from the formal definitions.
\begin{center}
\begin{tabular}{|l|c|l|}  \hline
name & formula & type \\ \hline
$\star$ & $(\true )_{\bot}$ & $\Oh$ \\
{\sf true} & $(\star \oplus \false )$ & {\sf B} \\
{\sf false} & $(\false \oplus \star )$ & {\sf B} \\
$\overline{0}$ & $(\star \oplus \false )$ & {\sf N} \\
$\overline{1}$ & $(\false \oplus \overline{0})$ & {\sf N} \\
$\overline{n+1}$ & $(\false \oplus \overline{n})$  & {\sf N} \\
{\sf nil} & $(\star \oplus \false )$ & {\sf List(N)} \\
$\overline{0} :: {\sf nil}$ & $(\false \oplus (\overline{0} \times {\sf nil}))$ & {\sf List(N)} \\
$\overline{0} :: \bot$ & $(\false \oplus (\overline{0} \times \true ))$ & {\sf List(N)} \\
{\sf parallel or} & $(({\sf true} \times \true ) \rightarrow {\sf true})$ & \\
& $\mbox{} \wedge ((\true \times {\sf true}) \rightarrow {\sf true})$ & \\
& $\mbox{} \wedge (({\sf false} \times {\sf false}) \rightarrow {\sf false})$ 
& $({\sf B} \times {\sf B}) \rightarrow {\sf B}$ \\ \hline
\end{tabular}
\end{center}
\subsection*{Auxiliary Predicates}
Before proceeding to the axiomatisation proper, we shall define some
auxiliary predicates on formulas. These will be used as side-conditions
on a number of axioms and rules (e.g. $(\rightarrow - \vee - R)$ below).
Thus it is important that they are recursive predicates, defined
syntactically on formulae. The main predicates we define are:
\begin{itemize}
\item {\sf PNF}($\phi$): $\phi$ is in {\em prime normal form},
defined by the condition that disjunctions only occur in $\phi$
immediately under $\Box$.
\end{itemize}
Then for $\phi$ in PNF, we shall define:
\begin{itemize}
\item {\sf C}($\phi$): $\phi$ is {\em consistent}, i.e. so that we have
\[ {\sf C}(\phi ) \; \Longleftrightarrow \; \neg (\phi \leq {\sl f})
\; \Longleftrightarrow \; \lsem \phi \rsem \neq \varnothing \]
(where $\lsem \cdot \rsem$ is the semantics to be introduced below).
\item {\sf T}($\phi$): $\phi$ requires {\em termination}, i.e. so that we have
\[ {\sf T}(\phi ) \; \Longleftrightarrow \; \neg ({\sl t} \leq \phi )
\; \Longleftrightarrow \; \bot \not\in \lsem \phi \rsem . \] 
\end{itemize}

Of these, the idea of formal consistency, and its definition for
function spaces, go back to \cite{Kre59}, and also play a major role in
\cite{Sco81,Sco82}.
The other predicates, as syntactic conditions on expressions, are
apparently new (and in the presence of the  type constructions
we are considering, specifically function space and coalesced sum,
the definitions of {\sf C} and {\sf T} are {\em mutually recursive}).
\[ \begin{array}{lll}
{\sf C}(\true ) & \equiv & {\sf true} \\
{\sf C}(\bigwedge_{i \in I}(\phi_{i} \times \psi_{i})) & \equiv & {\sf C}(\bigwedge_{i \in I}
\phi_{i}) \; \& \; {\sf C}(\bigwedge_{i \in I}\psi_{i}) \\
{\sf C}(\bigwedge_{i \in I}(\phi_{i} \rightarrow
\psi_{i})) & \equiv  & \forall J \subseteq I. \: 
{\sf C}
(\bigwedge_{j \in J}\phi_{j})
\; \Rightarrow \; {\sf C}(\bigwedge_{j \in J}\psi_{j}) \\
{\sf C}(\bigwedge_{i \in I}(\phi_{i} \oplus {\sl f}) & & \\
\mbox {} \wedge \bigwedge_{j \in J}({\sl f} \oplus \psi_{j})) & \equiv &
\neg ({\sf T}(\bigwedge_{i \in I}\phi_{i}) \; \& \; 
{\sf T}(\bigwedge_{j \in J}\psi_{j})) \\
& & \& \; {\sf C}(\bigwedge_{i \in I}\phi_{i}) \; \& \;
{\sf C}(\bigwedge_{j \in J}\psi_{j}) \\
{\sf C}(\bigwedge_{i \in I}(\phi_{i})_{\bot}) & \equiv & 
{\sf C}(\bigwedge_{i \in I}\phi_{i}) \\
{\sf C}(\bigwedge_{i \in I}\Diamond \phi_{i}) & \equiv & \forall i \in I. \:
{\sf C}(\phi_{i}) \\
{\sf C}(\bigwedge_{i \in I} \Box \bigvee_{j \in J_{i}} \phi_{ij}) & \equiv &
\exists f \in \prod_{i \in I} J_{i}. \: {\sf C}(\bigwedge_{i \in I} \phi_{i f(i)  } )
\end{array} \]
\[ \begin{array}{lll}
{\sf T}(\bigwedge_{i \in I}\phi_{i}) & \equiv & \exists i \in I. \: {\sf T}(\phi ) \\
{\sf T}(\phi \rightarrow \psi ) & \equiv & {\sf C}(\phi ) \; \& \; {\sf T}(\psi ) \\
{\sf T}(\phi \times \psi ) & \equiv & {\sf T}(\phi ) \; \mbox{or} \; {\sf T}(\psi ) \\
{\sf T}(\phi \oplus {\sl f}) & \equiv &
{\sf T}({\sl f} \oplus \phi ) \;\; \equiv \;\; {\sf T}(\phi ) \\
{\sf T}((\phi )_{\bot}) & \equiv & {\sf true} \\
{\sf T}(\Diamond \phi ) & \equiv & {\sf T}(\Box \phi ) \equiv T(\phi ) .
\end{array} \] 

Once we have defined {\sf C} and {\sf T}, we can introduce the following derived predicates:
\begin{eqnarray*}
{\sf CPNF}(\phi) & \equiv & {\sf PNF}(\phi ) \; \mbox{and for all sub-formulae
$\psi$ of $\phi$,} \\
& &  {\sf PNF}(\psi ) \; \Rightarrow \; {\sf C}(\psi ) . \\
{\sf CDNF}(\phi ) & \equiv & \phi = \bigvee_{i \in I} \phi_{i} \; \& \; \forall   i \in I. \, {\sf CPNF}(\phi_{i}) \\
\#(\phi ) & \equiv & \phi = \bigvee_{i \in I} \phi_{i} \; \& \; \forall i \in I. \, {\sf PNF}(\phi ) \: \& \: \neg {\sf C}(\phi ) \\ 
(\phi ) \converges  & \equiv & \phi = \bigvee_{i \in I} \phi_{i} \; \& \; \forall i \in I. \, {\sf PNF}(\phi ) \: \& \: {\sf T}(\phi ).
\end{eqnarray*}

Now we turn to the axiomatization.
The axioms of our logic are all ``polymorphic'' in character,
i.e. they arise from the type constructions uniformly over
the types to which the constructions are applied.
Thus we omit type subscripts.

The axioms fall into two main groups.
\subsection*{Logical Axioms}
These give each $\cal L (\sigma )$ the structure of a distributive
lattice.
\[ ({\leq}-{\rm ref}) \;\;\; \phi \leq \phi \;\;\;\;\;\;
({\leq}-{\rm trans}) \;\;\;  \frac{\phi \leq \psi , \; \psi \leq \chi}
{\phi \leq \chi} \]
\[ ({=}-I) \;\;\; \frac{\phi \leq \psi , \; \psi \leq \phi}{\phi = \psi} 
\;\;\;\;\;\;
({=}-E) \;\;\; \frac{\phi = \psi}{\phi \leq \psi , \; \psi \leq \phi} \]
\[ ({\true}-I) \;\;\; \phi \leq \true \;\;\;\;\;\;	
({\wedge}-I) \;\;\; \frac{\phi \leq \psi_{1} , \; \phi \leq \psi_{2}}
{\phi \leq \psi_1 \wedge \psi_2} \]
\[ ({\wedge}-E-L) \;\;\; \phi \wedge \psi \leq \phi \;\;\;\;\;\; 
({\wedge}-E-R) \;\;\; \phi \wedge \psi \leq \psi  \]
\[ ({\false}-E) \;\;\; \false \leq \phi \;\;\;\;\;\;
({\vee}-I) \;\;\; \frac{\phi_{1} \leq \psi , \; \phi_{2} \leq \psi}
{\phi_{1} \vee \phi_{2} \leq \psi} \]
\[ ({\vee}-E-L) \;\;\; \phi \leq \phi \vee \psi \;\;\;\;\;\;
({\vee}-E-R) \;\;\; \psi \leq \phi \vee \psi \]  
\[ ({\wedge}-{\rm dist}) \;\;\; \phi \wedge (\psi \vee \chi )  \leq  
(\phi \wedge \psi )
\vee (\psi \wedge \chi ) \]

\subsection*{Type-specific Axioms}
These articulate each type construction, by showing how its generators
interact with the logical structure.

\[ ({\times}-{\leq}) \;\;\; \frac{\phi \leq \phi' , \; \psi \leq \psi'}
{(\phi \times \psi ) \leq (\phi' \times \psi' )} \]
\[ ({\times}-{\wedge}) \;\;\; \bigwedge_{i \in I} (\phi_{i} \times \psi_{i}) = (\bigwedge_{i \in I} \phi_{i} \times \bigwedge_{i \in I} \psi_{i} ) \]
\[ ({\times}-{\vee}-L) \;\;\; ( \bigvee_{i \in I} \phi_{i} \times \psi ) = \bigvee_{i \in I} (\phi \times \psi ) \]
\[ ({\times}-{\vee}-R) \;\;\; (\phi \times \bigvee_{i \in I}\psi_{i}) = \bigvee_{i \in I} (\phi \times \psi_{i}) \]
\[ ({\rightarrow}-\leq ) \;\;\; \frac{\phi^{\prime} \leq \phi ,\; 
\psi \leq \psi^{\prime} }
{(\phi \rightarrow \psi )  \leq (\phi^{\prime} \rightarrow \psi^{\prime} )} \]
\[ ({\rightarrow}-\wedge ) \;\;\; 
(\phi \rightarrow \bigwedge_{i \in I} \psi_i ) = 
\bigwedge_{i \in I} (\phi \rightarrow \psi_i ) \]
\[ ({\rightarrow}-\vee-L) \;\;\; 
(\bigvee_{i \in I} \phi_{i} \rightarrow \psi ) =
\bigwedge_{i \in I} (\phi_{i} \rightarrow \psi ) \]
\[ ({\rightarrow}-\vee-R) \;\;\; 
(\phi \rightarrow \bigvee_{i \in I} \psi_{i}) =
\bigvee_{i \in I} (\phi \rightarrow \psi_{i})  \;\;\;\; ({\sf CPNF}(\phi )) \]
\[ ({\oplus}-{\leq}) \;\;\; \frac{\phi \leq \psi}{(\phi \oplus \false ) \leq (\psi \oplus \false ), \; (\false \oplus \phi ) \leq (\false \oplus \psi )} \]
\[ ({\oplus}-{\wedge}-L) \;\;\; (\bigwedge_{i \in I} \phi_{i} \oplus \false ) = 
\bigwedge_{i \in I} (\phi_{i} \oplus \false ) \]
\[ ({\oplus}-{\wedge}-R) \;\;\; (\false \oplus \bigwedge_{i \in I} \psi_{i}) =
\bigwedge_{i \in I} (\false \oplus \psi_{i}) \]
\[ ({\oplus}-{\vee}-R) \;\;\; (\bigvee_{i \in I} \phi_{i} \oplus \false ) =
\bigvee_{i \in I} (\phi_{i} \oplus \false ) \]
\[ ({\oplus}-{\vee}-L) \;\;\; (\false \oplus \bigvee_{i \in I} \psi_{i}) =
\bigvee_{i \in I} (\false \oplus \psi_{i})   \]
\[ ({( \cdot )_{\bot}}-{\leq}) \;\;\; \frac{\phi \leq \psi}{(\phi )_{\bot} \leq
(\psi )_{\bot}} \]
\[ ({( \cdot )_{\bot}}-{\wedge}) \;\;\; (\phi \wedge \psi )_{\bot} = (\phi )_{\bot} \wedge (\psi )_{\bot} \]
\[ ({( \cdot )_{\bot}}-{\vee}) \;\;\; (\bigvee_{i \in I} \phi_{i} )_{\bot} =
\bigvee_{i \in I} (\phi_{i})_{\bot} \]
\[ ({\Box}-{\leq}) \;\;\; \frac{\phi \leq \psi}{\Box \phi \leq \Box \psi} \]
\[ ({\Box}-{\wedge}) \;\;\; \Box \bigwedge_{i \in I} \phi_{i} =
\bigwedge_{i \in I} \Box \phi_{i} \]
\[ ({\Box}-{\sl f}) \;\;\; \Box {\sl f} = {\sl f} \]
\[ ({\Diamond}-{\leq}) \;\;\; \frac{\phi \leq \psi}{\Diamond \phi \leq \Diamond \psi} \]
\[ ({\Diamond}-{\vee}) \;\;\; \Diamond \bigvee_{i \in I} \phi_{i} =
\bigvee_{i \in I} \Diamond \phi_{i} \]
\[ ({\Diamond}-{\true}) \;\;\; \Diamond \true = \true \]
\[ (\#) \;\;\; \phi \leq \false \;\;\;\; (\# (\phi )) \]

The axiom $({\Box}-{\sl f})$ exemplifies the possibilities for
fine-tuning in our approach. It corresponds exactly to the
{\em omission} of the empty set from the upper powerdomain.

To make precise the sense in which this axiomatic presentation is
equivalent to the usual denotational construction of domains we define,
for each (closed) type expression $\sigma$, an interpretation function
\[ \lsem \cdot \rsem_{\sigma } : L(\sigma ) \longrightarrow K \Omega (\cal D 
(\sigma ) ) \]
by
\[ \begin{array}{lll}
\lsem \phi \wedge \psi \rsem_{\sigma} & = & \lsem \phi \rsem_{\sigma} \cap
\lsem \psi \rsem_{\sigma} \\
\lsem {\sl t} \rsem_{\sigma} & = & D(\sigma ) = 1_{K \Omega (\cal D (\sigma ))} 
\\
\lsem \phi \vee \psi \rsem_{\sigma} & = & \lsem \phi \rsem_{\sigma} \cup
\lsem \psi \rsem_{\sigma} \\
\lsem \false \rsem_{\sigma} & = & \varnothing = 0_{K \Omega (\cal D (\sigma ))}
\\
\lsem (\phi \times \psi ) \rsem_{\sigma \times \tau} & = & 
\{ \ltuple u, v \rtuple : u \in \lsem \phi \rsem_{\sigma}, \; v \in \lsem \psi
\rsem_{\tau} \} \\
\lsem (\phi \rightarrow \psi ) \rsem_{\sigma \rightarrow \tau} & = &
\{ f \in D(\sigma \rightarrow \tau ) : f(\lsem \phi \rsem_{\sigma })
\subseteq \lsem \psi \rsem_{\tau} \} \\
\lsem (\phi \oplus \false ) \rsem_{\sigma \oplus \tau} & = &
\{ \ltuple 0, u \rtuple : u \in \lsem \phi \rsem_{\sigma} - \{ \bot_{\sigma} \} \} \\
&  & \mbox{} \cup \{ \bot_{\sigma \oplus \tau} : \bot_{\sigma} \in \lsem \phi \rsem_{\sigma} \} \\
\lsem (\false \oplus \psi ) \rsem_{\sigma \oplus \tau} & = &
\{ \ltuple 1, v \rtuple : v \in \lsem \psi \rsem_{\tau} - \{ \bot_{\tau} \} \} \\
& & \mbox{} \cup \{ \bot_{\sigma \oplus \tau} : \bot_{\tau} \in \lsem \psi \rsem_{\tau} \} \\
\lsem (\phi )_{\bot} \rsem_{(\sigma )_{\bot}} & = & \{ \ltuple 0, u \rtuple :
u \in \lsem \phi \rsem_{\sigma} \} \\
\lsem \Box \phi \rsem_{P_{u} \sigma } & = & \{ S \in D(P_{u} \sigma ) :
S \subseteq \lsem \phi \rsem_{\sigma} \} \\
\lsem \Diamond \phi \rsem_{P_{l} \sigma } & = & \{ S \in D(P_{l} \sigma ) : 
S \cap \lsem \phi \rsem_{\sigma} \neq \varnothing \} \\
\lsem \phi \rsem_{{\sf rec} \: t. \, \sigma} & = & 
\{ {\foldalph}_{\sigma}(u) : u \in \lsem \phi \rsem_{\sigma [{\sf rec} \: t. \, {\sigma}/ t]} \}
\end{array} \] 
where ${\foldalph}_{\sigma} : \Dom (\sigma [{\sf rec} \: t. \, {\sigma}/t]) \cong \Dom ({\sf rec} \: t. \, \sigma )$\label{foldalph} is the isomorphism
arising from the initial solution to the domain equation $t = \sigma (t)$.

Then for $\phi , \psi \in L(\sigma )$, we define
\[ \cal D (\sigma ) \models \phi \leq \psi \; \equiv \; \lsem \phi 
\rsem_{\sigma}
\subseteq \lsem \psi \rsem_{\sigma}. \]

We now use the results of Chapter~3 to establish some fundamental properties
of our system of ``Domain Logic''.

Firstly, we note that operations on prelocales in the style of
Chapter~3 can be distilled from our definitions for product, lifting and
Hoare powerdomain.
The reader will find no difficulty in carrying out the same
programme for these constructions as that shown for function space,
Smyth powerdomain and coalesced sum in Chapter~3.
Now using~\ref{obs}, we see that, for each closed $\sigma$ and any $\rho \in {\sf LEnv}$:
\[ {\cal L} \lsem \sigma \rsem \rho = \Ell (\sigma ) . \]
The following results are then immediate consequences of our work
in Chapter~3.

\noindent {\bf Notation.} ${\sf PNF}(\sigma ) \equiv \{ \phi \in L(\sigma ) :
{\sf PNF}(\phi ) \}$, 
and similarly for ${\sf CPNF}(\sigma )$, ${\sf CDNF}(\sigma )$.
\begin{proposition}
\label{metap}
For all $\phi \in {\sf PNF}(\sigma )$:
\[ \begin{array}{rl}
(i) & \lsem \phi \rsem_{\sigma} \in {\sf pr}(K \Omega (\Dom (\sigma ))) \\
(ii) & {\sf C}(\phi ) \;\; \Longleftrightarrow \;\; \lsem \phi \rsem_{\sigma}
\not= \varnothing \\
(iii) & {\sf T}(\phi ) \;\; \Longleftrightarrow \;\; \bot_{\sigma} \not\in \lsem \phi \rsem .
\end{array} \]
\end{proposition}
\begin{lemma}[Normal Forms]
\label{dptnf}
For all $\phi \in L(\sigma )$, for some $\psi \in {\sf CDNF}(\sigma )$:
\[ \Ell (\sigma ) \vdash \phi = \psi . \]
\end{lemma}

Now we define a relation
\[ {\leftrightsquigarrow} \subseteq {\sf CPNF}(\sigma ) \times K( \Dom (\sigma )):\]
\[ \phi \leftrightsquigarrow u \equiv \lsem \phi \rsem_{\sigma} = \diverges u. \]
\begin{proposition}
\label{squig}
$\leftrightsquigarrow$ is a surjective total function.
\end{proposition}

Now we come to the main results of the section:
\begin{theorem}[Soundness and Completeness]
For all $\phi , \psi \in L(\sigma )$:
\[ {\cal L} (\sigma ) \vdash \phi \leq \psi \;\; \Longleftrightarrow
\;\; {\cal D} (\sigma ) \models \phi \leq \psi . \]
\end{theorem}
Now we define 
\[{\cal LA}(\sigma ) \; \equiv \; (L(\sigma )/{=}_{\sigma}, \:
\leq_{\sigma}/{=}_{\sigma}),\]
the {\em Lindenbaum algebra} of
${\cal L}(\sigma )$.
\begin{theorem}[Stone Duality]
\label{sdual}
${\cal LA}(\sigma )$ is the Stone dual of ${\cal D}(\sigma )$, i.e.
\[\begin{array}{rl}
(i)  &  {\cal D}(\sigma ) \; \cong \; {\sf Spec} \: {\cal LA}(\sigma ) \\
(ii) &  K\Omega ( {\cal D}(\sigma )) \; \cong \; {\cal LA}(\sigma ).
\end{array} \]
\end{theorem}

\section{Programs as Elements: Endogenous Logic}
We extend our meta-language for denotational semantics to include typed
terms.

\subsection*{Syntax}
For each type $\sigma$, we have a set of variables 
\[ {\sf Var}(\sigma ) = \{ x^{\sigma}, y^{\sigma}, z^{\sigma}, \ldots \} . \]
We give the term formation rules {\it via} an inference system for assertions
of the form $M : \sigma$, i.e. ``$M$ is a term of type $\sigma$''.
\[ ({\sf Var}) \;\;\; x^{\sigma} : \sigma \]
\[ ({\bf 1}-I) \;\;\; \star : {\bf 1} \]
\[ ({\times}-I) \;\; \frac{M : \sigma , \;\; N : \tau}{(M, N) : \sigma \times \tau}
\;\;\;\;\;\; 
   ({\times}-E) \;\; \frac{M : \sigma \times \tau , \;\; N : \upsilon}{\mylet{M}{x^{\sigma}}{y^{\tau}}{N} 
 : \upsilon} \]
\[ ({\rightarrow}-I) \;\; \frac{M : \tau}{\lambda x^{\sigma}. M:\sigma \rightarrow \tau} \;\;\;\;\;\;
   ({\rightarrow}-E) \;\; 
\frac{M:\sigma \rightarrow \tau  , \;\; N:\sigma}{MN:\tau} \]
\[ ({\oplus}-I-L) \;\; \frac{M : \sigma}{\imath_{\sigma \tau}(M) : \sigma \oplus \tau}
\;\;\;\;\;\;
   ({\oplus}-I-R) \;\; \frac{N : \tau}{\jmath_{\sigma \tau}(M) : \sigma \oplus \tau} \]
\[ ({\oplus}-E) \;\; \frac{M : \sigma \oplus \tau , \;\; N_{1}, N_{2} : 
\upsilon}{\mycases{M}{x^{\sigma}}{N_{1}}{y^{\tau}}{N_{2}} : \upsilon} \]
\[ ((\cdot )_{\bot}-I) \;\; \frac{M : \sigma}{{\sf up}(M) : (\sigma )_{\bot}}
\;\;\;\;\;\;
   ((\cdot )_{\bot}-E) \;\; \frac{M : (\sigma )_{\bot}, \;\; N : \tau}{\lift{M}{x^{\sigma}}{N} : \tau } \]
\[ ({\Diamond}-I) \;\; \frac{M : \sigma}{\lsing M \rsing_{l} : P_{l} \sigma}
\;\;\;\;\;\;
   ({\Box}-I) \;\; \frac{M:\sigma}{\lsing M \rsing_{u} : P_{u} \sigma} \] 
\[ ({\Diamond}-E) \;\; \frac{M : P_{l} \sigma , \;\; N : P_{l} \tau}{\lextend{M}{x^{\sigma}}{N} : P_{l} \tau} \]
\[ ({\Box}-E) \;\; \frac{M: P_{u} \sigma , \;\; N : P_{u} \tau}{\uextend{M}{x^{\sigma}}{N} : P_{u} \tau} \]
\[ ({\Diamond}-{+}) \;\; \frac{M, N : P_{l} \sigma}{M \uplus_{l} N : P_{l} \sigma}
\;\;\;\;\;\;
   ({\Box}-{+}) \;\; \frac{M, N: P_{u} \sigma}{M \uplus_{u} N : P_{u} \sigma} \] 
\[ ({\Diamond}-{\otimes}) \;\; \frac{M : P_{l} \sigma , \;\; N : P_{l} \tau}{M \otimes_{l} N : P_{l} (\sigma \times \tau )}
\;\;\;\;\;\;
   ({\Box}-{\otimes}) \;\; \frac{M:P_{u}
\sigma , \;\; N:P_{u} \tau}{M \otimes_{u} N:P_{u} (\sigma \times \tau )} \]
\[ ({\sf rec}-I) \;\; \frac{M : \sigma [ {\sf rec} \: t. \, \sigma / t]}{{\sf fold}_{t, \sigma}(M) : {\sf rec} \: t . \, \sigma}
\;\;\;\;\;\;
   ({\sf rec}-E) \;\; \frac{M : {\sf rec} \: t . \, \sigma}{{\sf unfold}_{t, \sigma}(M) : \sigma [{\sf rec} \: t . \, \sigma / t]} \]
\[ ({\mu}-I) \;\; \frac{M : \sigma}{\mu x^{\sigma} . \, M : \sigma} \]
We write $\Lambda (\sigma )$ for the set of terms of type $\sigma$. 
Note the systematic presentation of these constructs as {\em introduction}
and {\em elimination} rules for each of the type constructions,
following ideas of Martin-L\"{o}f \cite{M-L83} and Plotkin \cite{Plo85}.
Note
that $\lambda$, {\sf let}, {\sf cases}, {\sf lift}, {\sf extend}, $\mu$  are
all {\em variable binding} operations in the obvious way.
Also, note
that $\lsing . \rsing$, {\sf extend} arise from the adjunction defining the
powerdomain construction; $\uplus$ is the operation of the free algebras
for this adjunction; while $\otimes$ is the universal map for the tensor
product with respect to this operation   \cite{HP79}.

We now introduce an endogenous program logic with assertions of the form
\[M,\Gamma \vdash \phi \] where $M:\sigma$, $\phi \in L(\sigma )$, and
$\Gamma \in \prod_{\sigma} \{{\sf Var}(\sigma ) \rightarrow L(\sigma )\}$
gives {\em assumptions} on the free variables of $M$.

\noindent {\bf Notation}
\[ \Gamma \leq \Delta \equiv \forall x \in {\sf Var}. \, {\cal L} \vdash \Gamma
x \leq \Delta x . \]
For the remainder of this Chapter, we shall omit type subscripts and 
superscripts ``whenever we think we can get away with it'',
in the delightful formulation of Barr and Wells \cite[p.\ 1]{BW84}.
\subsection*{Axiomatisation}
\[ ({\vdash}-{\wedge}) \;\;\; \frac{ \{ M, \Gamma \vdash \phi_{i} \}_{i \in I}}{M, \Gamma \vdash \bigwedge_{i \in I} \phi_{i}} \;\;\;\;\;\;
   ({\vdash}-{\vee}) \;\;\; \frac{ \{ M, \Gamma [x \mapsto \phi_{i}] \vdash \psi \}_{i \in I}}{M, \Gamma [x \mapsto \bigvee_{i \in I}\phi_{i}] \vdash \psi} \]
\[ ({\vdash}-{\leq}) \;\;\; \frac{\Gamma \leq \Delta \;\; M, \Delta \vdash \phi \;\; \phi \leq \psi}{M, \Gamma \vdash \psi} \;\;\;\;\;\;
   x, \Gamma [x \mapsto \phi ] \vdash \phi \] 
\[ \frac{M, \Gamma \vdash \phi \;\;\; N, \Gamma \vdash \psi}{(M, N), \Gamma \vdash (\phi \times \psi )} \;\;\;\;\;\;
   \frac{M, \Gamma \vdash (\phi \times \psi ) \;\;\; N, \Gamma [x \mapsto \phi , y \mapsto \psi ] \vdash \theta}{\mylet{M}{x}{y}{N} , \Gamma \vdash \theta} \]
\[\frac{M, \Gamma [x \mapsto \phi ] \vdash \psi}{\lambda x.M, \Gamma
\vdash (\phi \rightarrow \psi )} \;\;\;\;\;\; 
  \frac{M, \Gamma \vdash (\phi
\rightarrow \psi ) \;\;\; N, \Gamma \vdash \phi}{MN, \Gamma \vdash \psi} \]
\[ \frac{M, \Gamma \vdash \phi}{\imath (M), \Gamma \vdash (\phi \oplus \false )}
\;\;\;\;\;\;
   \frac{M : (\phi \oplus \false ) \;\;\; (\phi \converges ) \;\;
N_{1}, \Gamma [x \mapsto \phi ] \vdash \theta}{\mycases{M}{x}{N_{1}}{y}{N_{2}}, \Gamma \vdash \theta} \]
\[ \frac{N, \Gamma \vdash \psi}{\jmath (N), \Gamma \vdash (\false \oplus \psi )}
\;\;\;\;\;\;
   \frac{M : (\false \oplus \psi ) \;\;\; (\psi \converges ) \;\;
N_{2}, \Gamma [y \mapsto \psi ] \vdash \theta}{\mycases{M}{x}{N_{1}}{y}{N_{2}}, \Gamma \vdash \theta} \]
\[ \frac{M, \Gamma \vdash \phi}{{\sf up}(M), \Gamma \vdash (\phi )_{\bot}} \;\;\;\;\;\;
   \frac{M, \Gamma \vdash (\phi )_{\bot} \;\;\; N, \Gamma [x \mapsto \phi ] \vdash \psi}{\lift{M}{x}{N}, \Gamma \vdash \psi} \]
\[\frac{M, \Gamma \vdash \phi}{\lsing M \rsing_{l}, \Gamma \vdash \Diamond
\phi} \;\;\;\;\;\;
  \frac{M, \Gamma \vdash \phi}{\lsing M \rsing_{u}, \Gamma \vdash \Box \phi} \]
\[ \frac{M, \Gamma \vdash \Diamond \phi \;\;\; N, \Gamma [x \mapsto \phi ] \vdash \Diamond \psi}{\lextend{M}{x}{N}, \Gamma \vdash \Diamond \psi} 
\;\;\;\;\;\;
   \frac{M, \Gamma \vdash \Box \phi \;\;\; N, \Gamma [x \mapsto \phi ] \vdash
\Box \psi}{\uextend{M}{x}{N}, \Gamma \vdash \Box \psi} \]
\[ \frac{M, \Gamma \vdash \Diamond \phi}{M \uplus_{l} N, \Gamma \vdash \Diamond \phi} \;\;\;\;\;\;
   \frac{N, \Gamma \vdash \Diamond \psi}{M \uplus_{l} N, \Gamma \vdash \Diamond \psi} \;\;\;\;\;\;
   \frac{M, \Gamma \vdash \Box \phi \;\;\; N, \Gamma \vdash \Box \phi}{M
\uplus_{u} N, \Gamma \vdash \Box \phi} \]
\[ \frac{M, \Gamma \vdash \Diamond \phi \;\;\; N, \Gamma 
\vdash \Diamond \psi}{M \otimes_{l} N, \Gamma \vdash \Diamond (\phi \times \psi )} \;\;\;\;\;\;
   \frac{M, \Gamma \vdash \Box \phi \;\;\; N, \Gamma
\vdash \Box \psi}{M \otimes_{u} N, \Gamma \vdash
\Box (\phi \times \psi )} \]
\[ \frac{M, \Gamma \vdash \phi}{{\sf fold}(M), \Gamma \vdash \phi} \;\;\;\;\;\;
   \frac{M, \Gamma \vdash \phi}{{\sf unfold}(M), \Gamma \vdash \phi} \]
\[ \frac{\mu x . \, M, \Gamma \vdash \phi \;\;\; M, \Gamma [x \mapsto \phi ]
\vdash \psi}{\mu x . \, M, \Gamma \vdash \psi} \]
Note that there is one inference rule for $\vdash$ per formation rule in
our syntax.
Thus we can refer {\it e.g.} to rule $({\vdash}-{\times}-E)$ without ambiguity.
Note the role of the convergence predicate $( \cdot ) \converges$ in 
$({\vdash}-{\oplus}-E)$; it plays a similar role in the elimination rules
for the other ``strict'' constructions of smash product \cite[Chapter 3 p.\  1]{PloLN} and strict 
function space \cite[Chapter 1 p.\ 11]{PloLN},
which we do not cover here.
\subsection*{Semantics}
Following standard ideas \cite{PloLN,SP82,Plo76}, we now give a denotational semantics for this
meta-language, in the form of a map
\[ \lsem \cdot \rsem_{\sigma} : \Lambda (\sigma ) \longrightarrow    {\sf Env}
\longrightarrow {\cal D}(\sigma ) \]
where ${\sf Env} \equiv \prod_{\sigma}\{{\sf Var}(\sigma ) \rightarrow {\cal
D}(\sigma ) \}$ is the set of {\em environments}. 
\[ \begin{array}{lcl}
\lsem x \rsem \rho & = & \rho x \\
\lsem (M, N) \rsem \rho & = & \ltuple \lsem M \rsem \rho , 
\lsem N \rsem \rho \rtuple \\
\lsem \mylet{M}{x}{y}{N} \rsem \rho & = & \lsem N \rsem \rho [ x \mapsto d,
y \mapsto e ] \\
& & {\rm where} \\
& & \ltuple d, e \rtuple = \lsem M \rsem \rho \\
\lsem \imath (M) \rsem \rho & = & \left\{ \begin{array}{ll}
\ltuple 0, \lsem M \rsem \rho \rtuple , & \lsem M \rsem \rho \not= \bot \\
\bot & \lsem M \rsem \rho = \bot  
\end{array}
\right. \\
\lsem \jmath (N) \rsem \rho & = & \left\{ \begin{array}{ll}
\ltuple 1, \lsem N \rsem \rho \rtuple , & \lsem N \rsem \rho \not= \bot \\
\bot & \lsem N \rsem \rho = \bot 
\end{array}
\right. \\
\lsem {\sf cases} \; M \; {\sf of} & & \\
\imath (x). \, N_{1} \; {\sf else} \; \jmath (y). \, N_{2} \rsem \rho & = & \left\{ \begin{array}{ll}
\lsem N_{1} \rsem \rho [x \mapsto d], & \lsem M \rsem \rho = \ltuple 0, d \rtuple \\
\lsem N_{2} \rsem \rho [x \mapsto e], & \lsem M \rsem \rho = \ltuple 1, e \rtuple \\
\bot , & \lsem M \rsem \rho = \bot
\end{array}
\right. \\
\lsem {\sf up}(M) \rsem \rho & = & \ltuple 0, \lsem M \rsem \rho \rtuple \\
\lsem \lift{M}{x}{N} \rsem \rho & = & \left\{ \begin{array}{ll}
\lsem N \rsem \rho [x \mapsto d], & \lsem M \rsem \rho = \ltuple 0, d \rtuple \\
\bot , & \lsem M \rsem \rho = \bot
\end{array}
\right. \\
\lsem \lsing M \rsing_{l} \rsem \rho & = & \converges ( \lsem M \rsem \rho ) \\
\lsem \lextend{M}{x}{N} \rsem \rho & = & \bigcup \{ \lsem N \rsem \rho [x
\mapsto d] : d \in \lsem M \rsem \rho \} \\
\lsem M \uplus_{l} N \rsem \rho & = & ( \lsem M \rsem \rho ) \cup ( \lsem N \rsem \rho ) \\
\lsem M \otimes_{l} N \rsem \rho & = & ( \lsem M \rsem \rho ) \times ( \lsem N
\rsem \rho ) \\
\end{array} \]
\[ \begin{array}{lcl}
\lsem \lsing M \rsing_{u} \rsem \rho & = & \diverges ( \lsem M \rsem \rho ) \\
\lsem \uextend{M}{x}{N} \rsem \rho & = & \bigcup \{ \lsem N \rsem \rho [x
\mapsto d] : d \in \lsem M \rsem \rho \} \\
\lsem M \uplus_{u} N \rsem \rho & = & ( \lsem M \rsem \rho ) \cup ( \lsem N
 \rsem \rho ) \\
\lsem M \otimes_{u} N \rsem \rho & = & ( \lsem M \rsem \rho ) \times ( \lsem N
\rsem \rho ) \\
\lsem {\sf fold}(M) \rsem \rho & = & {\foldalph}(\lsem M \rsem \rho ) \\
\lsem {\sf unfold}(M) \rsem \rho & = & {\foldalph}^{-1}(\lsem M \rsem \rho ) \\
\lsem \mu x. \, M \rsem \rho & = & \bigsqcup_{k \in \omega} d_{k} \\
& & {\rm where} \\
& & d_{0} = \bot , \;\; d_{k+1} = \lsem M \rsem \rho [x \mapsto d_{k}]
\end{array} \] 
Here $\foldalph$ is the initial algebra isomorphism as in Section~2 page~\pageref{foldalph}.
We can use this 
semantics to define a notion of validity for assertions:
\[ M, \Gamma \models \phi \; \equiv \;  \forall \rho \in {\sf Env} . \, \rho \models
\Gamma \Rightarrow \lsem M \rsem_{\sigma} \rho \models \phi \]
where
\[ \rho \models \Gamma \; \equiv \; \forall x \in {\sf Var} . \, 
\rho x \models \Gamma x
\]
and for $d \in D(\sigma )$, $\phi \in L(\sigma )$:
\[ d \models \phi \; \equiv \; d \in \lsem \phi \rsem_{\sigma} . \]
We can now state the main result of this section:
\begin{theorem}
\label{endogth}
The Endogenous logic is sound and complete:
\[ \forall M, \Gamma , \phi . \: M, \Gamma \vdash \phi \;\; \Longleftrightarrow \;\; 
M, \Gamma \models \phi . \]
\end{theorem}

We can state this result more sharply in terms of Stone Duality: it says
that
\[ \eta_{\sigma}^{-1}(\{ [\phi ]_{=_{\sigma}} :  M, \Gamma \vdash \phi \}) = 
\lsem M \rsem_{\sigma} \rho , \]
where 
\[ \eta_{\sigma} : {\cal D}(\sigma ) \; \cong \; {\sf Spec} \: {\cal LA}(\sigma ) \]
is the component of the natural isomorphism arising from Theorem~\ref{sdual};
i.e. that we recover the point of ${\cal D}(\sigma )$ given by the 
denotational semantics
of $M$ from the properties we can prove to hold of $M$ in our logic.

We now turn to the proof of Theorem~\ref{endogth}.
Our strategy is analogous to that of Chapter~3; we get Completeness {\it via} Prime Completeness.
Firstly, we have:
\begin{theorem}[Soundness]
\label{endogsoun}
For all $M$, $\Gamma$, $\phi$:
\[ M, \Gamma \vdash \phi \;\; \Longrightarrow \;\; M, \Gamma \models \phi . \]
\end{theorem}

\proof\ By a routine induction on the length of proofs in the endogenous logic.
We give two cases for illustration.

\noindent 1. Suppose the last step in the proof is an application of $({\vdash}-{\rightarrow}-I)$:
\[ \frac{M, \Gamma [x \mapsto \phi ] \vdash \psi}{\lambda x . M, \Gamma \vdash (\phi \rightarrow \psi )} \]
By induction hypothesis, $M, \Gamma [x \mapsto \phi ] \models \psi$, i.e for all $\rho \models \Gamma$, $d \in {\cal D}(\sigma )$, 
\[  d \in \lsem \phi \rsem \;\; \Longrightarrow \;\; \lsem M \rsem \rho [x \mapsto d] \in \lsem \psi \rsem , \]
which implies
\[ \lambda x . M, \Gamma \models (\phi \rightarrow \psi ) . \]

\noindent 2. Next we consider $({\vdash}-{\Box}-E)$:
\[ \frac{M, \Gamma \vdash \Box \phi \;\;\; N, \Gamma [x \mapsto \phi ] \vdash \Box \psi}{\uextend{M}{x}{N}, \Gamma \vdash \Box \psi} \]
By induction hypothesis, $M, \Gamma \models \Box \phi$ and $N, \Gamma [x \mapsto \phi ] \models \Box \psi$.
Hence for $\rho \models \Gamma$, $\lsem M \rsem \rho \subseteq \lsem \phi \rsem$,
and for $d \in {\cal D}(\sigma )$,
\[ d \in \lsem \phi \rsem \;\; \Longrightarrow \;\; \lsem N \rsem \rho [x \mapsto d] \subseteq \lsem \psi \rsem . \]
Thus
\[ \begin{array}{ll}
& \bigcup_{d \in \lsem M \rsem \rho} \lsem N \rsem \rho [x \mapsto d] \subseteq \lsem \psi \rsem \\
\Longrightarrow & \lsem \uextend{M}{x}{N} \rsem \rho \subseteq \lsem \psi \rsem \\
\Longrightarrow & \uextend{M}{x}{N}, \Gamma \models \Box \psi . \;\;\; \qed
\end{array} \]

Next, we shall need a technical lemma which  describes our program constructs under the denotational semantics.
\begin{lemma}
\label{techlem}
For $u \in {\cal K}({\cal D}(\sigma ))$, $v \in {\cal K}({\cal D}(\tau ))$, $w \in {\cal K}({\cal D}(\upsilon ))$, 
$X \in \wp_{\sf fne}({\cal K}({\cal D}(\sigma )))$, 
$Y \in \wp_{\sf fne}({\cal K}({\cal D}(\tau )))$, 
$Z \in \wp_{\sf fne}({\cal K}({\cal D}(\sigma \times \tau )))$,
$w_{1} \in {\cal K}({\cal D}({\sf rec} \: t. \, \sigma ))$, $w_{2} \in {\cal K}({\cal D}(\sigma [{\sf rec} \: t. \, \sigma /t]))$:
\[ \begin{array}{rl}
(i) & (u, v) \sqsubseteq \lsem (M, N) \rsem \rho \; \Leftrightarrow \; u \sqsubseteq \lsem M \rsem \rho \: \& \: v \sqsubseteq \lsem N \rsem \rho \\
(ii) & w \sqsubseteq \lsem \mylet{M}{x}{y}{N} \rsem \rho \; \Leftrightarrow \; \exists u, v.  \\
& (u, v) \sqsubseteq \lsem M \rsem \rho \: \& \: w \sqsubseteq \lsem N \rsem \rho [x \mapsto u, y \mapsto v] \\
(iii) & [u, v] \sqsubseteq \lsem \lambda x . M \rsem \rho \; \Leftrightarrow \; v \sqsubseteq \lsem M \rsem \rho [x \mapsto u] \\
(iv) & v \sqsubseteq \lsem MN \rsem \rho \;  \Leftrightarrow \; \exists u. [u, v] \sqsubseteq \lsem M \rsem \rho \: \& \: u \sqsubseteq \lsem N \rsem  \rho \\
(v) & \ltuple 0, u \rtuple \sqsubseteq \lsem \imath (M) \rsem \rho \; \Leftrightarrow \; u \sqsubseteq \lsem M \rsem \rho \\
&  \ltuple 1, v \rtuple \sqsubseteq \lsem \jmath (N) \rsem \rho \; \Leftrightarrow \; v \sqsubseteq \lsem N \rsem \rho \\
(vi) & w \not= \bot \;\; \Longrightarrow \;\; w \sqsubseteq \lsem \mycases{M}{x}{N_{1}}{y}{N_{2}} \rsem \rho \; \Leftrightarrow \\
& \exists u \not= \bot . \, \ltuple 0, u \rtuple \sqsubseteq \lsem M \rsem \rho \: \& \: w \sqsubseteq \lsem N_{1} \rsem \rho [x \mapsto u] \\
& \mbox{or} \\
& \exists v \not= \bot . \, \ltuple 1, v \rtuple \sqsubseteq \lsem M \rsem \rho \: \& \: w \sqsubseteq \lsem N_{2} \rsem \rho [x \mapsto v] \\ 
(vii) & \ltuple 0, u \rtuple \sqsubseteq \lsem {\sf up}(M) \rsem \rho \; \Leftrightarrow \; u \sqsubseteq \lsem M \rsem \rho \\
(viii) & v \not= \bot \;\; \Longrightarrow \;\; v \sqsubseteq \lsem \lift{M}{x}{N} \rsem \rho \; \Leftrightarrow \\
& \exists u. \, \ltuple 0, u \rtuple \sqsubseteq \lsem M \rsem \rho 
\: \& \: v \sqsubseteq \lsem N \rsem \rho [x \mapsto u] \\
(ix) & \converges X \sqsubseteq \lsem \lsing M \rsing_{l} \rsem \rho \; \Leftrightarrow \; \forall x \in X. \, x \sqsubseteq \lsem M \rsem \rho \\
(x) & \converges Y \sqsubseteq \lsem \lextend{M}{x}{N} \rsem \rho \; \Leftrightarrow \; \exists X. \, \converges X \sqsubseteq \lsem M \rsem \rho \\
& \: \& \: \converges Y \sqsubseteq \bigcup_{u \in X} \lsem N \rsem \rho [x \mapsto u] \\
(xi) & \converges X \sqsubseteq \lsem M \uplus_{l} N \rsem \rho \; \Leftrightarrow \; \converges X \sqsubseteq \lsem M \rsem \rho \; \mbox{or} \; \converges X \sqsubseteq \lsem N \rsem \rho \\
(xii) & \converges Z \sqsubseteq \lsem M \otimes_{l} N \rsem \rho \; \Leftrightarrow \; \exists X, Y. \: \converges Z \sqsubseteq \converges X \otimes_{l} \converges Y \\
& \: \& \: \converges X \sqsubseteq \lsem M \rsem \rho \: \& \: \converges Y \sqsubseteq \lsem N \rsem \rho  \\
(xiii) & \diverges X \sqsubseteq \lsem \lsing M \rsing_{u} \rsem \rho \; \Leftrightarrow \; \exists x \in X. \, x \sqsubseteq \lsem M \rsem \rho 
\end{array} \]
\[ \begin{array}{rl}
(xiv) & \diverges Y \sqsubseteq \lsem \uextend{M}{x}{N} \rsem \rho \; \Leftrightarrow \; \exists X. \, \diverges X \sqsubseteq \lsem M \rsem \rho \\
& \: \& \: \diverges Y \sqsubseteq \bigcup_{u \in X} \lsem N \rsem \rho [x \mapsto u] \\ 
(xv) & \diverges X \sqsubseteq \lsem M \uplus_{u} N \rsem \rho \; \Leftrightarrow \; \diverges X \sqsubseteq \lsem M \rsem \rho \; \& \; \diverges X \sqsubseteq \lsem N \rsem \rho \\
(xvi) & \diverges Z \sqsubseteq \lsem M \otimes_{u} N \rsem \rho \; \Leftrightarrow \; \exists X, Y . \: \diverges Z \sqsubseteq \diverges X \otimes_{u} \diverges Y \\
& \: \& \: \diverges X \sqsubseteq \lsem M \rsem \rho \: \& \: \diverges Y \sqsubseteq \lsem N \rsem \rho \\
(xvii) & w_{1} \sqsubseteq \lsem {\sf fold}(M) \rsem \rho \; \Leftrightarrow \; {\foldalph}^{-1}(w_{1}) \sqsubseteq \lsem M \rsem \rho \\
(xviii) & w_{2} \sqsubseteq \lsem {\sf unfold}(M) \rsem \rho \; \Leftrightarrow \; {\foldalph}(w_{2}) \sqsubseteq \lsem M \rsem \rho \\
(xix) & u \sqsubseteq \lsem \mu x .  M \rsem \rho \; \Leftrightarrow \; \exists k \in \omega , \, u_{0}, \ldots , u_{k}. \, u_{0} = \bot \: \& \: u_{k} = u \\
& \: \& \: \forall i: 0 \leq i < k. \, u_{i+1} \sqsubseteq \lsem M \rsem \rho [x \mapsto u_{i}] 
\end{array} \]
\end{lemma}

\proof\ The content of this Lemma is all quite standard, at least in the folklore.
It amounts to a description of the combinators underlying the denotational semantics of terms as {\em approximable mappings}.
Most of it can be found, couched in the language of information systems, in \cite{Sco82}, and for neighbourhood systems in \cite{Sco81}.
We shall just give a couple of the less familiar cases for illustration.

\noindent (xii).
\[ \begin{array}{ll}
\bullet & \converges Z \sqsubseteq \lsem M \otimes_{l} N \rsem \rho \\
\Leftrightarrow & \converges Z \subseteq \bigsqcup \{ \converges X \otimes_{l} \converges Y : \converges X \sqsubseteq \lsem M \rsem \rho \: \& \: \converges Y \sqsubseteq \lsem N \rsem \rho \} \\
& \mbox{since $\otimes_{l}$ is continuous} \\
\Leftrightarrow & \exists X, Y. \: \converges Z \sqsubseteq \converges X 
\otimes_{l} \converges Y \: \& \: \converges X \sqsubseteq \lsem M \rsem \rho 
\; \& \; \converges Y \sqsubseteq \lsem N \rsem \rho \\
& \mbox{since $\converges Z$ is finite.}  \\
\end{array} \]

\noindent (xiv).
\[ \begin{array}{ll}
\bullet & \diverges Y \sqsubseteq \lsem \uextend{M}{x}{N} \rsem \rho \\
\Leftrightarrow & \diverges Y \sqsubseteq \bigsqcup_{\diverges X \sqsubseteq \lsem M \rsem \rho} \bigcup \{ \lsem N \rsem \rho [x \mapsto u] : u \in \diverges X \} \\
& \mbox{since {\sf extend} is continuous} \\
\Leftrightarrow & \exists X. \, \diverges X \sqsubseteq \lsem M \rsem \rho \; \& \; \diverges Y \sqsubseteq \bigcup_{u \in \diverges X} \lsem N \rsem \rho [x \mapsto u] 
\end{array} \]
since $\diverges Y$ is finite.
The argument is completed by observing that
\[ \bigcup_{u \in \diverges X} \lsem N \rsem \rho [x \mapsto u] = \bigcup_{u \in X} \lsem N \rsem \rho [x \mapsto u] . \;\;\; \qed \]

Now for Prime Completeness.

{\bf Notation.} ${\sf CPNF}(\Gamma ) \equiv \forall x \in {\sf Var}. \, {\sf CPNF}(\Gamma x)$.

\begin{theorem}[Prime Completeness]
\label{pricom}
${\sf CPNF}(\Gamma )$ and ${\sf CPNF}(\phi )$ imply that
\[ M, \Gamma \models \phi \;\; \Longrightarrow \;\; M, \Gamma \vdash \phi \]
\end{theorem}

\proof\ We begin by establishing some useful notation.
Given $\Gamma$ with ${\sf CPNF}(\Gamma )$, we define an environment $\rho_{\Gamma}$ by:
\[ \forall x \in {\sf Var}. \, \Gamma x \leftrightsquigarrow \rho_{\Gamma} x . \]
This is well-defined by Proposition~\ref{squig}.
Similarly, let $\phi \leftrightsquigarrow u$.
Now we have:
\begin{equation}
\label{stareq}
M, \Gamma \models \phi \;\; \Longleftrightarrow \;\; u \sqsubseteq \lsem M \rsem \rho_{\Gamma} . 
\end{equation}
The proof proceeds by induction on $M$.
As the various cases all share a common pattern, we shall only give a selection of the more interesting for illustration.

Abstraction. We argue by induction on $\phi$. 
The inductive case, which can only be a conjunction, since
$\phi$ is in {\sf CPNF}, is trivial.
We are left with the case for a generator $(\phi \rightarrow \psi )$, where $\phi$, $\psi$ are in {\sf CPNF}.
Let $\phi \leftrightsquigarrow u$, $\psi \leftrightsquigarrow v$.
Then
\[ \begin{array}{llr}
\bullet & \lambda x . M, \Gamma \models (\phi \rightarrow \psi ) & \\
\Rightarrow & [u, v] \sqsubseteq \lsem \lambda x . M \rsem \rho_{\Gamma} & \ref{stareq} \\
\Rightarrow & v \sqsubseteq \lsem M \rsem \rho_{\Gamma}[x \mapsto u] & \mbox{\ref{techlem}(iii)} \\
\Rightarrow & M, \Gamma [x \mapsto \phi ] \models \psi & \ref{stareq} \\
\Rightarrow & M, \Gamma [x \mapsto \phi ] \vdash \psi & \mbox{ind. hyp.} \\
\Rightarrow & \lambda x . M, \Gamma \vdash (\phi \rightarrow \psi ) & ({\vdash}-{\rightarrow}-I)
\end{array} \]

Application.
\[ \begin{array}{llr}
\bullet & MN, \Gamma \models \phi  &\\
\Rightarrow & u \sqsubseteq \lsem MN \rsem \rho_{\Gamma} & \ref{stareq} \\
\Rightarrow & \exists v. \, [v, u] \sqsubseteq \lsem M \rsem \rho \: \& \: v \sqsubseteq \lsem N \rsem \rho & \mbox{\ref{techlem}(iv)} \\
\Rightarrow & M, \Gamma \models (\psi \rightarrow \phi ) \; \& \; N, \Gamma \models \psi & \ref{stareq} \\
& \mbox{where $\psi \leftrightsquigarrow v$} & \\
\Rightarrow & M, \Gamma \vdash (\psi \rightarrow \phi ) \; \& \; N, \Gamma \vdash \psi & \mbox{ind. hyp.} \\
\Rightarrow & MN, \Gamma \vdash \phi & ({\vdash}-{\rightarrow}-E).
\end{array} \]

Case expression.
\[ \begin{array}{llr}
& \mycases{M}{x}{N_{1}}{y}{N_{2}}, \Gamma \models \phi & \\
\Leftrightarrow & u \sqsubseteq \lsem \mycases{M}{x}{N_{1}}{y}{N_{2}} \rsem \rho_{\Gamma} & \ref{stareq}.
\end{array} \]
If $u = \bot$, then ${\cal L} \vdash \true \leq \phi$, and the required conclusion follows by $({\vdash}-{\wedge})$ and $({\vdash}-{\leq})$.
Otherwise, by~\ref{techlem}(vi), either
\[ (i) \;\; \exists u_{1} \not= \bot . \, \ltuple 0, u_{1} \rtuple \sqsubseteq \lsem M \rsem \rho_{\Gamma} \: \& \: u \sqsubseteq \lsem N_{1} \rsem \rho_{\Gamma} [x \mapsto u_{1}] \]
or
\[ (ii) \;\; \exists u_{2} \not= \bot . \, \ltuple 1, u_{2} \rtuple \sqsubseteq \lsem M \rsem \rho_{\Gamma} \: \& \: u \sqsubseteq \lsem N_{2} \rsem \rho_{\Gamma} [x \mapsto u_{2}]  . \]
We shall consider sub-case (i); (ii) is entirely similar.
Let $\phi_{1} \leftrightsquigarrow u_{1}$. Then
\[ \begin{array}{llr}
\bullet & \ltuple 0, u_{1} \rtuple \sqsubseteq \lsem M \rsem \rho_{\Gamma} \: \& \: u \sqsubseteq \lsem N_{1} \rsem \rho_{\Gamma} [x \mapsto u_{1}]  & \\
\Rightarrow & M, \Gamma \models (\phi_{1} \oplus \false ) \; \& \; N_{1}, \Gamma [x \mapsto \phi_{1}] \models \phi & \ref{stareq} \\
\Rightarrow & M, \Gamma \vdash (\phi_{1} \oplus \false ) \; \& \; N_{1}, \Gamma [x \mapsto \phi_{1}] \vdash \phi & \mbox{ind. hyp.} \\
\Rightarrow & \mycases{M}{x}{N_{1}}{y}{N_{2}}, \Gamma \vdash \phi & 
\mbox{by $({\vdash}-{\oplus}-E)$} \\
& \mbox{since $u_{1} \not= \bot$ implies $\phi_{1} \converges$ by \ref{metap}.} &
\end{array} \]

Tensor product.
We write $\phi \in {\sf CPNF}(P_{u}(\sigma \times \tau ))$ as $\Box \bigvee_{i \in I}(\phi \times \psi )$, and define $Z = \diverges \{ (u_{i}, v_{i}) : i \in I \}$, where
\[ \phi_{i} \leftrightsquigarrow u_{i}, \;\;\; \psi_{i} \leftrightsquigarrow v_{i} \;\;\; (i \in I). \]
Now
\[ \begin{array}{llr}
\bullet & M \otimes_{u} N, \Gamma \models \Box \bigvee_{i \in I}(\phi \times \psi ) & \\
\Rightarrow & Z \sqsubseteq \lsem M \otimes_{u} N \rsem \rho_{\Gamma} & \ref{stareq} \\
\Rightarrow & \exists X, Y. \: \diverges X \sqsubseteq \lsem M \rsem \rho_{\Gamma} \; \& \; \diverges Y \sqsubseteq \lsem N \rsem \rho_{\Gamma} & \\
& \& \; \diverges Z \sqsubseteq \diverges X \otimes_{u} \diverges Y = \diverges (X \times Y) & \mbox{\ref{techlem}(xvi)} 
\end{array} \]
Let $X = \{ u_{k} \}_{k \in K}$, $Y = \{ v_{l} \}_{l \in L}$, and define
\[ \phi_{k} \leftrightsquigarrow u_{k} \;\; (k \in K), \;\;\; \psi_{l} \leftrightsquigarrow v_{l} \;\; (l \in L). \]
Now
\[ \begin{array}{llr}
\bullet & \diverges X \sqsubseteq \lsem M \rsem \rho_{\Gamma} \; \& \; \diverges Y \sqsubseteq \lsem N \rsem \rho_{\Gamma} &  \\
\Rightarrow & M, \Gamma \models \Box \bigvee_{k \in K} \phi_{k} \;\; \& \;\; N, \Gamma \models \Box \bigvee_{l \in L} \psi_{l} & \ref{stareq} \\
\Rightarrow & M, \Gamma \vdash \Box \bigvee_{k \in K} \phi_{k} \;\; \& \;\; N, \Gamma \vdash \Box \bigvee_{l \in L} \psi_{l} & \mbox{ind. hyp.} \\
\Rightarrow & M \otimes_{u} N, \Gamma \vdash \Box ( \bigvee_{k \in K} \phi_{k} \times \bigvee_{l \in L} \psi_{l}) & ({\vdash}-{\Box}-{\otimes}).
\end{array} \]
Finally,
\[ \begin{array}{rclr}
{\cal L} \vdash ( \bigvee_{k \in K} \phi_{k} \times \bigvee_{l \in L} \psi_{l}) & = & \bigvee_{(k,l) \in K \times L} (\phi_{k} \times \psi_{l}) & ({\times}-{\vee}) \\
& \leq & \bigvee_{i \in I} (\phi_{i} \times \psi_{i}) &
\end{array} \]
since $Z \sqsubseteq \diverges X \otimes_{u} \diverges Y$ implies
\[ \forall k, l. \, \exists i. \: {\cal L} \vdash (\phi_{k} \times \psi_{l}) \leq (\phi_{i} \times \psi_{i}). \]
Hence by $({\vdash}-{\leq})$,
\[ M \otimes_{u} N, \Gamma \vdash \Box  \bigvee_{i \in I} (\phi_{i} \times \psi_{i}). \]

Extension.
As in the case for abstraction, it suffices to consider the case when $\phi$ is a generator $\Box \bigvee_{i \in I}\phi_{i}$.
We define $Y = \{ u_{i} \}_{i \in I}$, where $\phi_{i} \leftrightsquigarrow u_{i}$, $(i \in I)$.
Now
\[ \begin{array}{llr}
\bullet & \uextend{M}{x}{N}, \Gamma \models \Box \bigvee_{i \in I} \phi_{i} & \\
\Rightarrow & \diverges Y \sqsubseteq \lsem \uextend{M}{x}{N} \rsem \rho_{\Gamma} & \ref{stareq} \\
\Rightarrow & \exists X. \: \diverges X \sqsubseteq \lsem M \rsem \rho_{\Gamma} \: \& \: \diverges Y \sqsubseteq \bigcup_{u \in X} \lsem N \rsem \rho_{\Gamma}[x \mapsto u] & \mbox{\ref{techlem}(xiv)} \\
\Rightarrow & \exists X. \: \diverges X \sqsubseteq \lsem M \rsem \rho_{\Gamma} \: \& \: \forall u \in X. \, \diverges Y \sqsubseteq \lsem N \rsem \rho_{\Gamma}[x \mapsto u] & 
\end{array} \]
Let $X = \{ v_{j} \}_{j \in J}$, $\psi_{j} \leftrightsquigarrow v_{j}$, $(j \in J)$. Then
\[ \begin{array}{llr}
\bullet &  \diverges X \sqsubseteq \lsem M \rsem \rho_{\Gamma} \: \& \: \forall u \in X. \, \diverges Y \sqsubseteq \lsem N \rsem \rho_{\Gamma}[x \mapsto u] & \\
\Rightarrow & M, \Gamma \models \Box \bigvee_{j \in J} \psi_{j} \; \& \; \forall j \in J. \: N, \Gamma [x \mapsto \psi_{j}] \models \phi & \ref{stareq} \\
\Rightarrow & M, \Gamma \vdash \Box \bigvee_{j \in J} \psi_{j} \; \& \; \forall j \in J. \: N, \Gamma [x \mapsto \psi_{j}] \vdash \phi & \mbox{ind. hyp.} \\
\Rightarrow & M, \Gamma \vdash \Box \bigvee_{j \in J} \psi_{j} \; \& \; N, \Gamma [x \mapsto \bigvee_{j \in J} \psi_{j}] \vdash \phi & ({\vdash}-{\vee}) \\
\Rightarrow & \uextend{M}{x}{N}, \Gamma \vdash  \phi & ({\vdash}-{\Box}-E)
\end{array} \]

Recursive types.
Firstly, we note that for $\phi \in {\cal L}({\sf rec} \: t. \, \sigma )$,
\[ \phi \leftrightsquigarrow u \;\; \Leftrightarrow \;\; \phi \leftrightsquigarrow \alpha^{-1}(u) , \]
since ${\cal L}({\sf rec} \: t. \, \sigma ) = {\cal L}(\sigma 
[{\sf rec} \: t. \,
\sigma /t])$.
Now,
\[ \begin{array}{llr}
\bullet & {\sf fold}(M), \Gamma \models \phi & \\
\Rightarrow & u \sqsubseteq \lsem {\sf fold}(M) \rsem \rho_{\Gamma} & \ref{stareq} \\
\Rightarrow & \alpha^{-1}(u) \sqsubseteq \lsem M \rsem \rho_{\Gamma} & \mbox{\ref{techlem}(xvii)} \\
\Rightarrow & M, \Gamma \models \phi & \ref{stareq} \\
\Rightarrow & M, \Gamma \vdash \phi & \mbox{ind. hyp.} \\
\Rightarrow & {\sf fold}(M), \Gamma \vdash \phi & ({\vdash}-{\sf rec}-I)
\end{array} \]

Recursion.
\[ \begin{array}{llr}
\bullet & \mu x. M, \Gamma \models \phi & \\
\Rightarrow & u \sqsubseteq \lsem \mu x. M \rsem \rho_{\Gamma} & \ref{stareq} \\
\Rightarrow & \exists k \in \omega , u_{0}, \ldots , u_{k}. \: u_{0} = \bot \: \& \: u_{k} = u & \\
& \: \& \: \forall i: 0 \leq i < k. \, u_{i+1} \sqsubseteq \lsem M \rsem \rho_{\Gamma} [x \mapsto u_{i}]  & \mbox{\ref{techlem}(xix).}
\end{array} \]
Let $\| u \|$ be the least such $k$ (as a function of $u$ for $u \sqsubseteq \lsem \mu x. M \rsem \rho_{\Gamma}$, keeping $\mu x. M$, $\Gamma$ fixed).
We complete the proof for this case by induction on $\| u \|$, with $\phi \leftrightsquigarrow u$.

Basis:
\[ \| u \| = 0 \Rightarrow u = \bot \Rightarrow \; \vdash \true \leq \phi \Rightarrow \mu x. M, \Gamma \vdash \phi , \]
by $({\vdash}-{\wedge})$ and $({\vdash}-{\leq})$.

Induction step: $\| u \| = k+1$. Then by definition of $\| u \|$, for some $v$:
\[ u \sqsubseteq \lsem M \rsem \rho_{\Gamma} [x \mapsto v] \; \& \; \| v \| = k. \]
Let $\psi \leftrightsquigarrow v$. Then
\[ \begin{array}{llr}
\bullet & u \sqsubseteq \lsem M \rsem \rho_{\Gamma} [x \mapsto v] \; \& \; \| v \| = k & \\
\Rightarrow & M, \Gamma [x \mapsto \psi ] \models \phi & \ref{stareq} \\
& \mbox{and} \; \mu x. M, \Gamma \vdash \psi & \mbox{inner ind. hyp.} \\
\Rightarrow & M, \Gamma [x \mapsto \psi ] \vdash \phi \; \& \; \mu x. M, \Gamma \vdash \psi & \mbox{outer ind. hyp.} \\
\Rightarrow & \mu x. M, \Gamma \vdash \phi & ({\vdash}-{\mu}-I). \;\;\; \qed
\end{array} \]

Finally, we can prove Theorem~\ref{endogth}.
One half is Theorem~\ref{endogsoun}.
For the converse, suppose $M, \Gamma \models \phi$.
We can assume that $\Gamma x \not= \false$\footnote{ 
meaning $\lsem \Gamma x \rsem \not= \varnothing$, or, equivalently by Theorem~\ref{sdual}, ${\cal L} \nvdash \Gamma x = \false$}  
for all $x \in {\sf Var}$, since otherwise we could apply $({\vdash}-{\vee})$ to obtain $M, \Gamma \vdash \phi$.
Let $V = {\sf FV}(M)$, the {\em free variables} of $M$.
(We omit the formal definition, which should be obvious).
We define $\Gamma_{V}$ by
\[ \Gamma_{V} x = \left\{ \begin{array}{ll}
\Gamma x, & x \in V \\
\true & \mbox{otherwise.}
\end{array}
\right. \]
Then by standard arguments we have:
\begin{eqnarray}
M, \Gamma \models \phi & \Leftrightarrow & M, \Gamma_{V} \models \phi \label{semfv} \\
M, \Gamma \vdash \phi & \Leftrightarrow & M, \Gamma_{V} \vdash \phi  \label{synfv} 
\end{eqnarray}
Now by Lemma~\ref{dptnf}, we have
\[ {\cal L} \vdash \phi = \bigvee_{i \in I} \phi_{i}, \]
and for all $x \in V$,
\[ {\cal L} \vdash \Gamma x = \bigvee_{j \in J_{x}} \psi_{j}, \]
with each $\phi_{i}$, $\psi_{j}$ in {\sf CPNF}.
Moreover, our assumption that $\Gamma x \not= \false$ for all $x$ implies that
$J_{x} \not= \varnothing$ for all $x \in V$.
Given $f \in \prod_{x \in V} J_{x}$ (i.e. a {\em choice function} selecting one of the disjuncts $\psi_{f x}$, $fx \in J_{x}$, for each $x \in V$), we define $\Gamma_{f}$ by:
\[ \Gamma_{f} \: x = \left\{ \begin{array}{ll}
\psi_{f x}, & x \in V \\
\true & \mbox{otherwise.}
\end{array}
\right. \]
Then
\[ \begin{array}{llr}
\bullet & M, \Gamma \models \phi & \\
\Rightarrow & M, \Gamma_{V} \models \phi & \ref{semfv} \\
\Rightarrow & \forall f \in \prod_{x \in V} J_{x}. \: M, \Gamma_{f} \models \bigvee_{i \in I} \phi_{i} & \mbox{$({\vdash}-{\leq})$, Soundness} \\
\Rightarrow & \forall f \in \prod_{x \in V} J_{x}. \, \exists i \in I. \: M, \Gamma_{f} \models \phi_{i} & \\
\Rightarrow & \forall f \in \prod_{x \in V} J_{x}. \, \exists i \in I. \: M, \Gamma_{f} \vdash \phi_{i} &  \mbox{Prime Completeness} \\
\Rightarrow & \forall f \in \prod_{x \in V} J_{x}.  \: M, \Gamma_{f} \vdash \phi & ({\vdash}-{\leq}) \\
\Rightarrow & M, \Gamma_{V} \vdash \phi & ({\vdash}-{\vee}) \\
\Rightarrow & M, \Gamma \vdash \phi & \ref{synfv} \;\;\; \qed 
\end{array} \]

\section{Programs as Morphisms: Exogenous Logic}
We now introduce a second extension of our denotational meta-language,
which provides a syntax of terms denoting {\em morphisms between},
rather than elements of, domains.
This is an extended version of the algebraic meta-language for cartesian closed
categories \cite{Poi86,LS86}, just as the language of the previous section was an extended typed
$\lambda$-calculus.
Terms are sorted on {\em morphism types} $(\sigma , \tau )$, with notation
$f : (\sigma , \tau )$.
We shall give the formation rules in ``polymorphic'' style, with type
subscripts omitted.
\subsection*{Syntax of morphism terms}
\[ \indic {\sf id} : (\sigma , \sigma ) \;\;\;\;\;\; 
\indic \frac{f : (\sigma , \tau ) \;\;\; g : (\tau , \upsilon )}{f ; g : (\sigma , \upsilon )} \]
\[ \indic 1 : (\sigma , {\bf 1}) \]
\[ \indic \frac{f : (\upsilon , \sigma ) \;\;\; g : (\upsilon , \tau )}{\ltuple f, g \rtuple : (\upsilon , \sigma \times \tau )} \;\;\;\;\;\;
   \indic {\sf p} : (\sigma \times \tau , \sigma ) \;\;\;\;\;\;
\indic {\sf q} : (\sigma \times \tau , \tau ) \]
\[ \indic \frac{f : (\sigma \times \tau , \upsilon )}{\curry{f} : (\sigma , \tau \rightarrow \upsilon )} \;\;\;\;\;\;
\indic {\sf Ap} : ((\sigma \rightarrow \tau ) \times \sigma , \tau ) \]
\[ \indic {\sf l} : (\sigma , \sigma \oplus \tau ) \;\;\;\;\;\;
   \indic {\sf r} : (\tau , \sigma \oplus \tau ) \;\;\;\;\;\;
   \indic \frac{f : (\sigma , \upsilon ) \;\;\; g : \tau , \upsilon )}{\stup{f}{g} : (\sigma \oplus \tau , \upsilon )} \]
\[ \indic {\sf up} : (\sigma , (\sigma )_{\bot}) \;\;\;\;\;\;
 \indic \frac{f : (\sigma , \tau )}{{\sf lift}(f) : ((\sigma)_{\bot}, \tau )} \;\;\;\;\;\;
\indic \frac{f : (\sigma , \tau )}{{\sf strict}(f) : (\sigma , \tau )} \]
\[ \indic \lsing \cdot \rsing_{l} : (\sigma , P_{l} \sigma ) \;\;\;\;\;\;
    \indic \lsing \cdot \rsing_{u} : (\sigma , P_{u} \sigma ) \]
\[ \indic \frac{f : (\sigma , P_{l} \tau )}{f^{\dagger}_{l} : (P_{l} \sigma , P_{l} \tau )} \;\;\;\;\;\;
   \indic \frac{f : (\sigma , P_{u} \tau )}{f^{\dagger}_{u} : (P_{u} \sigma , P_{u} \tau )} \]
\[ \indic +_{l} : (P_{l} \sigma \times P_{l} \sigma , P_{l} \sigma ) \;\;\;\;\;\;
   \indic +_{u} : (P_{u} \sigma \times P_{u} \sigma , P_{u} \sigma ) \]
\[ \indic \otimes_{l} : (P_{l} \sigma \times P_{l} \tau , P_{l} (\sigma \times \tau )) \;\;\;\;\;\;
   \indic \otimes_{u} : (P_{u} \sigma \times P_{u} \tau , P_{u} (\sigma \times \tau )) \]
\[ \indic {\sf fold} : (\sigma [{\sf rec} \: t. \, \sigma /t], {\sf rec} \: t. \, \sigma ) \;\;\;\;\;\;
\indic {\sf unfold} : ( {\sf rec} \: t. \, \sigma , \sigma [{\sf rec} \: t. \, \sigma /t]) \]
\[ \indic {\sf Y} : (\sigma \rightarrow \sigma , \sigma ) \]
We now form an exogenous logic $\DDL$ (for {\em dynamic domain logic}, because of the evident analogy with dynamic logic \cite{Pra79,Har79}).
$\DDL$ is an extension of ${\cal L}$, the basic domain logic described in Section~2.
\subsection*{Formation Rules}
We define the set of formulas ${\rm DDL}(\sigma )$ for each type $\sigma$.
\[ \indic L(\sigma ) \subseteq {\rm DDL}(\sigma ) \;\;\;\;\;\;
\indic \frac{f : (\sigma , \tau ) \;\;\; \psi \in {\rm DDL}(\tau )}{[f]\psi \in {\rm DDL}(\sigma )} \]
\[ \indic \true , \false \in {\rm DDL}(\sigma ) \;\;\;\;\;\;
\indic \frac{\phi , \psi \in {\rm DDL}(\sigma )}{\phi \wedge \psi , \phi \vee \psi \in {\rm DDL}(\sigma )} \]
\subsection*{Axiomatization}
The following axioms and rules are added to those of $\cal L$.
\[ \indic \frac{\phi \leq \psi}{[f]\phi \leq [f]\psi} \;\;\;\;\;\;
   \indic [f]\bigwedge_{i \in I} \phi_{i} = \bigwedge_{i \in I} [f] \phi_{i} \;\;\;\;\;\;
   \indic [f]\bigvee_{i \in I} \phi_{i} = \bigvee_{i \in I} [f] \phi_{i} \]
\[ \indic [{\sf id}]\phi = \phi \;\;\;\;\;\; \indic [f ; g] \phi = [f][g]\phi \]
\[ \indic [\ltuple f, g \rtuple ] (\phi \times \psi ) = [f]\phi \wedge [g]\psi \]
\[ \indic [{\sf p}]\phi = (\phi \times \true ) \;\;\;\;\;\;
\indic [{\sf q}]\psi = (\true \times \psi ) \]
\[ \indic \frac{(\phi \times \psi ) \leq [f]\theta}{\phi \leq [\curry{f}](\psi \rightarrow \theta )} \;\;\;\;\;\;
\indic (\phi \rightarrow \psi ) \times \phi \leq [{\sf Ap}]\psi \]
\[ \indic [{\sf l}](\phi \oplus \false ) = \phi \;\;\;\;\;\;
\indic [{\sf l}](\false \oplus \psi ) = \false \;\; (\psi \converges ) \]
\[ \indic [{\sf r}](\phi \oplus \false ) = \false \;\; (\phi \converges ) \;\;\;\;\;\;
\indic [{\sf r}](\false \oplus \psi ) = \psi \]
\[ \indic [\stup{f}{g}]\phi = ([{\sf strict}(f)]\phi \oplus \false ) \vee (\false \oplus [{\sf strict}(g)]\phi ) \]
\[ \indic \frac{\phi \leq [f]\psi}{\phi \leq [{\sf strict}(f)]\psi} \;\; (\phi \converges ) \]
\[ \indic [{\sf up}](\phi )_{\bot} = \phi \;\;\;\;\;\;
\indic [{\sf lift}(f)]\phi = ([f]\phi )_{\bot} \;\; (\phi \converges ) \]
\[ \indic [\lsing \cdot \rsing_{l}] \Diamond \phi = \phi \;\;\;\;\;\;
\indic [\lsing \cdot \rsing_{u}] \Box \phi = \phi \]
\[ \indic \frac{\phi \leq [f]\Diamond \psi}{\Diamond \phi \leq [f^{\dagger}_{l}]\Diamond \psi} \;\;\;\;\;\;
\indic \frac{\phi \leq [f]\Box \psi}{\Box \phi \leq [f^{\dagger}_{u}]\Box \psi} \]
\[ \indic [+_{l}]\Diamond \phi = (\Diamond \phi \times \true ) \vee (\true \times \Diamond \phi ) \;\;\;\;\;\;
\indic [+_{u}]\Box \phi = (\Box \phi \times \Box \phi )  \]
\[ \indic [\otimes_{l}]\Diamond (\phi \times \psi ) = (\Diamond \phi \times \Diamond \psi ) \;\;\;\;\;\;
\indic [\otimes_{u}]\Box (\phi \times \psi ) = (\Box \phi \times \Box \psi ) \]
\[ \indic [{\sf fold}]\phi = \phi \;\;\;\;\;\; 
\indic [{\sf unfold}]\phi = \phi \;\;\;\;\;\; 
\indic \frac{\phi \leq [{\sf Y}]\psi}{\phi \wedge (\psi \rightarrow \theta ) \leq [{\sf Y}]\theta} \]
At this point, we could proceed to give a direct treatment of the semantics and meta-theory of $\DDL$, just as we did for the endogenous logic in Section~3.
This would ignore the salient fact that  our morphism term language and the
typed $\lambda$-calculus presented in Section~3 are essentially 
{\em equivalent}.
Instead, we shall give a translation of morphism terms into $\lambda$-terms.
The idea is that a morphism term $f : (\sigma , \tau )$ is translated into
a $\lambda$-term $\trans{f} : \sigma \rightarrow \tau$.
\subsection*{Translation}
\[ \begin{array}{rcl}
\trans{{\sf id}} & = & \lambda x.x  \\
\trans{f;g} & = & \lambda x. \trans{g}(\trans{f} x) \\
\trans{1} & = & \lambda x. \star  \\
\trans{\ltuple f, g \rtuple} & = & \lambda x. (\trans{f} x, \trans{g} x) \\
\trans{{\sf p}} & = & \lambda z. \mylet{z}{x}{y}{x} \\
\trans{{\sf q}} & = & \lambda z. \mylet{z}{x}{y}{y}  \\
\trans{\curry{f}} & = & \lambda x. \lambda y. \trans{f} (x, y)  \\
\trans{{\sf Ap}} & = & \lambda f. \lambda x. f x \\
\trans{{\sf l}} & = & \lambda x. \imath (x) \\
\trans{{\sf r}} & = & \lambda y. \jmath (y) \\
\trans{\stup{f}{g}} & = & \lambda z. \mycases{z}{x}{\trans{f} x}{y}{\trans{g} y} \\
\trans{{\sf strict}(f)} & = & \lambda z. \mycases{\imath (\trans{f} x)}{x}{\trans{f} x}{y}{y} \\
\trans{{\sf up}} & = & \lambda x. {\sf up}(x) \\
\trans{{\sf lift}(f)} & = & \lambda y. \lift{y}{x}{\trans{f} x} \\
\end{array} \]
\[ \begin{array}{rcl}
\trans{\lsing \cdot \rsing_{l}} & = & \lambda x. \lsing x \rsing_{l} \\
\trans{\lsing \cdot \rsing_{u}} & = & \lambda x. \lsing x \rsing_{u} \\
\trans{f^{\dagger}_{l}} & = & \lambda z. \lextend{z}{x}{\trans{f} x} \\
\trans{f^{\dagger}_{u}} & = & \lambda z. \uextend{z}{x}{\trans{f} x} \\
\trans{+_{l}} & = & \lambda z. \mylet{z}{x}{y}{x \uplus_{l} y} \\
\trans{+_{u}} & = & \lambda z. \mylet{z}{x}{y}{x \uplus_{u} y} \\
\trans{\otimes_{l}} & = & \lambda z. \mylet{z}{x}{y}{x \otimes_{l} y} \\
\trans{\otimes_{u}} & = & \lambda z. \mylet{z}{x}{y}{x \otimes_{u} y} \\
\trans{{\sf fold}} & = & \lambda x. {\sf fold}(x) \\
\trans{{\sf unfold}} & = & \lambda x. {\sf unfold}(x) \\
\trans{{\sf Y}} & = & \lambda f. \mu x. f x 
\end{array} \]
\subsection*{Semantics}
Let ${\cal M}(\sigma , \tau )$ be the set of morphism terms of sort $(\sigma , \tau )$.
Since 
\[ {\bf SDom}({\cal D}(\sigma ), {\cal D}(\tau )) \cong {\cal D}(\sigma \rightarrow \tau ) \]
by cartesian closure, we can get a semantics
\[ \lsem \cdot \rsem_{\sigma \tau} : {\cal M}(\sigma , \tau ) \longrightarrow {\bf SDom}({\cal D}(\sigma ), {\cal D}(\tau )) \]
for morphism terms from the above translation.
We use this to extend our semantics for $\cal L$ from Section~2 to $\DDL$:
\[ \lsem [f]\phi \rsem = (\lsem f \rsem )^{-1} (\lsem \phi \rsem ) \]
(the other clauses being handled in the obvious way).
Note that the denotations of formulas in $\DDL$ are still {\em open} sets (continuity!),
but need no longer be compact-open, since compactness is not preserved under inverse image in general.

This semantics yields a notion of validity for $\DDL$ assertions:
\[ \models \phi \leq \psi \;\; \equiv \;\; \lsem \phi \rsem \subseteq \lsem \psi \rsem . \]
\begin{theorem} $\DDL$ is sound:
\[ \DDL \vdash \phi \leq \psi \;\; \Longrightarrow \;\; \models \phi \leq \psi \]
\end{theorem}

\proof\ The usual routine induction on the length of proofs.
We give a few cases for illustration.

Left injection.
\begin{eqnarray*}
(i) \;\; \lsem [{\sf l}]( \phi  \oplus \false ) \rsem & = & ( \lsem {\sf l} \rsem )^{-1}(\lsem ( \phi \oplus \false ) \rsem ) \\
& = & \{ d : \ltuple 0, d \rtuple \in \lsem (\phi \oplus \false ) \rsem \} \cup \{ \bot : \bot \in \lsem ( \phi \oplus \false ) \rsem \} \\
& = & \lsem \phi \rsem .
\end{eqnarray*}
\[ (ii) \;\; \psi \converges \Rightarrow \bot \not\in \lsem \psi \rsem \Rightarrow
(\lsem {\sf l} \rsem )^{-1} (\lsem ( \false \oplus \psi ) \rsem ) 
= \varnothing . \] 

Strictification. Note that
\[ \lsem {\sf strict}(f) \rsem d = \left\{ \begin{array}{ll}
\bot , & d = \bot \\
f d  & \mbox{otherwise}
\end{array}
\right. \]
Now,
\[ \phi \converges \Rightarrow \bot \not\in \lsem \phi \rsem \Rightarrow
\forall d \in \lsem \phi \rsem. \: \lsem {\sf strict}(f) \rsem d = fd, \]
which implies
\[ \lsem \phi \rsem \subseteq \lsem [f]\psi \rsem \; \Leftrightarrow \; \lsem
\phi \rsem \subseteq \lsem [{\sf strict}(f)]\psi \rsem . \]

Union.
\begin{eqnarray*}
(i) \;\; \lsem [+_{l}] \Diamond \phi \rsem & = &
\{ (X, Y) : (X \cup Y) \cap \lsem \phi \rsem \not= \varnothing \} \\
& = & \{ (X, Y) : X \cap \lsem \phi \rsem \not= \varnothing \; \mbox{or} \;
Y \cap \lsem \phi \rsem \not= \varnothing \} \\
& = & \{ (X, Z) : X \cap \lsem \phi \rsem \not= \varnothing \} \\
&   & \mbox{} \cup \{ (Z, Y) : Y \cap \lsem \phi \rsem \not= \varnothing \} \\
& = & \lsem (\Diamond \phi \times \true ) \vee (\true \times \Diamond \phi ) \rsem
\end{eqnarray*}
\begin{eqnarray*}
(ii) \;\; \lsem [+_{u}]\Diamond \phi \rsem & = &
\{ (X, Y) : X \cup Y \subseteq \lsem \phi \rsem \} \\
& = & \{ (X, Y) : X \subseteq \lsem \phi \rsem \: \& \: Y \subseteq \lsem \phi \rsem \} \\
& = & \lsem (\Box \phi \times \Box \phi ) \rsem .
\end{eqnarray*}

Recursion.
\[ \begin{array}{ll}
\bullet & \lsem \phi \rsem \subseteq \lsem [{\sf Y}]\psi \rsem \\
\Rightarrow & \forall f \in \lsem \phi \rsem . \: {\sf Y}f \in \lsem \psi \rsem \\
\Rightarrow & \forall f \in \lsem \phi \rsem \cap \lsem (\psi \rightarrow \theta ) \rsem . \: {\sf Y}f = f ({\sf Y} f) \in \lsem \theta \rsem . \;\;\; \qed
\end{array}  \]

Next, we turn to what can be proved in the way of completeness.
A {\em Hoare triple} in $\DDL$ is a formula $\phi \leq [f]\psi$
such that $\phi$ and $\psi$ are formulas of $\cal L$, i.e. do not contain any program modalities.
\begin{theorem}[Completeness For Hoare Triples]
\label{Htrip}
Let $\phi \leq [f]\psi$ be a Hoare triple. Then
\[ \DDL \vdash \phi \leq [f]\psi \;\; \Longleftrightarrow \;\; \models \phi \leq [f]\psi. \]
\end{theorem}
This result can either be proved directly, in similar fashion to Theorem~\ref{endogth}; or it can be reduced to that result, since
\[ \models \phi \leq [f]\psi \;\; \Longleftrightarrow \;\;
\trans{f}, \Gamma_{\true} \models (\phi \rightarrow \psi ) \;\; 
\Longleftrightarrow \;\; \trans{f}, \Gamma_{\true} \vdash (\phi \rightarrow \psi ) \]
(where $\Gamma_{\true}$ is the constant map $x \mapsto \true$).
It thus suffices to prove:
\[ \trans{f}, \Gamma_{\true} \vdash (\phi \rightarrow \psi ) \;\; \Longrightarrow \;\; \DDL \vdash \phi \leq [f]\psi . \]
In either approach, the argument is a straightforward variation on our work in
section~3, which we omit since it adds nothing new.

Finally, we come to a limitative result, which differentiates $\DDL$ from the
endogenous logic of Section~3, and shows that the restricted form of \ref{Htrip} is necessary.
The result is of course not ``surprising'', since 
$\DDL$ is semantically more expressive than the endogenous logic,  allowing
the description of non-compact open sets.
\begin{theorem}
The validity problem for $\DDL$ is $\Pi^{0}_{2}$-complete.
\end{theorem}

\proof\ 
We will need some notions on effectively given domains; see \cite[Chapter 7]{PloLN}.
Firstly, each type expression in our meta-language has an effectively given 
domain as its denotation (since effectively given domains are closed under recursive definitions and all our type constructions \cite[Chapter 7 pp.\ 16, 21, Chapter 8 pp.\ 16, 54]{PloLN}).
Similarly, each term $f : (\sigma , \tau )$ denotes a computable morphism from
${\cal D}(\sigma )$ to ${\cal D}(\tau )$.
Moreover, each $\phi \in {\cal L}(\sigma )$ denotes a compact-open, and hence
computable open set in ${\cal D}(\sigma )$; and computable open sets are closed
under inverse images of computable maps \cite[Chapter 7 p.\ 9]{PloLN},
and under finite unions and intersections \cite[Chapter 7 p.\ 7]{PloLN}.
Thus each formula of $\DDL$ denotes a computable open set, and the problem
of deciding the validity of the assertion $\phi \leq \psi$ can be reduced
to that of deciding the inclusion of r.e. sets $\lsem \phi \rsem \subseteq \lsem \psi \rsem$, which as is well-known \cite[IV.1.6]{Soa87} is  $\Pi^{0}_{2}$.

To complete the argument, we take a standard $\Pi_{2}^{0}$-complete problem,
and reduce it to validity in $\DDL$.
The problem we choose is
\[ {\sf Tot} = \{ x : W_{x} = \Nat \} \]
i.e. the set of codes of {\em total} recursive functions \cite[IV.3.2]{Soa87}.
To perform the reduction, we proceed as follows:
\begin{itemize}
\item The type ${\Nat}_{\bot} \equiv {\sf rec} \: t. \, ({\bf 1})_{\bot} \oplus t$ is
used to model the flat domain of natural numbers.
\item We can show that every partial recursive function $\varphi : \Nat \rightarrow \Nat$, thought of as a strict continuous function of type 
${\Nat}_{\bot} \rightarrow {\Nat}_{\bot}$, can be defined by a morphism term.
This is quite standard: the numerals are constructed from the injections,
lifting, 
and {\sf fold} and {\sf unfold}; the conditional and basic predicates from source tupling; and primitive recursion from general recursion ({\sf Y}) and conditional.
We omit the details.
\item In particular, we can define a morphism term 
$N : ({\Nat}_{\bot}, {\Nat}_{\bot})$ such that:
\[ \lsem N \rsem d = \left\{ \begin{array}{ll}
\bot , & d = \bot \\
0  & \mbox{otherwise}
\end{array}
\right. \]
\item Now given a partial recursive function $\varphi$, represented
by a morphism term $f$, the totality of $\varphi$ is equivalent to the $\DDL$-validity of
\[ N \leq [f][N]\bar{0}  \]
where $\bar{0} \equiv ((\true )_{\bot} \oplus \false)$ (so $\lsem \bar{0} \rsem = \{ 0 \}$). \qed
\end{itemize}

\section{Applications: The Logic of a Domain Equation}
A denotational analysis of a computational situation results in the
description of a domain which provides an appropriate semantic universe
for this situation.
Canonically, domains are specified by type expressions in a metalanguage.
We can then use our approach to ``turn the handle'', and generate 
a logic for this situation in a quite mechanical way.

We shall now go on to develop two case studies of this kind, in the areas of concurrency (Chapter~5) and the $\lambda$-calculus (Chapter~6).

\chapter{Applications to Concurrency:  
A Domain Equation for Bisimulation}
\section{Introduction}
Our aim in this Chapter is to treat some basic topics in the theory
of concurrency from the point of view of domain logic. 
This will serve as a major case study for the general theory developed in the previous two Chapters; and will also weave another of the strands mentioned in Chapter~1 into our narrative. 
Our aim is not only to exemplify the general theory, but to {\em apply} it in order to shed some new light on concurrency.
In particular, we shall study {\em bisimulation} \cite{Par81,Mil83,HM85}. 
This notion has emerged as one of the more stable and mathematically natural concepts to have been formulated in the study of concurrency over the past decade.
It is commonly accepted as the {\em finest} extensional or behavioural equivalence on processes one would want to impose.
To date, bisimulation has been studied almost exclusively from the operational and logical points of view.
Our aim is to show that this notion can be captured elegantly in the setting of domain theory, using Plotkin's powerdomain construction \cite{Plo76}.
Moreover, we shall make extensive use of the logical form of domain theory developed in the previous Chapter.
Thus our motivation can be summarised as follows:
\begin{itemize}

\item To show that more can be done in the sphere of concurrency using domain-theoretic and denotational methods than seems to be commonly realised.

\item To analyze the apparently {\it ad hoc} and ``application oriented'' notions of bisimulation over labelled transition systems and Hennessy-Milner logic by means of the general, mathematically basic, and ``reusable'' notions of domain theory, specifically type constructions and the solution of recursive domain equations.

\item To form part of our general programme of connecting
\begin{enumerate}

\item Domain theory and operational notions of observability

\item Denotational semantics and program logics.
\end{enumerate}

This programme is made systematic by using the information conveyed in the syntactic description of domains by type expressions.
It can be argued that a full domain-theoretic analysis of some computational situation is only obtained when we have written down an explicit type expression, rather than using some {\it ad hoc} construction of a cpo.
At any rate, the benefits which flow from having such a description are very considerable.
Using the ideas developed in the previous Chapter, we can derive a propositional theory from the type expression, and use this to explore the ``observational logic'' of the computational situation.
\end{itemize}

We now summarise the further contents of the Chapter.
After reviewing some basic notions on transition systems etc., we introduce a domain of synchronisation trees defined by means of a domain equation (recursive type expression).
Then we present a domain logic for transition systems, which is derived from this domain equation in the sense of Chapter 3.
The main result of section 4 is that the finitary part of this logic is the Stone dual of our domain of synchronisation trees.

In section 5, we present a number of applications of this logic.
It is shown to be equivalent to Hennessy-Milner logic in the infinitary case, and hence to characterise bisimulation.
In the finitary case, it more powerful than Hennessy-Milner logic, and we obtain a more satisfactory characterisation result for it; namely, it is shown to characterise the ``finitary part'' of bisimulation for {\em all} transition systems.

We also develop an extension of Hennessy-Milner logic which is equivalent to the finitary domain logic.
The infinitary domain logic is then used to {\em axiomatize} a suitable notion of ``finitary transition system''.
These systems are shown indeed to be finitary in a strong sense --- their bisimulation preorders are algebraic.
Finally, the domain of synchronisation trees (i.e. the spectral space of the logic) is shown to be finitary {\it qua} transition system, and moreover to be {\em final} in a suitable category of such systems.
This yields a syntax-free ``universal semantics'' for transition systems, which is fully abstract with respect to bisimulation.

In section 6, we give a conventional (syntax-directed) denotational semantics for the concurrent calculus SCCS \cite{Mil83}, based on our domain of synchronisation trees.
A full abstraction result is proved for this semantics; as a by-product, our domain is shown to be isomorphic to Hennessy's term model \cite{Hen81}.

\section{Transition Systems and Related Notions}

We begin with the basic notion of a labelled transition system (with divergence), which abstracts from the operational semantics of many concurrent calculi.

\begin{definition}
{\rm A {\em transition system} is a structure
\[ ({\rm Proc}, {\sf Act}, \rightarrow, \diverges ) \]
where:
\begin{itemize}

\item ${\rm Proc}$ is a set of {\em processes} or {\em agents}.

\item ${\sf Act}$ is a set of atomic {\em actions} or {\em experiments}.

\item ${\rightarrow} \subseteq {\rm Proc} \times {\sf Act} \times {\rm Proc}$ (notation: $p \stackrel{a}{\rightarrow} q$).

\item $\diverges \subseteq {\rm Proc}$ (notation: $p \diverges$).
\end{itemize}}
\end{definition}

We write
\[ p \converges \equiv \neg (p \diverges ) . \]
We read $p \stackrel{a}{\rightarrow} q$ as ``$p$ has the capability to do $a$ and become (i.e. change state to) $q$''; $p \diverges$ as ``$p$ may diverge''; and $p \converges$ as ``$p$ definitely converges''.
We define
\[ {\sf sort}(p) \equiv \{ a \in {\sf Act} \: | \: \exists q, r . \, p \rightarrow^{\star} q \stackrel{a}{\rightarrow} r \} \]
where $p \rightarrow q \equiv \exists a \in {\sf Act} . \, p \stackrel{a}{\rightarrow} q$, and $\rightarrow^{\star}$ is the reflexive, transitive closure of $\rightarrow$.

We now define a number of finiteness conditions on transition systems:
\begin{center}
\begin{tabular}{ll}
{\bf image-finiteness} & $\forall p \in {\rm Proc} , a \in {\sf Act} . \, \{ q \: | \: p \stackrel{a}{\rightarrow} q \}$ is finite. \\
{\bf sort-finiteness} & $\forall p \in {\rm Proc} . \, {\sf sort}(p)$ is finite. \\
{\bf finite-branching} & $\forall p \in {\rm Proc} . \, \{ q \: | \: p \rightarrow q \}$ is finite. \\
{\bf initials-finiteness} & $\forall p \in {\rm Proc} . \, \{ a \in {\sf Act} \: | \: \exists q . \, p \stackrel{a}{\rightarrow} q \}$ is finite.
\end{tabular}
\end{center}

Each of these properties has a weak form, obtained by making it conditional on convergence. For example:
\begin{center}
\begin{tabular}{ll}
{\bf weak image-finiteness} &
$\forall p \in {\rm Proc} , a \in {\sf Act} . \, p \converges \; \Rightarrow \; \{ q \: | \: p \stackrel{a}{\rightarrow} q \}$ is finite. 
\end{tabular}
\end{center}

We now introduce a particularly useful source of examples for transition systems, the {\em synchronisation trees}.
Given a set ${\sf Act}$ of actions, ${\sf ST}_{\infty}({\sf Act})$, the synchronisation trees over ${\sf Act}$, are defined as the (proper) class of infinitary terms generated by the following inductive definition:
\begin{equation}
\frac{ \{ a_{i} \in {\sf Act}, t_{i} \in {\sf ST}_{\infty}({\sf Act}) \}_{i \in I}}{\sum_{i \in I} a_{i}t_{i} \; [ + \Omega ] \in {\sf ST}_{\infty}({\sf Act})} \label{st}
\end{equation}
where $[ + \Omega ]$ means optional inclusion of $\Omega$ as a summand (i.e. there are really two clauses in this definition). We write
\begin{eqnarray*}
\Oh & \equiv & \sum_{i \in \varnothing} a_{i}t_{i} \\
\Omega & \equiv & \sum_{i \in \varnothing} a_{i}t_{i} + \Omega .
\end{eqnarray*}

The subclass of terms formed using only finite sums is denoted 
${\sf ST}_{\omega}({\sf Act})$. 
Given a synchronisation tree $t$ formed according to \ref{st}, we stipulate:
\begin{itemize}
\item $t \diverges$ iff $\Omega$ is included as a summand.
\item $t \stackrel{a_{i}}{\rightarrow} t_{i}$ for each summand $a_{i}t_{i}$ $(i \in I)$.
\end{itemize}

This defines a (large) transition system $( {\sf ST}_{\infty}({\sf Act}) , {\sf Act}, \rightarrow , \diverges )$; restriction to a subset of synchronisation trees yields a small transition system. 
In particular, by choosing a canonical system of representatives for ${\sf ST}_{\omega}({\sf Act})$ which is closed under subtrees 
we obtain a countable transition system of finite synchronisation trees, which by abuse of notation we refer to also as ${\sf ST}_{\omega}({\sf Act})$.

We are now ready to introduce the main concept we will study.

\begin{definition}
{\rm (\cite{Par81,Mil80,Mil81})
A relation $R \subseteq {\rm Proc} \times {\rm Proc}$ is a {\em prebisimulation} if, for all $p, q \in {\rm Proc}$:
\[ \begin{array}{lrl}
p R q & \Longrightarrow & \forall a \in {\sf Act} . \\
& \bullet & p \stackrel{a}{\rightarrow} p' \;\; \Longrightarrow \;\; \exists q' . \, q \stackrel{a}{\rightarrow} q' \: \& \: p' R q' \\
& \bullet & p \converges \;\; \Longrightarrow \;\; q \converges \; \& \; [q \stackrel{a}{\rightarrow} q' \; \Rightarrow \; \exists p' . \, p \stackrel{a}{\rightarrow} p' \: \& \: p' R q'] .
\end{array} \]
We write
\[ p \preord^{B} q \equiv \exists R . \, R \; \mbox{is a prebisimulation and} \; p R q . \] }
\end{definition}

For an alternative description of $\preord^{B}$, let $Rel({\rm Proc})$ be the set of all binary relations over ${\rm Proc}$; this is a complete lattice under set inclusion. Now define
\[ F : Rel({\rm Proc}) \rightarrow Rel({\rm Proc}) \]
\[ \begin{array}{lrl}
F(R) & = & \{ (p, q) \: | \: \forall a \in {\sf Act} . \\
& & \bullet \; p \stackrel{a}{\rightarrow} p' \; \Rightarrow \; \exists q' . \, q \stackrel{a}{\rightarrow} q' \: \& \: p' R q' \\
& & \bullet \; p \converges \; \Rightarrow \; q \converges \: \& \:  [q \stackrel{a}{\rightarrow} q' \; \Rightarrow \; \exists p' . \, p \stackrel{a}{\rightarrow} p' \: \& \: p' R q'] \} .
\end{array} \]
Clearly, $R$ is a prebisimulation iff $R \subseteq F(R)$, i.e. $R$ is a {\em pre-fixed point of $F$}. Since $F$ is monotone, by Tarski's Theorem it has a maximal fixpoint, given by $\bigcup \{ R \: | \: R \subseteq F(R) \}$, i.e. $\preord^{B}$. Thus $\preord^{B}$ is itself a prebisimulation, and evidently the largest one. Moreover, it is reflexive and transitive; the corresponding equivalence is denoted $\sim^{B}$.

We can also describe $\preord^{B}$ more explicitly, in terms of iterations of $F$. We define relations $\preord_{\alpha}$, $(\alpha \in {\sf Ord})$ (the class of ordinals), by the following ordinal recursion:
\begin{itemize}
\item $p \preord_{0} q$ always (i.e. $\preord_{0} = {\rm Proc} \times {\rm Proc}$, the top element in the lattice $Rel({\rm Proc})$).
\item $p \preord_{\alpha + 1} q$ iff
\[ \begin{array}{l}
\forall a \in {\sf Act} . \\
\bullet \; p \stackrel{a}{\rightarrow} p' \;\; \Longrightarrow \;\; \exists q' . \, q \stackrel{a}{\rightarrow} q' \: \& \: p' \preord_{\alpha} q' \\
\bullet \; p \converges \;\; \Longrightarrow \;\; q \converges \; \& \;  [q \stackrel{a}{\rightarrow} q' \; \Rightarrow \; \exists p' . \, p \stackrel{a}{\rightarrow} p' \: \& \: p' \preord_{\alpha} q' ]  .
\end{array} \]
(i.e. $\preord_{\alpha + 1} = F(\preord_{\alpha})$).

\item For limit $\lambda$, $p \preord_{\lambda} q$ iff $\forall \alpha < \lambda . \, p \preord_{\alpha} q$ (i.e. $\preord_{\lambda} = \bigcap_{\alpha < \lambda} \preord_{\alpha}$).
\end{itemize}

This sequence of relations is decreasing, and bounded below by $\preord^{B}$; i.e. for all $\alpha$
\[ {\preord_{\alpha}} \supseteq {\preord_{\alpha + 1}} \supseteq {\preord^{B}} . \]
For any (small) transition system the sequence is eventually stationary; for some $\lambda$, for all $\alpha > \lambda$, $\preord_{\alpha} = \preord_{\lambda}$.
The least ordinal $\lambda$ for which this holds is called the {\em closure ordinal} \cite{Mos74}; and we have ${\preord_{\lambda}} = {\preord^{B}}$.
Note that each $\preord_{\alpha}$ is relexive and transitive.

The relations $\preord^{B}$ and $\sim^{B}$ have been defined in the context of a given transition system. However, we frequently want to use them to compare processes from different transition systems.
This is easily accomplished by forming the disjoint union of the two systems, and then using $\preord^{B}$ as defined above. In the sequel, we will do this without further comment.

We now introduce a program logic due to Hennessy and Milner \cite{HM85}. The idea is to obtain a characterisation of $\preord^{B}$ in terms of a suitable notion of {\em property} of process; $p \preord^{B} q$ iff every property satisfied by $p$ is satisfied by $q$.

\begin{definition}
{\rm Given a set of actions ${\sf Act}$, the language ${\rm HML}_{\infty}({\sf Act})$ (we henceforth elide the parameter ${\sf Act}$) is defined by the following inductive clauses:
\[ \frac{a \in {\sf Act} , \: \phi \in {\rm HML}_{\infty}}{[a] \phi ,  \ltuple a \rtuple \phi \in {\rm HML}_{\infty}} \]
\[ \frac{\phi_{i} \in {\rm HML}_{\infty} \: (i \in I)}{\bigwedge_{i \in I} \phi_{i} , \bigvee_{i \in I} \phi_{i} \in {\rm HML}_{\infty}} \]}
\end{definition}
In particular, we write:
\begin{eqnarray*}
\true & \equiv & \bigwedge_{i \in \varnothing} \phi_{i} \\
\false & \equiv & \bigvee_{i \in \varnothing} \phi_{i} .
\end{eqnarray*}
We use the subscript $\infty$ to indicate the presence of infinite conjunctions and disjunctions.
We write ${\rm HML}_{\omega}$ for the sublanguage obtained by restricting the formation rules to finite conjunctions and disjunctions.

We now define a satisfaction relation ${\models} \subseteq {\rm Proc} \times {\rm HML}_{\infty}$.
\[ \begin{array}{lcl}
p \models \bigwedge_{i \in I} \phi_{i} & \equiv & \forall i \in I . \, p \models \phi_{i} \\
p \models \bigvee_{i \in I} \phi_{i} & \equiv & \exists i \in I . \, p \models \phi_{i} \\
p \models {\ltuple a \rtuple} \phi & \equiv & \exists q . \,  \labarrow{p}{a}{q} \; \& \; q \models \phi \\
p \models {[ a ]} \phi & \equiv & \forall q . \,  \labarrow{p}{a}{q} \;\; \Longrightarrow \;\; q \models \phi . \\
\end{array} \]
We write
\begin{eqnarray*}
{\rm HML}_{\infty}(p) & \equiv & \{ \phi \in {\rm HML}_{\infty} : p \models \phi \}
\end{eqnarray*}
plus obvious variations on this notation.

We define two useful assignments of ordinals to formulas in ${\rm HML}_{\infty}$, the {\em modal depth}:
\[ \begin{array}{lllll}
{\sf md} (\bigwedge_{i \in I}\phi_{i} ) & \equiv & {\sf md} (\bigvee_{i \in I}\phi_{i} ) & \equiv & \sup \{ {\sf md}  ( \phi_{i} ) : i \in I \} \\
{\sf md} ( [ a ] \phi ) & \equiv & {\sf md}  ( \ltuple a \rtuple \phi ) & \equiv & {\sf md} ( \phi ) + 1 
\end{array} \]
and the {\em height}:
\[ \begin{array}{lllll}
{\sf ht}(\bigwedge_{i \in I}\phi_{i} ) & \equiv & {\sf ht} (\bigvee_{i \in I}\phi_{i} ) & \equiv & \sup \{ {\sf ht} ( \phi_{i} ) : i \in I \} + 1 \\
{\sf ht}( [ a ] \phi ) & \equiv & {\sf ht} ( \ltuple a \rtuple \phi ) & \equiv & {\sf ht}( \phi ) + 1 .
\end{array} \]

We define ${\sf sort}( \phi )$ to be the set of action symbols which occur in $\phi$.

Now given a set $A \subseteq {\sf Act}$ and an ordinal $\lambda$, we define a sublanguage of ${\rm HML}_{\infty}$:
\[ {\rm HML}^{(A, \lambda )}_{\infty} = \{ \phi \in {\rm HML}_{\infty} : {\sf sort}( \phi ) \subseteq A \; \& \; {\sf md} ( \phi ) \leq \lambda \} . \]

We are now ready to prove a generalised and strengthened version of the Modal Characterisation Theorem \cite{Mil81,Mil85,HM85}.

\begin{theorem}[Modal Characterisation Theorem]
\label{mct}
Suppose that $A \subseteq {\sf Act}$ satisfies
\[ {\sf sort}(p) \cup {\sf sort}(q) \subseteq A \not= \varnothing ; \]
then 
\[ p \preord_{\lambda} q \;\; \Longleftrightarrow \;\; {\rm HML}^{(A, \lambda )}_{\infty} (p) \subseteq  {\rm HML}^{(A, \lambda )}_{\infty} (q). \]
As an immediate consequence we obtain
\[  p \preord^{B} q \;\; \Longleftrightarrow \;\; {\rm HML}_{\infty} (p) \subseteq  {\rm HML}_{\infty} (q). \]
\end{theorem}

\proof\ The left-to-right implication is proved by induction on $\lambda$. 
The cases for $\lambda = 0$, $\lambda$ a limit ordinal are trivial. 
For $\lambda = \alpha + 1$, we argue by induction on ${\sf ht}(\phi )$. 
The cases for $\bigwedge_{i \in I} \phi_{i}$, $\bigvee_{i \in I} \phi_{i}$ are trivial. 
Suppose $p \models {\ltuple a \rtuple} \phi$. 
Then for some $p'$, $\labarrow{p}{a}{p'}$ and $p \models \phi$. 
Since $p \preord_{\lambda} q$, for some $q'$, $\labarrow{q}{a}{q'}$ and $p' \preord_{\alpha} q'$. 
By the outer induction hypothesis, $q' \models \phi$, hence $q \models {\ltuple a \rtuple} \phi$, as required. 
The case for ${[a]} \phi$ is similar.

For the converse, we argue by induction on $\lambda$. 
Suppose $p \npreord_{\lambda} q$: we must find $\phi \in  {\rm HML}^{(A, \lambda )}_{\infty} (p) -  {\rm HML}^{(A, \lambda )}_{\infty} (q)$. \\
Case 1: $\labarrow{p}{a}{p'}$ and for all $q'$, $\labarrow{q}{a}{q'}$ implies $p' \npreord_{\alpha} q'$ for some $\alpha < \lambda$. 
By induction hypothesis, for each such $q'$ there is $\phi \in  {\rm HML}^{(A, \alpha )}_{\infty} (p') -  {\rm HML}^{(A, \alpha )}_{\infty} (q')$. 
Now take
\[ \phi = {\ltuple a \rtuple} \bigwedge \{ \phi_{q'} : \labarrow{q}{a}{q'} \} . \]
Case 2: $p \converges$ and $p \diverges$. Take $\phi \equiv {[a]} \true$, for any $a \in A$. \\
Case 3: $p \converges$, $q \converges$, $\labarrow{q}{a}{q'}$, and for all $p'$, $\labarrow{p}{a}{p'}$ implies $p' \npreord_{\alpha} q'$ for some $\alpha < \lambda$. 
Defining $\phi_{p'}$ analogously to Case 1,
\[ \phi = {[a]} \bigvee \{ \phi_{p'} : \labarrow{p}{a}{p'} \} . \;\;\; \qed \]
The reader familiar with infinitary logic will recognise the strong similarity between this result and Karp's Theorem \cite{Barw75}.
Similar remarks apply to ``Master Formula Theorems'' as in \cite{Rou85}, {\em vis a vis}  the
Scott Isomorphism Theorem \cite{Barw75}.

Note that, if $A$ is a finite set and $\lambda$ a finite ordinal, then (up to logical equivalence) ${\rm HML}^{(A, \lambda )}_{\infty}$ is finite. 
It follows easily from this observation that each formula in ${\rm HML}^{(A, \lambda )}_{\infty}$ is equivalent to one in ${\rm HML}^{(A, \lambda )}_{\omega}$. 
Hence as a Corollary to the Characterisation Theorem we obtain
\begin{theorem}
\cite{Abr87b} If the transition system is sort-finite, then
\[ p \preord_{\omega} q \;\; \Longleftrightarrow \;\; {\rm HML}_{\omega}(p) \subseteq {\rm HML}_{\omega}(q) . \]
\end{theorem}

Moreover, we have the following result from \cite{HM85}:
\begin{theorem}
If the transition system is image-finite, then
\[ \begin{array}{rl}
(i) & {\preord_{\omega}} = {\preord^{B}} \\
(ii) & p \preord_{\omega} q \;\; \Longleftrightarrow \;\; {\rm HML}_{\omega}(p) \subseteq {\rm HML}_{\omega}(q) . 
\end{array} \]
\end{theorem}

Unfortunately, if unguarded recursion is allowed in any of the standard concurrent calculi (SCCS, CCS, CSP, etc.) 
they are neither image-finite nor sort-finite (though sort-finiteness may be regained e.g. for CCS by imposing fairly mild restrictions on the relabelling operators). 
Thus these two Theorems cannot be applied. 
To see how weak finitary Hennessy-Milner logic is when the set of actions is finite, consider the following

\noindent {\bf Example.}
\begin{eqnarray*}
p & \equiv & a \Oh + \Omega \\
q & \equiv & \sum_{n \in \omega} a b_{n} \Oh + \Omega
\end{eqnarray*}
where we assume $b_{m} \not= b_{n}$ for $m \not= n$. Now $p \npreord_{2} q$, but we have

\begin{proposition}
\label{bex}
${\rm HML}_{\omega}(p) \subseteq {\rm HML}_{\omega}(q).$
\end{proposition}

In order to prove this Proposition we need a lemma.

\begin{lemma}
Every formula in ${\rm HML}_{\omega}(\Oh )$ is satisfied by cofinitely many of the $b_{n}\Oh$.
\end{lemma}

\proof\ By induction on formulas in  ${\rm HML}_{\omega}(\Oh )$. 
For conjunctions and disjunctions, the intersection and union of finitely many cofinite sets are cofinite. 
(It is the case for conjunction which necessitates the strength of statement of the Lemma).
The case for ${\ltuple b \rtuple}\phi$ is vacuous. 
For ${[b]}\phi$, cofinitely many (in fact, all but at most one) of the $b_{n}\Oh$ do not have a $b$-action, hence satisfy ${[b]}\phi$. \qed

The Proposition can now be proved by induction on formulas in ${\rm HML}_{\omega}$. 
The only non-trivial case is ${\ltuple a \rtuple}\phi$, which follows from the Lemma.

The deficiency of Hennessy-Milner logic illustrated by this example is disturbing, 
because processes generated by a finitary calculus (including $p$ and $q$ above) should be adequately modelled by a finitary semantics and logic. 
This suggests that Hennessy-Milner logic is not quite right as it stands.

\section{A Domain Equation for Synchronisation Trees}

In this section, we shall define a domain of synchronisation trees, and establish some of its basic properties.
Since our definitions will use the Plotkin powerdomain, we need to work in a category which is closed under this construction.
This means that we cannot use {\bf SDom}, as we did in the previous two Chapters.
Instead, we will use {\bf SFP}.
The only facts about {\bf SFP} which we will need are that it is a category of algebraic domains closed under the following type constructions:
\subsection*{Separated Sum}
Let $A$ be a countable set, and $\{ D_{a} \}_{a \in A}$ an $A$-indexed family of domains. Then $\sum_{a \in A} D_{a}$ is formed by taking the disjoint union of the $D_{a}$ and adjoining a bottom element.
We shall write elements of the disjoint union as $\ltuple a, d \rtuple$ ($a \in A$, $d \in D_{a}$).
Note that the ordering is defined so that
\[ \ltuple a, d \rtuple \sqsubseteq \ltuple a' , d' \rtuple \;\; \Longleftrightarrow \;\; a = a' \: \& \: d \sqsubseteq_{D_{a}} d' . \]
\begin{itemize}
\item For each $a \in A$, the function
\[ D_{a} \rightarrow \sum_{a \in A} D_{a} \]
\[ d \mapsto \ltuple a, d \rtuple \]
is continuous.

\item Separated sum is functorial; given a family
\[ f_{a} : D_{a} \rightarrow E_{a} \;\; (a \in A) , \]
\[ \sum_{a \in A} f_{a} : \sum_{a \in A} D_{a} \rightarrow \sum_{a \in A} E_{a} \]
is defined by:
\[ \begin{array}{lll}
(\sum_{a \in A} f_{a}) \bot & = & \bot \\
(\sum_{a \in A} f_{a}) \ltuple a, d \rtuple & = & \ltuple a, f_{a} d \rtuple .
\end{array} \]
\end{itemize}

\subsection*{The Plotkin Powerdomain}
We write $P[D]$ for the Plotkin powerdomain over $D$.
Although this construction is best {\em characterised} abstractly, as in 
\cite{HP79}, for purposes of comparison with more concrete operational notions a good representation is invaluable.
This is provided in \cite{Plo76,PloLN}.
\begin{definition}
{\rm For an algebraic domain $D$ the {\em Lawson topology} on $D$ is generated by the sub-basic sets
\[ \diverges b , \;\; D - \diverges b \]
for finite $b \in D$ (so the Lawson topology refines the Scott topology). 
We will write the closure operator associated with the Lawson topology as $Cl$. (NB: in \cite{Plo76}, the Lawson topology is called the Cantor topology).}
\end{definition}

\begin{definition}
{\rm For $X \subseteq D$,
\[ \begin{array}{rlcl}
(i) & Con(X) & \equiv & \{ d : \exists d_{1}, d_{2} \in X . \, d_{1} \sqsubseteq d \sqsubseteq d_{2} \} \\
(ii) & X^{\star} & \equiv & Con \circ Cl .
\end{array} \]
$X$ is said to be
\begin{itemize}
\item {\em Lawson-closed} if $X = Cl \; X$
\item {\em Convex-closed} if $X = Con \; X$
\item {\em Closed} if $X = X^{\star}$.
\end{itemize}}
\end{definition}

\begin{definition}
{\rm The {\em Egli-Milner order}. For $X, Y \subseteq D$:}
\begin{eqnarray*}
X \sqsubseteq_{EM} Y & \equiv & \forall x \in X . \, \exists y \in Y . \, x \sqsubseteq y \; \& \; \forall y \in Y . \, \exists x \in X . \, x \sqsubseteq y .
\end{eqnarray*}
\end{definition}

The representation of the Plotkin powerdomain can now be defined as follows:
\begin{eqnarray*}
P[D] & \equiv & ( \{ X \subseteq D : X \not= \varnothing , X = X^{\star} \} , \sqsubseteq_{EM} ) .
\end{eqnarray*}

There are also a number of (continuous) operations associated with the Plotkin powerdomain, which we shall describe in terms of our representation of $P[D]$.

\begin{itemize}

\item Firstly, $P$ is {\em functorial}: given $f \: D \rightarrow E$,
\[ P f : P[D] \rightarrow P[E] \]
is defined by
\begin{eqnarray*}
P f (X) & \equiv & \{ f(x) | x \in X \}^{\star} .
\end{eqnarray*}

\item {\em Singleton:}
\[ \lsing . \rsing : D \rightarrow P[D] \]
is defined by
\begin{eqnarray*}
\lsing d \rsing & \equiv & \{ d \}^{\star} = \{ d \} .
\end{eqnarray*}

\item {\em Union:}
\[ \uplus : P[D]^{2} \rightarrow P[D] \]
is defined by
\begin{eqnarray*}
X \uplus Y & \equiv & (X \uplus Y)^{\star} = Con(X \cup Y) .
\end{eqnarray*}

\item {\em Big Union:}
\[ \biguplus : P[P[D]] \rightarrow P[D] \]
is defined by
\begin{eqnarray*}
\biguplus ( \Theta ) & \equiv & ( \bigcup \Theta )^{\star} = Con ( \bigcup \Theta ) .
\end{eqnarray*}

\item {\em Tensor Product} \cite{HP79}. We will only need the following: given
\[ f : D^{n} \rightarrow D \]
the {\em multilinear extension}
\[ f^{\dagger} \: P[D]^{n} \rightarrow P[D] \]
is defined by
\begin{eqnarray*}
f^{\dagger} (X_{1}, \ldots , X_{n} ) & \equiv & \{ f(x_{1}, \ldots , x_{n} ) : x_{i} \in X_{i} \}^{\star} .
\end{eqnarray*}
(Note that for $n = 1$, $f^{\dagger} = P f$.) This extension has the property
\begin{eqnarray*}
f^{\dagger} (X_{1}, \ldots , X_{i} \uplus X'_{i}, \ldots , X_{n}) & = & f^{\dagger} (X_{1}, \ldots , X_{i}, \ldots , X_{n}) \\
& & \mbox{} \uplus f^{\dagger} (X_{1}, \ldots ,  X'_{i}, \ldots , X_{n})  
\end{eqnarray*}
for $(1 \leq i \leq n)$.
\end{itemize}

\subsection*{Adjoining the empty set}

To the best of my knowledge, the only significant precursor of our work in this Chapter is \cite{MM79}.
The main reason that something like our present programme could not have been carried through in their framework is that, because of a technical problem, they used the Smyth rather than the Plotkin powerdomain.
This rules out any hope of gaining a correspondence with bisimulation.
The technical problem is that of adjoining the empty set to the
powerdomain to model the convergent process with no actions (NIL in CCS
\cite{Mil80}, $\Oh$ in SCCS \cite{Mil83}, STOP in CSP \cite{Hoa85},
$\delta$ in ACP \cite{BK84}, etc.).
If we add the empty set to our representation of $P[D]$, it is not related to
anything except itself under $\sqsubseteq_{EM}$; in category-theoretic
terms, the problem is the non-existence of a certain free construction
(\cite{PloLN} ).
Fortunately, we do not need these non-existent solutions.
We shall adjoin the empty set to the Plotkin powerdomain in a way which
has two advantages:
\begin{enumerate}
\item There is no theoretical overhead, since it is definable as a derived
operation from standard type constructions.
\item It works, i.e. is exactly suited to our semantic purposes, as the
results to follow will show.
\end{enumerate}

For motivation, consider a transition system $({\rm Proc}, {\sf Act}, \rightarrow ,
\diverges )$ and processes $p, r \in {\rm Proc}$ such that
\[ \begin{array}{rl}
(i) & p\diverges , \; r \converges \\
(ii) & p \nrightarrow , \; r \nrightarrow .
\end{array} \]
Then it is easy to see that, for all $q \in {\rm Proc}$:
\[ \begin{array}{rlcl}
(i) & r \preord^{B} q & \Longleftrightarrow & r \sim^{B} q \\
(ii) & q \preord^{B} r & \Longleftrightarrow & q \nrightarrow \\
& & \Longleftrightarrow & q \sim^{B} p \; \mbox{or} \; q \sim^{B} r .
\end{array} \]

This suggests the following
\begin{definition}
{\rm $P^{0}[D]$, the Plotkin powerdomain with empty set.

\noindent Representation of $P^{0}[D]$:
\begin{center}
\begin{tabular}{ll}
{\bf Elements} & $\{ X \subseteq D : X = X^{\star} \} = P[D] \cup \{
\varnothing \}$. \\
{\bf Ordering} & $X \sqsubseteq Y \; \equiv \; X = \{ \bot \} \; \mbox{or} \;
X \sqsubseteq _{EM} Y$.
\end{tabular}
\end{center}}
\end{definition}

\begin{observation}
$P^{0}[D] \; \cong \; ({\bf 1})_{\bot} \oplus P[D]$.
\end{observation}

In principle, we could work throughout with 3.5 as the {\em definition} of
$P^{0}[D]$; in practice, it is much more convenient to work with the
representation given by 3.4.
This requires that we extend our definitions of the powerdomain
operations to work on $P^{0}[D]$.
In fact, all of the definitions following 3.3 still make sense for $P^{0}[D]$.
It is easily checked that $\uplus$, $\biguplus$ and $\lsing \cdot \rsing$
are continuous on $P^{0}[D]$.
For $P^{0} f$ and $f^{\dagger}$ a technical point arises, which is not
specific to 3.4, but stems from the use of coalesced sum in 3.5.
As is well known,  coalesced sum is functorial only on the category of
{\em strict} functions.
Hence we can only use $P^{0} f$ if $f$ is strict, and $f^{\dagger}$ if $f$ is
strict in each argument separately.
With these provisos, the extended operations are continuous.

{\bf Notation}. We use $\emptyset$ to denote the empty set in $P^{0}[D]$;
if $I$ is a finite index set, we write
\[ \biguplus_{i \in I} X_{i} \]
meaning the iterated use of $\uplus$ (which is associative, commutative
and idempotent on $P^{0}[D]$, just as it is on $P[D]$) if $I \not=
\varnothing$, and $\emptyset$ otherwise.
Also, we write
\[ \lsing d : A \rsing \]
where $d \in D$ and $A$ is some sentence, meaning $\lsing d \rsing$ if
$A$ is true, and $\emptyset$ otherwise.

We are now ready for the main definition of the section.
\begin{definition}
{\rm Let ${\sf Act}$ be a {\em countable} set of actions.
Then $\Dom ({\sf Act})$, the domain of synchronisation trees over ${\sf Act}$ (we
henceforth omit the parameter ${\sf Act}$), is defined to be the initial solution
of the domain equation}
\begin{equation}
\Dom \; \cong \; P^{0}[ \sum_{a \in {\sf Act}} \Dom ]  . \label {domeq} 
\end{equation}
\end{definition}
Here the sum $\sum_{a \in {\sf Act}} \Dom$ is the ``copower'' of ${\sf Act}$ copies of 
 $\Dom$.
The equation is essentially that of \cite{MM79}, minus the value passing
and with a different powerdomain.

How can we relate this domain equation to the formalism of Chapter~4?
Suppose we extend the metalanguage of types introduced there with a
constructor $P_{p}( \cdot )$ for the Plotkin powerdomain.
Then we can write
\[ \Dom \; \equiv \; {\sf rec} \: t . ({\bf 1})_{\bot} \oplus P_{p}[\sum_{a \in {\sf Act}}
t] \]
using 3.5 to eliminate $P^0$.
This is not yet a valid type expression because of the sum
\begin{equation}
\sum_{a \in {\sf Act}} t \label{sumeq}
\end{equation}
Let us take the main case of interest, where ${\sf Act}$ is countably infinite,
say ${\sf Act} = \{ a_{n} \}_{n \in \omega}$.
Then we can replace~\ref{sumeq} by the recursive expression
\begin{equation}
{\sf rec} \: u . (t)_{\bot} \oplus u  \label{recsum} 
\end{equation}
yielding the overall expression
\begin{equation}
\Dom \; \equiv \; {\sf rec} \: t. ({\bf 1})_{\bot} \oplus P_{p}[{\sf rec} \: u . (t)_{\bot}
\oplus u  ]  \label{typeexp} 
\end{equation}
the intention being that the $i$'th summand as we unfold~\ref{recsum}
corresponds to $a_{i} \in {\sf Act}$.

The reader will by now probably appreciate our efforts to streamline the
presentation.
Nevertheless, we regard the ``closed form'' expression \ref{typeexp} as
fundamental, and the logic we shall introduce in the next section could be
derived mechanically from it in the manner detailed in Chapter 4.

In the remainder of this section, we shall apply some standard
domain-theoretic methods to elucidate the structure of $\Dom$.

{\bf Notation.} We write $\bot$ for the bottom element of
$\sum_{a \in {\sf Act}} \Dom$; $\lsing \bot \rsing$ is then the bottom element
of $ P^{0}[ \sum_{a \in {\sf Act}} \Dom ]$.

How can we unpack the structure of $\Dom$ from the domain equation
\ref{domeq}?
This is best done in two parts:
\begin{enumerate}
\item A {\em specified} isomorphism pair
\[ \Dom \begin{array}{c}
\eta \\
\rightleftarrows \\
\theta
\end{array} P^{0}[ \sum_{a \in {\sf Act}} \Dom ] . \]
In fact, we shall elide $\eta$ and $\theta$, and treat \ref{domeq} as an
identity; this is only a notational convenience, and the reader can put
$\eta$ and $\theta$ back without encountering any difficulties.

\item {\em Initiality}. The categorical framework is clumsy to work with
for our purposes.
Instead, we will use an ``intrinsic'' (or in the terminology of
\cite{SP82} a ``local'' or ``{\bf O}-notion'') formulation.
\end{enumerate}

\begin{definition}
{\rm We define a sequence of functions
\[ \pi_{k} : \Dom \rightarrow \Dom \]
as follows:
\[ \begin{array}{lll}
\pi_{0} & \equiv & \lambda x \in \Dom . \lsing \bot \rsing \\
\pi_{k+1} & \equiv & P^{0} \sum_{a \in {\sf Act}} \pi_{k} .
\end{array} \] }
\end{definition}
Note that $\sum_{a \in {\sf Act}}$ always produces a strict function, so this is
well-defined.

Now the following proposition is standard (\cite[Chapter 5 Theorem 3]{PloLN}):
\begin{proposition}
\label{icolim}
\Dom\ is the ``internal colimit'' of the $\pi_{k}$:
\[ \begin{array}{rl}
(i) & \mbox{Each $\pi_{k}$ is continuous and $\pi_{k} \sqsubseteq
\pi_{k+1}$} \\
(ii) & \bigsqcup_{k} \pi_{k} = {\sf id}_{\Dom} \\
(iii) & \pi_{k} \circ \pi_{k} = \pi_{k} \\
(iv) & \forall d_{1}, d_{2} \in \Dom . \, d_{1} \sqsubseteq d_{2} \;\;
\Longleftrightarrow \;\; \forall k . \, \pi_{k} d_{1} \sqsubseteq \pi_{k} d_{2} .
\end{array} \]
\end{proposition}
In particular, we will use part $(iv)$ of this Proposition as the
cutting edge of initiality.

Next, it will be useful to have an explicit description of the finite
elements of $\Dom$, which, as already noted, is in {\bf SFP}, and hence
algebraic.

\begin{definition}
\label{felts}
{\rm $K(\Dom ) \subseteq \Dom$ is defined inductively as follows:}
\begin{itemize}
\item $\emptyset \in K(\Dom )$
\item $\lsing \bot \rsing \in K(\Dom )$
\item $a \in {\sf Act}, d \in K(\Dom ) \; \Rightarrow \; \lsing {<}a, d{>} \rsing
\in K(\Dom )$
\item $d_{1}, d_{2} \in K(\Dom ) \; \Rightarrow \; d_{1} \uplus d_{2} \in
K(\Dom )$.
\end{itemize}
\end{definition}
The following is again standard:
\begin{proposition}
\label{feltp}
$ K(\Dom )$ is exactly the set of finite elements of $\Dom$.
\end{proposition}

Finally, we consider $\Dom$ as a {\em transition system} $(\Dom , {\sf Act},
\rightarrow , \diverges )$ defined by:
\[ \begin{array}{clcl}
\bullet & \labarrow{d}{a}{d'}  & \equiv & {<}a, d' {>} \in d \\
\bullet & d \diverges & \equiv & \bot \in d .
\end{array} \]

\begin{proposition}
\label{ifs}
$\Dom$ is ``internally fully abstract'', i.e.
\[ \forall d_{1}, d_{2} \in \Dom\ . \, d_{1} {\preord}^{B} d_{2} \;\;
\Longleftrightarrow \;\; d_{1} \sqsubseteq d_{2} . \]
\end{proposition}

\proof\ We shall prove
\[ (1) \;\; \forall k . \; d_{1} \preord_{k} d_{2} \;\; \Longrightarrow \;\; \pi_{k}
d_{1} \sqsubseteq \pi_{k} d_{2} \]
and
\[ (2) \;\; {\sqsubseteq} \subseteq {\preord^{B}} . \]
Clearly (1) implies
\[ (3) \;\; {\preord_{\omega}} \subseteq {\sqsubseteq}  \]
by \ref{icolim}$(iv)$, and since
\[ (4) \;\; {\preord^{B}} \subseteq {\preord_{\omega}} , \]
we obtain ${\preord^{B}} = {\sqsubseteq}$, as required.

\noindent (1). By induction on $k$. The basis is trivial. 
For the inductive step, assume $d \preord_{k+1} e$.
Now $d = \emptyset$ and $d \preord_{k+1} e$ implies $e = \emptyset$,
while $d = \lsing \bot \rsing$ implies $d \sqsubseteq e$, so we may
assume $d \not= \emptyset \not= e$, and it suffices to prove $d
\sqsubseteq_{EM} e$.

From the definitions we have $\pi_{k+1} d = X^{\star}$, where
\[ X = \{ {<}a, \pi_{k} d' {>} : {<}a, d' {>} \in d \} \cup \{ \bot : \bot \in d \} , 
\]
and similarly $\pi_{k+1} e = Y^{\star}$. Now
\[ \begin{array}{ll}
\bullet & {<}a, \pi_{k} d' {>} \in X \\
\Longrightarrow & \labarrow{d}{a}{d'} \\
\Longrightarrow & \exists e' . \, \labarrow{e}{a}{e'} \: \& \: d' \preord_{k} e' \\
\Longrightarrow & \exists e' . \, {<}a, e' {>} \in e \: \& \: \pi_{k} d' \sqsubseteq
\pi_{k} e' 
\;\; \mbox{by induction hypothesis} \\
\Longrightarrow & \exists {<}a, \pi_{k} e' {>} \in Y . \, {<}a, \pi_{k} d' {>}
\sqsubseteq {<}a, \pi_{k} e' {>} .
\end{array} \]
Again,
\[ \begin{array}{ll}
\bullet & \bot \not\in X \\
\Longrightarrow & \bot \not\in d \\
\Longrightarrow & \bot \not\in e \: \& \: [ \labarrow{e}{a}{e'}  \; \Rightarrow \;
\exists d' . \, \labarrow{d}{a}{d'} \: \& \: d' \preord_{k} e' ] \\
\Longrightarrow & \bot \not\in Y \: \& \: \forall {<}a, \pi_{k} e' {>} \in Y . \,
\exists {<}a, \pi_{k} d' {>} \in X . \, \pi_{k} d' \sqsubseteq \pi_{k} e' 
\end{array} \]
by the induction hypothesis again,
and we have shown $X \sqsubseteq_{EM} Y$, which implies $X^{\star}
\sqsubseteq_{EM} Y^{\star}$, as required.

\noindent (2). It suffices to show that $\sqsubseteq$ is a prebisimulation.
This is a simple calculation:
\[ \begin{array}{ll}
\bullet & d \sqsubseteq e \\
\Longrightarrow &  \forall {<}a, d' {>} \in d . \, \exists {<}a,
e' {>} \in e . \, d' \sqsubseteq e' \\
& \& \: \bot \not\in d \; \Rightarrow \; \bot \not\in e 
\; \& \; [ \forall {<}a, e' {>} \in e . \, \exists {<}a, d' {>} \in d . \, d'
\sqsubseteq e'] \\
\Longrightarrow & \forall a \in {\sf Act} . \, \labarrow{d}{a}{d'} \; \Rightarrow \;
\exists e' . \, \labarrow{e}{a}{e'} \: \& \: d' \sqsubseteq e' \\
& \& \: d \converges \; \Rightarrow \; e \converges 
\; \& \; [ \labarrow{e}{a}{e'} \; \Rightarrow \; \exists d' . \,
\labarrow{d}{a}{d'} \: \& \: d' \sqsubseteq e' ] . \;\;\; \qed
\end{array} \]

We finish with some examples to illustrate the richness of $\Dom$ as a
transition system.

\subsection*{Examples}

(1). $\Dom$ is not sort-finite.
\begin{eqnarray*}
d_{0} & \equiv & \lsing {<} a_{0} , \lsing \bot \rsing {>}  \rsing \\
d_{1} & \equiv & \lsing {<}a_{0} , \lsing {<} a_{1}, \lsing \bot \rsing {>}
\rsing {>} \rsing \\
& \vdots & \\
{\sf sort}( \bigsqcup d_{k} ) & = &  \{ a_{0}, a_{1} , \ldots \} 
\end{eqnarray*}

\noindent (2). $\Dom$ is not weakly image-finite.
\begin{eqnarray*}
c_{k} & \equiv & {\sum_{i \leq k} a^{i} \Oh } + a^{k} \Omega \;\;\; (k \in
\omega ) \\
\bigsqcup c_{k} & = & { \sum_{k \in \omega} a^{k} \Oh } + a^{\omega} .
\end{eqnarray*}

\section{A Domain Logic for Transition Systems} 
We now introduce our domain logic in an infintary version \Ellinfty, with
a finitary subset $\Ellomega$. 
We show how \Ellinfty\ can be interpreted in any transition system,
present a proof system, and establish its soundness. 
We then turn to \Ellomega\ , and prove the main result of the section:
\Ellomega\ is the Stone dual of \Dom. That is, \Dom\ is isomorphic to
the spectral space of \Ellomega, while \Ellomega\ is isomorphic to the
lattice of compact-open subsets of \Dom. 
This duality will be crucial to our work in the next section. 

\begin{definition} 
{\rm The language \Ellinfty\ has two {\em sorts}: $\pi$ (process) and
$\kappa$ (capability). We write $\Ell_{\infty \pi}$ ($\Ell_{\infty
\kappa}$) for the class of formulae of sort $\pi$ ($\kappa$), which are
defined inductively as follows:} 
\[ \bullet \;\;\; \frac{\{\phi_{i} \in \Ell_{\infty \sigma} \}_{i \in
I}}{\bigvee_{i \in I}\phi_{i} , \bigwedge_{i \in I}\phi_{i} \in \Ell_{\infty
\sigma}} \;\;\; (\sigma \in \{ \pi , \kappa \}) \] 
\[ \bullet \;\;\; \frac{a \in {\sf Act}, \;\; \phi \in \Ell_{\infty \pi}}{a(\phi ) \in
\Ell_{\infty \kappa}} \]
\[ \bullet \;\;\; \frac{\phi \in \Ell_{\infty \kappa}}{\Box \phi , \Diamond
\phi \in \Ell_{\infty \pi}} . \] 
\end{definition} 

\noindent {\bf Notation.} We write $\true \; \equiv \;  \bigwedge_{i \in
\varnothing} \phi_{i}$, $\false \; \equiv \; \bigvee_{i \in \varnothing}
\phi_{i}$. 

The sublanguage of \Ellinfty\ obtained by the restriction to {\em finite}
conjunctions and disjunctions is denoted \Ellomega\ . {\em Height}, {\em
modal depth} and {\em sort} are defined for \Ell\ in entirely analogous
fashion to HML. 
For example: 
\[ \begin{array}{llllll} 
\bullet & {\sf md}(\bigwedge_{i \in I}\phi_{i}) & \equiv & {\sf md}(\bigwedge_{i \in
I}\phi_{i}) & \equiv & \sup \; \{ {\sf md}(\phi_{i} : i \in I \} \\ 
\bullet & {\sf md}(a(\phi )) & \equiv & {\sf md}(\phi ) & & \\ 
\bullet & {\sf md}(\Box \phi ) & \equiv & {\sf md}(\Diamond \phi ) & \equiv & {\sf md}(\phi
) + 1 . 
\end{array} \] 
For each $A \subseteq {\sf Act}$ and ordinal $\lambda$: 
\[ {\Ell}^{(A, \lambda )}_{\infty} \; \equiv \; \{ \phi \in \Ellinfty :
{\sf sort}(\phi ) \subseteq A \: \& \: {\sf md}(\phi ) \leq \lambda \} . \] 

It should be clear how the form of our language is derived from the type
expression 
\[ {\sf rec} \: t. \, P^{0}[\sum_{a \in {\sf Act}} t] . \] 
The two-sorted structure of \Ell\ corresponds to the type constructions
$P^0$~($\pi$) and $\sum_{a \in {\sf Act}}$~($\kappa$). 
The recursion in the type expression is mirrored by the mutual recursion
between the two sorts. 
Note that the Plotkin powerdomain is built from the combination of the
{\em must} modality $\Box$ of the Smyth powerdomain and the {\em may}
modality $\Diamond$ of the Hoare powerdomain ({\it cf.} \cite{Abr83a,Win83}). 
\subsection*{Interpretation of $\cal L$\/  in transition systems} 
Given a transition system $({\rm Proc}, {\sf Act}, {\rightarrow} , \diverges )$, we
define 
\[ {\rm Cap} \; \equiv \; \{ \bot \} \cup {({\sf Act} \times {\rm Proc})} \] 
\[ C : {\rm Proc} \rightarrow \wp ({\rm Cap}) \] 
\[ C(p) = \{ \bot : p \diverges \} \cup \{ \ltuple a, q \rtuple : \labarrow{p}{a}{q} \} . \] 
$C(p)$ is the set of {\em capabilities} of $p$. 
We can now define satisfaction relations 
\[ {\models_{\pi}} \subseteq {\rm Proc} \times \Ell_{\infty \pi} , \] 
\[ {\models_{\kappa}} \subseteq {\rm Proc} \times \Ell_{\infty \kappa} : \] 
For $\sigma  \in \{ \pi , \kappa \}$:
\[ \begin{array}{lll}
w \models_{\sigma} \bigwedge_{i \in I} \phi_{i} & \equiv & \forall i \in I .
\, w \models_{\sigma} \phi_{i} \\
w \models_{\sigma} \bigvee_{i \in I} \phi_{i} & \equiv & \exists i \in I . \,
w \models_{\sigma} \phi_{i} \\
p \models_{\pi} \Box \phi & \equiv & \forall c \in C(p) . \, c
\models_{\kappa} \phi \\
p \models_{\pi} \Diamond \phi & \equiv & \exists c \in C(p) \cup \{ \bot \}
. \, c \models_{\kappa} \phi \\
c \models_{\kappa} a(\phi ) & \equiv & c = {<} a, q {>} \: \& \: q
\models_{\pi} \phi .
\end{array} \]
The {\em assertions} over \Ell\ have the form
\[ \phi \leq_{\sigma} \psi , \;\; \phi =_{\sigma} \psi \;\;\;\; (\sigma \in \{
\pi , \kappa \} , \phi , \psi \in \Ell_{\infty \sigma}) . \]
The satisfaction relation between transition systems and assertions is
defined by:
\begin{eqnarray*}
{\cal T} \models \phi \leq_{\sigma} \psi & \equiv & \forall w \in
S_{\sigma} . \, w \models_{\sigma} \phi \;\; \Longrightarrow \;\; w
\models_{\sigma} \psi \\
{\cal T} \models \phi =_{\sigma} \psi & \equiv & \forall w \in S_{\sigma}
. \, w \models_{\sigma} \phi \;\; \Longleftrightarrow \;\; w \models_{\sigma}
\psi .
\end{eqnarray*}
\[ ( \sigma \in \{ \pi , \kappa \} , S_{\pi} = {\rm Proc}, S_{\kappa} = {\rm Cap} ). \]
This is extended to a class of transition systems {\bf C} by:
\[ {\bf C} \models A \;\; \equiv \;\; \forall {\cal T} \in {\bf C} . \, {\cal T}
\models A . \]
If {\bf C} is the class of all transition systems, we simply write $\models
A$.

\subsection*{A Proof System For $\Ellinfty$}
Firstly, we define a predicate $( \cdot ){\converges}$ on \Ellinfty\ :
\[ \begin{array}{lll}
(\bigwedge_{i \in I} \phi_{i} ) {\converges} & \equiv & \exists i \in I . \,
\phi_{i} {\converges} \\
(\bigwedge_{i \in I} \phi_{i} ) {\converges} & \equiv & \forall i \in I . \,
\phi_{i} {\converges} \\
a(\phi ) {\converges} & \equiv & {\sf true} \\
(\Box \phi ) {\converges} & \equiv & \phi {\converges} \\
(\Diamond \phi ) {\converges} & \equiv & \phi {\converges} .
\end{array} \]
Intuitively, $\phi {\converges}$ means that at least the completely
undefined process does {\em not} satisfy $\phi$ (i.e. $\phi \not= \true$).
We will use it to restrict one of our axiom schemes.

We now present a proof system for assertions over \Ellinfty\ .
Sort subscripts are omitted.

\subsection*{Logical Axioms}
Exactly as in Chapter 4, except that the restriction to finite index sets on
conjunctions and disjunctions is lifted.

\subsection*{Modal Axioms}
\begin{center}
\[ (a-{\leq}) \;\;\;  \frac{\phi \leq \psi}{a(\phi ) \leq a(\psi )} \]
\[ (a-{\wedge})(i) \;\;\; a(\bigwedge_{i \in I} \phi_{i} ) = \bigwedge_{i \in
I} a(\phi_{i} ) \;\;\;\; (I \not= \varnothing ) \]
\[ (a-{\wedge})(ii) \;\;\; a(\phi ) \wedge b(\psi ) = \false \;\;\;\; (a \not=
b) \]
\[ (a-{\vee}) \;\;\; a(\bigvee_{i \in I}\phi_{i} ) = \bigvee_{i \in I}
a(\phi_{i})  \]
\[ ({\Box}-{\leq}) \;\;\; \frac{\phi \leq \psi}{\Box \phi \leq \Box \psi} \]
\[ ({\Box}-{\wedge}) \;\;\; \Box \bigwedge_{i \in I} \phi_{i} =
\bigwedge_{i \in I} \Box \phi_{i} \]
\[ ({\Diamond}-{\leq}) \;\;\;  \frac{\phi \leq \psi}{\Diamond \phi \leq
\Diamond \psi} \]
\[ ({\Diamond}-{\vee}) \;\;\; \Diamond \bigvee_{i \in I} \phi_{i} =
\bigvee_{i \in I} \Diamond \phi_{i}  \]
\[ ({\Box}-{\vee}) \;\;\; \Box (\phi \vee \psi ) \leq \Box \phi \vee
\Diamond \psi  \]
\[ ({\Diamond}-{\wedge}) \;\;\;  \Box \phi \wedge \Diamond \psi \leq
\Diamond ( \phi \wedge \psi ) \;\;\;\; ( \psi {\converges}) \]
\[ ({\Diamond}-{\true}) \;\;\; \Diamond \true\ = \true . \]
\end{center}

The form of our axiomatisation follows the same pattern as that of
Chapter 4, of (the general approach exemplified by) which it is of course a
special case.
The first group of axioms and rules give the logical structure of
entailment, conjunction and disjunction.
They give (the Lindenbaum algebra of) \Ellinfty\ the structure of a (large)
{\em completely distributive lattice} \cite{Joh82}.
We then articulate the modal structure by showing how the 
constructors interact with the logical structure.
The axioms for the $a( \cdot )$ constructor correspond to those for coalesced
sum given in Chapter 4; the fact that {\em separated} sum is intended here
is reflected by the side-condition on $(a-{\wedge})(i)$.
The axioms for $\Box$ and $\Diamond$ individually correspond to those
presented for the upper and lower powerdomains in Chapter 4; however,
these two modalities interact in the Plotkin powerdomain, resulting in its
greater complexity; these interactions are expressed in logical terms by
$({\Box}-{\vee})$ and $({\Diamond}-{\wedge})$.
Our surgery on the ordering to keep a least element while adding the empty
set is reflected by the presence of $({\Diamond}-\true )$ and the side
condition on $({\Diamond}-{\wedge})$.

We write $\Ell  \vdash  A$ or just $\vdash  A$ if an assertion $A$ is
derivable from the above rules and axioms.
It will be convenient to have equational versions of $({\Box}-{\vee})$ and
$({\Diamond}-{\wedge})$, which can be obtained as theorems of \Ell\ :
\[ \begin{array}{clr}
(D1) & \vdash  \Box ( \phi \vee \psi ) = \Box \phi \vee ( \Box (\phi \vee
\psi ) \wedge \Diamond \psi )  & \\
(D2) & \vdash  \Box \phi \wedge \Diamond \psi = \Box \phi \wedge
\Diamond (\phi \wedge \psi )  & (\psi {\converges}) .
\end{array} \]

We now turn to the question of soundness for our system.
As a first step, we show that our auxiliary predicate $(){\converges}$
works as intended.

\begin{proposition}
\label{auxp}
(i)  $\forall \phi \in \Ell_{\infty \kappa}. \: \phi \converges \;\;
\Longleftrightarrow \;\;  \bot \nvDash_{\kappa} \phi$. 

\noindent (ii) $\forall \phi \in \Ell_{\infty \pi}. \: \phi \converges \;\;
\Longleftrightarrow \;\;  p \models_{\pi} \phi \; \Rightarrow \; C(p) \not=
\{ \bot \}$.
\end{proposition}

\proof\ We prove (i) and (ii) simultaneously by induction on $\phi$.
We consider the two non-trivial cases:

\noindent $\Box \phi$: Assume $(\Box \phi )\converges  \equiv  \phi
\converges$, and $p \models_{\pi} \Box \phi$.
$C(p) = \{ \bot \}$ would then imply $\bot \models_{\kappa} \phi$, but
this is impossible by the induction hypothesis.
For the converse, suppose $(\Box \phi ) \diverges$, i.e. $\phi \diverges$.
Then by induction hypothesis, $\bot \models_{\kappa} \phi$, and hence
$\Omega\ \models_{\pi} \Box \phi$ with $C(\Omega ) = \{ \bot \}$.

\noindent $\Diamond \phi$: Assume $\phi \converges$ and $p
\models_{\pi} \Diamond \phi$.
Then $\bot \nvDash_{\kappa} \phi$, and so there must be $c \in C(p) - \{
\bot \}$ with $c \models_{\kappa} \phi$.
The converse is proved by the same argument as for $\Box \phi$. \qed

\begin{theorem}[Soundness of $\Ell$]
$\vdash  A \;\; \Longrightarrow \;\; \models  A$.
\end{theorem}

\proof\ By a routine induction over proofs.
For illustration, we consider $({\Diamond}-{\wedge})$.
Assume $\psi \converges$ and $p \models_{\pi} \Box \phi \wedge
\Diamond \psi$.
Then $p \models_{\pi} \Diamond \psi$, and so by~\ref{auxp}, $C(p) \not= \{ \bot
\}$ and $\bot \nvDash_{\kappa} \psi$, and there must be $c \in C(p) - \{
\bot \}$ such that $c \models_{\kappa} \psi$.
But then $p \models_{\pi} \Box \phi$ implies that $c \models_{\kappa}
\phi$, and so $p \models_{\pi} \Diamond ( \phi \wedge \psi )$ as required.
\qed

We now turn to the finitary logic $\Ellomega$.
Henceforth we assume that ${\sf Act}$ is countable.
It is then clear that $\Ellomega$ can be made into a countable set by a
suitable choice of canonical representatives of logical equivalence
classes.

Recall that ${\sf Spec} \; \Ellomega$ is the set of {\em prime filters} over
$\Ell_{\omega \pi}$, i.e. subsets $x \subseteq \Ell_{\omega \pi}$
satisfying
\[ \begin{array}{cc}
\bullet & \phi \in x \: \& \: \vdash  \phi \leq \psi \;\; \Rightarrow \;\;
\psi \in x \\
\bullet & \true\ \in x \\
\bullet & \phi , \psi \in x \;\; \Rightarrow \;\; \phi \wedge \psi \in x \\
\bullet & \false\ \not\in x \\
\bullet & \phi \vee \psi \in x \;\; \Rightarrow \;\; \phi \in x \; \mbox{or}
\; \psi \in x .
\end{array} \]

${\sf Spec} \; \Ellomega$ is topologised by taking as basic opens
\[ U_{\phi} \; \equiv \;  \{ x \in {\sf Spec} \; \Ellomega\ : \phi \in x \}
\;\;\; (\phi \in \Ell_{\omega \pi}) , \]
or, equivalently in our context, by taking the Scott topology over the
specialisation order on ${\sf Spec} \; \Ellomega$, which is simply set
inclusion.

Our aim is to prove the following fundamental result, which ahows that
the logic $\Ellomega$ does indeed correspond exactly to the domain 
\Dom\ :

\begin{theorem}[Stone Duality]
\label{bdual}
\Dom\ and \Ellomega\ are Stone duals, i.e.
\[ \begin{array}{rl}
(i) & \Dom\ \; \cong \; {\sf Spec} \; \Ellomega \\
(ii) & K \Omega (\Dom\ ) \; \cong \; (\Ell_{\omega \pi}/{=_{\pi}},
{\leq_{\pi}}/{=_{\pi}}) .
\end{array} \]
\end{theorem}

Here $K \Omega (D)$ is the lattice of compact-open subsets of $\Dom$, while 
\[ ( {\Ell}_{\omega \pi}/{=_{\pi}}, \leq_{\pi}/{=_{\pi}}) \]
is the {\em Lindebaum algebra} of $\Ellomega$. 
Since $\Dom$ is coherent, (i) and (ii) are indeed equivalent (\cite{Joh82}).

The Stone Duality Theorem is entirely analogous to Theorem \ref{sdual}, and our proof strategy is identical.
However, some of the technical details are more complex; in particular, the syntactic identification of primes is less obvious than for Scott domains, since primes are no longer preserved under meets.

We begin by defining a normal form for $\Ellomega$.
\begin{definition}
{\rm (i) $\phi$ is in {\em strong disjunctive normal form} (SDNF) if it has the form $\bigvee_{i \in I} \phi_{i}$, where each $\phi_{i}$ is in {\em prime normal form} (PNF).

\noindent (ii) $\phi$ is in PNF if it has one of the forms
\begin{itemize}
\item $\bigwedge_{i \in I} \Diamond a_{i} (\phi_{i})$, where each $\phi_{i}$ is in PNF.
\item $\Box \bigvee_{i \in I} a_{i} ( \phi_{i}) \; \wedge \; \bigwedge_{j \in J} \Diamond b_{j}(\psi_{j})$, where
\begin{enumerate}
\item Each $\phi_{i}$ and $\psi_{j}$ is in PNF.
\item $\forall i \in I. \, \exists j \in J. \, \vdash  b_{j}(\psi_{j}) \leq 
a_{i}(\phi_{i})$.
\item $\forall j \in J. \, \exists i \in I. \, \vdash  b_{j}(\psi_{j}) \leq 
a_{i}(\phi_{i})$.
\end{enumerate}
\end{itemize}}
\end{definition}
We call (2) and (3) the {\em convexity conditions} (note the resemblance to the Egli--Milner ordering).

The combinatorics are concentrated in the following 
\begin{theorem}[SDNF]
\label{SDNF}
For every $\phi \in \Ell_{\omega \pi}$, there is (effectively) a $\psi$ in SDNF such that 
\[ \vdash  \phi =_{\pi} \psi . \]
\end{theorem}

\proof\ By induction on ${\sf md}(\phi )$.
The idea is to form a sequence of ``transformations''
\[ \phi \equiv \phi_{0} \rightsquigarrow \phi_{1} \rightsquigarrow \cdots 
\rightsquigarrow \phi_{n} \]
such that
\[ \begin{array}{ll}
(1) & \vdash  \phi_{i} = \phi_{i+1} \;\;\; (0 \leq i < n) \\
(2) & {\sf md}(\phi_{i+1}) \leq {\sf md}(\phi_{i}) \;\;\; (0 \leq i < n) \\
(3) & \phi_{n} \; \mbox{is in SDNF.}
\end{array} \]
(Condition (2) is needed to keep the induction going.)
To keep the notation bearable, we shall omit indices in conjunctions and disjunctions, writing e.g. $\bigvee \{ \phi \}$.

Firstly, using the distributive lattice laws we can transform $\phi_{0}$ into
\begin{equation}  
\label{SDNF1}
\bigvee \{ \bigwedge \{ \Box \bigwedge \{ \bigvee \{ a(\phi ) \} \} \} \; \wedge \; \bigwedge \{ \Diamond \bigwedge \{ \bigvee \{ b(\psi ) \} \} \} \}
\end{equation}
Using $({\Box}-{\wedge})$ in the outwards direction for each $\Box$-conjunct in \ref{SDNF1}, and the distributive law and then $({\Diamond}-{\vee})$, followed by the distributive law again, in each $\Diamond$-conjunct, we otain
\begin{equation} 
\label{SDNF2}
\bigvee \{ \bigwedge \{ \Box \bigvee \{ a(\phi ) \} \} \; \wedge \; \bigwedge \{ \Diamond \bigwedge \{ b(\psi ) \} \} \}
\end{equation}
Now for each non-empty conjunction
\[ \bigwedge \{ \Box \bigvee \{ a(\phi ) \} \} \]
in \ref{SDNF2}, we can use $({\Box}-{\wedge})$, the distributive law, and $(a-{\wedge})$ $(i)$ or $(ii)$; similarly, inside each $\Diamond \bigwedge \{ b(\psi ) \}$ we can use $({\Diamond}-{\true})$ if the conjunction is empty, and otherwise $(b-{\wedge})$ $(i)$ or $(ii)$ (with further applications of $({\Diamond}-{\vee})$ and the distributive laws as in the previous step if $(b-{\wedge})(ii)$ is applicable), to obtain
\begin{equation} 
\label{SDNF3}
\bigvee \{ \theta \}
\end{equation}
where each $\theta$ is in one of the forms
\begin{equation} 
\label{SDNF4}
\bigwedge \{ \Diamond b(\psi ) \}
\end{equation}
or
\begin{equation} 
\label{SDNF5}
\Box \bigvee \{ a(\phi ) \} \; \wedge \; \bigwedge \{ \Diamond b(\psi ) \}
\end{equation}
Since we have not increased modal depth in obtaining \ref{SDNF3}, we can apply the inductive hypothesis to each $\phi$ and $\psi$ to obtain $\bigvee \{ \phi' \}$, $\bigvee \{ \psi' \}$ with each $\phi'$ and $\psi'$ in PNF.
Using  $(a-{\vee})$, $({\Diamond}-{\vee})$ and the distributive laws, we can thus obtain a formula of the same form as \ref{SDNF3}, in which each $\phi$ and $\psi$ in \ref{SDNF4} and \ref{SDNF5} is in PNF.

At this point, our formula \ref{SDNF3} can only fail to be in SDNF because of disjuncts \ref{SDNF5} which do not satisfy the convexity conditions
\begin{itemize}
\item For each $a(\phi )$, for some $b(\psi )$: $\vdash  b(\psi ) \leq a(\phi )$.
\item For each $b(\psi )$, for some $a(\phi )$: $\vdash   b(\psi ) \leq a(\phi )$.
\end{itemize}

Our strategy is to remove any failures of these two conditions, using our derived equations $(D1)$ and $(D2)$ respectively. We begin with the first condition.
We argue by induction on $(m, n)$ in the lexicographic ordering on $\omega \times \omega$, where:
\begin{itemize}
\item $m$ is the maximum number of $a(\phi )$ occurring in one of the disjuncts \ref{SDNF5} of our formula \ref{SDNF3} such that there is no $b(\psi )$ with $\vdash  b(\psi ) \leq a(\phi )$.
\item $n$ is the number of disjuncts attaining this maximum.
\end{itemize}
If $m = 0$, there is nothing to prove.
Otherwise, choose such an $a(\phi )$ in one of the maximal disjuncts.
We can apply $(D1)$ to 
\[ \Box \bigvee \{ a' (\phi' ) \} \; \vee \; a(\phi ) \]
to obtain
\begin{equation} 
\label{SDNF6}
\Box \bigvee \{ a' ( \phi' ) \} \;\; \vee \;\; [ \Box ( \bigvee \{ a' ( \phi' ) \} \vee a(\phi )) \wedge \Diamond a(\phi ) ]
\end{equation}
We can then use the distributive law to obtain a new formula of the form \ref{SDNF3} to which the inner induction hypothesis can be applied, since the first disjunct in \ref{SDNF6} has jettisoned $a(\phi )$, while the second disjunct evidently contains a $\Diamond b(\psi )$ such that $\vdash  b(\psi ) \leq a(\phi )$, namely $a(\phi )$ itself.

The final stage is to remove failures of the second condition.
We argue by induction in the same way as for the previous stage.
Suppose we are given a $b(\psi )$ in \ref{SDNF5} with no $a(\phi )$ such that $\vdash  b(\psi ) \leq a(\phi )$.
Firstly, we note that
$\psi \diverges$  implies $\vdash  \psi = \true$, 
which is easily proved by induction on $\psi$.
Hence if $\psi \diverges$, we can use $({\Diamond}-{\true})$ to eliminate the conjunct $\Diamond b(\psi )$.
Otherwise, we can use $(D2)$ to obtain
\begin{equation} 
\label{SDNF7}
\Box \bigvee \{ a(\phi ) \} \; \wedge \; \Diamond [ b(\psi ) \wedge \bigvee \{ a(\phi ) \} ] \; \wedge \; \bigwedge \{ \Diamond b' (\psi' ) \}
\end{equation}
Now we can use the distributive law inside the second main conjunct in \ref{SDNF7}, followed by $(a-{\wedge})$, $({\Diamond}-{\vee})$, and the distributive law again.
In this way, the disjunct \ref{SDNF7} of our main formula is replaced by the disjunction of all those formulae
\begin{equation} 
\label{SDNF8}
\Box \bigvee \{ a(\phi ) \} \; \wedge \; \Diamond b(\phi' \wedge \psi ) \; \wedge \; \bigwedge \{ \Diamond b' ( \psi' ) \}
\end{equation}
for $a'(\phi' ) \in \{ a(\phi ) \}$ with $a' = b$.
For each such $\phi' \wedge \psi$, we can apply the outer induction hypothesis to obtain $\bigvee \{ \theta' \}$ with each $\theta'$ in PNF.
Applying $(b-{\vee})$, $({\Diamond}-{\vee})$ and the distributive laws as before, we obtain disjuncts of the form
\begin{equation} 
\label{SDNF9}
\Box \bigvee \{ a(\phi ) \} \; \wedge \; \Diamond b(\theta' ) \; \wedge \;  \bigwedge \{ \Diamond b' ( \psi' ) \}
\end{equation}
Since
\[ \vdash \; \theta' \leq \bigvee \{ \theta' \} = \phi' \wedge \psi \leq \phi' , \]
we can apply the inner induction hypothesis to \ref{SDNF9}.
This completes the process of transforming $\phi$ into SDNF. \qed

We shall now prove that formulae in PNF denote primes in $K \Omega (\Dom )$.
\begin{proposition}
\label{psoun}
For all $\phi$ in PNF there exsists $k(\phi ) \in {\cal K}(\Dom )$ such that:
\[ \forall d \in \Dom . \; d \models \phi \;\; \Longleftrightarrow \;\; k(\phi ) \sqsubseteq d . \]
\end{proposition}

\proof\ We define $k(\phi )$ (which must clearly be unique) by induction on $\phi$:
\[ \bullet \;\; k(\bigwedge_{i \in I} \Diamond a_{i}(\phi_{i} )) \; \equiv \; \biguplus_{i \in I} \lsing \ltuple a_{i} , k(\phi_{i}) \rtuple \rsing \uplus \lsing \bot \rsing \]
\[ \bullet \;\; k(\Box \bigvee_{i \in I} a_{i}(\phi_{i}) \; \wedge \; \bigwedge_{j \in J} \Diamond b_{j}(\psi_{j})) \; \equiv \]
\[  \;\;\;\; \biguplus_{i \in I} \lsing \ltuple a_{i}, k(\phi_{i} ) \rtuple \rsing \; \uplus \; \biguplus_{j \in J} \lsing \ltuple b_{j}, k(\psi_{j}) \rtuple \rsing . \]

We shall prove the proposition by induction on $\phi$.
Note that in the statement of the proposition, we are viewing $\Dom$ as a transition system, according to~\ref{ifs}.
With our convention of eliding the isomorphisms between $\Dom$ and $P^{0} [ \sum_{a \in {\sf Act}} \Dom ]$, we have:
$d = C(d)$, $(d \in \Dom )$. 

\noindent Case 1:  $\phi \equiv \bigwedge_{i \in I} \Diamond a_{i}(\phi_{i} )$.
\[ \begin{array}{ll}
\bullet & d \models \bigwedge_{i \in I} \Diamond a_{i}(\phi_{i}) \\
\Longleftrightarrow & \forall i \in I . \, \exists \ltuple a_{i}, d_{i} \rtuple \in d . \, d_{i} \models \phi_{i} \\
\Longleftrightarrow & \forall i \in I . \, \exists \ltuple a_{i}, d_{i} \rtuple \in d . \, k(\phi_{i}) \sqsubseteq d_{i} \;\; \mbox{by induction hypothesis} \\
\Longleftrightarrow & k(\phi ) \sqsubseteq d .
\end{array} \]

\noindent Case 2: $\phi  \equiv  \Box \bigvee_{i \in I} a_{i}(\phi_{i})  \wedge  \bigwedge_{j \in J} \Diamond b_{j}(\psi_{j})$.
Let $\Phi = \{ a_{i}(\phi_{i}) : i \in I \} \cup \{ b_{j}(\psi_{j}) : j \in J \}$.
\[ \begin{array}{ll}
\bullet & d \models \phi \\
\Longleftrightarrow & \forall \ltuple a, d' \rtuple \in d. \, \exists i \in I . \, a = a_{i} \: \& \: d' \models \phi_{i} \\
& \& \: \bot \not\in d \: \& \: \forall j \in J . \, \exists \ltuple b_{j}, d_{j} \rtuple \in d . \, d_{j} \models \psi_{j} \\
\Longleftrightarrow & \forall \ltuple a, d' \rtuple \in d. \, \exists  a(\theta ) \in \Phi . \, d' \models \theta \\
& \& \: \bot \not\in d \: \& \: \forall a(\theta ) \in \Phi . \, \exists \ltuple a, d' \rtuple \in d . \, d \models \theta \\
& \mbox{by the convexity conditions and the Soundness Theorem,} \\
\Longleftrightarrow & k(\phi ) \sqsubseteq d, \; \mbox{by induction hypothesis.} \;\;\; \qed
\end{array} \]

\begin{theorem}[Prime Completeness]
For all $\phi$, $\phi'$ in PNF:
\[ \Dom  \models  \phi \leq \phi' \;\; \Longrightarrow \;\; \Ell  \vdash  \phi \leq \phi' . \]
\end{theorem}

\proof\ By 4.7,
\[ \Dom  \models  \phi \leq \phi' \;\; \Longleftrightarrow \;\; k(\phi' ) \sqsubseteq k(\phi ) . \]
Suppose then that $k(\phi' ) \sqsubseteq k(\phi )$.
We argue by induction on $\phi$.
There are a number of cases, according to the forms of $\phi$ and $\phi'$.
We consider the case
\begin{eqnarray*}
\phi & \equiv & \Box \bigvee_{i \in I} a_{i}(\phi_{i}) \; \wedge \; \bigwedge_{j \in J} \Diamond b_{j}(\psi_{j}) , \\
\phi' & \equiv & \Box \bigvee_{i' \in I'} a_{i'}(\phi_{i'}) \; \wedge \; \bigwedge_{j' \in J'} \Diamond b_{j'}(\psi_{j'}) .
\end{eqnarray*}
\[ \begin{array}{ll}
\bullet & k(\phi' ) \sqsubseteq k(\phi ) \\
\Longleftrightarrow & \forall j' \in J' . \, \exists j \in J . \, b_{j} = b_{j'} \: \& \: k(\psi_{j'}) \sqsubseteq k(\psi_{j}) \\
& \mbox{}\& \; \forall i \in I . \, \exists i' \in I' . \, a_{i} = a_{i'} \: \& \: k(\phi_{i'}) \sqsubseteq k(\phi_{i}), \\
& \mbox{by the convexity conditions, Soundness, and \ref{psoun}} \\
\Longrightarrow & \forall j' \in J' . \, \exists j \in J . \, \vdash  b_{j}(\psi_{j}) \leq b_{j'}(\psi_{j'}) \\
& \mbox{} \& \; \forall i \in I . \, \exists i' \in I' . \, \vdash \: a_{i}(\phi_{i}) \leq a_{i'}(\phi_{i'}), \\
& \mbox{by the induction hypothesis,} \\
\Longrightarrow & \vdash \: \phi \leq \phi' . \;\;\; \qed
\end{array} \]

We can now use the same arguments as in Chapter~3 T7 to prove

\begin{theorem}[Completeness]
For all $\phi , \psi \in \Ellomega$:
\[ \Dom  \models  \phi \leq \psi \;\; \Longrightarrow \;\; \Ellomega  \vdash  \phi \leq \psi . \]
\end{theorem}

We now establish a converse to \ref{psoun}.
\begin{theorem}[Definability]
For all $d \in {\cal K}(\Dom )$, for some $\phi$ in PNF, $k(\phi ) = d$.
\end{theorem}

\proof\ We define $\phi (d)$ by induction on the construction of $d$ according to~\ref{felts}:
\begin{eqnarray*}
\phi (\biguplus_{i \in I} \lsing \ltuple a_{i}, d_{i} \rtuple \rsing \; \uplus \; \lsing \bot \rsing ) & \equiv & \bigwedge_{i \in I} \Diamond a_{i}(\phi (d_{i} )) \\
\phi ( \biguplus_{i \in I} \lsing \ltuple a_{i}, d_{i} \rtuple \rsing ) & \equiv & \Box \bigvee_{i \in I} a_{i} (\phi (d_{i} )) \; \wedge \; \bigwedge_{i \in I} \Diamond a_{i}(\phi (d_{i})).
\end{eqnarray*} 
Note in particular that $\phi (\emptyset ) = \Box \false$.
It is easily verified that $\phi (d)$ is in PNF and that $k(\phi (d)) = d$. \qed

The Duality Theorem is an immediate consequence of Soundness, Completeness and Definability, just as in Chapter~3 T8.

Combining Soundness and Completeness we obtain
\begin{theorem}[Completeness for $\Ellomega$]
Let {\bf C} be any class of transition systems containing $\Dom$.
Then for $\phi , \psi \in \Ellomega$:
\[ {\bf C}  \models \phi \leq \psi \;\; \Longleftrightarrow \;\; \Dom  \models  \phi \leq \psi \;\; \Longleftrightarrow \;\; \Ell  \vdash  \phi \leq \psi . \]
\end{theorem}

\section{Applications of the Domain Logic}
We shall now use domain logic to study bisimulation.
Our results in this section can be grouped under four main headings:
\begin{enumerate}
\item Comparisons with Hennessy-Milner logic 
\item Characterisation Theorems
\item Finitary Transition Systems
\item Universal Semantics
\end{enumerate}
Of these, (1) and (2) will confirm the appropriateness of our definitions, while (3) and (4) will represent a distinctive payoff for our approach.

\subsection*{Comparison with Hennessy-Milner logic}
We begin with some technicalities on normal forms.
\begin{definition}
{\rm We define a class of normal forms ${\sf N} \Ellinfty \subseteq \Ell_{\infty \pi}$ inductively as follows:}
\[ \bullet \;\; \frac{ \{\phi_{i} \in {\sf N} \Ellinfty \}_{i \in I}}{\bigwedge_{i \in I} \phi_{i}, \bigvee_{i \in I} \phi_{i} \in {\sf N} \Ellinfty} \]
\[ \bullet \;\; \frac{\phi \in {\sf N} \Ellinfty\ , \;\; a \in {\sf Act}}{\Diamond a(\phi ) \in {\sf N} \Ellinfty} \]
\[ \bullet \;\; \frac{ \{ \phi_{i} \in {\sf N} \Ellinfty \}_{i \in I}, \;\; \{ a_{i} \in {\sf Act} \}_{i \in I} \;\; \{ i \not= j \; \Rightarrow \; a_{i} \not= a_{j} \}_{i, j \in I}}
{\Box \bigvee_{i \in I} a_{i} ( \phi_{i}) \in {\sf N} \Ellinfty} \]
\end{definition}

\begin{lemma}[Normal Forms]
\label{hmnf}
For all $\phi \in \Ell_{\infty \pi}$, for some $\psi \in {\sf N} \Ellinfty$:
\[ \Ellinfty  \vdash  \phi = \psi . \]
\end{lemma}

\proof\ By induction on ${\sf md}(\phi )$.
We consider the two non-trivial cases.

\noindent $\Diamond \phi$: In this case, using the distributive lattice laws there is $\phi'$ of the form
\[ \bigvee_{i \in I} \bigwedge_{j \in J_{i}} a_{ij}(\phi_{ij}) \]
such that $\vdash  \phi = \phi'$, and ${\sf md}(\phi' ) \leq {\sf md}(\phi)$.
By the induction hypothesis, for each $\phi_{ij}$ there is $\phi'_{ij} \in {\sf N} \Ellinfty$ such that $\vdash  \phi_{ij} = \phi'_{ij}$.
Using $(a-{\leq})$ and $({\Diamond}-{\leq})$, we have
\begin{equation} 
\label{nf1}
\vdash  \Diamond \phi = \Diamond \bigvee_{i \in I} \bigwedge_{j \in J_{i}} a_{ij}(\phi_{ij}) .
\end{equation}
Now for each $i \in I$, there are three cases:
\begin{enumerate}
\item $J_{i} = \varnothing$. In this case, $\vdash \Diamond \phi = \Diamond \true$, and we can use $({\Diamond}-\true\ )$ to obtain a normal form.
\item $\exists j_{1}, j_{2} \in J_{i}$, $a_{j_{1}} \not= a_{j_{2}}$.
In this case, we can use $(a-{\wedge})$ to delete the $i$'th disjunct in the RHS of~\ref{nf1}.
\item $\{ a_{ij} : j \in J_{i} \} = \{ a \}$, for some $a \in {\sf Act}$. In this case, we can use $(a-{\wedge})(i)$.
\end{enumerate}
In this way, we obtain {\em either}
\[ \vdash \Diamond \phi = \true , \]
if case (1) is ever applicable, {\em or}
\[ \vdash \Diamond \phi = \Diamond \bigvee_{i' \in I'} a_{i'}(\psi_{i'}) \;\;\; (\psi_{i'} \in {\sf N} \Ellinfty ). \]
In the latter case, we can apply $({\Diamond}-{\vee})$ to get a normal form.

\noindent $\Box \phi$: Similarly to the previous case, we have
\[ \vdash \Box \phi = \Box \bigwedge_{i \in I} \bigvee_{j \in J_{i}} a_{ij} (\phi_{ij}) \;\;\; (\phi_{ij} \in {\sf N} \Ellinfty ). \]
We can then use $({\Box}-{\wedge})$ to get
\[ \vdash \Box \phi =  \bigwedge_{i \in I} \Box \bigvee_{j \in J_{i}} a_{ij} (\phi_{ij}) . \]
Now if we partition each $J_{i}$ by $\sim_{i}$, with
\[ j \sim_{i} k \;\; \Longleftrightarrow \;\; a_{ij} = a_{ik} \;\;\; (j, k \in J_{i}),  \]
we have
\[ \vdash \Box \phi =  \bigwedge_{i \in I} \Box \bigvee_{[j] \in J_{i}/{\sim_{i}}} ( \bigvee_{k \in [j]} a_{ij} (\phi_{ik}))  \]
using the lattice laws; we can then apply $(a-{\vee})$ to get a normal form. \qed

\begin{definition}
\label{tfuns}
{\rm We define translation functions}
\[ ( \cdot )^{\ast} : {\rm HML}_{\infty} \longrightarrow {\sf N} \Ellinfty\ , \]
\[ ( \cdot )^{\dagger} : {\sf N} \Ellinfty  \longrightarrow {\rm HML}_{\infty} . \]
\[ \begin{array}{lll}
( \bigwedge_{i \in I} \phi_{i})^{\ast} & = & \bigwedge_{i \in I} (\phi_{i})^{\ast} \\
( \bigvee_{i \in I} \phi_{i})^{\ast} & = & \bigvee_{i \in I} (\phi_{i})^{\ast} \\
( \ltuple a \rtuple \phi )^{\ast} & = & \Diamond a(\phi^{\ast}) \\
( [ a ] \phi )^{\ast} & = & \Box  a((\phi )^{\ast}) \vee \bigvee \{ b(\true ) : b \in {\sf Act} - \{ a \} \} ) \\
( \bigwedge_{i \in I} \phi_{i})^{\dagger} & = & \bigwedge_{i \in I} (\phi_{i})^{\dagger} \\
( \bigvee_{i \in I} \phi_{i})^{\dagger} & = & \bigvee_{i \in I} (\phi_{i})^{\dagger} \\
( \Diamond a(\phi ))^{\dagger} & = &  \ltuple a \rtuple (\phi )^{\dagger} \\
( \Box \bigvee_{i \in I} a_{i}(\phi_{i}))^{\dagger} & = & \bigwedge_{i \in I} [ a_{i} ] (\phi_{i})^{\dagger} \; \wedge \; \bigwedge \{ [ b ] \false : b \in {\sf Act} - \{ a_{i} : i \in I \} \}
\end{array} \]
\end{definition}
The following is easily verified.
\begin{proposition}
\label{faitht}
For all $\phi \in {\rm HML}_{\infty}, \psi \in {\sf N} \Ellinfty$:
\[ \begin{array}{rrcl}
(i) & {\sf md}(\phi ) & = & {\sf md}(\phi^{\ast}) \\
(ii) & {\sf md}(\psi ) & = & {\sf md}(\psi^{\dagger}) \\
(iii) & p \: \models \: \phi & \Longleftrightarrow & p \: \models \phi^{\ast} \\
(iv) &  p \: \models \: \psi & \Longleftrightarrow & p \: \models \psi^{\dagger} .
\end{array} \]
\end{proposition}

As an immediate consequence of this Proposition together with~\ref{hmnf}, we have

\begin{theorem}[Comparison Theorem (Infinitary Case)]
For $p, q \in {\rm Proc}$ in any transition system,  $A \subseteq {\sf Act}$ and $\lambda \in {\sf Ord}$:
\[ \Ell^{(A, \lambda )}_{\infty}(p) \subseteq \Ell^{(A, \lambda )}_{\infty}(q) \;\; \Longleftrightarrow \;\; {\rm HML}^{(A, \lambda )}_{\infty}(p) \subseteq {\rm HML}^{(A, \lambda )}_{\infty}(q) . \]
\end{theorem}

Thus in the infinitary case, $\Ellinfty$ determines the same preorder on processes as ${\rm HML}_{\infty}$.
However, when {\sf Act} is infinite this does {\em not} cut down to a corresponding result for the finitary case, since our
translation functions introduce infinite disjunctions in translating $[a]$, and infinite conjunctions in translating $\Box$, even for finite formulas.
Our general considerations on observability in Chapter 2 suggest that the introduction of infinite conjunctions is more serious, and indicates a weakness of expressive power in ${\rm HML}_{\infty}$ as an ``observational logic''.
This is in keeping with our remarks at the end of Section~2.
In fact, our translation functions suggest an appropriate way of extending ${\rm HML}_{\infty}$ so as to render it equivalent to $\Ellomega$.
This will be the content of a second Comparison Theorem which we will prove later in this section, when we have some additional machinery at our disposal.

\subsection*{Characterisation Theorems}
Combining the Comparison Theorem with the Modal Characterisation Theorem~\ref{mct}, we have:
\begin{theorem}[Characterisation Theorem for \Ellinfty]
\label{linfct}
With notation as in the previous Theorem,
\[ p {\preord}_{\lambda} q \;\; \Longleftrightarrow \;\; {\Ell}^{({\sf Act}, \lambda )}_{\infty}(p) \subseteq {\Ell}^{({\sf Act}, \lambda )}_{\infty}(q) \]
and therefore
\[ p \preord^{B} q \;\; \Longleftrightarrow \;\; {\Ellinfty}(p) \subseteq {\Ellinfty}(q) . \]
\end{theorem}

We now turn to the question of finding a Characterisation Theorem for $\Ellomega$.
Intuitively, \Ellomega\ represents finitely observable properties of processes, hence should correspond to the ``finitely observable part'' of bisimulation.
If we accept the finite synchronisation trees ${\sf ST}_{\omega}$ as a suitable notion of {\em finite process}, we can use them to determine the {\em algebraic} part of the bisimulation preorder, in the sense e.g. of \cite{Gue81}.
\begin{definition}
{\rm The {\em finitary preorder} $\preord^{F}$ is defined on any transition system by:}
\[ p \preord^{F} q \; \equiv \; \forall t \in {\sf ST}_{\omega}. \: t \preord^{B} p \; \Rightarrow \; t \preord^{B} q . \]
\end{definition}
Our aim is to prove
\begin{theorem}[Characterisation Theorem for $\Ellomega$]
\label{lct}
With notation as in the previous Theorem,
\[ p \preord^{F} q \;\; \Longleftrightarrow \;\; \Ellomega (p) \subseteq \Ellomega (q). \]
\end{theorem}

We will need a few auxiliary results which also have some independent interest.
\begin{definition}
{\rm The {\em height} of a synchronisation tree is defined by:}
\[ {\sf ht}(\sum_{i \in I} a_{i}t_{i} \; [+ \Omega ] ) = \sup \: \{ {\sf ht}(t_{i}) : i \in I \} + 1 \]
\end{definition}

\begin{lemma}
\label{htl}
For any synchronisation tree $T \in {\sf ST}_{\infty}$, ${\sf ht}(T) < \lambda$ implies
\[ T \preord^{B} p \;\; \Longleftrightarrow \;\; T \preord_{\lambda} p . \]
\end{lemma}

\proof\ The left-to-right implication is immediate; the converse is an easy induction on ${\sf ht}(T)$. \qed

In particular, we see that for a finite synchronisation tree $t \in {\sf ST}_{\omega}$, $t \preord^{B} p \; \Leftrightarrow \; t\preord_{\omega} p$.  
Thus we have the inclusions
\[ {\preord^{B}} \subseteq {\preord_{\omega}} \subseteq {\preord^{F}} . \]
In general, these inclusions are strict.
\subsection*{Examples}
(1) ${\preord^{B}} \not= {\preord_{\omega}}$.
\[ p \equiv a^{\omega} + \Omega , \;\;\;\; q \equiv \sum_{k \in \omega} a^{k} \Oh\ + \Omega \]
Then $p \preord_{\omega} q$, but $p \npreord_{\omega + 1} q$.

\noindent (2) ${\preord_{\omega}} \not= {\preord^{F}}$.
\begin{eqnarray*}
p & \equiv & a( \sum_{n \in \omega} b_{n} \Oh + \Omega ) + \Omega \\
q & \equiv & \sum_{n \in \omega} a(\sum_{m \in \omega - \{ n \} } b_{n} \Oh + \Omega ) + \Omega
\end{eqnarray*}
Then $p \preord^{F} q$, but $p \npreord_{2} q$.

These examples gain in significance because all the processes involved can be defined in finitary calculi, in particular SCCS, as we shall see in the next section.

\begin{lemma}[Sort Lemma]
\label{stl}
In any transition system, let $p, q \in {\rm Proc}$, ${\sf sort}(p) \subseteq A \subseteq {\sf Act}$, $\lambda \in {\sf Ord}$. Then
\[ p \npreord_{\lambda} q \;\; \Longrightarrow \;\; {\Ell}^{(A, \lambda )}_{\infty}(p) \not\subseteq {\Ell}^{(A, \lambda )}_{\infty}(q) . \]
\end{lemma}

\proof\ By induction on $\lambda$.
We assume $ p \npreord_{\lambda} q$, and must construct $\phi \in {\Ell}^{(A, \lambda )}_{\infty}(p) - {\Ell}^{(A, \lambda )}_{\infty}(q)$.
There are three cases.

\noindent (1) $\labarrow{p}{a}{p'}$ and for all $q'$, $\labarrow{q}{a}{q'}$ implies $p' \npreord_{\alpha} q'$ for some $\alpha < \lambda$.
By induction hypothesis, for each such $q$ there is $\phi_{q'} \in {\Ell}^{(A, \alpha )}_{\infty}(p') - {\Ell}^{(A, \alpha )}_{\infty}(q')$.
Now define
\[ \phi \; \equiv \; \Diamond a ( \bigwedge \{ \phi_{q'} : \labarrow{q}{a}{q'} \} ) . \]

\noindent (2) $p \converges$ and $q \diverges$.
Let $\phi \equiv \Box \bigvee \{ a(\true ) : \exists p' . \, \labarrow{p}{a}{p'} \}$.

\noindent (3) $p \converges$, $q \converges$, $\labarrow{q}{a}{q'}$, and for all $p'$, $\labarrow{p}{a}{p'}$ implies $p' \npreord_{\alpha} q'$ for some $\alpha < \lambda$.
Define $\phi_{p'}$ similarly to case (1). Then we define
\[ \phi \; \equiv \; \Box ( \bigvee \{ a(\phi_{p'}) : \labarrow{p}{a}{p'} \} \; \vee \; \bigvee \{ b(\true ) : b \not= a \: \& \: \exists r . \, \labarrow{p}{a}{r} \} ) . \;\;\; \qed \]
Note that this result is stronger than the Modal Characterisation Theorem~\ref{mct} for Hennessy-Milner logic, since we only require ${\sf sort}(p) \subseteq A$.
This is significant in the light of the example at the end of Section 2.

\begin{proposition}
\label{finbis}
For all $t \in {\sf ST}_{\omega}$:
\[ t \preord^{B} p \;\; \Longleftrightarrow \;\; {\Ellomega}(t) \subseteq {\Ellomega}(p) . \]
\end{proposition}

\proof\ Combining \ref{htl} and \ref{stl}, we see that
\[ t \preord^{B} p \;\; \Longleftrightarrow \;\; {\Ell}^{(A, k )}_{\infty}(t)  \subseteq {\Ell}^{(A, k )}_{\infty}(p) , \]
where $A = {\sf sort}(t)$ and $k = {\sf ht}(t)$.
Since $A$ and $k$ are both finite, $ {\Ell}^{(A, k )}_{\infty}$ is finite up to logical equivalence (i.e. the Lindenbaum algenbra is finite).
Thus each formula in $ {\Ell}^{(A, k )}_{\infty}$ is equivalent to one in $\Ellomega$, and the proposition is proved. \qed

We need one more auxiliary result, which will in fact be a consequence of our work on SCCS in the next section.
Firstly, we define a map from prime normal forms to finite synchronisation trees
\[ {\sf st} : \mbox{PNF} \rightarrow {\sf ST}_{\omega} \]
as follows:
\[ \begin{array}{lll}
{\sf st}( \bigwedge_{i \in I} \Diamond a_{i}(\phi_{i})) & \equiv & {\sum_{i \in I} a_{i} {\sf st}(\phi_{i}) } + \Omega \\
{\sf st}( \Box \bigvee_{i \in I} a_{i}(\phi_{i}) \; \wedge \; \bigwedge_{j \in J} \Diamond b_{j}(\psi_{i})) & \equiv & {\sum_{i \in I} a_{i} {\sf st}(\phi_{i}) } + { \sum_{j \in J} b_{j} {\sf st}(\psi_{j}) }  .
\end{array} \]
Now analogously to~\ref{psoun} we have
\begin{proposition} 
\label{stprop}
For all $\phi$ in PNF, and $p \in {\rm Proc}$ in any transition system:
\[ p  \models  \phi \;\; \Longleftrightarrow \;\; {\sf st}(\phi ) \preord^{B} p . \]
\end{proposition}
The proof is entirely analogous to \ref{psoun}.

We can now prove~\ref{lct}.
Firstly, ${\Ellomega}(p) \subseteq {\Ellomega}(q)$ implies $p \preord^{F} q$, by \ref{finbis}.
For the converse, assume $p \preord^{F} q$ and $p \models \phi$, $(\phi \in \Ellomega\ )$.
By the SDNF Theorem~\ref{SDNF},
\[ \begin{array}{llr}
\bullet & \vdash \phi = \bigvee_{i \in I} \phi_{i} & (\phi_{i} \in \mbox{PNF}) \\
\Longrightarrow & \exists i \in I . \, p \models \phi_{i} & \\
\Longrightarrow & {\sf st}(\phi_{i}) \preord^{B} p & \ref{stprop} \\
\Longrightarrow & {\sf st}(\phi_{i}) \preord^{B} q & p \preord^{F} q \\
\Longrightarrow & q \models \phi_{i} & \ref{stprop} \\
\Longrightarrow & q \models \phi . & \qed\ 
\end{array} \]
\subsection*{Finitary Transition Systems}

We now embark on our next topic.
The various finiteness conditions on transition systems defined in section~2 reflect attempts to capture features of finitary processes.
Nowever, none of these conditions seems to capture exactly the right class of systems unless we make some unwelcome assumptions such as that the set of actions is finite.
We shall adopt what seems to be a novel approach, of using our program logic to axiomatize a class of systems which we propose as the finitary ones.
Our axiomatisation consists of two schemes over $\Ellinfty$.

\noindent {\bf Notation.} ${\sf Fin}(I)$ is the set of finite subsets of $I$.
\begin{itemize}
\item The axiom scheme of {\em bounded non-determinacy}:
\[ \mbox{(BN)} \;\; \Box \bigvee_{i \in I} \phi_{i} \leq \bigvee_{J \in {\sf Fin}(I)} \Box \bigvee_{j \in J} \phi_{j} \;\;\; (\phi_{i} \in \Ellomega\ ). \]
\item The axiom scheme of {\em finite approximability}:
\[ \mbox{(FA)} \;\; \bigwedge_{J \in {\sf Fin}(I)} \Box \bigwedge_{j \in J} \phi_{j}  \leq \Diamond \bigwedge_{i \in I} \phi_{i} \;\;\; (\phi_{i} \in \Ellomega\ ). \]
\end{itemize}

Note that these axioms are duals.
Since the opposite entailments are theorems of {\Ellinfty}, we shall in fact use (BN) and (FA) to denote the corresponding {\em equations}.
The axioms could equivalently be formulated as: $\Box$~preserves directed joins, $\Diamond$~preserves filtered meets.

What are the intuitions behind these axioms?
(BN) is (thinking of each process as the set of its capabilities and each $\phi_{i}$ as an open set) exactly a statement of {\em compactness}; the link between compactness and the computational notion of bounded non-determinacy is well-known from the literature on powerdomains \cite{PloLN,Smy83}.

The axiom of finite approximability is less familiar from either the topological or the computer science literature.
It is best understood as a logical (or localic) expression of the idea that only {\em closed} sets are taken as elements of a finitary powerdomain construction (or, better put, that from the point of view of finite observability we cannot distinguish between a set and its closure).
The best way to get a more precise understanding is probably to read the proof of the next Theorem.

The duality between the two axioms is reminiscent of the discussion of finite {\em breadth} (BN) and finite {\em length} (FA) limitations of testing in \cite{Abr83a}.

\begin{definition}
\label{fts}
{\rm A transition system is {\em finitary} if it satisfies (all instances of) (BN) and (FA).
The class of finitary transition systems is denoted {\bf FTS}.}
\end{definition}

As a first step, we shall give a substantive example of a finitary transition system.
As we will see, it is actually the best possible example.
\begin{theorem}
\Dom\ is a finitary transition system.
\end{theorem}

\proof\ By the Duality Theorem~ \ref{bdual}, we have a map
\[ \lsem \cdot \rsem : \Ell_{\omega \pi} \longrightarrow K \Omega (\Dom ) \]
\[ \lsem \phi \rsem  \equiv  \{ d \in \Dom : d \models \phi \} . \]
Now for $d \in \Dom$,
\[ d \models \Box \bigvee_{i \in I} \phi_{i}  \;\; \Longrightarrow \;\; d \models \bigvee_{J \in {\sf Fin}(I)} \Box \bigvee_{j \in J} \phi_{j}  \]
is just the statement
\[ d \subseteq \bigcup_{i \in I} O_{i} \;\; \Longrightarrow \;\; \exists J \in {\sf Fin}(I) . \, d \subseteq \bigcup_{j \in J} O_{j} , \]
where $O_{i} = \lsem \phi_{i} \rsem$, i.e. that $d$ is compact as a subset of $\sum_{a \in {\sf Act}} \Dom$.
Since $d \in \Dom \cong P^{0} [ \sum_{a \in {\sf Act}} \Dom ]$, and elements of the Plotkin powerdomain are Scott-compact subsets of the base domain (\cite{PloLN}), this proves that \Dom\ satisfies (BN).

Next we show that \Dom\ satisfies (FA).
Since there are only countably many distinct formulae in $\Ellomega$, it suffices to prove the following:
\begin{itemize}
\item Given a sequence $\{ U_{n} \}$ of compact-open subsets of $\Dom$, 
with $U_{n} \supseteq U_{n + 1} \; (n \in \omega )$, and an element $d \in \Dom$ such that $d \cap U_{n} \not= \varnothing$ $(n \in \omega)$, then $d \cap \bigcap_{n \in \omega} U_{n} \not= \varnothing$.
\end{itemize}
(The alternative case for $d \models U_{n}$, namely $\bot \in U_{n}$ for all $n$, is trivial.)

Since each $U_{n}$ is compact-open, it has the form $\diverges B_{n}$, where $B_{n}$ is a finite subset of ${\cal K}(\Dom\ )$.
Also, $B_{n} \sqsubseteq_{u} B_{n + 1}$, where
\[ X \sqsubseteq_{u} Y \; \equiv \; \forall y \in Y. \, \exists x \in X . \, x \sqsubseteq y \;\;\; (X, Y \subseteq \Dom\ ) . \]
Now define
\[ C_{n} \; \equiv \; \{ b \in B_{n} : \exists x \in d . \, b \sqsubseteq x \} \;\;\; (n \in \omega ) . \]
Since $d \cap U_{n} \not= \varnothing$, $C_{n} \not= \varnothing$ for all $n$.
Also, $C_{n} \sqsubseteq_{u} C_{n + 1}$.
Thus by K\"{o}nig's Lemma in the form given e.g. in~\cite{Niv81}, there is a sequence $\{ c_{n} \}$ with $c_{n} \sqsubseteq c_{n+1}$ and $c_{n} \in C_{n}$.
Now define
\[ e_{n} \; \equiv \; \lsing c_{n} \rsing \uplus \lsing \bot \rsing \;\;\; (n \in \omega ). \]
Clearly $e_{n} \sqsubseteq e_{n + 1}$ and $e_{n} \sqsubseteq d$ for all $n$, whence $\bigsqcup e_{n} \sqsubseteq d$.
But $\bigsqcup c_{n} \in \bigsqcup e_{n}$ (using the description of least 
upper bounds of chains in the Plotkin powerdomain given in \cite[Theorem 8]{Plo76}), 
and so for some $x \in d$, $\bigsqcup c_{n} \sqsubseteq x$.
Since $\bigsqcup c_{n} \in U_{n}$ for all $n$, $d \cap \bigcap_{n \in \omega} U_{n} \not= \varnothing$, and the proof is complete. \qed

We now draw some striking consequences from the finitary axioms.
\begin{definition}
{\rm A formula $\phi \in \Ellinfty$ is in {\em finitary normal form} if it has the form}
\[ \bigwedge_{i \in I} \bigvee_{j \in J_{i}} \phi_{ij} \;\;\; (\phi_{ij} \in \Ellomega ). \]
\end{definition}

\begin{lemma} \label{fnf}
For each $\phi \in \Ellinfty$, for some finitary normal form $\psi$:
\[ \mbox{(BN) + (FA)} \vdash \phi = \psi . \]
\end{lemma}

\proof\ An easy induction on ${\sf ht}(\phi)$. \qed

\begin{proposition}
\label{p18}
In any finitary transition system ${\cal T}$, for all $p, q \in {\rm Proc}$:
\[ \Ellinfty (p) \subseteq \Ellinfty (q) \;\; \Longleftrightarrow \;\; \Ellomega (p) \subseteq \Ellomega (q) . \]
\end{proposition}

\proof\ The left to right implication is immediate.
For the converse, suppose $ \Ellomega (p) \subseteq \Ellomega (q)$, and $p \models \phi$, $(\phi \in \Ellinfty )$.
By~\ref{fnf},
\[ \mbox{(BN) + (FA)} \vdash \phi = \bigwedge_{i \in I} \bigvee_{j \in J_{i}} \phi_{ij} \;\;\;  (\phi_{ij} \in \Ellomega ) \]
hence since ${\cal T} \models \mbox{(BN) + (FA)}$, ${\cal T} \models \phi = \bigwedge_{i \in I} \bigvee_{j \in J_{i}} \phi_{ij}$, and
\[ \begin{array}{ll}
\bullet & p \models \bigwedge_{i \in I} \bigvee_{j \in J_{i}} \phi_{ij} \\
\Longrightarrow & \forall i \in I . \, \exists j \in J_{i} . \, p \models \phi_{ij} \\
\Longrightarrow &  \forall i \in I . \, \exists j \in J_{i} . \, q \models \phi_{ij} \\
\Longrightarrow & q \models \bigwedge_{i \in I} \bigvee_{j \in J_{i}} \phi_{ij} \\
\Longrightarrow & q \models \phi . \;\;\; \qed
\end{array} \]

\begin{theorem}[Finitary Characterisation Theorem]
\label{fct}
With notation as in the previous Proposition:
\[ p \preord^{B} q \;\; \Longleftrightarrow \;\; p \preord_{\omega} q \;\; \Longleftrightarrow \;\; p \preord^{F} q \;\; \Longleftrightarrow \;\; \Ellomega (p) \subseteq \Ellomega (q) . \]
\end{theorem}

\proof\ Combine Theorems~\ref{linfct}, \ref{lct} and \ref{p18}. \qed

In order to continue our study of finitary transition systems, we need to introduce some notions from our final topic of this section.

\subsection*{Universal Semantics}

Given any transition system and $p \in {\rm Proc}$, it is easy to see that $\Ellomega (p) \subseteq \Ellomega$ satisfies the axioms of a prime filter; hence we have a map
\[ \Ellomega (\cdot ) : {\rm Proc} \longrightarrow {\sf Spec} \; \Ellomega\ . \]
If we compose this with the isomorphism $ {\sf Spec} \; \Ellomega \cong \Dom$ from the Duality Theorem~\ref{bdual}, we get a map
\[ \lsem \cdot \rsem : {\rm Proc} \longrightarrow \Dom \]
which takes each process to an element of our domain.
This map can be regarded as a {\em syntax-free denotational semantics}; it is {\em universal} since it is defined on every transition system.

\begin{theorem}[Universal Semantics] 
\label{usem}
For any transition system ${\cal T}$ with $p, q \in {\rm Proc}$:
\[  \begin{array}{rl}
(i)  & p \preord^{F} q \;\; \Longleftrightarrow \;\; \lsem p \rsem \sqsubseteq \lsem q \rsem \\
(ii) & p \sim^{F} \lsem p \rsem .
\end{array} \]
If ${\cal T}$ is finitary, then:
\[ \begin{array}{rl}
(iii)  & p \preord^{B} q \;\; \Longleftrightarrow \;\; \lsem p \rsem \sqsubseteq \lsem q \rsem \\
(iv) & p \sim^{B} \lsem p \rsem .
\end{array} \] 
\end{theorem}

\proof\ Clearly (i) follows from (ii), and (iii) from (iv).
Now $\Ellomega (p) = \Ellomega (\lsem p \rsem )$; 
and so (ii) follows from \ref{lct}; while (iv) follows from \ref{fct}. \qed

We can think of \ref{usem} as a {\em full abstraction theorem} 
\cite{Mil75,Plo77,Mil77} for our semantics; it says that every transition system (finitary transition system) can be embedded in \Dom\ with as much identification as possible modulo the finitary equivalence (bisimulation).

Since \Dom\ can itself be viewed as a transition system, we can tie things up even more neatly.
Let {\bf TS} be the category with objects the transition systems, and morphisms ${\cal T}_{1} \rightarrow {\cal T}_{2}$ maps
\[ f : {\rm Proc}_{1} \rightarrow {\rm Proc}_{2} \]
for which
\[ \Ellomega (p) = \Ellomega (f(p)) \;\;\; (p \in {\rm Proc}_{1}). \]
It is clear that for such $f$
\[ p \preord^{F} q \;\; \Longleftrightarrow \;\; f(p) \preord^{F} f(q) , \]
and if ${\cal T}_{1}$ and  ${\cal T}_{2}$ are finitary,
\[ p \preord^{B} q \;\; \Longleftrightarrow \;\; f(p) \preord^{B} f(q) . \]
Now we have

\begin{theorem}[Final Algebra Theorem]
\label{falg}
\Dom\ is final in {\bf TS}, and also in the subcategory {\bf FTS} of finitary transition systems.
\end{theorem}

\proof\ All we need to show is that the semantic map $\lsem \cdot \rsem$ is the unique morphism from a transition system to $\Dom$.
But for $d_{1}, d_{2} \in \Dom$,
\begin{Eqarray}
\Ellomega (d_{1}) \subseteq \Ellomega (d_{2}) & \Longleftrightarrow & K \Omega (d_{1}) \subseteq K \Omega (d_{2}) & \mbox{by~\ref{bdual}} \\
& \Longleftrightarrow & d_{1} \sqsubseteq d_{2} & \mbox{since \Dom\ is coherent,} \\
\end{Eqarray}
which gives uniqueness. \qed

\subsection*{Finitary Transition Systems Resumed}
Firstly, some conditions equivalent to finitariness.
\begin{proposition}
\label{ftsequiv}
For any transition system ${\cal T}$, the following conditions are equivalent:
\begin{center}
\begin{tabular}{rl}
(i) & ${\cal T}$ is finitary \\
(ii) & $\forall p \in {\rm Proc} . \, p \sim^{B} \lsem p \rsem$ \\
(iii) & ${\preord^{B}} = {\preord^{F}}$ in the combined system ${\cal T} + \Dom$ (disjoint union).
\end{tabular}
\end{center}
\end{proposition}

\proof\  $(i) \; \Longrightarrow \; (ii)$ is~\ref{usem} (iv); $(ii) \; \Longrightarrow \; (iii)$ since \Dom\ is finitary.

\noindent $(ii) \; \Longrightarrow \; (i)$. Suppose that ${\cal T}$ is not finitary, in particular that (BN) fails; i.e. that for some $p \in {\rm Proc}$,
\[ p \models \Box \bigvee_{i \in I} \phi_{i} \;\;\; ( \phi_{i} \in \Ellomega ) \]
and $\forall J \in {\sf Fin}(I). \, p \nvDash \bigvee_{j \in J} \phi_{j}$.
Since $\Ellomega (p) = \Ellomega ( \lsem p \rsem )$, and each $\bigvee_{j \in J} \phi_{j} \in \Ellomega$, 
$\lsem p \rsem \nvDash \bigvee_{j \in J} \phi_{j}$ for all $J \in {\sf Fin}(I)$;
hence since $\lsem p \rsem \in \Dom$ and $\Dom$ is finitary, $\lsem p \rsem \nvDash \Box \bigvee_{i \in I} \phi_{i}$.
Thus $\Ellinfty (\lsem p \rsem ) \not= \Ellinfty (p)$, and so by~\ref{linfct} $p \sim^{B} \lsem p \rsem$.
The case when (FA) fails is similar. 

\noindent $(iii) \; \Longrightarrow \; (ii)$. Suppose for some $p$, $p \nsim^{B} \lsem p \rsem$.
Then since $p \sim^{F} \lsem p \rsem$ by~\ref{usem} (ii), ${\preord^{B}} \not= {\preord^{F}}$. \qed

Note that in part (iii) of this Proposition we have ``added in'' \Dom\ to the given transition system ${\cal T}$.
This is to overcome the problem that there may not be enough processes in ${\cal T}$ alone to cause $\preord^{B} = \preord^{F}$ to fail.

Now we relate some of the finitariness conditions of Section~2 to our axioms.

\begin{proposition}
\label{finots}
(i) Weakly finite branching is equivalent to weakly image finite plus weakly initials finite.

\noindent (ii) Weakly finite branching implies (BN).

\noindent (iii) (BN) implies weakly initials finite.

\noindent (iv) (BN) + (FA) do {\em not} imply weakly image finite.
\end{proposition}

\proof\ (i). Easy.

\noindent (ii). Suppose $p \models \Box \bigvee_{i \in I} \phi_{i}$.
$(\bigvee_{i \in I} \phi_{i}) \diverges  \Leftrightarrow  \exists i \in I. \, \phi_{i} \diverges$, in which case $\vdash \phi_{i} = \true$, and the conclusion is trivial.
Otherwise, $p \converges$, and so $C(p)$ is finite, say 
\[ C(p) = \{ \ltuple a_{1}, p_{1} \rtuple\ , \ldots , \ltuple a_{n}, p_{n} \rtuple \} . \]
Then for each $k$ with $1 \leq k \leq n$, $\ltuple a_{k}, p_{k} \rtuple \models \phi_{i_{k}}$ for some $i_{k} \in I$, and so $p \models \Box \bigvee_{j \in J} \phi_{j}$, where $J = \{ i_{1}, \ldots , i_{n} \}$.

\noindent (iii). Assume (BN) and $p \converges$.
Then $p \models \Box \bigvee_{a \in {\sf Act}} a(\true\ )$, and so by (BN)
\[ p \models \bigvee_{J \in {\sf Fin}({\sf Act})} \Box \bigvee_{a \in J} a(\true\ ) , \]
which says exactly that $p$ has a finite set of initial actions.

\noindent (iv). $\sum_{n \in \omega} a^{n} + a^{\omega}$ is in $\Dom$. \qed

All the usual finitary  calculi are weakly finite branching, and so satisfy (BN).
However, in general these calculi do {\em not} satisfy (FA) (analogously to the fact that generating trees over domains do not yield closed sets, although they always yield compact ones; cf. \cite{PloLN}).
As a standard counterexample, define
\[ \begin{array}{lll}
p & \equiv & \sum_{n \in \omega} a^{n} \Oh + \Omega \\
\phi_{0} & \equiv & \true\  \\
\phi_{k + 1} & \equiv & a(\Diamond \phi_{k}) .
\end{array} \]
Then for all $J \in {\sf Fin}(\omega )$, $p \models \Diamond \bigwedge_{j \in J} \phi_{j}$, but $p \nvDash \Diamond \bigwedge_{i \in \omega} \phi_{i}$.

Thus if $p$ can be defined in our calculus, it does not satisfy (FA).
Since $p$ {\em can} be defined in CCS, SCCS (see next section), etc., these calculi are not finitary transition systems according to Definition~\ref{fts}.
However, we can take the view that if we only take account of {\em observable} 
information via the semantics $\lsem \cdot \rsem$, we have collapsed the 
given system into a finitary one which will actually, 
by Theorems~\ref{usem} and~\ref{falg}, be isomorphic to a subsystem (or, topologically, a subspace) of $\Dom$.

\subsection*{Comparison Theorems Resumed}
We now return to the question of finding a suitable correspondence between the finitary parts of HML and $\Ell$.
As confirmation of our claim that ${\rm HML}_{\omega}$ is unsatisfactory, we have:

\noindent {\bf Observation.} ${\rm HML}_{\omega}$ does not characterise $\preord^{F}$.

In fact, \ref{bex} provides a counter-example since, with the notation used there, $p \npreord^{F} q$ while $ {\rm HML}_{\omega}(p) \subseteq {\rm HML}_{\omega}(q)$.

We can get an idea of how to extend $ {\rm HML}_{\omega}$ by inspection of the translation functions~\ref{tfuns}.
Although $( \cdot )^{\dagger}$ introduces infinitary conjunctions, these are of a special kind, for which a finitary counterpart can be found.

\begin{definition} 
{\rm ${\rm HML}^{+}$ is the extension of ${\rm HML}_{\omega}$ with additional atomic fomulae of the form
\[ {\sf init}(A) \;\;\; (A \in {\sf Fin}({\sf Act})). \]
The definition of the satisfaction relation is extended by
\[ p \models {\sf init}(A)  \equiv  p \converges \; \& \; \{ a \in {\sf Act} : \exists q . \, \labarrow{p}{a}{q} \} \subseteq A . \] }
\end{definition}

We can now modify the translation function $( \cdot )^{\dagger}$ as follows:
\[ ( \Box \bigvee_{i \in I} a_{i} ( \phi_{i} ))^{\dagger} \; \equiv \; \bigwedge_{i \in I} [a_{i}] ( \phi_{i})^{\dagger} \; \wedge \; {\sf init}(\{ a_{i} : i \in I \} ). \]
Proposition~\ref{faitht} clearly still holds with this modification, and $( \cdot )^{\dagger}$ now cuts down to a function
\[ {\sf N} \Ellomega \longrightarrow {\rm HML}^{+}. \]
There is still a mismatch in the other direction, since $( \cdot )^{\ast}$ introduces infinite disjunctions.
To overcome this, we have to make the assumption that the transition system satisfies (BN)---a mild one, as~\ref{finots} and the ensuing discussion shows.

Let $\Ell_{\bigvee \infty}$ be the sublanguage of \Ellinfty\ obtained by the restriction to finite {\em conjunctions} (but with infinite disjunctions still allowed).
\begin{proposition} 
In any transition system satisfying (BN), for all $p, q \in {\rm Proc}$:
\[ \Ell_{\bigvee \infty}(p) \subseteq \Ell_{\bigvee \infty}(q) \;\; \Longleftrightarrow \;\; \Ellomega(p) \subseteq \Ellomega(q) . \]
\end{proposition}

\proof\ Just like~\ref{p18}. \qed

Clearly, $( \cdot )^{\ast}$, extended by the clause
\[ ({\sf init}(A))^{\ast} \; \equiv \; \Box \bigvee \{ a(\true\ : a \in A \} \]
cuts down to a function
\[ {\rm HML}^{+}  \longrightarrow  {\sf N} \Ell_{\bigvee \infty} . \]
We thus arrive at our
\begin{theorem}[Comparison Theorem (Finitary Case)]
With notation as in the previous Proposition:
\[ {\rm HML}^{+}(p) \subseteq {\rm HML}^{+}(q) \;\; \Longleftrightarrow \;\; \Ellomega (p) \subseteq \Ellomega (q) . \]
\end{theorem}

\section{Full Abstraction for SCCS}
So far, we have worked with abstract transition systems, in a syntax-free fashion.
This degree of abstraction carries a price; we lose compositionality.
Indeed, we need syntax to {\em define} compositionality.
Accordingly, in this Section we turn to a particular transition system specified by an algebraic syntax, namely Milner's SCCS \cite{Mil83}.
We equip our domain \Dom\ with a continuous algebraic structure corresponding to the signature of SCCS.
Our main result is that the resulting denotational semantics for SCCS is {\em fully abstract} \cite{Mil75,Plo77} with respect to bisimulation for finite terms, and with respect to the finitary preorder for recursive terms.
As a by-product we will show that \Dom\ is isomorphic to Hennessy's term model \cite{Hen81}, and hence obtain a complete axiomatisation of its equational theory as an immediate consequence of Hennessy's results.

Our choice of SCCS is for illustrative purposes, because it is simple and yet expressive.
Similar accounts could be given for CCS \cite{Mil80}, MEIJE \cite{AB84}, ACP \cite{BK84}, etc.
Note, however, that our semantics is fully abstract with respect to the {\em strong} congruence in Milner's terminology \cite{Mil83}, where all actions are observable.
A corresponding treatment of {\em observation equivalence} \cite{HM85}, where unobservable actions are factored out, is still an open problem as far as I know; some hints of a possible approach may be gleaned from \cite{Abr87b}.

We begin by recalling some basic definitions on SCCS from \cite{Mil83,Hen81}.
We assume familiarity with basic notions of universal algebra; see e.g. \cite{ADJ78,EM85}.

We fix a set of actions {\sf Act}, which we assume comes equipped with an {\em abelian monoid} structure comprising
\begin{itemize}
\item an associative, commutative binary operation which we denote by juxtaposition, e.g. $ab$
\item a unit 1.
\end{itemize}
The (one-sorted) signature $\Sigma$ of SCCS is then defined as follows:

\begin{definition} 
{\rm $\Sigma = \{ \Sigma_{n} \}_{n \in \omega}$, where $\Sigma_{n}$ is the set of operation symbols of arity $n$ in $\Sigma$.
\begin{eqnarray*}
\Sigma_{0} & \equiv & \{ \Oh , \Omega \} \\
\Sigma_{1} & \equiv & \{ a\_ : a \in {\sf Act} \} \cup \{ \_{\restriction} A : A \subseteq {\sf Act} \} \\
& & \mbox{} \cup \{ \_ [S] : S \; \mbox{is a monoid endomorphism on {\sf Act}} \}  \\
\Sigma_{2} & \equiv & \{ {+}, {\times} \} \\
\Sigma_{n} & \equiv & \varnothing , \;\; n > 2 .
\end{eqnarray*} }
\end{definition}

Thus our version of SCCS only has {\em finite} sums (in contrast with \cite{Mil83}), and has a constant for the undefined process as in \cite{Hen81}.

We define the subsignature $\Sigma' \subseteq \Sigma$ to be obtained by omitting the {\it restriction} operators $\_{\restriction} A$, 
the {\it relabelling} operators $\_ [S]$, and the {\it synchronous product} operator $\times$, leaving only the {\it nullary sum} $\Oh$, the {\it binary sum} $+$, {\it prefixing} $a\_$, and the {\it undefined} process $\Omega$.

We take the {\em finite processes} of SCCS to be the terms over the signature $\Sigma$, i.e. the elements of the term algebra $T_{\Sigma}$.
Evidently, we can take the elements of $T_{\Sigma'}$ as notations for the finite synchronisation trees ${\sf ST}_{\omega}$.

\begin{definition}[Operational Semantics]
\label{opdef}
{\rm We make $T_{\Sigma}$ into a transition system by defining the transition relation
and divergence predicate in a syntax-directed way, as the {\em least}
relations satisfying the following axioms and rules:}
\[ (D \Omega ) \;\; \Omega \diverges \]
\[ (D+L) \;\; \frac{t_{1} \diverges}{(t_{1} + t_{2}) \diverges} \;\;\;\;\;\;\;\; (D+R) \;\; \frac{t_{2} \diverges}{(t_{1} + t_{2}) \diverges} \]
\[ (D{\restriction} ) \;\; \frac{t \diverges}{(t {\restriction} A) \diverges} \;\;\;\;\;\;\;\; (DS) \;\; \frac{t \diverges}{t [S] \diverges} \]
\[ (D \times L) \;\; \frac{t_{1} \diverges}{t_{1} \times t_{2} \diverges} \;\;\;\;\;\;\;\; (D \times R) \;\; \frac{t_{2} \diverges}{t_{1} \times t_{2} \diverges} \]
\[ (Ta) \;\;  \labarrow{at}{a}{t} \]
\[ (T+L) \;\; \frac{\labarrow{t_{1}}{a}{t'_{1}}}{\labarrow{t_{1} + t_{2}}{a}{t'_{1}}} \;\;\;\;\;\;\;\; 
(T+R) \;\; \frac{\labarrow{t_{2}}{a}{t'_{2}}}{\labarrow{t_{1} + t_{2}}{a}{t'_{2}}} \]
\[ (T {\restriction} ) \;\; \frac{\labarrow{t}{a}{t'}, \; a \in A}{\labarrow{t {\restriction} A}{a}{t' {\restriction} A}} \;\;\;\;\;\;\;\; (TS) \;\; \frac{\labarrow{t}{a}{t'}}{\labarrow{t[S]}{S a}{t'[S]}} \]
\[ (T \times ) \;\; \frac{\labarrow{t_{1}}{a}{t'_{1}} \;\; \labarrow{t_{2}}{b}{t'_{2}}}{\labarrow{t_{1} \times t_{2}}{ab}{t'_{1} \times t'_{2}}} \] 
\end{definition}

For an illuminating discussion of the conceptual basis for these and related axioms, see \cite{Mil86}.

We now have a transition system $(T_{\Sigma}, {\sf Act}, {\rightarrow}, \diverges )$ implicitly defined by~\ref{opdef}.
The following proposition gives a more explicit description of this system.

\begin{proposition}
\label{tops}
For all $t, t_{1}, t_{2} \in T_{\Sigma}$:
\[ \begin{array}{rlcl}
(i) (a) & \Oh \converges & & (b) \;\; \nlabarrow{\Oh}{a}{} \\
(ii) (a) & \Omega \diverges & & (b) \;\; \nlabarrow{\Omega}{a}{} \\
(iii) (a) & at \converges \\
(b) & \labarrow{at_{1}}{b}{t_{2}} & \Longleftrightarrow & b = a \: \& \: t_{1} = t_{2} \\
(iv) (a) & (t_{1} + t_{2}) \diverges & \Longleftrightarrow & t_{1} \diverges \;  {\rm or} \; t_{2} \diverges \\
(b) & \labarrow{(t_{1} + t_{2})}{a}{t} & \Longleftrightarrow & \labarrow{t_{1}}{a}{t} \; {\rm or} \; \labarrow{t_{2}}{a}{t} \\
(v) (a) & (t {\restriction} A) \diverges & \Longleftrightarrow & t \diverges \\
(b) & \labarrow{t_{1} {\restriction} A}{a}{t_{2}} & \Longleftrightarrow & \exists t . \, \labarrow{t_{1}}{a}{t} \: \& \: t_{2} = t {\restriction} A \: \& \: a \in A \\
(vi) (a) & t[S] \diverges & \Longleftrightarrow & t \diverges \\
(b) & \labarrow{t_{1}[S]}{a}{t_{2}} & \Longleftrightarrow & \exists b, t . \, \labarrow{t_{1}}{b}{t} \: \& \: t_{2} = t[S] \: \& \:  a = Sb \\
(vii) (a) & (t_{1} \times t_{2}) \diverges & \Longleftrightarrow & t_{1} \diverges \;  {\rm or} \; t_{2} \diverges \\
(b) & \labarrow{t_{1} \times t_{2}}{a}{t} & \Longleftrightarrow & \exists t'_{1}, t'_{2}, b_{1}, b_{2} . \, \labarrow{t_{i}}{b_{i}}{t'_{i}} \; (i = 1, 2) \\
& & & \& \: t = t'_{1} \times t'_{2} \: \& \: a = b_{1}b_{2} .
\end{array} \]
\end{proposition}

\proof\ By induction on the length of proofs of $t \diverges$ and $\labarrow{t_{1}}{a}{t_{2}}$. \qed

Now given any $\Sigma$-algebra ${\cal A}$, by initiality of $T_{\Sigma}$ there is a unique $\Sigma$-homomorphism
\[ \lsem \cdot \rsem^{\cal A} : T_{\Sigma} \; \longrightarrow \; {\cal A} , \]
which is just another notation for a compositional denotational semantics as 
in \cite{MS76,Sto77,Gor79}.
Thus to form a denotational semantics $\lsem \cdot \rsem^{\cal D}$ based on our domain $\Dom$, it suffices to define each operation in $\Sigma$ as a function of the appropriate arity over $\Dom$.
We shall in fact define the operations so that they are {\em continuous} over $\Dom$.

\begin{definition} 
\label{sigd}
{\rm We specify a $\Sigma$-structure on $\Dom$:
\[ \begin{array}{rlcl}
(i) & \Oh^{\Dom} & \equiv & \emptyset \\
(ii) & \Omega^{\Dom} & \equiv & \lsing \bot \rsing \\
(iii) & {a\_}^{\Dom} & \equiv & \lambda d \in {\Dom}. \lsing \ltuple a, d \rtuple \rsing \\
(iv) & +^{\Dom} & \equiv & \uplus \\
\end{array} \]
Restriction:
\[ (v) \;\; (\_ {\restriction} A)^{\Dom}  \equiv  \mu \Phi \in [\Dom \rightarrow \Dom ] . \, \biguplus \circ P^{0} (g_{A} \Phi ) \]
where
\[ g_{A} : [ \Dom \rightarrow \Dom ] \rightarrow [ \sum_{a \in {\sf Act}} \Dom \rightarrow \Dom ] \]
is defined by
\[ \begin{array}{rcl}
g_{A} \Phi \bot & = & \lsing \bot \rsing \\
g_{A} \Phi \ltuple a, d \rtuple & = & \left\{ \begin{array}{ll} 
\lsing \ltuple a, \Phi d \rtuple \rsing & \mbox{if $a \in A$} \\
\emptyset & \mbox{otherwise}
\end{array}
\right. \\
\end{array} \]
(i.e. 
\[ g_{A} \Phi = {\coprod_{a \in A} \lambda d \in \Dom . \lsing \ltuple a, \Phi d \rtuple \rsing} \; {\textstyle \coprod} \; {\coprod_{a \in {\sf Act} - A} \lambda d \in \Dom . \emptyset } , \]
where $\coprod$ is ``source tupling'' \cite{ADJ85}).

\noindent Relabelling:
\[ (vi) \;\; (\_ [S])^{\Dom}  \equiv  \mu \Phi \in [\Dom \rightarrow \Dom ] . \,  P^{0} (g_{S} \Phi ) \]
where
\[ g_{S} : [ \Dom \rightarrow \Dom ] \rightarrow [ \sum_{a \in {\sf Act}} \Dom \rightarrow \sum_{a \in {\sf Act}} \Dom  ] \]
is defined by
\[ \begin{array}{rcl}
g_{S} \Phi \bot & = &  \bot  \\
g_{S} \Phi \ltuple a, d \rtuple & = & \ltuple Sa, \Phi d \rtuple
\end{array} \]
Product: 
\[ (vii) \;\; \times^{\Dom}  \equiv  \mu \Phi \in [\Dom^{2} \rightarrow \Dom ] . \,   (f \Phi )^{\dagger} \]
where
\[ f : [ \Dom^{2} \rightarrow \Dom ] \rightarrow [ ( \sum_{a \in {\sf Act}} \Dom )^{2} \rightarrow \sum_{a \in {\sf Act}} \Dom  ] \]
is defined by
\[ \begin{array}{rcl}
f \Phi (x, \bot ) = f \Phi (\bot , x)  & = &  \bot  \\
f \Phi (\ltuple a, d \rtuple  , \ltuple b, e \rtuple ) & = &  \ltuple ab, \Phi (d, e) \rtuple 
\end{array} \] }
\end{definition}

The only point which needs to be checked to ensure that this definition yields well-defined continuous functions is that $g_{A} \Phi$, $g_{S} \Phi$ and $f \Phi$ are (bi)strict and continuous, which is immediate from the definitions.
Note that restriction, relabelling and product are defined recursively, while sum and prefixing are interpreted by the basic operations derived from the domain equation for $\Dom$.
This corresponds to the fact that restriction, relabelling and product can be {\em eliminated} (for finite terms) in the equational theory of SCCS modulo bisimulation.

The continuous $\Sigma$-algebra defined by \ref{sigd} is denoted $\Dom_{\Sigma}$.
The following is an easy consequence of~\ref{sigd} and~\ref{feltp}.

\begin{proposition}
\label{surs}
The semantic function
\[ \lsem\cdot \rsem^{\Dom} : T_{\Sigma} \; \longrightarrow \; \Dom_{\Sigma} \]
cuts down to surjections
\[ T_{\Sigma} \twoheadrightarrow {\cal K}(\Dom ), \;\;\;\; T_{\Sigma'} \twoheadrightarrow {\cal K}(\Dom ) . \]
\end{proposition}
Thus the finite synchronisation trees provide a notation for the finite elements of $\Dom$.

We now relate our definitions of the SCCS operations on \Dom\ to the transition system view of $\Dom$.
\begin{proposition}
\label{dops}
For all $d, d_{1}, d_{2} \in {\cal K}(\Dom )$:
\[ \begin{array}{rlcl}
(i) (a) & \Oh^{\Dom} \converges & &  (b) \;\; \nlabarrow{\Oh^{\Dom}}{a}{} \\
(ii) (a) & \Omega^{\Dom} \diverges & & (b) \;\; \nlabarrow{\Omega^{\Dom}}{a}{} \\
(iii) (a) & a^{\Dom} d \converges & & \\
(b) & \labarrow{a^{\Dom} d_{1}}{b}{d_{2}} & \Longleftrightarrow & b = a \: \& \: d_{1} = d_{2} \\
(iv) (a) & (d_{1} +^{\Dom} d_{2}) \diverges & \Longleftrightarrow & d_{1} \diverges \; \mbox{or} \; d_{2} \diverges \\
(b) & \labarrow{d_{1} +^{\Dom} d_{2}}{a}{d} & \Longleftrightarrow & \labarrow{d_{1}}{a}{d} \; \mbox{or} \; \labarrow{d_{2}}{a}{d} \\
\end{array} \]
Restriction:
\[ \begin{array}{rlcl}
(v) (a) & (d {\restriction}^{\Dom} A) \diverges & \Longleftrightarrow & d \diverges \\
(b) & \labarrow{d_{1} {\restriction}^{\Dom} A}{a}{d_{2}} & \Longleftrightarrow & \exists e_{1}, e_{2} . \, \labarrow{d_{1}}{a}{e_{i}}, \; (i = 1, 2) \\
& & & \mbox{} \& \; e_{1} {\restriction}^{\Dom} A \sqsubseteq d_{2} \sqsubseteq e_{2} {\restriction}^{\Dom} A \\
& & &  \& \; a \in A 
\end{array} \]
Relabelling:
\[ \begin{array}{rlcl}
(vi) (a) & (d [S]^{\Dom}) \diverges & \Longleftrightarrow & d \diverges \\
(b) & \labarrow{d_{1} [S]^{\Dom}}{a}{d_{2}} & \Longleftrightarrow & \exists e_{1}, e_{2}, b_{1}, b_{2} . \, \labarrow{d_{1}}{a}{e_{i}}, \; (i = 1, 2) \\
& & & \mbox{} \& \; e_{1} [S]^{\Dom} \sqsubseteq d_{2} \sqsubseteq e_{2} [S]^{\Dom} \\
& & &  \& \; S b_{1} = a = S b_{2} 
\end{array} \]
Product:
\[ \begin{array}{rlcl}
(vii) (a) & (d_{1} \times^{\Dom} d_{2}) \diverges & \Longleftrightarrow & d_{1} \diverges \; \mbox{or} \; d_{2} \diverges \\
(b) & \labarrow{d_{1} \times^{\Dom} d_{2}}{a}{d} & \Longleftrightarrow & \exists u_{i}, v_{i}, b_{i}, c_{i} \; (i = 1, 2). \\
& & & \labarrow{d_{1}}{b_{i}}{u_{i}} \: \& \: \labarrow{d_{2}}{c_{i}}{v_{i}} \; (i = 1, 2) \\
& & & \mbox{} \& \; (u_{1} \times^{\Dom} v_{1}) \sqsubseteq d \sqsubseteq (u_{2} \times^{\Dom} v_{2}) \\
& & &  \& \; b_{i}c_{i} = a \; (i = 1, 2).
\end{array} \]
\end{proposition}

\proof\ We give two cases for illustration.

\noindent (v). We define
\begin{eqnarray*}
\Theta & \equiv & \{ \{ \ltuple a, d' {\restriction}^{\Dom} A \rtuple \} : \ltuple a, d' \rtuple \in d, a \in A \} \\
&  & \mbox{} \cup \{ \varnothing : d = \emptyset \; \mbox{or} \; \exists \ltuple a, d' \rtuple \in d. \, a \not\in A \} \\
&  & \mbox{} \cup \{ \{ \bot \} : \bot \in d \} .
\end{eqnarray*}
Now
\begin{eqnarray*}
d {\restriction}^{\Dom} A & = & Con( \bigcup \Theta^{\star})  \\
& = & Con((\bigcup \Theta )^{\star}) \;\; \mbox{by \cite{Plo76} p. 477} \\
& = & Con(\bigcup \Theta ) \;\; \mbox{since $d \in {\cal K}(\Dom )$} \\
& = & Con( \{ \ltuple a, d' {\restriction}^{\Dom} A \rtuple : \ltuple a, d' \rtuple \in d \: \& \: a \in A \} \\
&   & \mbox{} \cup \{ \bot : \bot \in d \} ), 
\end{eqnarray*}
and (v) is readily derived from this description.

\noindent (vii). Similarly to (v),
\begin{eqnarray*}
d_{1} \times^{\Dom} d_{2} & = & Con( \{ \ltuple b_{1}b_{2}, e_{1} \times^{\Dom} e_{2} \rtuple : \ltuple b_{i}, e_{i} \rtuple \in d_{i}, \; i = 1, 2 \} \\
& & \mbox{} \cup \{ \bot : \bot \in d_{1} \; \mbox{or} \; \bot \in d_{2} \} ). \;\;\; \qed
\end{eqnarray*}

\begin{proposition}
\label{sbis}
For all $t \in T_{\Sigma}$, $t \sim^{B} \lsem t \rsem^{\Dom}$.
\end{proposition}

\proof\ Firstly, we define a height function on $T_{\Sigma}$ in the obvious way:
\[ {\sf ht}( \sigma (t_{1}, \ldots , t_{n}) = \sup \; \{ {\sf ht}(t_{i} : 1 \leq i \leq n \} + 1 . \]
As an easy consequence of~\ref{tops}, we have:
\[ \labarrow{t}{a}{t'} \;\; \Longrightarrow \;\; {\sf ht}(t' ) < {\sf ht}(t) . \]
The proposition is proved by induction on ${\sf ht}(t)$, and cases on the construction of $t$.
The cases arising from operations in $\Sigma'$ are immediate in the light of the parallelism between~\ref{tops} and~\ref{dops}.
We give one of the remaining cases for illustration.

\noindent $t \equiv t_{1} {\restriction}^{\Dom} A$.
Firstly, 
\begin{Eqarray}
t \diverges & \Longleftrightarrow & t_{1} \diverges & \mbox{by \ref{tops}(v)} \\
& \Longleftrightarrow & \lsem t_{1} \rsem^{\Dom} \diverges & \mbox{by induction hypothesis} \\
& \Longleftrightarrow & ( \lsem t_{1} \rsem^{\Dom} {\restriction}^{\Dom} A) \diverges & \mbox{by \ref{dops}(v)} \\
& \Longleftrightarrow & \lsem t_{1} {\restriction}  A) \rsem^{\Dom} \diverges . &
\end{Eqarray}
Next,
\[ \begin{array}{llr}
\bullet & \labarrow{t}{a}{t'} & \\
\Longrightarrow & \labarrow{t_{1}}{a}{t'_{1}} \: \& \: t' = t'_{1} {\restriction}  A \: \& \: a \in A & \mbox{by \ref{tops}(v)} \\
\Longrightarrow & \exists d' . \, \labarrow{\lsem t_{1} \rsem^{\Dom}}{a}{d'} \; \& \; t'_{1} \preord^{B} d' & \mbox{ind. hyp. on $t_{1}$} \\
\Longrightarrow & t'_{1} {\restriction}  A \sim^{B} \lsem t'_{1} {\restriction}  A \rsem^{\Dom} & \mbox{ind. hyp. on $t'_{1} {\restriction}  A$} \\
& \mbox{} = \lsem t'_{1} \rsem^{\Dom}  {\restriction}^{\Dom}  A & \\
& \; \preord^{B} d' {\restriction}^{\Dom} A & \mbox{by \ref{ifs}} \\
& \mbox{(since ${\restriction}^{\Dom}$ is monotone)} & \\
\Longrightarrow & \exists u . \, \labarrow{\lsem t \rsem^{\Dom}}{a}{u} \; \& \; t' \preord^{B} u  & \mbox{by \ref{dops}(v).}
\end{array} \]
Similarly, we can show
\[ \labarrow{t}{a}{t'} \; \Rightarrow \; \exists u. \, \labarrow{\lsem t \rsem^{\Dom}}{a}{u} \: \& \: u \preord^{B} t' . \]
Again,
\[ \begin{array}{llr}
\bullet & \labarrow{\lsem t \rsem^{\Dom}}{a}{d} & \\
\Longrightarrow & \exists d_{1}, d_{2}. \: \labarrow{\lsem t_{1} \rsem^{\Dom}}{a}{d_{i}}, \; i = 1, 2  & \\
& \mbox{} \& \; d_{1} {\restriction}^{\Dom} A \sqsubseteq d \sqsubseteq d_{2} {\restriction}^{\Dom} A \\
& \mbox{} \& \; a \in A & \mbox{by \ref{dops}(v)} \\
\Longrightarrow & \exists t'_{1}, t'_{2}. \: \labarrow{t_{1}}{a}{t'_{i}}, \; i = 1, 2 & \\
& \mbox{} \& \; t'_{1} \preord^{B} d_{1}, \; d_{2} \preord^{B} t'_{2} & \mbox{by induction hypothesis} \\
\Longrightarrow & \labarrow{t}{a}{t'_{i} {\restriction} A}, \; i = 1, 2 & \\
& \mbox{} \& \; t'_{1} {\restriction} A \sim^{B} \lsem t'_{1} {\restriction} A \rsem^{\Dom} &  \mbox{by induction hypothesis} \\
& \mbox{} = \lsem t'_{1} \rsem^{\Dom} {\restriction}^{\Dom} A \; \preord^{B} \; d_{1} {\restriction}^{\Dom} A \; \preord^{B} d, & 
\end{array} \]
and similarly $d \preord^{B} t'_{2} {\restriction} A$.
Altogether, we have $t \sim^{B} \lsem t \rsem^{\Dom}$. \qed

As an immediate consequence of this Proposition and~\ref{ifs} we have
\begin{theorem}[Full Abstraction for Finite Terms]
\label{faft}
For all $t_{1}, t_{2} \in T_{\Sigma}$: 
\[ t_{1} \preord^{B} t_{2} \;\; \Longleftrightarrow \;\; \lsem t_{1} \rsem^{\Dom} \sqsubseteq \lsem t_{2} \rsem^{\Dom} . \]
\end{theorem}
As further consequences of \ref{faft} we have
\begin{itemize}
\item $\lsem \cdot \rsem^{\Dom}$ agrees with the syntax-free map $\lsem \cdot \rsem$ defined in Section 5.
Indeed, $t \sim^{B} \lsem t \rsem^{\Dom}$ implies $\Ellomega (\lsem t \rsem^{\Dom}) = \Ellomega (t) = \Ellomega (\lsem t \rsem )$, which implies $\lsem t \rsem^{\Dom} = \lsem t \rsem$.
\item $T_{\Sigma}$ is a finitary transition system, by \ref{ftsequiv}.
\end{itemize}
Moreover, we can derive two further characterisations of \Dom.
\begin{theorem}
(i) ${\cal K}(\Dom ) \; \cong \; (T_{\Sigma'}/{\sim^{B}}, {\preord^{B}}/{\sim^{B}})$, and therefore

\noindent (ii) $D \; \cong \; {\sf Idl} \: (T_{\Sigma'}/{\sim^{B}}, {\preord^{B}}/{\sim^{B}})$.
\end{theorem}

\proof\ Immediate from \ref{surs} and \ref{faft}. \qed

We recall the notion of {\em continuous $\Sigma$-algebra} \cite{ADJ78,Gue81}.
This is just a $\Sigma$-algebra whose carrier is a cpo, and whose operations are continuous.
A homomorphism of such algebras which is continuous on the carriers is 
a {\em continuous $\Sigma$-homomorphism}.
The category of these algebras and homomorphisms is denoted ${\bf CAlg}(\Sigma )$.
\begin{definition}
{\rm {\bf SCCS-Alg} is the full subcategory of ${\bf CAlg}(\Sigma )$ of those algebras ${\cal A}$ satisfying
\[ \forall t_{1}, t_{2} \in T_{\Sigma}. \: t_{1} \preord^{B} t_{2} \; \Longrightarrow \; \lsem t_{1} \rsem^{\cal A} \sqsubseteq \lsem t_{2} \rsem^{\cal A} . \] }
\end{definition}

\begin{theorem}
\label{binit}
$\Dom_{\Sigma}$ is initial in {\bf SCCS-Alg}.
\end{theorem}

\proof\ We begin by recalling a useful fact about continuous algebras (\cite{Gue81} Proposition 3.12).
Suppose $\cal A$ is a continuous algebra whose carrier $A$ is an algebraic domain, such that the finite elements ${\cal K}(A)$ form a $\Sigma$-subalgebra.
Then, given any monotonic $\Sigma$-homomorphism
\[ f : {\cal K}(A)  \longrightarrow  {\cal B} \]
to a continuous $\Sigma$-algebra $\cal B$, there is a unique extension
\[ \hat{f} : {\cal A}  \longrightarrow  {\cal B} \]
to a continuous $\Sigma$-homomorphism on $\cal A$.

By \ref{surs}, ${\cal K}(\Dom )$ is closed under the $\Sigma$-operations.
Hence it suffices to construct a unique monotone $\Sigma$-homomorphism
\[ f : {\cal K}(\Dom )  \longrightarrow  {\cal A} \]
to any $\cal A$ in {\bf SCCS-Alg}.
Given $d \in {\cal K}(\Dom )$, by \ref{surs} there is $t \in T_{\Sigma}$ with 
$\lsem t \rsem^{\Dom} = d$, and the only possible definition for $f$ giving a $\Sigma$-homomorphism is
\[ f : d \mapsto \lsem t \rsem^{\cal A} . \]
This establishes uniqueness.
For existence,
\begin{Eqarray}
\lsem t_{1} \rsem^{\Dom} = \lsem t_{2} \rsem^{\Dom} & \Longleftrightarrow & \lsem t_{1} \rsem^{\Dom} \sim^{B} \lsem t_{2} \rsem^{\Dom} &\mbox{by \ref{ifs}} \\
& \Longleftrightarrow & t_{1} \sim^{B} t_{2} & \mbox{by \ref{faft}} \\
& \Longrightarrow & \lsem t_{1} \rsem^{\cal A} = \lsem t_{2} \rsem^{\cal A} &
\end{Eqarray}
since $\cal A$ is in {\bf SCCS-Alg}, and so $f$ is well-defined.
Similarly,
\[ \lsem t_{1} \rsem^{\Dom} \sqsubseteq \lsem t_{2} \rsem^{\Dom} \; \Rightarrow \; t_{1} \preord^{B} t_{2} \; \Rightarrow \; \lsem t_{1} \rsem^{\cal A} \sqsubseteq \lsem t_{2} \rsem^{\cal A}, \]
and so $f$ is monotone. \qed

The purely algebraic part of SCCS which we have developed so far only allows the description of {\em finite} processes.
We now extend the calculus with recursion.
\begin{definition}
{\rm We fix a set of variables {\sf Var}, ranged over by $x, y, z$.
The syntax of {\em recursive terms} ${\rm REC}_{\Sigma}$, is then defined by
\[ t \;\; ::= \;\; \sigma (t_{1}, \ldots , t_{n}) \;\; (\sigma \in \Sigma_{n}) \; | \; x \; | \; {\sf rec} \: x.t \] }
\end{definition}
In an obvious way, we can take $T_{\Sigma}$ as a subset of ${\rm REC}_{\Sigma}$. Note that $ {\sf rec} \: x.t$ is a variable-binding construct.
The set of {\em closed recursive terms} is denoted ${\rm CREC}_{\Sigma}$.

We now extend the definition of the operational semantics to ${\rm CREC}_{\Sigma}$:
\[ (D{\sf rec}) \;\; \frac{t[{\Omega}/x] \diverges}{{\sf rec} \: x.t \diverges} 
\;\;\;\;\;\;\;\; (T {\sf rec}) \;\; \frac{\labarrow{t[{\sf rec} \: x.t/x]}{a}{t'}}{\labarrow{{\sf rec} \: x.t }{a}{t'}} \]

We thus obtain a transition system $({\rm CREC}_{\Sigma}, {\sf Act}, \rightarrow , \diverges )$.
It is not too hard to see that this system is weakly finite-branching, and therefore by~\ref{finots} satisfies (BN).
However, most of the other finiteness conditions on transition systems fail, as the following examples show.

\subsection*{Examples}
(1) {\bf Failure of sort-finiteness.} Assume {\sf Act} is infinite, in 
particular that $\{ a_{n} \}$ is a sequence of distinct actions, and that $S$ is a relabelling such that
\[ S a_{n} = a_{n + 1} \;\; (n \in \omega ). \]
Then
\[ {\sf rec} \: x. \: a_{0} \Oh + x[S] \]
has the behaviour described by the synchronisation tree
\[ {\sum_{n \in \omega} a_{n} \Oh } + \Omega . \]

\noindent (2) {\bf Failure of (FA), and ${\preord_{\omega}} \not= {\preord^{B}}$.} By the example following~\ref{finots}, it suffices to show that the synchronisation tree
\[ p \equiv {\sum_{n \in \omega} a^{n} \Oh } + \Omega \]
can be defined in SCCS to disprove (FA); while the same example shows that $\preord_{\omega} \not= \preord^{B}$, since
\[ p \sim_{\omega} p + a^{\omega}, \;\; p \nsim_{\omega + 1} p + a^{\omega} , \]
and we can define $a^{\omega} \equiv {\sf rec} \: x. \: ax$. 
But using {\em unguarded} recursion (cf. \cite{Mil83}), we can define
\[ p \equiv ({\sf rec} \: x. \: ( \Delta a + ( \Delta a \times x)))\restriction \{ a \} \]
where $\Delta a \equiv {\sf rec} \: y. \: a 1^{\omega} + 1 y$.

\noindent (3) ${\preord^{F}} \not= {\preord_{\omega}}$. Again, following the examples after~\ref{htl}, it suffices to show that the synchronisation trees
\begin{eqnarray*}
p & \equiv & a(\sum_{n \in \Nat} b_{n}\Oh ) + \Omega \\
q & \equiv & {\sum_{n \in \Nat} a({\sum_{m \in \Nat - \{ n \}} b_{m} \Oh} + \Omega )} + \Omega
\end{eqnarray*}
are definable in  SCCS.
Clearly $p$ is definable in the same way as Example~(1).
For $q$, we need some additional assumptions on {\sf Act}:
\begin{itemize}
\item There are $c, \{ c_{n} \} \in {\sf Act}$ such that,
for $k, m \in \Nat$:
\begin{eqnarray*}
c^{(k)} c_{m} & = & b_{m} \;\; (k \not= m) \\
c^{(m)} c_{m} & = & b_{m + 1} 
\end{eqnarray*}
where $c^{(k)} \equiv \underbrace{c \ldots c}_{k}$, i.e. the product in the monoid {\sf Act}.
\item There is a relabelling $S$ such that
\[ S c_{n} = c_{n + 1} \;\; (n \in \Nat ). \]
\end{itemize}
(To see that these requirements can be met, let {\sf Act} be the free abelian
monoid over the generators $0, a, b_{k}, c, c_{k}$ $(k \in \Nat)$ subject
to the relations
\[ 0x = x0 = 0, \;\;\;\; c^{(k)}c_{m} = b_{m} \;\;(k \not= m), \;\;\;\; c^{(m)}c_{m} = b_{m+1} \]
for $k, m \in \Nat$.
Let $S$ be the endomorphism induced by
\[ S 0 = S a = S b_{k} = S c = 0, \;\;\;\; S c_{k} = c_{k+1} , \]
which is well-defined since $S$ preserves the relations.)

Then we can define
\begin{eqnarray*}
q & \equiv & {\sf rec} \: x. \: ar + (1 c \Oh \times x) \\
r & \equiv & {\sf rec} \: y. \: c_{1} \Oh + x[S] ,
\end{eqnarray*}
and calculate:
\begin{eqnarray*}
r & = & {\sum_{n \in \Nat} c_{n} \Oh } + \Omega , \\
q & = & {\sum_{n \in \Nat} ( \prod_{i=1}^{n} 1 c \Oh \times ar) } + \Omega \\
& = & {\sum_{n \in \Nat} a(c^{(n)} \Oh \times \sum_{m \in \Nat} c_{m} \Oh + \Omega )} + \Omega \\
& = & {\sum_{n \in \Nat} a( \sum_{m \in \Nat} (c^{(n)} c_{m}) \Oh + \Omega )} + \Omega \\
& = &  {\sum_{n \in \Nat} a( \sum_{m \in \Nat - \{ n \}} b_{m} \Oh + \Omega )} + \Omega 
\end{eqnarray*}
as required.

By contrast with Example (3), Hennessy claims in~\cite{Hen81} Theorem~4.1 that ${\preord^{F}} = {\preord_{\omega}}$ for SCCS.
The defect in his argument occurs in the definition of $p^{(n)}$ at the start of section~4 of \cite{Hen81}; 
there appears to be an implicit assumption that SCCS is sort-finite.
Indeed, as an easy consequence of our work in the previous Section, we have
\begin{proposition}
In any sort-finite transition system satisfying (BN):
\[ {\preord^{F}} = {\preord_{\omega}} . \]
\end{proposition}

\proof\ Let $p, q \in {\rm Proc}$ in such a system.
\begin{Eqarray}
p \preord^{F} q & \Longrightarrow & \Ellomega (p) \subseteq \Ellomega (q) & \\
& \Longrightarrow & \Ell_{\bigvee \infty}(p) \subseteq \Ell_{\bigvee \infty}(q) & \mbox{(BN)} \\
& \Longrightarrow & {\rm HML}_{\omega}(p) \subseteq {\rm HML}_{\omega}(q) & \\
& \Longrightarrow & p \preord_{\omega} q & \mbox{sort-finiteness. \qed}
\end{Eqarray}

Nevertheless, Hennessy's results on full abstraction are valid when $\preord_{\omega}$ is replaced by $\preord^{F}$, and we shall make use of them shortly.

Firstly, we need to extend our denotational semantics $\lsem \cdot \rsem^{\Dom}$ to recursive terms.
This is done in the standard way; we introduce environments to deal with variables, and interpret recursion by least fixed points.
\begin{definition}
{\rm Denotational semantics of recursive terms:}
\[ {\sf Env} \equiv \Dom^{\sf Var} \]
\[ \lsem \cdot \rsem^{\Dom} : {\rm REC}_{\Sigma}  \longrightarrow {\sf Env} \longrightarrow \Dom \]
\[ \begin{array}{lll}
\lsem x \rsem^{\Dom} \rho & \equiv & \rho x \\
\lsem \sigma (t_{1}, \ldots , t_{n}) \rsem^{\Dom} \rho & \equiv & \sigma^{\Dom} (\lsem t_{1} \rsem^{\Dom} \rho , \ldots , \lsem t_{n} \rsem^{\Dom} \rho ) \\
\lsem {\sf rec} \: x. \, t \rsem^{\Dom} \rho & \equiv & \mu d \in \Dom . \: \lsem t \rsem^{\Dom} \rho [ x \mapsto d ] .
\end{array} \]
\end{definition}

We now want to extend our Full Abstraction Theorem to recursive terms.
We can use Hennessy's results in~\cite{Hen81} to get a cheap proof.
In that paper, Hennessy constructs a term model $\cal I$ with the following properties:
\begin{enumerate}
\item $\cal I$ is an algebraic continuous $\Sigma$-algebra all finite elements of which are definable in $T_{\Sigma}$.
\item $\cal I$ is fully abstract for recursive terms with repect to the finitary preorder; for all $t_{1}, t_{2} \in {\rm CREC}_{\Sigma}$:
\[ t_{1} \preord^{F} t_{2} \;\; \Longleftrightarrow \;\; \lsem t_{1} \rsem^{\cal I} \sqsubseteq  \lsem t_{2} \rsem^{\cal I}. \]
\end{enumerate}
Combining (1) and (2) with Theorem~\ref{binit}, we obtain
\begin{theorem}
\label{isoalg}
$\Dom_{\Sigma}$ and $\cal I$ are isomorphic as continuous $\Sigma$-algebras.
\end{theorem}
Let $h : \Dom_{\Sigma} \rightarrow {\cal I}$ be the isomorphism given by Theorem~\ref{isoalg}.
It is immediate that $h$ preserves denotations of terms in $T_{\Sigma}$:
\[ \forall t \in T_{\Sigma}. \: h(\lsem t \rsem^{\Dom}) = \lsem t \rsem^{\cal I} . \]
To extend this to recursive terms we need one further piece of machinery.
\begin{definition}
{\rm Let $\simeq$ be the least $\Sigma$-congruence over ${\rm REC}_{\Sigma}$ generated by
\[ {\sf rec} \: x. \, t \simeq t[{\sf rec} \: x. \, t/x] . \]
Let $t_{\Omega}$ be the term obtained from $t$ by replacing each subexpression of the form ${\sf rec} \: x. \, t'$ by $\Omega$.
The {\em syntactic approximants} of $t$ are defined by:
\[ SA(t) \equiv \{ t'_{\Omega} : t' \simeq t \} . \] } 
\end{definition}
Note that $SA(t) \subseteq T_{\Sigma}$ for all $t \in {\rm CREC}_{\Sigma}$.

Now the following is standard (cf. e.g. \cite{ADJ77}):
\begin{lemma}[Syntactic Approximation]
\label{sapprox}
For all $t \in {\rm CREC}_{\Sigma}$:
\[ \lsem t \rsem^{\Dom} = \bigsqcup \{ \lsem t' \rsem^{\Dom} : t' \in SA(t) \} . \]
\end{lemma}
Hennessy proves the corresponding result for $\lsem \cdot \rsem^{\cal I}$ as his Lemma~3.4.

\begin{proposition}
\label{hpres}
For all $t \in {\rm CREC}_{\Sigma}$:
\[ h(\lsem t \rsem^{\Dom}) = \lsem t \rsem^{\cal I} . \]
\end{proposition}

\proof\ 
\begin{Eqarray}
h(\lsem t \rsem^{\Dom}) & = & h( \bigsqcup \{ \lsem t' \rsem^{\Dom} : t' \in SA(t) \} ) & \mbox{by \ref{sapprox}} \\
& = & \bigsqcup \{ h( \lsem t' \rsem^{\Dom}) : t' \in SA(t) \}  & \mbox{$h$ is continuous} \\
& = & \bigsqcup \{ \lsem t' \rsem^{\cal I} : t' \in SA(t) \}  & \mbox{by \ref{isoalg}} \\
& = & \lsem t \rsem^{\cal I} . & \qed
\end{Eqarray}

\begin{theorem}[Full Abstraction for Recursive Terms]
For all $t_{1}, t_{2} \in {\rm CREC}_{\Sigma}$:
\[ t_{1} \preord^{F} t_{2} \;\; \Longleftrightarrow \;\;  \lsem t_{1} \rsem^{\Dom} \sqsubseteq \lsem t_{2} \rsem^{\Dom} . \]
\end{theorem}

\proof\ 
\begin{eqnarray*}
t_{1} \preord^{F} t_{2} & \Longleftrightarrow & \lsem t_{1} \rsem^{\cal I} \sqsubseteq \lsem t_{2} \rsem^{\cal I} \\
& \Longleftrightarrow & \lsem t_{1} \rsem^{\Dom} \sqsubseteq \lsem t_{2} \rsem^{\Dom} ,
\end{eqnarray*}
by~\ref{hpres} and since $h$ is an order-isomorphism. \qed

Since $\Dom$ is algebraic, this result extends to terms with variables in the obvious way.
It follows that the axiomatisation of the order and equality relations between 
terms of SCCS presented in \cite{Hen81} is sound and complete for $\Dom_{\Sigma}$.

\chapter{Applications to Functional Programming: 
The Lazy Lambda-Calculus}
\section{Introduction}
In this Chapter, we turn to our second case study, which concerns the 
foundations of functional programming. 
Once again, we aim not merely to exemplify our theory, 
but to use it in order to break some new ground.

The commonly accepted basis for functional programming is the 
$\lambda$-calculus; 
and it is folklore that the $\lambda$-calculus {\em is} the prototypical 
functional language in purified form. But what is the $\lambda$-calculus? 
The syntax is simple and classical; variables, abstraction and application 
in the pure calculus, with applied calculi obtained by adding constants. 
The further elaboration of the theory, covering conversion, reduction, 
theories and models, is laid out in Barendregt's already classical treatise 
\cite{Bar}. 
It is instructive to recall the following crux, which occurs rather early 
in that work (p.\  39):
\subsection*{Meaning of $\lambda$-terms: first attempt}
\begin{itemize}
\item The meaning of a $\lambda$-term is its normal form (if it exists).
\item All terms without normal forms are identified.
\end{itemize}
This proposal incorporates such a simple and natural interpretation of the 
$\lambda$-calculus as a programming language, that if it worked 
there would surely be no doubt that it was the right one. 
However, it gives rise to an inconsistent theory! (see the above reference).
\subsection*{Second attempt}
\begin{itemize}
\item The meaning of $\lambda$-terms is based on head normal forms 
via the notion of {\em Bohm tree}.
\item All unsolvable terms (no head normal form) are identified.
\end{itemize}
This second attempt forms the central theme of Barendregt's book, 
and gives rise to a very beautiful and successful theory 
(henceforth referred to as the ``standard theory''), as that work shows.

This, then, is the commonly accepted foundation for functional programming; 
more precisely, for the {\em lazy} functional languages, 
which represent the mainstream of current functional programming practice. 
Examples: MIRANDA \cite{Tur85}, LML \cite{Aug84}, LISPKIT \cite{Hen80}, 
ORWELL \cite {Wad85}, PONDER \cite{Fai85}, 
TALE \cite{BvL86}. 
But do these languages as defined and implemented actually evaluate terms 
to head normal form? 
To the best of my knowledge, {\em not a single one of them does so}. 
Instead, they evaluate to {\em weak head normal form}, i.e. they do not 
evaluate under abstractions. 
\subsection*{Example} 
$\lambda x. (\lambda y . y)M$ is in weak head normal form, but not in head normal form, since it contains the head redex $(\lambda y . y) M$.

So we have a mismatch between theory and practice. 
Since current practice is well-motivated by efficiency considerations and 
is unlikely to be abandoned readily, it makes sense to see if a good 
modified theory can be developed for it. 
To see that the theory really does need to be modified:
\subsection*{Example}
Let $\Omega \equiv (\lambda x . xx)(\lambda x . xx)$ be the standard 
unsolvable term. Then
\[ \lambda x . \Omega = \Omega \]
in the standard theory, since $\lambda x . \Omega$ is also unsolvable; 
but $\lambda x. \Omega$ is in weak head normal form, hence should be 
distinguished from $\Omega$ in our ``lazy'' theory.

We now turn to a second point in which the standard theory is not completely 
satisfactory.
\subsection*{Is the $\lambda$-calculus a programming language?}
In the standard theory, the $\lambda$-calculus may be regarded as being 
characterised by the type equation
\[ D = [D \rightarrow D] \]
(for justification of this in a general categorical framework, see e.g. 
\cite{Sco80a}, \cite{Koy82,LS86}).

It is one of the most remarkable features of the various categories of 
domains used in denotational semantics that they admit non-trivial 
solutions of this equation. 
However, there is no {\em canonical} solution in any of these categories 
(in particular, the initial solution is trivial -- the one-point domain).

I regard this as a symptom of the fact that the pure $\lambda$-calculus in 
the standard theory {\em is not a programming language}. 
Of course, this is to some extent a matter of terminology, but I feel that 
the expression ``programming language'' should be reserved for a formalism 
with a definite computational interpretation (an operational semantics). 
The pure $\lambda$-calculus as ordinarily conceived is too schematic to qualify.

A further indication of the same point is that studies such as Plotkin's 
``LCF Considered as a Programming Language'' \cite{Plo77} have not been 
carried over to the pure $\lambda$-calculus, for lack of any convincing way of 
doing do in the standard theory. 
This in turn impedes the development of a theory which integrates the 
$\lambda$-calculus with concurrency and other computational notions.

We shall see that by contrast with this situation, the lazy $\lambda$-calculus 
we shall develop does have a canonical model; that Plotkin's ideas can be 
carried over to it in a very natural way; 
and that the theory we shall develop will run quite strikingly in parallel 
with our treatment of concurrency in the previous Chapter.

The plan of the remainder of the Chapter is as follows. 
In the next section, we introduce the intuitions on which our theory is based, 
in the concrete setting of $\lambda$-terms. 
We then set up the axiomatic framework for our theory, 
based on the notion of {\em applicative transition systems}. 
This forms a bridge both to the standard theory, and to concurrency and other 
computational notions. 
Just as in Chapter 4, we introduce a domain equation for applicative 
transition systems, and the corresponding domain logic. 
We prove Duality, Characterisation, and Final Algebra theorems.

We then show how the ideas of \cite{Plo77} can be formulated in our setting. 
Two distinctive features of our approach are:
\begin{itemize}
\item the axiomatic treatment of concepts and results usually presented 
concretely in work on programming language semantics
\item the use of our domain logic as a tool in studying the equational theory 
over our ``programs'' ($\lambda$-terms).
\end{itemize}
Our results can also be interpreted as settling a number of questions and 
conjectures concerning the Domain Interpretation of Martin-Lof's 
Intuitionistic Type Theory raised at the 1983 Chalmers University Workshop on 
Semantics of Programming Languages \cite{Cha83}.

Finally, we consider some extensions and variations of the theory.

\section{The Lazy Lambda-Calculus}
We begin with the syntax, which is standard.
\begin{definition}
{\rm We assume a set {\sf Var} of variables, ranged over by $x, y, z$. The set ${\bf \Lambda}$ of $\lambda$-terms, ranged over by {\mit M, N, P, Q, R} is defined by}
\[ M \;\; ::= \;\; x \; | \; \lambda x . M \; | \; M N . \]
\end{definition}
For standard notions of free and bound variables etc. we refer to \cite{Bar}. The reader should also refer to that work for definitions of notation such as: 
${\sf FV}(M)$, $C[\cdot ]$, $\Lambda^{0}$. 
Our one point of difference concerns substitution; we write $M[N/x]$ rather than $M[x := N]$.
\begin{definition}
\label{convdef}
{\rm The relation $M \Converges N$ (``$M$ converges to principal weak head normal form $N$'') is defined inductively over $\Lambda^{0}$ as follows:}
\[ \bullet \;\; \lambda x . M \Converges \lambda x . M \]
\[ \bullet \;\; \frac{M \Converges \lambda x . P \;\; P[N/x] \Converges Q}{M N \Converges Q} \]
\end{definition}
{\bf Notation} 
\begin{Eqarray} 
M \Converges & \equiv & \exists N . M \Converges N & 
\mbox{(``$M$ converges'')} \\
M \Diverges & \equiv & \neg (M \Converges ) & \mbox{(``$M$ diverges'')}
\end{Eqarray}
It is clear that $\Converges$ is a partial function, i.e. evaluation is deterministic.

We now have an (unlabelled) transition system $(\Lambda^{0}, 
\underline{\ }\Converges \underline{\ } )$. The relation $\Converges$ by itself is too ``shallow'' to yield information about the behaviour of a term under all experiments. However, just as in  the study of concurrency, we shall use it as a building block for a deeper relation, which we shall call {\em applicative bisimulation}. To motivate this relation, let us spell out the observational scenario we have in mind.

Given a closed term $M$, the only experiment of depth 1 we can do is to evaluate $M$ and see if it converges to some abstraction (weak head normal form) $\lambda x . M_{1}$. If it does so, we can continue the experiment to depth 2 by supplying a term $N_{1}$ as input to $M_{1}$, and so on. Note that what the experimenter can observe at each stage is only the {\em fact} of convergence, not which term lies under the abstraction. 
We can picture matters thus:
\begin{center}
\begin{tabular}{cl}
Stage 1 of experiment: & $M \Converges \lambda x . M_{1}$; \\
& environment ``consumes'' $\lambda$, \\
& produces $N_{1}$ as input \\
Stage 2 of experiment: & $M_{1}[N_{1}/x] \Converges \ldots$ \\
$\vdots$ &
\end{tabular}
\end{center}
\begin{definition}[Applicative Bisimulation]
{\rm We define a sequence of relations $\{ \preord_{k} \}_{k \in \omega}$ on $\Lambda^{0}$:
\[ M \preord_{0} N \;\;\;\; {\rm always} \]
\begin{eqnarray*}
M \preord_{k+1} n \;\; \Longleftrightarrow \;\; M \Converges \lambda x . M_{1} & \Rightarrow & \exists N_{1}. \, N \Converges \lambda y . N_{1} \; \& \; \forall P \in \Lambda^{0}. \\
& & M_{1}[P/x] \preord_{k} N_{1}[P/x]
\end{eqnarray*}
\[ M \preord^{B} N \;\; \equiv \;\; \forall k \in \omega . \, M \preord_{k} N \]
Clearly each $\preord_{k}$ and $\preord^{B}$ is a preorder. We extend $\preord^{B}$ to $\Lambda$ by:
\[ M \preord^{B} N \;\; \equiv \;\; \forall \sigma : {\sf Var} \rightarrow \Lambda^{0}. \, M \sigma \preord^{B} N \sigma \]
(where e.g. $M \sigma$ means the result of substituting $\sigma x$ for each $x \in FV(M)$ in $M$). Finally,}
\[ M \bisim^{B} N \;\; \equiv \;\; M \preord^{B} N \; \& \; N \preord^{B} M. \]
\end{definition}
Analogously to our treatment of bisimulation in the previous Chapter, $\preord^{B}$ can be shown to be the maximal fixpoint of a certain function, and hence to satisfy:
\begin{eqnarray*}
M \preord^{B} N \;\; \Longleftrightarrow \;\; M \Converges \lambda x . M_{1} & \Rightarrow & \exists N_{1}. \, N \Converges \lambda y . N_{1} \; \& \; \forall P \in \Lambda^{0}. \\
& & M_{1}[P/x] \preord^{B} N_{1}[P/y]
\end{eqnarray*}
Further details are given in the next section.

The applicative bisimulation relation can be dexcribed in a more traditional 
way (from the point of view of $\lambda$-calculus) as a ``Morris-style 
contextual congruence'' \cite{Mor68,Plo77,Mil77,Bar}.
\begin{definition}
{\rm The relation $\preord^{C}$ on $\Lambda^{0}$ is defined by
\[ M \preord^{C} N \;\; \equiv \;\; \forall C[\cdot ] \in \Lambda^{0}. \, C[M] \Converges \; \Rightarrow \; C[N] \Converges . \]
This is extended to $\Lambda$ in the same way as $\preord^{B}$.}
\end{definition}
\begin{proposition}
\label{cont}
${\preord^{B}} = {\preord^{C}}$.
\end{proposition}
This is a special case of a result we will prove later. Our proof will make essential use of domain logic, despite the fact that the {\em statement} of the result does not mention domains at all. The reader who may be sceptical of our approach is invited to attempt a direct proof.

We now list some basic properties of the relation $\preord^{B}$ (superscript omitted).
\begin{proposition}
\label{lazycong}
For all $M, N, P \in \Lambda$:
\[\begin{array}{rl}
(i) & M \preord M \\
(ii) & M \preord N \; \& \; N \preord P \;\; \Rightarrow \;\; M \preord P \\
(iii) & M \preord N \;\; \Rightarrow \;\; M[P/x] \preord N[P/x] \\
(iv) & M \preord N \;\; \Rightarrow \;\; P[M/x] \preord P[N/x] \\
(v) & \lambda x . M \bisim \lambda y . M[y/x] \\
(vi) & M \preord N \;\; \Rightarrow \;\; \lambda x . M \preord \lambda x . N \\
(vii) & M_{i} \preord N_{i} \; (i=1,2) \;\; \Rightarrow \;\; M_{1}M_{2} \preord N_{1}N_{2} .
\end{array} \]
\end{proposition}

\proof\ $(i)$--$(iii)$ and $(v)$--$(vi)$ are trivial; 
$(vii)$ follows from $(ii)$ and $(iv)$, since taking $C_{1} \equiv [\cdot ]M_{2}$, 
$M_{1}M_{2} \preord N_{1}M_{2}$, and taking $C_{2} \equiv N_{1}[\cdot ]$, $N_{1}M_{2} \preord N_{1}N_{2}$, whence $M_{1}M_{2} \preord N_{1}N_{2}$. It remains to prove $(iv)$, which by 2.5 is equivalent to
\[ M \preord^{C} N \;\; \Rightarrow \;\; P[M/x] \preord^{C} P[N/x]. \]
We rename all bound variables in $P$ to avoid clashes with $M$ and $N$, 
and replace $x$ by $[\cdot ]$ to obtain a context $P[\cdot ]$ such that
\[ P[M/x] = P[M], \;\;\;\; P[N/x] = P[N]. \]
Now let $C[\cdot ] \in \Lambda^{0}$ and $\sigma \in {\sf Var} \rightarrow \Lambda^{0}$ be given. 
Let $C_{1}[\cdot ] \equiv C[P[\cdot ] \sigma ]$. $M \preord^{C} N$ implies
\[ C_{1}[M \sigma ] \Converges \;\; \Rightarrow \;\; C_{1}[N \sigma ] \Converges \]
which, since $(P[M/x])\sigma = (P[\cdot ] \sigma )[M \sigma ]$, yields
\[ C[(P[M/x]) \sigma ] \Converges \;\; \Rightarrow \;\; C[(P[N/x]) \sigma ] \Converges , \]
as required. \qed

This Proposition can be summarised as saying that $\preord^{B}$ is a {\em precongruence}. We thus have an (in)equational theory $\lambda \ell = (\Lambda , \sqsubseteq , = )$, where:
\[ \lambda \ell \; \vdash \; M \sqsubseteq N \;\;\; \equiv \;\;\; M \preord^{B} N \]
\[ \lambda \ell \; \vdash \; M = N \;\;\; \equiv \;\;\; M \bisim^{B} N. \]
What does this theory look like?
\begin{proposition}
\label{lazyprop}
(i) The theory $\lambda$ \cite{Bar} is included in $\lambda \ell$; in particular,
\[ \lambda \ell \; \vdash \; (\lambda x . M)N = M[N/x] \;\;\;\; (\beta ).\]
(ii) ${\bf \Omega} \equiv (\lambda x . xx)(\lambda x . xx)$ is a least element for $\sqsubseteq$, i.e.
\[ \lambda \ell \; \vdash \; {\bf \Omega} \sqsubseteq x . \]
(iii) $(\eta )$ is not valid in $\lambda \ell$, e.g.
\[ \lambda \ell \; \not{\vdash} \; \lambda x . {\bf \Omega} x = {\bf \Omega} , \]
but we do have the following conditional version of $\eta$:
\[ (\Converges \eta ) \;\; \lambda \ell \; \vdash \; \lambda x . M x = M \;\;\;\; (M \Converges , \; x \not\in FV(M)) \]
\[ (M \Converges \; \equiv \; \forall \sigma \in {\sf Var} \rightarrow \Lambda^{0}
. \, (M \sigma ) \Converges). \]
(iv) {\bf YK} is a greatest element for $\sqsubseteq$, i.e.
\[ \lambda \ell \; \vdash \; x \sqsubseteq {\bf YK} . \]
\end{proposition}

\proof\ {\em (i)} is an easy consequence of \ref{lazycong}. \\
{\em (ii)}. ${\bf \Omega} \Diverges$, hence ${\bf \Omega} \preord^{B} M$ for all $M \in \Lambda^{0}$. \\
{\em (iii)}. $\lambda x . {\bf \Omega} x \npreord_{1} {\bf \Omega}$, since $(\lambda x . {\bf \Omega} x)\Converges $. 
Now suppose $M\Converges$, and let $\sigma : {\sf Var} \rightarrow \Lambda^{0}$ be given. 
Then $(M \sigma ) \Converges \lambda y . N$, and $(\lambda x . {\bf \Omega} x)\sigma \Converges \lambda x . {\bf \Omega} x$. 
For any $P \in \Lambda^{0}$,
\begin{Eqarray}
(M \sigma )P \Converges Q & \Leftrightarrow & ((M \sigma )x) [P/x] \Converges Q
& \mbox{since $x \not\in FV(M)$,} \\
& \Leftrightarrow & ((\lambda x . M x) \sigma ) P \Converges Q , & 
\end{Eqarray}
and so $M \bisim^{B} \lambda x . M x$, as required. \\
{\em (iv)}. Note that ${\bf YK} \Converges \lambda y . N$, where 
$N \equiv (\lambda x . {\bf K} (x x))(\lambda x . {\bf K} (x x))$, and that for all $P$,
\[ N[P/y] \Converges \lambda y . N . \]
Hence for all $P_{1}, \ldots , P_{n} \;\; (n \geq 0)$,
\[ {\bf YK} P_{1} \ldots P_{n} \Converges , \]
and so $M \preord^{B} {\bf YK}$ for all $M \in \Lambda^{0}$. \qed

To understand {\em (iv)}, we can think of {\bf YK} as the infinite process \[ \stackrel{\lambda}{\circlearrowleft} \]
solving the equation
\[ \xi = \lambda x . \xi . \]
This is a top element in our applicative bisimulation ordering because it converges under all finite stages of evaluation for all arguments---the experimenter can always observe convergence (or ``consume an infinite $\lambda$-stream'').

We can make some connections between the theory $\lambda \ell$ and \cite{Lon83}, as pointed out to me by Luke Ong. Firstly, \ref{lazyprop}(ii) can be generalised to:
\begin{itemize}
\item The set of terms in $\Lambda^{0}$ which are least in $\lambda \ell$ 
are exactly the $PO_{0}$ terms in the terminology of \cite{Lon83}.
\end{itemize}
Moreover, {\bf YK} is an $O_{\infty}$ term in the terminology of \cite{Lon83}, although it is {\em not} a greatest element in the ordering proposed there.

\section{Applicative Transition Systems}
The theory $\lambda \ell$ defined in the previous section was derived from 
a particular operational model, the transition system $(\Lambda^{0}, \Converges )$. What is the general concept of which this is an example?
\begin{definition}
{\rm A {\em quasi-applicative transition system} is a structure $(A, ev)$ where}
\[ ev : A \rightharpoonup (A \rightarrow A) . \]
\end{definition}
{\bf Notations:}
\[ \begin{array}{rrcl}
(i) & a \Converges f & \equiv & a \in {\sf dom} \, ev \; \& \; ev(a) = f \\
(ii) & a \Converges & \equiv & a \in {\sf dom} \, ev \\
(iii) & a \Diverges & \equiv & a \not\in {\sf dom} \, ev
\end{array} \]
\begin{definition}[Applicative Bisimulation]
\label{apsim}
{\rm Let $(A, ev)$ be a quasi-ats. We define
\[ F : Rel(A) \rightarrow Rel(A) \]
by
\[ F(R) = \{ (a,b) : a \Converges f \;\; \Longrightarrow \;\; b \Converges g \; \& \; \forall c \in A . \, f(c) R g(c) \} . \]
Then $R \in Rel(A)$ is an {\em applicative bisimulation} iff $R \subseteq F(R)$; and ${\preord^{B}} \in Rel(A)$ is defined by
\[ a \preord^{B} b \; \equiv \; a R b \; \mbox{for some applicative bisimulation} \; R. \]}
\end{definition}
Thus ${\preord^{B}} = \bigcup \{ R \in Rel(A) : R \subseteq F(R) \}$, and hence is the maximal fixpoint of the monotone function $F$. 
Since the relation $\Converges$ is a partial function, it is easily shown that the closure ordinal of $F$ is $\leq \omega$, and we can thus describe $\preord^{B}$ more explicitly as follows:
\[ \bullet \;\; a \preord^{B} b \;\; \equiv \;\; \forall k \in \omega . \, a \preord_{k} b \]
\[ \bullet \;\; a \preord_{0} b \;\; {\rm always} \]
\[ \bullet \;\; a \preord_{k+1} b \;\; \equiv \;\; a \Converges f \;\; \Longrightarrow \;\; b \Converges g \; \& \; \forall c \in A. \, f(c) \preord_{k} g(c) \]
\[ \bullet \;\; a \bisim^{B} b \;\; \equiv \;\; a \preord^{B} b \; \& \; b \preord^{B} a . \]
It is easily seen that $\preord^{B}$, and also each $\preord_{k}$, is a preorder; $\bisim^{B}$ is therefore an equivalence.

We now come to our main definition.
\begin{definition}
{\rm An {\em applicative transition system} (ats)  is a quasi-ats $(A, ev)$ satisfying:}
\[ \forall a, b, c \in A. \, a \Converges f \; \& \; b \preord^{B} c \; \Rightarrow \; f(b) \preord^{B} f(c) . \]
\end{definition}
An ats has a well-defined quotient $(A/{\bisim^{B}}, ev/{\bisim^{B}})$, where
\[ ev/{\bisim^{B}} ([a]) = \left\{ \begin{array}{ll}
[b] \mapsto [f(b)], & a \Converges f \\
\mbox{undefined} & \mbox{otherwise.}
\end{array}
\right. \]

The reader should now refresh her memory of such notions as {\em applicative structure, combinatory algebra {\rm and} lambda model} from \cite[Chapter 5]{Bar}.
\begin{definition}
{\rm A {\em quasi-applicative structure with divergence} is a structure $(A, \appl\ , \Diverges )$ such that $(A, \appl\ )$ is an applicative structure, and $\Diverges \subseteq A$ is a divergence predicate satisfying}
\[ x \Diverges \;\; \Longrightarrow \;\; (x\appl\ y)\Diverges . \]
\end{definition}
Given $(A, \appl\ , \Diverges )$, we can define
\[ a \preord^{A} b \;\; \equiv \;\; a \Converges \;\; \Longrightarrow \;\; 
b \Converges \: \& \: \forall c \in A. \, a\appl\ c \preord^{A} b\appl\ c \]
as the maximal fixpoint of a monotone function along identical lines to \ref{apsim}.

Applicative transition systems and applicative structures with divergence are not quite equivalent, but are sufficiently so for our purposes:
\begin{proposition}
\label{appstruct}
Given an ats ${\cal B} = (A, ev)$, we define ${\cal A} = (A, \appl\ , \Diverges)$ by
\[ a\appl\ b \; \equiv \; \left\{ \begin{array}{ll}
a, & a \Diverges \\
f(b) & a \Converges f .
\end{array} \right. \]
Then 
\[ a \preord^{A} b \;\; \Longleftrightarrow \;\; a \preord^{B} b , \]
and moreover we can recover ${\cal B}$ from ${\cal A}$ by
\[ ev(a) =  \left\{ \begin{array}{ll}
b \mapsto a\appl\ b, & a \Converges \\
\mbox{undefined} & \mbox{otherwise.}
\end{array} \right. \]
Furthermore, $\appl\ $ is compatible with $\preord^{B}$, i.e.
\[ a_{i} \preord^{B} b_{i} \; (i = 1,2) \; \Rightarrow \; a_{1}\appl\ a_{2} \preord^{B} b_{1}\appl\ b_{2} . \;\;\; \qed \]
\end{proposition}
We now turn to a language for talking about these structures.
\begin{definition}
{\rm We assume a fixed set of variables {\sf Var}. Given an applicative structure ${\cal A} = (A, \appl\ )$, we define $CL({\cal A})$, the {\em combinatory terms over ${\cal A}$}, by
\[ \bullet \;\; {\sf Var} \subseteq CL({\cal A}) \]
\[ \bullet \;\; \{ c_{a} : a \in A \} \subseteq CL({\cal A}) \]
\[ \bullet \;\; M, N \in CL({\cal A}) \; \Rightarrow \; MN \in CL({\cal A}) . \]
Let $Env({\cal A}) \; \equiv \; {\sf Var} \rightarrow A$. Then the {\em interpretation function}
\[ \lsem \rsem^{{\cal A}} : CL({\cal A}) \rightarrow Env({\cal A}) \rightarrow A \]
is defined by:}
\begin{eqnarray*}
\lsem x \rsem^{{\cal A}}_{\rho} & = & \rho x \\
\lsem c_{a} \rsem^{{\cal A}}_{\rho} & = & a \\
\lsem MN \rsem^{{\cal A}}_{\rho} & = & (\lsem M \rsem^{{\cal A}}_{\rho})\appl\ (\lsem N \rsem^{{\cal A}}_{\rho}) .
\end{eqnarray*}
\end{definition}
Given an ats ${\cal A} = (A, ev)$, with derived applicative structure $(A, \appl\ )$, the satisfaction relation between ${\cal A}$ and atomic formulae over $CL({\cal A})$, of the forms
\[ M \sqsubseteq N, \;\; M = N, \;\; M \Converges \;\; M \Diverges \]
is defined by:
\begin{eqnarray*}
{\cal A}, \rho \models M \sqsubseteq N & \equiv & \lsem M \rsem^{{\cal A}}_{\rho} \preord^{B} \lsem N \rsem^{{\cal A}}_{\rho} \\
{\cal A}, \rho \models  M = N & \equiv & \lsem M \rsem^{{\cal A}}_{\rho} \bisim^{B} \lsem N \rsem^{{\cal A}}_{\rho} \\
{\cal A}, \rho \models M \Converges & \equiv & \lsem M \rsem^{{\cal A}}_{\rho} \Converges \\
{\cal A}, \rho \models M \Diverges & \equiv & \lsem M \rsem^{{\cal A}}_{\rho} \Diverges 
\end{eqnarray*}
while
\[ {\cal A} \models \phi \; \equiv \; \forall \rho \in Env({\cal A}). \, {\cal A}, \rho \models \phi . \]
This is extended to first-order formulae in the usual way.

Note that equality in $CL({\cal A})$ is being interpreted by bisimulation in ${\cal A}$. 
We could have retained the standard notion of interpretation as in \cite{Bar} by working in the quotient structure $(A/{\bisim^{B}}, \appl\  /{\bisim^{B}})$. 
This is equivalent, in the sense that the same sentences are satisfied.
\begin{definition}
{\rm A {\em lambda transition system} (lts) is a structure $(A, ev, k, s)$, where:
\begin{itemize}
\item $(A, ev)$ is an ats
\item $k, s \in A$, and $A$ satisfies the following axioms 
(writing {\bf K, S} for $c_{k}, c_{s}$):
\[ \bullet \;\; {\bf K} \Converges , \;\;\; {\bf K} x \Converges \]
\[ \bullet \;\; {\bf K} x y = x \]
\[ \bullet \;\; {\bf S} \Converges , \;\;\; {\bf S} x \Converges , \;\;\; {\bf S} x y \Converges \]
\[ \bullet \;\; {\bf S} x y z = (x z)(y z) \]
\end{itemize}}
\end{definition}

We now check that these definitions do indeed capture our original example.
\subsection*{Example}
{\rm We define $\ell = (\Lambda^{0}, ev)$, where
\[ ev(M) = \left\{ \begin{array}{ll}
P \mapsto N[P/x] , & M \Converges \lambda x . N \\
\mbox{undefined} & \mbox{otherwise.}
\end{array} \right. \]
$\ell$ is indeed an ats by $\ref{lazycong}(iv)$. Moreover, it is an lts via the definitions
\[ k \; \equiv \; \lambda x . \lambda y . x \]
\[ s \; \equiv \; \lambda x . \lambda y . \lambda z . (x z) (y z) . \]

We now see how to interpret $\lambda$-terms in any lts.
\begin{definition}
{\rm Given an lts ${\cal A}$, we define $\Lambda ({\cal A})$, the $\lambda$-terms over ${\cal A}$, by the same clauses as for $CL({\cal A})$, plus the additional one:
\[ \bullet \;\; x \in {\sf Var}, M \in \Lambda ({\cal A}) \; \Rightarrow \; \lambda x . M \in \Lambda({\cal A}) . \]}
\end{definition}
We define a translation
\[ ( \cdot )_{CL} : \Lambda ({\cal A}) \rightarrow CL({\cal A}) \]
by
\begin{eqnarray*}
(x)_{CL} & \equiv & x \\
(c_{a})_{CL} & \equiv & c_{a} \\
(MN)_{CL} & \equiv & (M)_{CL}(N)_{CL} \\
(\lambda x . M)_{CL} & \equiv & \lambda^{\ast} x . (M)_{CL}
\end{eqnarray*}
where
\begin{eqnarray*}
\lambda^{\ast} x . x & \equiv & {\bf I} \; ( \equiv {\bf SKK}) \\
\lambda^{\ast} x . M & \equiv & {\bf K} M \;\; (x \not\in FV(M)) \\
\lambda^{\ast} x . MN & \equiv & {\bf S} (\lambda^{\ast} x . M) (\lambda^{\ast} x . N) .
\end{eqnarray*}
We now extend $\lsem \cdot \rsem$ to $\Lambda ({\cal A})$ by:
\[ \lsem M \rsem^{{\cal A}}_{\rho} \; \equiv \; \lsem (M)_{CL}\rsem^{{\cal A}}_{\rho} . \]
\begin{definition}
{\rm We define two sets of formulae over $\Lambda$:
\begin{itemize}
\item {\em Atomic formulae:}
\begin{eqnarray*}
{\sf AF} & \equiv & \{ M \sqsubseteq N, \: M = N, \: M \Diverges , \: N \Diverges \; | \;   M, N \in \Lambda \}
\end{eqnarray*} 
\item {\em Conditional formulae:}
\begin{eqnarray*}
{\sf CF} & \equiv & \{ \bigwedge_{i \in I} M_{i} \Converges \wedge \bigwedge_{j \in J} N_{j} \Diverges \Rightarrow F : F \in {\sf AF}, M_{i}, N_{i} \in \Lambda, \\
& &  I, J \; {\rm finite} \}
\end{eqnarray*} 
\end{itemize}
Note that, taking $I = J = \varnothing$, ${\sf AF} \subseteq {\sf CF}$. 
Now given an lts ${\cal A}$, ${\Im}({\cal A})$, the {\em theory} of ${\cal A}$, is defined by
\[  {\Im}({\cal A}) \; \equiv \; \{ C \in {\sf CF} : {\cal A} \models C \} . \]
We also write ${\Im}^{0}({\cal A})$ for the restriction of 
${\Im}({\cal A})$ to closed formulae; 
and given a set {\sf Con} of constants and an interpretation 
${\sf Con} \rightarrow A$, we write ${\Im}({\cal A}, {\sf Con})$ 
for the theory of conditional formulae built from terms in $\Lambda ({\sf Con})$.}
\end{definition}

\noindent {\bf Example (continued)}. 
We set $\lambda \ell = {\Im}(\ell )$. 
This is consistent with our usage in the previous section. 
We saw there that $\lambda \ell$ satisfied much stronger properties than the 
simple combinatory algebra axioms in our definition of lts. 
It might be expected that these would fail for general lts; 
but this is to overlook the powerful extensionality principle built into our 
definition of the theory of an ats through the applicative bisimulation 
relation.
\begin{proposition}
\label{extprop}
Let ${\cal A}$ be an ats. The axiom scheme of {\rm conditional extensionality} over $CL({\cal A})$:
\[ (\Converges {\rm ext}) \;\;\; M\Converges \: \& \: N \Converges \; \Rightarrow \; ([\forall x . M x = N x] \; \Rightarrow \; M = N) \]
\begin{flushright}
$(x \not\in FV(M) \cup FV(N))$
\end{flushright}
is valid in ${\cal A}$.
\end{proposition}

\proof\ Let $\rho \in Env({\cal A})$.
\[ {\cal A}, \rho \; \models \; M \Converges \: \& \: N \Converges \: \& \: \forall x . \, M x = N x \]
\[ \Rightarrow \;\; \lsem M \rsem^{{\cal A}}_{\rho} \Converges \: \& \:  
\lsem N \rsem^{{\cal A}}_{\rho} \Converges \: \& \: \forall a \in A . \, \lsem M \rsem^{{\cal A}}_{\rho} \appl\  a = \lsem N \rsem^{{\cal A}}_{\rho} \appl\  a \]
\begin{flushright}
since $x \not\in FV(M) \cup FV(N)$
\end{flushright}
\[ \Rightarrow \;\; \lsem M \rsem^{{\cal A}}_{\rho} \bisim^{A} \lsem N \rsem^{{\cal A}}_{\rho} \]
\[ \Rightarrow \;\; \lsem M \rsem^{{\cal A}}_{\rho} \bisim^{B} \lsem N \rsem^{{\cal A}}_{\rho} \]
\[ \Rightarrow \;\; {\cal A}, {\rho} \; \models \; M = N . \;\;\; \qed \]
Using this Proposition, we can now generalise most of \ref{lazyprop} to an arbitrary lts.
\begin{theorem}
\label{ltsprop}
Let ${\cal A} = (A, ev, k, s)$ be an lts. Then \\
(i) $(A, ., k, s)$ is a lambda model, and hence $\lambda \subseteq {\Im}({\cal A})$. \\
(ii) ${\cal A}$ satisfies the conditional $\eta$ axiom scheme:
\[ (\Converges \eta ) \;\; M \Converges \; \Rightarrow \; \lambda x . M x = M \;\;\;\; (x \not\in FV(M)) \]
(iii) For all $M \in \Lambda^{0}$:
\[ \lambda \ell \; \vdash \; M \Converges \;\; \Rightarrow \;\; {\cal A} \; \models \; M \Converges \]
(iv) ${\cal A} \; \models \; x \sqsubseteq {\bf YK}$. \\
(v) $\sqsubseteq$ is a precongruence in ${\Im}({\cal A})$.
\end{theorem}

\proof\ (i). Firstly, by the very definition of lts, ${\cal A}$ is a combinatory algebra. We now use the following result due to Meyer and Scott, cited from \cite[Theorem 5.6.3, p.\ 117]{Bar}:
\begin{itemize}
\item Let ${\cal M}$ be a combinatory algebra. Define
\[ {\bf 1} \: \equiv \: {\bf 1}_{1} \: \equiv \: {\bf S(KI)}, \]
\[ {\bf 1}_{k+1} \: \equiv \: {\bf S(K} {\bf 1}_{{\rm k}}) . \]
Then ${\cal M}$ is a lambda model iff it satisfies
\[ \begin{array}{rl}
\mbox{(I)} & \forall x . \, a x = b x \; \Rightarrow \; {\bf 1} a = {\bf 1} b \\
\mbox{(II)} & {\bf 1}_{2} {\bf K} = {\bf K} \\
\mbox{(III)} & {\bf 1}_{3} {\bf S} = {\bf S} .
\end{array} \]
\end{itemize}
Thus it is sufficient to check that ${\cal A}$ satisfies (I)--(III). 
For (I), note firstly that ${\cal A} \; \models \; {\bf 1} a 
\Converges 
x \: \& \: {\bf 1} b \Converges$ by the convergence axioms for an lts. Hence we can apply \ref{extprop} to obtain
\[ {\cal A} \; \models \; [\forall x . \, {\bf 1} a x = {\bf 1} b x] \; \Rightarrow \; {\bf 1} a = {\bf 1} b . \]
We now assume $\forall x . \, a x =  b x$ and prove $\forall x . \, {\bf 1} a x = {\bf 1} b x$:
\begin{eqnarray*}
{\bf 1} a x & = & {\bf S(KI)} a x \\
& = & {\bf (KI)} x (a x) \\
& = & {\bf (KI)} x (b x) \\
& = & {\bf S (KI)} b x \\
& = & {\bf 1} b x .
\end{eqnarray*}
(II) and (III) are proved similarly.

\noindent (ii). Let $\rho \in Env({\cal A})$, and assume ${\cal A}, \rho \; \models \; M \Converges$. We must prove that
\[ {\cal A}, \rho \; \models \; \lambda x . M x = M . \]
Firstly, note that for any abstraction $\lambda z . P$,
\[ {\cal A} \; \models \; \lambda z . P \Converges \]
by the definition of $\lambda^{\ast} z . P$ and the convergence axioms for an lts. Thus since $x \not\in FV(M)$, we can apply $(\Converges {\rm ext})$ to obtain
\[ {\cal A}, \rho \; \models \; [ \forall x . \, (\lambda x . M x) x = M x ] \; \rightarrow \; \lambda x . M x = M . \]
It is thus sufficient to show
\[ {\cal A} \; \models \; (\lambda x . M x) x = M x . \]
But this is just an instance of $(\beta )$, which ${\cal A}$ satisfies by (i).

\noindent (iii). We calculate:
\begin{eqnarray*}
\lambda \ell \; \vdash \; M \Converges & \Rightarrow & M \Converges \lambda x . N \\
& \Rightarrow & \lambda \; \vdash \; M = \lambda x . N \\
& \Rightarrow & {\cal A} \; \models \; M = \lambda x . N \\
& \Rightarrow & {\cal A} \; \models \; M \Converges , 
\end{eqnarray*}
since ${\cal A} \; \models \; \lambda x . N \Converges$, as noted in (ii).

\noindent (iv). By (i) and (iii),
\[ {\cal A} \; \models \; {\bf YK} \Converges \: \& \: \forall x . \, {\bf (YK)} x = {\bf YK} . \]
Hence we can use the same argument as in \ref{lazyprop}(iv) to prove that
\[ {\cal A} \; \models \; x \sqsubseteq {\bf YK} . \]

\noindent (v). This assertion amounts to the same list of properties as Proposition \ref{lazycong}, 
but with respect to ${\Im}({\cal A})$. 
The only difference in the proof is that \ref{lazycong}(vii) follows immediately from \ref{appstruct} and the fact that ${\cal A}$ is an ats, and can then be used to prove \ref{lazycong}(iv) by induction on $P$. \qed

Part (iii) of the Theorem tells us that all the closed terms which we expect to converge must do so in any lts. What of the converse? For example, do we have
\[ {\cal A} \; \models \; {\bf \Omega} \Diverges \]
in every lts? This is evidently not the case, since we have not imposed any axioms which require {\em anything} to be divergent.

\begin{observation}
Let ${\cal A} = (A, ev)$ be an ats in which $ev$ is {\em total}, i.e. ${\sf dom} \; ev = A$. Then ${\Im}({\cal A})$ is {\em inconsistent}, in the sense that
\[ {\cal A} \; \models \; x = y . \]
\end{observation}
This is of course because the distinctions made by applicative bisimulation are based on divergence.

In the light of this observation and \ref{ltsprop}, it is natural to make the following definition in analogy with that in \cite{Bar}:
\begin{definition}
{\rm An lts ${\cal A}$ is {\em sensible} if the converse to \ref{ltsprop}(iii) holds, i.e. for all $M \in \Lambda^{0}$:
\[ {\cal A} \; \models \; M \Converges \;\; \Longleftrightarrow \;\; 
\lambda \ell \; \vdash \; M \Converges \;\; \Longleftrightarrow \;\; \exists x, N. \, \; \lambda \; \vdash \; M = \lambda x . N . \]
(The second equivalence is justified by an appeal to the Standardisation Theorem \cite{Bar}.)}
\end{definition}

\section{A Domain Equation for Applicative Bisimulation}
We now embark on the same programme as in the previous Chapter; 
to obtain a domain-theoretic analysis of our computational notions, 
based on a suitable domain equation. 
What this should be is readily elicited from the definition of ats. 
The structure map
\[ ev : A \rightharpoonup (A \rightarrow A) \]
is {\em partial}; the standard approach to partial maps in domain theory 
({\em pace} Plotkin's recent work on predomains \cite{Plo85}) is to make 
them into total ones by sending undefined arguments to a ``bottom'' element, 
i.e. changing the type of $ev$ to
\[ A \rightarrow (A \rightarrow A)_{\bot} . \]
This suggests the domain equation
\[ D = (D \rightarrow D)_{\bot} \]
i.e. the denotation of the type expression 
${\sf rec} \, t. (t \rightarrow t)_{\bot}$. 
This equation is composed from the function space and lifting constructions. 
Since {\bf SDom} is closed under these constructions, $D$ is a Scott domain. 
Indeed, by the same reasoning it is an algebraic lattice. 
The crucial point is that this equation has a 
{\em non-trivial initial solution}, and thus there is a good candidate 
for a canonical model. 
To see this, consider the ``approximants'' $D_{k}$, with 
$D_{0} \equiv {\bf 1}$, $D_{k+1} \equiv (D_{k} \rightarrow D_{k})_{\bot}$. 
Then
\begin{eqnarray*}
D_{1} & = &  ({\bf 1} \rightarrow {\bf 1})_{\bot} \cong ({\bf 1})_{\bot} \cong \sierp \\
D_{2} & \cong & (\sierp \rightarrow \sierp )_{\bot}, \;\;\; \mbox{with four elements} \\
& \vdots & 
\end{eqnarray*} 
etc. 
We now unpack the structure of $D$. Our treatment will be rather cursory, 
as it proceeds along similar lines to our work in the previous Chapter. 
Firstly, there is an isomorphism pair 
\[ {\sf unfold} : D \rightarrow (D \rightarrow D)_{\bot}, \]
\[ {\sf fold} : (D \rightarrow D)_{\bot} \rightarrow D . \]
Next, we recall the categorical description of lifting, as the left adjoint 
to the forgetful functor
\[ U : {\bf Dom}_{\bot} \rightarrow {\bf Dom} \]
where ${\bf Dom}_{\bot}$ is the sub-category of strict functions. 
Thus we have:
\begin{itemize}
\item A natural transformation ${\sf up} : I_{\bf Dom} \rightarrow U \circ 
( \cdot )_{\bot}$.
\item For each continuous map $f : D \rightarrow U E$ its adjoint
\[ {\sf lift}(f) : (D)_{\bot} \rightarrow_{\bot} E . \]
\end{itemize}
Concretely, we can take
\begin{eqnarray*}
(D)_{\bot} & \equiv & \{ \bot \} \; \cup \; \{ {<}0, d{>} \: | \: d \in D \} \\
x \sqsubseteq y & \equiv & x = {\bot} \\
& & \mbox{or} \; x = {<}0, d{>} \: \& \: y = {<}0, d'{>} \: \& \: d \sqsubseteq_{D} d' \\
{\sf up}_{D}(d) & \equiv & {<}0, d{>} \\
{\sf lift}(f)(\bot ) & \equiv & \bot_{E} \\
{\sf lift}(f) {<}0, d{>} & \equiv & f(d) .
\end{eqnarray*}
We can now define
\[ ev : D \rightharpoonup (D \rightarrow D) \]
by
\[ ev(d) = \left\{ \begin{array}{ll}
f, & {\sf unfold}(d) = {<}0, f{>} \\
{\rm undefined} & {\sf unfold}(d) = \bot .
\end{array} \right. \]
Thus $(D, ev)$ is a quasi-ats, and we write $d \Converges f$, $d \Diverges$ etc. Note that we can recover $d$ from $ev(d)$ by
\[ d = \left\{ \begin{array}{ll}
{\sf fold}({<}0, f{>}),  & d \Converges f \\
\bot_{D} & d \Diverges .
\end{array} \right. \]
The final ingredient in the definition of $D$ is initiality. 
The only direct consequence of this which we will use is contained in
\begin{theorem}
$D$ is internally fully abstract, i.e.
\[ \forall d, d' \in D . \, d \sqsubseteq d' \;\; \Longleftrightarrow \;\; d \preord^{B} d' . \]
\end{theorem}

\proof\ Unpacking the definitions, we see that for all $d, d' \in D$:
\[ d \sqsubseteq d' \;\; \Longleftrightarrow \;\; d \Converges f \; \Rightarrow \; d' \Converges g \: \& \: \forall d'' \in D . \, f(d'' ) \sqsubseteq g(d'' ) . \]
Thus the domain ordering is an applicative bisimulation, and so is included in $\sqsubseteq^{B}$. For the converse, we need some additional notions. We define $d_{k}$, $f_{k}$ for $d \in D$, $f \in [D \rightarrow D]$, $k \in \omega$ by:
\[ d_{0} \Diverges \]
\[ d \Diverges \; \Rightarrow \; d_{k} \Diverges \]
\[ d \Converges f \; \Rightarrow \; d_{k+1} \Converges f_{k} \]
\[ f_{k} : d \mapsto (f d)_{k} . \]
We can use standard techniques to prove, from the initiality of $D$:
\[ \bullet \;\; \forall d \in D . \, d = \bigsqcup_{k \in \omega} d_{k} . \]
The proof is completed with a routine induction to show that:
\[ \forall k \in \omega . \, d \preord_{k} d' \; \Rightarrow \; d_{k} \sqsubseteq d'_{k} . \;\;\; \qed \]
As an immediate corollary of this result, we see that $D$ is an ats. We thus have an interpretation function
\[ \lsem \cdot \rsem^{D} : CL(D) \rightarrow Env(D) \rightarrow \rightarrow D . \]
We extend this to $\Lambda (D)$ by:
\[ \lsem \lambda x . M \rsem^{D}_{\rho} = {\sf fold}({\sf up}(\lambda d \in D . \lsem M \rsem^{D}_{\rho [ x \mapsto d ]} )) . \]
Note that the application  induced from $(D, ev)$ can be described by
\[ d \appl\  d' = {\sf lift}(Ap) \: {\sf unfold}(d) \: d' \]
where
\[ Ap : [D \rightarrow D] \rightarrow D \rightarrow D \]
is the standard application function; and is therefore continuous. This together with standard arguments about environment semantics guarantees that our extension of $\lsem \rsem^{D}$ is well-defined. Note also that $\lsem \lambda x . M \rsem^{D}_{\rho} \neq \bot_{D}$, as expected.

We can now define
\[ k \equiv \lsem \lambda x . \lambda y . x \rsem^{D}_{\rho} , \]
\[ s \equiv \lsem \lambda x . \lambda y . \lambda z . (x z) (y z) \rsem^{D}_{\rho} \]
for $D$. It is straightforward to verify
\begin{proposition}
$D$ is an lts. \qed
\end{proposition}

Thus far, we have merely used our domain equation to construct a particular lts $D$. However, its ``categorical'' or ``absolute'' nature should lead us to suspect that we can use $D$ to study the whole class of lts. The medium we will use for this purpose is once again a suitable domain logic.

\section{A Domain Logic for Applicative Transition Systems}
\begin{definition}
{\rm The syntax of our domain logic ${\cal L}$ is defined by}
\[ \phi \;\; ::= \;\; {\sl t} \; | \; \phi \wedge \psi \; | \; (\phi \rightarrow \psi)_{\bot} \]
\end{definition}
\begin{definition}[Semantics of ${\cal L}$]
{\rm Given a quasi ats ${\cal A}$, we define the satisfaction relation ${\models_{{\cal A}}} \subseteq {{\cal A} \times {\cal L}}$:}
\[ a \; \models_{{\cal A}} \; {\sl t}  \;\; {\rm always} \]
\[ a \; \models_{{\cal A}} \; \phi \wedge \psi \; \; \equiv \;\;   a \; \models_{{\cal A}} \; \phi \; \& \;  a \; \models_{{\cal A}} \; \psi \]
\[ a \; \models_{{\cal A}} \; (\phi \rightarrow \psi )_{\bot} \;\; \equiv \;\; a \Converges f \; \& \; \forall b \in A . \, b \; \models_{{\cal A}} \; \phi \; \Rightarrow \; f(b) \; \models_{{\cal A}} \; \psi . \]
\end{definition}
{\bf Notation:}
\begin{eqnarray*}
{\cal L}(a) & \equiv &  \{ \phi \in {\cal L} \; : \; a \; \models_{{\cal A}} \; \phi \} \\
{\cal A} \; \models \; \phi \leq \psi  & \equiv & \forall a \in A . \, a \; \models_{{\cal A}} \; \phi \;\; \Longrightarrow \;\; a \; \models_{{\cal A}} \; \psi \\
{\cal A} \; \models \; \phi = \psi   & \equiv & \forall a \in A . \, a \; \models_{{\cal A}} \; \phi \;\; \Longleftrightarrow \;\; a \; \models_{{\cal A}} \; \psi \\
\models \; \phi \leq \psi  & \equiv & \forall {\cal A} . \, {\cal A} \; \models \; \phi \leq \psi \\
\lambda  & \equiv & ({\sl t} \rightarrow {\sl t})_{\bot} \\
a \sqsubseteq^{\cal L} b & \equiv & {\cal L}(a) \subseteq {\cal L}(b) .
\end{eqnarray*}
Note that: $\forall a \in A . \, a \Converges \;\; \Longleftrightarrow \;\; a \models_{{\cal A}} \lambda$.
\begin{lemma}
Let ${\cal A}$ be a quasi ats. Then
\[ \forall a, b \in A . \, a \sqsubseteq^{B} b \;\; \Longrightarrow \;\; a \sqsubseteq^{{\cal L}} b . \]
\end{lemma}

\proof\ We assume $a \sqsubseteq^{B} b$ and prove $\forall \phi \in {\cal L} . \,  a \; \models_{{\cal A}} \; \phi \; \Rightarrow \;  b \; \models_{{\cal A}} \; \phi$ by induction on $\phi$. The non-trivial case is $(\phi \rightarrow \psi )_{\bot}$.
\[ \begin{array}{clr}
\bullet & a \; \models_{{\cal A}} \; (\phi \rightarrow \psi )_{\bot} & \\
\Longrightarrow & a \Converges f & \\ 
\Longrightarrow & b \Converges g \: \& \: \forall c . \, f(c) \sqsubseteq^{B} g(c) & \\ 
\Longrightarrow & \forall c . \,  c \; \models_{{\cal A}} \; \phi \; 
\Longrightarrow \;  f(c) \sqsubseteq^{B} g(c) \: \& \: f(c) \; \models_{{\cal A}} \; \psi & \\ 
\Longrightarrow & \forall c . \,  c \; \models_{{\cal A}} \; \phi \; \Rightarrow \;  g(c) \; \models_{{\cal A}} \; \psi & \mbox{ind. hyp.} \\ 
\Longrightarrow & b \; \models_{{\cal A}} \; (\phi \rightarrow \psi )_{\bot} . \;\;\; \qed 
\end{array} \]
To get a converse to this result, we need a condition on ${\cal A}$.
\begin{definition}
{\rm A quasi ats {\cal A} is {\em approximable} iff}
\[ \forall a, b_{1}, \ldots , b_{n} \in A . \, a b_{1} \ldots b_{n} \Converges \; \Rightarrow \; \exists \phi_{1} , \cdots , \phi_{n} . \]
\[ \;\; a \; \models_{{\cal A}} \; (\phi_{1} \rightarrow \cdots ( \phi_{n} \rightarrow \lambda )_{\bot} \cdots )_{\bot} \;\; \& \;\; b_{i} \; \models_{\cal A} \; \phi_{i} , \;\; 1 \leq i \leq n . \]
\end{definition}
This is a natural condition, which says that convergence of a function application is caused by some finite amount of information (observable properties) of its arguments.

As expected, we have
\begin{theorem}[Characterisation Theorem]
Let ${\cal A}$ be an approximable quasi ats. Then
\[ {\preord^{B}} = {\preord^{\cal L}} . \]
\end{theorem}

\proof\ By 5.3, ${\preord^{B}} \subseteq {\preord^{\cal L}}$. For the converse, suppose $a \notpreord^{B} b$. Then for some $k$, $a \notpreord^{B}_{k} b$, and so for some $c_{1}, \cdots , c_{k} \in A$:
\[ a c_{1} \cdots c_{k} \Converges \; \& \; b c_{1} \cdots c_{k} \Diverges . \]
By approximability, for some $\phi_{1} , \cdots , \phi_{k} \in {\cal L}$,
\[ a \; \models_{{\cal A}} \; (\phi_{1} \rightarrow \cdots ( \phi_{k} \rightarrow \lambda )_{\bot} \cdots )_{\bot} \; \& \; b_{i} \; \models_{\cal A} \; \phi_{i} , \;\; 1 \leq i \leq k . \]
Clearly $b \; \nvDash_{{\cal A}} \; (\phi_{1} \rightarrow \cdots ( \phi_{k} \rightarrow \lambda )_{\bot} \cdots )_{\bot}$, and so $a \notpreord^{\cal L} b$. \qed

As a further consequence of approximability, we have:
\begin{proposition}
An approximable quasi ats is an ats.
\end{proposition}

\proof\ Suppose $a \Converges f$ and $b \preord^{B} c$. We must show $f(b) \preord^{B} f(c)$. It is sufficient to show that for all $k \in \omega$, $d_{1}, \ldots , d_{k} \in A$:
\[ f(b) d_{1} \ldots d_{k} \Converges \;\; \Rightarrow \;\; f(c) d_{1} \ldots d_{k} \Converges . \]
Now $f(b) d_{1} \ldots d_{k} \Converges$ implies $a b d_{1} \ldots d_{k} \Converges$; hence by approximability, for some $\phi, \phi_{1}, \ldots \phi_{k} \in {\cal L}$:
\[ a \; \models_{\cal A} \; (\phi_{1} \rightarrow \cdots ( \phi_{k} \rightarrow \lambda )_{\bot} \cdots )_{\bot} \]
and
\[ b \; \models_{\cal A} \; \phi , \;\;  b_{i} \; \models_{\cal A} \; \phi_{i} , \;\; 1 \leq i \leq k . \]
By 5.5, $c \; \models_{\cal A} \phi$, and so $a b d_{1} \ldots d_{k} \; \models_{\cal A} \; \lambda$, and $f(c) d_{1} \ldots d_{k} \Converges$ as required. \qed

We now introduce a proof system for assertions of the form $\phi \leq \psi$, $\phi = \psi$ ($\phi , \psi \in {\cal L})$.
\subsection*{Proof System For $\cal L$}
\[ ({\rm REF}) \;\;\; \phi \leq \phi \]
\[ ({\rm TRANS}) \;\;\; \frac{\phi \leq \psi \;\; \psi \leq \xi}{\phi \leq \xi} \]
\[ (=-I) \;\;\; \frac{\phi \leq \psi \;\; \psi \leq \phi}{\phi = \psi} \]
\[ (=-E) \;\;\; \frac{\phi = \psi}{\phi \leq \psi \;\; \psi \leq \phi} \]
\[(\true\ -I) \;\;\; \phi \leq \true \]
\[ (\wedge -I) \;\;\; \frac{\phi \leq \phi_{1} \;\; \phi \leq \psi_{2}}{\phi \leq \phi_{1} \wedge \phi_{2}} \]
\[ (\wedge -E) \;\;\; \phi \wedge \psi \leq \phi \;\;\;\; \phi \wedge \psi \leq \psi \]
\[ ((\rightarrow )_{\bot}-\leq ) \;\;\; \frac{\phi_{2} \leq \phi_{1} \;\; \psi_{1} \leq \psi_{2}}{(\phi_{1} \rightarrow \psi_{1})_{\bot} \leq (\phi_{2} \rightarrow \psi_{2})_{\bot}} \]
\[ ((\rightarrow )_{\bot}-\wedge ) \;\;\; (\phi \rightarrow \psi_{1} \wedge \psi_{2} )_{\bot} = (\phi \rightarrow \psi_{1})_{\bot} \wedge (\phi \rightarrow \psi_{2})_{\bot} \]
\[ ((\rightarrow )_{\bot}-\true ) \;\;\; (\phi \rightarrow \true )_{\bot} \leq (\true \rightarrow \true )_{\bot} . \]
We write ${\cal L} \; \vdash \; A$ or just $\vdash \; A$ to indicate that an assertion $A$ is derivable from these axioms and rules.
Note that the converse of $((\rightarrow )_{\bot}-\true )$ is derivable from 
$(\true -I)$ and $((\rightarrow )_{\bot}-\leq )$; 
by abuse of notation we refer to the corresponding equation by the same name.

\begin{theorem}[Soundness Theorem]
\label{lsoun}
$\vdash \; \phi \leq \psi \;\; \Longrightarrow \;\; \models \; \phi \leq \psi$.
\end{theorem}

\proof\ By a routine induction on the length of proofs. \qed

So far, our logic has been presented in a syntax-free fashion so far as the elements of the ats are concerned. Now suppose we have an lts ${\cal A}$. $\lambda$-terms can be interpreted in $\cal A$, and for $M \in \BLambda^{0}$, $\rho \in Env(\cal A )$, we can define:
\[ M, \, \rho \; \models_{\cal A} \; \phi \;\; \equiv \;\; \lsem M \rsem^{\cal A}_{\rho} \; \models_{\cal A} \; \phi . \]
We can extend this to arbitrary terms $M \in \BLambda$ in the presence of {\em assumptions} $\Gamma :  {\sf Var} \rightarrow {\cal L}$ on the variables:
\[ M, \, \Gamma \; \models_{\cal A} \; \phi \;\; \equiv \;\; \forall \rho \in Env({\cal A}). \, \rho \; \models_{\cal A} \; \Gamma \;\; \Rightarrow \;\; \lsem M \rsem^{\cal A}_{\rho} \; \models_{\cal A} \; \phi  \]
where
\[ \rho \; \models_{\cal A} \; \Gamma \;\; \equiv \;\; \forall x \in {\sf Var}. \, \rho x \; \models_{\cal A} \; \Gamma x . \]
We write
\[ M, \, \Gamma \; \models \; \phi \;\; \equiv \;\; \forall {\cal A} . \, M, \, \Gamma \; \models_{\cal A} \; \phi . \]

We now introduce a proof system for assertions of the form $M, \, \Gamma \; \vdash \; \phi$.
\subsection*{Proof System For Program Logic}
\[ (TR) \;\;\; M, \, \Gamma \vdash \; \true \]
\[ (AND) \;\;\; \frac{M, \, \Gamma \; \vdash \; \phi \;\; M, \, \Gamma \; \vdash \; \psi}{M, \, \Gamma \; \vdash \; \phi \wedge \psi} \]
\[ (LEQ) \;\;\; \frac{\Gamma \leq \Delta \;\; M, \, \Delta \; \vdash \; \phi \;\; \phi \leq \psi}{M, \, \Gamma \; \vdash \; \psi} \]
\[ (VAR) \;\;\; x, \, \Gamma [ x \mapsto \phi ] \; \vdash \; \phi \]
\[ (ABS) \;\;\; \frac{M, \, \Gamma [ x \mapsto \phi ] \; \vdash \; \psi}{\lambda x . M, \, \Gamma \; \vdash \; (\phi \rightarrow \psi )_{\bot}} \]
\[ (APP) \;\;\; \frac{M, \, \Gamma \; \vdash \; ( \phi \rightarrow \psi )_{\bot} \;\; N, \, \Gamma \; \vdash \; \phi}{MN, \, \Gamma \; \vdash \; \psi} . \]
\begin{theorem}[Soundness of Program Logic]
For all $M$, $\Gamma$, $\phi$:
\[ M, \, \Gamma \: \vdash \: \phi \;\; \Longrightarrow \;\; M, \, \Gamma \: 
\models \: \phi . \;\;\; \qed \]
\end{theorem}
The proof is again routine. 
Note the striking similarity of our program logic with type inference, in particular with the intersection type discipline and Extended Applicative Type Structures of \cite{CDHL84}. 
The crucial {\em difference} lies in the entailment relation $\leq$, and in particular the fact that their axiom (in our notation)
\[ \true \leq (\true \rightarrow \true )_{\bot} \]
is {\em not} a theorem in our logic; instead, we have the weaker $((\rightarrow )_{\bot})$. This reflects a different notion of ``function space''; we discuss this further in section 7.

We now come to the expected connection between the domain logic $\cal L$ 
and the domain $D$. 
Once again, the connecting link is the domain equation used to define $D$, 
and from which $\cal L$ is derived. 
Since this equation corresponds to the type expression 
$\sigma \; \equiv \; {\sf rec} \, t. (t \rightarrow t)_{\bot}$, 
it falls within the scope of the general theory developed in Chapter 4. 
The logic $\cal L$ presented in this section is a streamlined version of 
${\cal L}(\sigma )$ as defined in Chapter 4. 
Once we have shown that $\cal L$ is equivalent to  ${\cal L}(\sigma )$, we 
can apply the results of Chapter 4 to obtain the desired relationships between 
${\cal L} \simeq {\cal L}(\sigma )$ and $D \simeq D(\sigma )$.

Firstly, note that $\cal L$ as presented contains no disjunctive structure, 
while the constructs $\rightarrow$, $(\cdot )_{\bot}$ appearing in $\sigma$ 
generate no inconsistencies according to the definition of {\sf C} in Chapter 4. 
Thus (the Lindenbaum algebra of) ${\cal L}_{\wedge}(\sigma )$, 
the purely conjunctive part of ${\cal L}(\sigma )$, is a meet-semilattice, 
and applying Theorem~\ref{msl}, we obtain
\[ {\sf Spec} \; ({\cal L}(\sigma )/{=_{\sigma}}, {\leq_{\sigma}}/{=_{\sigma}}) \; \cong \; {\sf Filt}({\cal L}_{\wedge}(\sigma )/{=_{\sigma}}, {\leq_{\sigma}}/{=_{\sigma}}) . \]
It remains to show that $\cal L$ is pre-isomorphic to 
${\cal L}_{\wedge}(\sigma )$. 
We can describe the syntax of ${\cal L}_{\wedge}(\sigma )$ as follows:
\begin{itemize}
\item  $L_{\wedge}(\sigma )$:
\[ \phi \;\; ::= \;\; \true \; | \; \phi \wedge \psi \; | \; (\phi )_{\bot} \;\; (\phi \in L(\sigma \rightarrow \sigma )) \]
\item $L_{\wedge}(\sigma \rightarrow \sigma )$:
\[ \phi \;\; ::= \;\; \true \; | \; \phi \wedge \psi \; | \; (\phi \rightarrow \psi) \;\; (\phi , \psi \in L(\sigma )). \] 
\end{itemize}
Using $(()_{\bot}-\wedge )$ and $(\rightarrow -\true )$ (i.e. the nullary instances of $(\rightarrow - \wedge )$) from Chapter 4, we obtain the following normal forms for $L_{\wedge}(\sigma )$:
\[ \phi \;\; ::= \;\; \true \; | \; \phi \wedge \psi \; | \; (\phi \rightarrow \psi )_{\bot} . \]
In this way we see that $L \subseteq L_{\wedge}(\sigma )$, and that each $\phi \in L_{\wedge}(\sigma )$ is equivalent to one in $L$. 
Moreover, the axioms and rules of $\cal L$ are easily seen to be derivable in ${\cal L}_{\wedge}(\sigma )$. 
For example, $((\rightarrow )_{\bot}-\true )$ is derivable, since
\[ {\cal L}_{\wedge}(\sigma ) \; \vdash \; (\phi \rightarrow \psi )_{\bot} = (\true )_{\bot} = (\true \rightarrow \true )_{\bot} . \]
It remains to show the converse, i.e. that for $\phi , \psi \in {\cal L}$:
\[ {\cal L}_{\wedge}(\sigma ) \; \vdash \; \phi \leq \psi \;\; \Longrightarrow \;\; {\cal L} \; \vdash \; \phi \leq \psi . \]
For this purpose, we use $((\rightarrow )_{\bot}-\wedge )$ and $((\rightarrow )_{\bot}-\true )$ to get normal forms for $\cal L$.
\begin{lemma}[Normal Forms]
Every formula in $\cal L$ is equivalent to one in $N{\cal L}$, where:
\[ \bullet \;\; N{\cal L} = \{ \bigwedge_{i \in I} \phi_{i} : I \; {\rm finite}, \; \phi_{i} \in SN{\cal L}, \; i \in I \}\]
\[ \bullet \;\; SN{\cal L} = \{ (\phi_{1} \rightarrow \cdots (\phi_{k} \rightarrow \lambda )_{\bot} \cdots )_{\bot} : k \geq 0, \phi_{i} \in N{\cal L}, \; 1 \leq i \leq k \} . \;\; \qed \]
\end{lemma}
Now by the semantic arguments of Chapter 3, we have
\begin{lemma}
For $\phi$, $\psi$ with
\[ \phi \; \equiv \; \bigwedge_{i \in I} (\phi_{i} \rightarrow \phi'_{i})_{\bot}, \]
\[ \psi \; \equiv \; \bigwedge_{j \in J} (\psi_{j} \rightarrow \psi'_{j})_{\bot} : \]
\[ {\cal L}(\sigma ) \; \vdash \; \phi \leq \psi \;\; \Longleftrightarrow \;\; \forall j \in J.\, 
{\cal L}(\sigma ) \; \vdash \; \bigwedge \{ \phi'_{i} \; : \; {\cal L}(\sigma ) \; \vdash \; \psi_{j} \leq \phi_{i} \} \leq \psi'_{j} . \] 
\end{lemma}
\begin{proposition}
For $\phi , \psi \in N{\cal L}$, if ${\cal L}(\sigma ) \; \vdash \; \phi \leq \psi$ then there is a proof of $\phi \leq \psi$ using only the meet-semilattice laws and the derived rule $((\rightarrow )_{\bot})$.
\end{proposition}

\proof\ By induction on the complexity of $\phi$ and $\psi$, and the preceding Lemma. \qed

We have thus shown that 
\[ {\cal L}(\sigma ) \; \cong \; {\cal L}_{\wedge}(\sigma ) \; \cong \;
{\cal L} , \]
and we can apply the Duality Theorem of Chapter 4 to obtain
\begin{theorem}[Stone Duality]
$\cal L$ is the Stone dual of $\cal D$:
\[ \begin{array}{rl}
(i) & {\cal D} \; \cong \; {\sf Filt} \: {\cal L} \\
(ii) & (K({\cal D}))^{op} \; \cong \; (L/{=}, {\leq}/{=}). 
\end{array} \]
\end{theorem}
\begin{corollary}
${\cal D} \; \models \; \phi \leq \psi \;\; \Longleftrightarrow \;\; {\cal L} \; \vdash \; \phi \leq \psi$. 
\end{corollary}

We can now deal with the program logic over $\lambda$-terms in a similar fashion. The denotational semantics for $\BLambda$ in $\cal D$ given in the precious section can be used to define a translation map
\[ ( \cdot )^{\ast} : \BLambda \rightarrow \BLambda (\sigma ) . \]
The logic presented in this section is equivalent to the endogenous logic of Chapter 4 in the sense that
\[ M,\, \Gamma \; \vdash \; \phi \;\; \Longleftrightarrow \;\; M^{\ast}, \, \Gamma \; \vdash \; \phi \]
where $M \in \BLambda$, $\Gamma : {\sf Var} \rightarrow L$, $\phi \in L \subseteq L(\sigma )$. We omit the details, which by now should be routine. As a consequence of this result, we can apply the Completeness Theorem for Endogenous Logic from Chapter 4, to obtain:
\begin{theorem}
$\cal D$ is $\cal L$-complete, i.e. for all $M \in \BLambda$, $\Gamma : {\sf Var} \rightarrow L$, $\phi \in L \subseteq L(\sigma )$:
\[ M, \, \Gamma \; \vdash \; \phi \;\; \Longleftrightarrow \;\; M, \, \Gamma \; \models_{\cal L} \; \phi . \]
\end{theorem}

In the previous section, we defined an lts over $\cal D$; 
and we have now shown that $\cal D$ is isomorphic to ${\sf Filt} \: {\cal L}$. 
We can in fact describe the lts structure over ${\sf Filt} \: {\cal L}$ directly; 
and this will show how $\cal D$, defined by a domain equation reminiscent of 
the $D_{\infty}$ construction, can also be viewed as a graph model or 
``PSE algebra'' in the terminology of \cite{Lon83}.

\noindent {\bf Notation.} For $X \subseteq L$, $X^{\dag}$ is the filter generated by $X$. This can be defined inductively by:
\begin{itemize}
\item $X \subseteq X^{\dag}$
\item $\true \in X^{\dag}$
\item $\phi , \psi \in X^{\dag} \; \Rightarrow \; \phi \wedge \psi \in X^{\dag}$
\item $\phi \in X^{\dag}, \; {\cal L} \: \vdash \: \phi \leq \psi \;\; \Rightarrow \;\; \psi \in X^{\dag}$ .
\end{itemize}
\begin{definition}
{\rm The quasi-applicative structure with divergence 
\[ ({\sf Filt} \: {\cal L}, \appl\ , \Diverges ) \]
is defined as follows:
\[ \bullet \;\; x \Diverges \;\; \equiv \;\; x = \{ \true \} \]
\[ \bullet \;\; x \appl\  y \;\; \equiv \;\; \{ \psi : \exists \phi . \, (\phi \rightarrow \psi )_{\bot} \in x \: \& \: \phi \in y \} \cup \{ \true \} . 
\]}
\end{definition}
It is easily verified that in this structure
\[ x \preord^{B} y \;\; \Longleftrightarrow \;\; x \subseteq y , \]
and hence that application is monotone in each argument, and ${\sf Filt} \: {\cal L}$ is an ats. Thus we have an interpretation function
\[ \lsem \cdot \rsem^{{\sf Filt} \: {\cal L}} : CL({\sf Filt} \: {\cal L}) \rightarrow Env({\sf Filt} \: {\cal L}) \rightarrow {\sf Filt} \: {\cal L} \]
which is extended to $\BLambda({\sf Filt} \: {\cal L})$ by
\[ \lsem \lambda x . M \rsem^{{\sf Filt} \: {\cal L}}_{\rho} = \{ (\phi \rightarrow \psi )_{\bot} : \psi \in \lsem M \rsem^{{\sf Filt} \: {\cal L}}_{\rho [ x \mapsto \diverges \psi ]} \}^{\dag} . \]
We then define
\begin{definition}
\begin{eqnarray*}
s & \equiv & \lsem \lambda x . \lambda y . \lambda z . (xz) (yz) \rsem^{{\sf Filt} \: {\cal L}} \\
k & \equiv & \lsem \lambda x . \lambda y . x \rsem^{{\sf Filt} \: {\cal L}} . 
\end{eqnarray*}
\end{definition}
\begin{proposition}
\label{isocalg}
${\sf Filt} \: {\cal L}$ is an lts. Moreover, ${\sf Filt} \: {\cal L}$ and $\cal D$ are isomorphic as combinatory algebras.
\end{proposition}

\proof\ It is sufficient to show that the isomorphism of the Duality Theorem 
preserves application, divergence and the denotation of $\lambda$-terms, 
since it then preserves $s$ and $k$ and so is a combinatory isomorphism, and 
${\sf Filt} \: {\cal L}$ is an lts, since $\cal D$ is.

Firstly, we show that application is preserved, i.e. for 
$d_{1}, d_{2} \in {\cal D}$:
\[ (\star ) \;\; {\cal L}(d_{1} \appl\ d_{2}) = {\cal L}(d_{1}) \appl\ {\cal L}(d_{2}) \]
The right to left inclusion follows by the same argument as the soundness of 
$(APP)$ in \ref{lsoun}. 
For the converse, suppose $\psi \in {\cal L}(d_{1} \appl\  d_{2})$, 
${\cal L} \; \nvdash \; \psi = \true$. 
By the Duality Theorem, each $\psi$ in ${\cal L}$ corresponds to a unique 
$c \in K({\cal D}$ with ${\cal L}(c) = {\diverges} \psi$. 
Since application is continuous in $\cal D$, $c \sqsubseteq d_{1}\appl\ d_{2}$, 
$c \neq \bot$ implies that for some $b \in K({\cal D})$, 
${\sf fold}({<}0, [b,c]{>}) \sqsubseteq d_{1}$ and 
$b \sqsubseteq d_{2}$. Let ${\cal L}(b) = {\diverges} \phi$, 
then $(\phi \rightarrow \psi )_{\bot} \in {\cal L}(d_{1})$ and 
$\phi \in {\cal L}(d_{2})$, as required.

Next, we show that denotations of $\lambda$-terms are preserved, i.e. for all 
$M \in \BLambda$, $\rho \in Env({\cal D})$:
\[ (\star \star ) \;\; {\cal L}( \lsem M \rsem^{\cal D}_{\rho}) = \lsem M \rsem^{{\sf Filt} \: {\cal L}}_{{\cal L} \circ \rho} . \]
This is proved by induction on $M$. The case when $M$ is a variable is trivial; 
the case for application uses $(\star )$. For abstraction, 
we argue by structural induction over ${\cal L}$. 
We show the non-trivial case. 
Let $\phi$, $b$ be paired in the isomorphism of the Duality Theorem. Then
\[ \begin{array}{clr}
& \lambda x . M , \, \rho \; \models_{\cal D} \; (\phi \rightarrow \psi )_{\bot} & \\
\Longleftrightarrow &  M , \, \rho [ x \mapsto b ] \; \models_{\cal D} \; \psi & \\
\Longleftrightarrow &  M , \, {\cal L}() \circ (\rho [ x \mapsto b ]) \; \models_{{\sf Filt} \: {\cal L}} \; \psi & \mbox{ind. hyp.} \\
\Longleftrightarrow &  M , \, ({\cal L}() \circ \rho ) [ x \mapsto \diverges \phi ] \; \models_{{\sf Filt} \: {\cal L}} \; \psi  & \\
\Longleftrightarrow & \lambda x . M , \, {\cal L}() \circ \rho \; \models_{{\sf Filt} \: {\cal L}} \; (\phi \rightarrow \psi )_{\bot} . &
\end{array} \]

Finally, divergence is trivially preserved, since the only divergent 
elements in 
$\cal D$, ${\sf Filt} \: {\cal L}$ are $\bot$, $\{ \true \}$, are these are in 
bi-unique correspondence under the isomorphism of the Duality Theorem. \qed

We can now proceed in exact analogy to Chapter 5, and use Stone Duality to convert the Characterisation Theorem into a Final Algebra Theorem.
\begin{definition}
{\rm We define a number of categories of transition systems:
\begin{description}
\item[ATS] Objects: applicative transition systems; morphisms ${\cal A} \rightarrow {\cal B}$: maps $f : A \rightarrow B$ satisfying
\[ a \; \models_{\cal A} \; \phi \;\; \Longleftrightarrow \;\; f(a) \; \models_{\cal B} \; \phi . \]
\item[LTS] The subcategory of {\bf ATS} of lts and morphisms which preserve application, $s$ and $k$.
\item[CLTS] The full subcategory of {\bf LTS} of those $\cal A$ satisfying {\em continuity}:
\[ \psi \neq \true , \; a b \; \models_{\cal A} \; \psi \;\; \Longrightarrow \;\; \exists \phi . \, a \; \models_{\cal A} \; (\phi \rightarrow \psi )_{\bot} \: \& 
\: b \; \models_{\cal A} \; \phi , \]
and also
\[ {\cal L}(s) = \lsem s \rsem^{{\sf Filt} \: {\cal L}} , \;\; {\cal L}(k) = \lsem k \rsem^{{\sf Filt} \: {\cal L}} . \]
Note that continuity implies approximability.
\end{description}}
\end{definition}
\begin{theorem}[Final Algebra]
(i) $\cal D$ is final in {\bf ATS}. \\
(ii) Let $\cal A$ be an approximable lts. The map
\[ {\sf t}_{\cal A} : {\cal A} \rightarrow {\cal D} \]
from (i) is an {\bf LTS} morphism iff $\cal A$ is continuous. \\
(iii) $\cal D$ is final in {\bf CLTS}.
\end{theorem}

\proof\  (i). Given $\cal A$ in {\bf ATS}, define
\[ {\sf t}_{\cal A} : {\cal A} \rightarrow {\cal D} \]
by
\[ {\sf t}_{\cal A} \;\; \equiv \;\; {\cal A} \stackrel{{\cal L}()}{\rightarrow} {\sf Filt} \: {\cal L} \stackrel{\eta}{\rightarrow} {\cal D} \]
where $\eta$ is the isomorphism from the Stone Duality Theorem. For $a \in A$,
\[ {\cal L}(a) = {\cal L} \circ \eta \circ {\cal L}(a) = {\cal L} \circ {\sf t}_{\cal A}(a) , \]
and so ${\sf t}_{\cal A}$ is an {\bf ATS} morphism; moreover, it is unique, since for $d, d' \in D$:
\[ {\cal L}(d) = {\cal L}(d') \; \Rightarrow \; {\cal K}(d) = {\cal K}(d') \; \Rightarrow \; d = d' . \]
(ii). That ${\cal L} ()$ is a combinatory morphism iff $\cal A$ is in {\bf CLTS} is an immediate consequence of the definitions; the result then follows from the fact that $\eta$ is a combinatory isomorphism. \\
(iii). Immediate from (ii). \qed

Note that if $\cal A$ is approximable, we have:
\[ a \preord^{B} b \;\; \Longleftrightarrow \;\; {\sf t}_{\cal A}(a) \preord^{B} {\sf t}_{\cal A}(b) . \]
Thus we can regard the Final Algebra Theorem as giving a syntax-free fully abstract semantics for approximable ats. 
However, from the point of view of applications to programming language semantics, this is not very useful.  
In the next section, we shall study full abstraction in a syntax-directed framework, using our domain logic as a tool.

\section{Lambda Transition Systems considered as Programming Languages}
The classical discussion of full abstraction in the  $\lambda$-calculus 
\cite{Plo77,Mil77} is set in the typed $\lambda$-calculus with ground data. 
As remarked in the Introduction, this material has not to date been 
transferred successfully to the pure untyped $\lambda$-calculus. 
To see why this is so, let us recall some basic notions from \cite{Plo77,Mil77}.

Firstly, there is a natural notion of {\em program}, namely closed term of 
ground type. Programs either diverge, or yield a ground constant as result. 
This provides a natural notion of observable behaviour for programs, and 
hence an operational order on them. This is extended to arbitrary terms 
via ground contexts; in other words, the point of view is taken that only 
program behaviour is directly observable, and the meaning of a higher-type 
term lies in the observable behaviour of the programs into which it can be 
embedded. 
Thus both the presence of ground data, and the fact that terms are typed, 
enter into the basic definitions of the theory.

By contrast, we have a notion of atomic observation for the lazy 
$\lambda$-calculus in the absence of types or ground data, namely convergence 
to weak head normal form. This leads to the applicative bisimulation relation, 
and hence to a natural operational ordering. 
We can thus develop a theory of full abstraction in the pure untyped 
$\lambda$-calculus. 
Our results will correspond recognisably to those in \cite{Plo77}, although 
the technical details contain many differences. One feature of 
our development is that we work axiomatically with classes of lts under 
various hypotheses, rather than with particular languages. 
(Note that operational transition systems and ``programming languages'' such as $\lambda \ell$ actually {\em are} lts under our definitions.)
\begin{definition}
{\rm Let $\cal A$ be an lts. $\cal D$ is {\em fully abstract} for $\cal A$ if ${\Im}({\cal A}) = {\Im}({\cal D})$. }
\end{definition}
This definition is consistent with that in \cite{Plo77,Mil77}, provided 
we accept the applicative bisimulation ordering on $\cal A$ as 
the appropriate operational preorder. The argument for doing so 
is made highly plausible by Proposition~\ref{cont}, which characterises 
applicative bisimulation as a contextual preorder analogous to those used 
in \cite{Plo77,Mil77}. We shall prove \ref{cont} later in this section.

We now turn to the question of conditions under which $\cal D$ is 
fully abstract for $\cal A$. 
As emerges from \cite{Plo77,Mil77}, this is essentially a question of definability.
\begin{definition}
{\rm An ats $\cal A$ is $\cal L$-{\em expressive} if for all $\phi \in {\cal L}$, for some $a \in {\cal A}$:
\[ {\cal L}(a) = {\uparrow} \phi \; \equiv \; \{ \psi \in {\cal L} \; : \; {\cal L} \: \vdash \: \phi \leq \psi \} . \]
}
\end{definition}
In the light of Stone Duality, $\cal L$-expressiveness can be read as: ``all finite elements of $\cal D$ are definable in $\cal A$''.
\begin{definition}
\label{Cdef}
{\rm Let $\cal A$ be an ats.
\begin{itemize}
\item {\em Convergence testing} is definable in $\cal A$ if for some $c \in  A$, $\cal A$ satisfies:
\begin{itemize}
\item $c {\Converges}$
\item $x {\Diverges} \; \Rightarrow \; c x {\Diverges}$
\item $x {\Converges} \; \Rightarrow \; cx = {\bf I}$.
\end{itemize}
In this case, we use {\sf C} as a constant to denote $c$.
\item {\em Parallel convergence} is definable in $\cal A$ if for some $p \in  A$, $\cal A$ satisfies:
\begin{itemize}
\item $p {\Converges} , \;\; p x {\Converges}$
\item $x {\Converges} \; \Rightarrow \; p x y {\Converges}$
\item $y {\Converges} \; \Rightarrow \; p x y {\Converges}$
\item $x {\Diverges} \: \& \: y {\Diverges} \; \Rightarrow \; p x y {\Diverges}$ .
\end{itemize}
In this case, we use {\sf P} to denote such a $p$.
\end{itemize}}
\end{definition}
Note that if {\sf C} is definable, it is unique (up to bisimulation); this is not so for {\sf P}.

The notion of parallel convergence is reminiscent of Plotkin's parallel or, 
and will play a similar role in our theory. 
(A sharper comparison will be made later in this section.) 
The notion of convergence testing is less expected. 
We can think of the combinator {\sf C} as a sort of ``1-strict'' version of {\bf K}:
\[ {\sf C} x y = {\bf K} x y = y \;\;\;\; {\rm if} \; x {\Converges} \]
\[ {\sf C} x y {\Diverges} \;\;\;\; {\rm if} \; x {\Diverges} . \]
This 1-strictness allows us to test, sequentially, a number of expressions 
for convergence. 
Under the hypothesis that {\sf C} is definable, we can give a very satisfactory picture of the relationship between all these notions.
\begin{theorem}[Full Abstraction]
Let $\cal A$ be a sensible, approximable lts in which {\sf C} is definable. 
The following conditions are equivalent:
\begin{center}
\begin{tabular}{rl}
(i) &  Parallel convergence is definable in $\cal A$. \\ 
(ii) &  $\cal A$ is $\cal L$-expressive. \\
(iii) &  $\cal A$ is $\cal L$-complete. \\
(iv)  & ${\sf t}_{\cal A}$ is a combinatory embedding with ${\sl K}({\cal D}) \subseteq {\sl Im} \; {\sf t}_{\cal A}$. \\ 
(v)  & $\cal D$ is fully abstract for $\cal A$. 
\end{tabular}
\end{center}
\end{theorem}

\proof\ We shall prove a sequence of implications to establish the theorem, indicating in each case which hypotheses on $\cal A$ are used.

\noindent $(i) \; \Longrightarrow \; (ii)$ ($\cal A$ sensible, {\sf C} definable).

Since $\cal A$ is sensible, $\BOmega$ diverges in $\cal A$.
\\
{\bf Notation.} Given a set {\sf Con} of constants, $\BLambda ({\sf Con})$ is the set of $\lambda$-terms over Con.

For each $\phi \in N{\cal L}$ we shall define terms $M_{\phi}, T_{\phi} \in \BLambda(\{{\sf P, C}\})$ such that:
\[ \bullet \;\; M_{\phi} \; \models_{\cal A} \; \psi \;\; \Longleftrightarrow \;\; {\cal L} \; \vdash \; \phi \leq \psi \]
\[ \bullet \;\; \forall a \in A. \, \left\{ \begin{array}{ll}
T_{\phi} a \Converges & \mbox{if $a \; \models_{\cal A} \; \phi$,} \\
T_{\phi} a \Diverges & \mbox{otherwise.}
\end{array} \right. \]
The definition is by induction on the complexity of
\[ \phi \; \equiv \; \bigwedge_{i \in I} (\phi_{i, 1} \rightarrow \cdots (\phi_{i, k_{i}} \rightarrow \lambda )_{\bot} \cdots )_{\bot} . \]
If $I = \varnothing$, $M_{\phi} \equiv \BOmega$. Otherwise, we define $M_{\phi} \; \equiv \; M(\phi , k)$, where $k = {\rm max} \: \{ k_{i} \: | \: i \in I \}$:
\begin{eqnarray*}
M( \phi , 0 ) & \equiv & {\bf K} \BOmega \\
M( \phi , i+1 ) & \equiv & \lambda x_{j} . \, {\sf C} N M(\phi , i ) 
\end{eqnarray*}
where
\begin{eqnarray*}
j & \equiv & k - i \\
N & \equiv & \sum \{ N_{i} : j \leq k_{i} \} \\
N_{i} & \equiv & {\sf C} (T_{\phi_{i, 1}} x_{1}) ({\sf C}(T_{\phi_{i, 2}} x_{2})( \ldots ({\sf C} ( T_{\phi_{i, j}} x_{j} )) \ldots )) \\
\sum \varnothing & \equiv & \BOmega \\
\sum \{ N \} \cup \Theta & \equiv & {\sf P} N ( \sum \Theta ) .
\end{eqnarray*}
\begin{eqnarray*}
T_{\phi} & \equiv & \lambda x . \, \prod \{ x M_{\phi_{i, 1}} \ldots M_{\phi_{i, k_{i}}} : i \in I \} \\
\prod \varnothing & \equiv & {\bf K} \BOmega \\
\prod \{ N \} \cup \Theta & \equiv & {\sf C} N ( \prod \Theta ) .
\end{eqnarray*}
We must show that these definitions have the required properties. Firstly, we prove for all $\phi \in N{\cal L}$:
\[ (1) \;\; M_{\phi} \; \models_{\cal A} \; \phi \]
\[ (2) \;\; a \; \models_{\cal A} \; \phi \;\; \Rightarrow \;\; T_{\phi} a \Converges \]
by induction on $\phi$:
\[ \begin{array}{clr}
\bullet & \forall i \in I. \, a_{j} \; \models_{\cal A} \; \phi_{i, j} \;\; (1 \leq j \leq k_{i}) & \\
\Rightarrow & M_{\phi} a_{1} \ldots a_{k_{i}} {\Converges} & \mbox{by induction hypothesis (2),} \\
\therefore & M_{\phi} \; \vdash_{\cal A} \; \phi . & 
\end{array} \]
\[ \begin{array}{clr}
\bullet & a \; \models_{\cal A} \; \phi & 
\mbox{by induction hypothesis (1)} \\
\Rightarrow & T_{\phi} a \Converges . &
\end{array} \]
We complete the argument by proving, for all $\phi , \psi \in N{\cal L}$:
\[ \begin{array}{cccc}
(3)  & M_{\phi} \; \models_{\cal A} \; \psi & \Rightarrow & {\cal L} \; \vdash \; \phi \leq \psi \\
(4)  & M_{\psi} \; \models_{\cal A} \; \phi & \Rightarrow & {\cal L} \; \vdash \; \psi \leq \phi \\
(5) & T_{\phi} M_{\psi} \Converges & \Rightarrow &  M_{\psi} \; \models_{\cal A} \; \phi \\
(6) & T_{\psi} M_{\phi} \Converges & \Rightarrow &  M_{\phi} \; \models_{\cal A} \; \psi .
\end{array} \]
The proof is by induction on $n + m$, where $n, m$ are the number of sub-formulae of $\phi , \psi$ respectively. Let
\[  \phi \; \equiv \; \bigwedge_{i \in I} (\phi_{i, 1} \rightarrow \cdots (\phi_{i, k_{i}} \rightarrow \lambda )_{\bot} \cdots )_{\bot} , \]
\[  \psi \; \equiv \; \bigwedge_{j \in J} (\psi_{j, 1} \rightarrow \cdots (\psi_{j, k_{j}} \rightarrow \lambda )_{\bot} \cdots )_{\bot} . \] 
(3):
\[ \begin{array}{clr}
\bullet & M_{\phi} \; \models_{\cal A} \; \psi & \\ 
\Rightarrow & \forall j \in J . \, M_{\phi}  M_{\psi_{j, 1}} \ldots M_{\psi_{j, k_{j}}} {\Converges} & \mbox{by (1) ,} \\
\Rightarrow & \forall j \in J . \, \exists i \in I . \, k_{j} \leq k_{i} \: \& \: T_{\phi_{i, l}} M_{\psi_{j, l}} \Converges , \;\; 1 \leq l \leq k_{j} & \\
\Rightarrow &  M_{\psi_{j, l}} \; \models_{\cal A} \; \phi_{i, l} , \;\; 1 \leq l \leq k_{j} & \mbox{ind. hyp. (5)} \\
\Rightarrow & {\cal L} \; \vdash \; \psi_{j, l} \leq \phi_{i, l} , \;\; 1 \leq l \leq k_{j} & \mbox{ind. hyp. (4)} \\
\Rightarrow & {\cal L} \; \vdash \; \phi \leq \psi . &
\end{array} \]

\noindent (4): Symmetrical to (3).

\noindent (5):
\[  \begin{array}{clr}
\bullet & T_{\phi} M_{\psi} {\Converges} & \\ 
\Rightarrow & \forall i \in I . \, M_{\psi} M_{\phi_{i, 1}} \ldots M_{\phi_{i, k_{i}}} {\Converges} & \\
\Rightarrow & \forall i \in I . \, \exists j \in J . \, k_{i} \leq k_{j} \: \& \: T_{\psi_{j, l}} M_{\phi_{i, l}} \Converges , \;\; 1 \leq l \leq k_{i} & \\
\Rightarrow &  M_{\phi_{i, l}} \; \models_{\cal A} \; \psi_{j, l} , \;\; 1 \leq l \leq k_{i} & \mbox{ind. hyp. (6)} \\
\Rightarrow & {\cal L} \; \vdash \; \phi_{i, l} \leq \psi_{j, l} , \;\; 1 \leq l \leq k_{i} & \mbox{ind. hyp. (3)} \\
\Rightarrow & {\cal L} \; \vdash \; \psi \leq \phi  & \\ 
\Rightarrow &  M_{\psi} \; \models_{\cal A} \; \phi & \mbox{by (1). } 
\end{array} \]

\noindent (6): Symmetrical to (5).

\noindent $(ii) \; \Longrightarrow \; (iii)$ ($\cal A$ approximable). 

\noindent {\bf Notation.} For each $\phi \in {\cal L}$, $a_{\phi} \in A$ is the element representing $\phi$. Given $\Gamma : {\sf Var} \rightarrow {\cal L}$, $\rho_{\Gamma} \in Env({\cal A})$ is defined by
\[ \rho_{\Gamma} x = a_{\Gamma x} . \]
Finally, $\Gamma_{\true} : {\sf Var} \rightarrow {\cal L}$ is the constant map $x \mapsto \true$.

We begin with some preliminary results.
\[ (1) \;\; {\cal A} \models \phi \leq \psi \;\; \Longleftrightarrow \;\; {\cal L} \vdash \phi \leq \psi . \]
One half is the Soundness Theorem for $\cal L$. For the converse, note that
\begin{eqnarray*}
{\cal A} \models \phi \leq \psi & \Rightarrow & a_{\phi} \models_{\cal A} \psi \\
& \Rightarrow & {\cal L} \vdash \phi \leq \psi . 
\end{eqnarray*}
\[ (2) \;\; \forall \psi \in N{\cal L} . \, \psi \neq \true \: \& \: ab \models_{\cal A} \psi \; \Rightarrow \; \exists \phi . \, a \models_{\cal A} (\phi \rightarrow \psi )_{\bot} \: \& \: b \models_{\cal A} \phi . \]
This is shown by induction on $\psi$.
\[ \begin{array}{ll}
\bullet & a b \models_{\cal A} \bigwedge_{i \in I} \psi_{i} \;\; (I \neq \varnothing )  \\
\Rightarrow & \forall i \in I . \, a b \models_{\cal A}  \psi_{i} \\
\Rightarrow & \forall i \in I . \, \exists \phi_{i} . \, a \models_{\cal A} (\phi_{i} \rightarrow \psi_{i})_{\bot} \: \& \: b \models_{\cal A} \phi_{i} \;\;\; \mbox{by ind. hyp.} \\
\Rightarrow & \forall i \in I . \, a \models_{\cal A} (\bigwedge_{i \in I} \phi_{i} \rightarrow \psi_{i})_{\bot} \: \& \: b \models_{\cal A} \bigwedge_{i \in I} \phi_{i} \\
\Rightarrow & a \models_{\cal A} (\bigwedge_{i \in I} \phi_{i} \rightarrow \bigwedge_{i \in I} \psi_{i})_{\bot} \: \& \: b \models_{\cal A} \bigwedge_{i \in I} \phi_{i} .
\end{array} \]
\[ \begin{array}{ll}
\bullet & a b \models_{\cal A} (\psi_{1} \rightarrow \cdots ( \psi_{k} \rightarrow \lambda )_{\bot} \cdots )_{\bot} \\
\Rightarrow & a b a_{\psi_{1}} \ldots a_{\psi_{k}} {\Converges} \\
\Rightarrow & \exists \phi , \phi_{1} , \ldots , \phi_{k} . \, b \models_{\cal A} \phi \: \& \: a_{\psi_{i}} \models_{\cal A} \phi_{i} \; (1 \leq i \leq k) \\ 
& \mbox{} \& \: a \models_{\cal A} (\phi \rightarrow ( \phi_{1} \rightarrow \cdots (\phi_{k} \rightarrow \lambda )_{\bot} \cdots )_{\bot}, \\
& \mbox{since {\cal A} is approximable} \\
\Rightarrow & {\cal L} \vdash \psi_{i} \leq \phi_{i} \; (1 \leq i \leq k) \\
\Rightarrow & {\cal L} \vdash (\phi \rightarrow (\phi_{1} \rightarrow \cdots (\phi_{k} \rightarrow \lambda )_{\bot} \cdots )_{\bot} \\
& \mbox{} \leq (\phi \rightarrow (\psi_{1} \rightarrow \cdots (\psi_{k} \rightarrow \lambda )_{\bot} \cdots )_{\bot} \\
\Rightarrow & a \models_{\cal A} (\phi \rightarrow \psi )_{\bot} \: \& \: b \models_{\cal A} \phi .
\end{array} \]

\[ (3) \;\; \forall M \in \BLambda . \, M, \Gamma \models_{\cal A} \phi \; \Longleftrightarrow \; M, \rho_{\Gamma} \models_{\cal A} \phi . \]
The right to left implication is clear, since $\rho_{\Gamma} \models_{\cal A} \Gamma$. We prove the converse by induction on $M$.
\begin{eqnarray*}
x, \Gamma \models_{\cal A} \phi & \Longleftrightarrow & {\cal A} \models \Gamma x \leq \phi \\
& \Longleftrightarrow & {\cal L} \vdash \Gamma x \leq \phi \;\; {\rm by (1)} \\
& \Longleftrightarrow & a_{\Gamma x} \models_{\cal A} \phi \\
& \Longleftrightarrow & x, \rho_{\Gamma} \models_{\cal A} \phi .
\end{eqnarray*} 

The case for $\lambda x . M$ is proved by induction on $\phi$. We show the non-trivial case.
\[ \begin{array}{llr}
\bullet & \lambda x . M, \rho_{\Gamma} \models_{\cal A} (\phi \rightarrow \psi )_{\bot} & \\
\Longrightarrow & M, \rho_{\Gamma}[x \mapsto a_{\phi}] \models_{\cal A} \psi & \\
\Longrightarrow & M, \Gamma [x \mapsto \phi ] \models_{\cal A} \psi &  \mbox{by (outer) induction hypothesis} \\
\Longrightarrow & \lambda x . M, \Gamma \models_{\cal A} (\phi \rightarrow \psi )_{\bot} . &
\end{array} \]

\[ \begin{array}{llr}
\bullet & MN, \rho_{\Gamma} \models_{\cal A} \psi & \\
\Longrightarrow & \lsem M \rsem^{\cal A}_{\rho_{\Gamma}} \lsem N \rsem^{\cal A}_{\rho_{\Gamma}} \models_{\cal A} \psi & \\
\Longrightarrow & \exists \phi . \, \lsem M \rsem^{\cal A}_{\rho_{\Gamma}} \models_{\cal A} (\phi \rightarrow \psi )_{\bot} \: \& \: \lsem N \rsem^{\cal A}_{\rho_{\Gamma}} \models_{\cal A} \phi & \mbox{by (2)} \\
\Longrightarrow & M, \Gamma \models_{\cal A} (\phi \rightarrow \psi )_{\bot} \: \& \: N, \Gamma \models_{\cal A} \phi & \mbox{ind. hyp.} \\
\Longrightarrow & M N , \Gamma \models_{\cal A} \psi . &
\end{array} \]

\noindent (4):
\[ \begin{array}{rrcl}
(i) & x, \Gamma [x \mapsto \phi ] \models_{\cal A} \psi & \Longleftrightarrow & {\cal L} \vdash \phi \leq \psi \\
(ii) & \lambda x . M , \Gamma \models_{\cal A} (\phi \rightarrow \psi )_{\bot} & \Longleftrightarrow & M, \Gamma [ x \mapsto \phi ] \models_{\cal A} \psi \\
(iii) & MN, \Gamma \models_{\cal A} \psi & \Longleftrightarrow & \exists \phi . \, M, \Gamma \models_{\cal A} (\phi \rightarrow \psi )_{\bot} \\
& & & \mbox{} \& \: N, \Gamma \models_{\cal A} \phi .
\end{array} \]

$4(i)$ is proved using (1).

$4(ii)$:
\[ \begin{array}{ll}
\bullet & \lambda x . M , \Gamma \models_{\cal A} (\phi \rightarrow \psi )_{\bot} \\
\Rightarrow & \forall \rho , a . \, \rho \models_{\cal A} \Gamma \: \& \: a \models_{\cal A} \phi \; \Rightarrow \; \lsem \lambda x . M \rsem^{\cal A}_{\rho} . a \models_{\cal A} \psi \\
\Rightarrow & \forall \rho . \, \rho \models_{\cal A} \Gamma [ x \mapsto \phi ] \; \Rightarrow \; M, \rho \models_{\cal A} \psi \\ 
& \;\; \mbox{since} \; \lsem \lambda x \appl\ M \rsem^{\cal A}_{\rho} . a = \lsem  M \rsem^{\cal A}_{\rho [ x \mapsto a ]} , \\
\Rightarrow & M, \Gamma [ x \mapsto \phi ] \models_{\cal A} \psi .
\end{array} \]
The converse follows from the soundness of $\cal L$.

$4(iii)$:
\begin{Eqarray}
MN, \Gamma \models_{\cal A} \psi & \Longleftrightarrow & MN, \rho_{\Gamma} \models_{\cal A} \psi & \mbox{by (3)} \\
& \Longleftrightarrow & \lsem M \rsem^{\cal A}_{\rho_{\Gamma}} \lsem N \rsem^{\cal A}_{\rho_{\Gamma}} \models_{\cal A} \psi & \\
& \Longleftrightarrow & \exists \phi . \, \lsem M \rsem^{\cal A}_{\rho_{\Gamma}} \models_{\cal A} (\phi \rightarrow \psi )_{\bot} \: \& \: \lsem N \rsem^{\cal A}_{\rho_{\Gamma}} \models_{\cal A} \phi & \mbox{by (2)} \\
& \Longleftrightarrow & \exists \phi . \, M, \Gamma \models_{\cal A} (\phi \rightarrow \psi )_{\bot} \: \& \: N, \Gamma \models_{\cal A} \phi & \mbox{by (3)}
\end{Eqarray}

We can now prove
\[ M, \Gamma \models_{\cal A} \phi \; \Rightarrow \; M, \Gamma \vdash \phi \]
by induction on $M$, using (4).

\noindent $(iii) \; \Longrightarrow \; (i)$.

Firstly, note that $(iii)$ implies
\[ {\cal A} \models \phi \leq \psi \; \Longleftrightarrow \; {\cal L} \vdash \phi \leq \psi . \]
One half is the Soundness Theorem. For the converse, suppose ${\cal A} \models \phi \leq \psi$ and ${\cal L} \nvdash \phi \leq \psi$. Then ${\bf I} \models_{\cal A} (\phi \rightarrow \psi )_{\bot}$ but ${\bf I} \nvdash (\phi \rightarrow \psi )_{\bot}$, and so $\cal A$ is not $\cal L$-complete.

Now suppose that {\sf P} is not definable in $\cal A$, and consider
\[ \phi \equiv (\lambda \rightarrow (\true \rightarrow \lambda )_{\bot})_{\bot} \wedge (\true \rightarrow ( \lambda \rightarrow \lambda )_{\bot})_{\bot} , \]
\[ \psi \equiv (\true \rightarrow (\true \rightarrow \lambda )_{\bot})_{\bot} . \]
Clearly, ${\cal L} \nvdash \phi \leq \psi$. 
However, for $a \in {\cal A}$, if $a \models_{\cal A} \phi$, then $x \Converges$ or $y \Converges$ implies $a x y \Converges$; 
since {\sf P} is not definable in $\cal A$, and in particular, $a$ does not define {\sf P}, we must have $a x y \Converges$ even if $x \Diverges$ and $y \Diverges$, and hence $a \models_{\cal A} \psi$. 
Thus ${\cal A} \models \phi \leq \psi$ and so by our opening remark, $\cal A$ is not $\cal L$-complete.

\noindent $(ii) \; \Longrightarrow \; (iv)$ ($\cal A$ approximable).

Clearly ${\sf Im} \; t_{\cal A} \supseteq  {\cal K}(D)$, by 5.14(ii). 
Also, since $\cal A$ is approximable, we can apply the Characterisation Theorem 
to deduce that $t_{\cal A}$ is injective (modulo bisimulation). 
To show that $t_{\cal A}$ is a combinatory morphism, we argue as in 
\ref{isocalg}. 
Application is preserved by $t_{\cal A}$ using (2) from the proof of 
$(ii)  \Rightarrow  (iii)$ and \ref{isocalg}. 
The proof is completed by showing that $t_{\cal A}$ preserves denotations of 
$\lambda$-terms, i.e.
\[ \forall M \in \BLambda , \rho \in Env({\cal A}) . \, t_{\cal A} (\lsem M \rsem^{\cal A}_{\rho}) = \lsem M \rsem^{D}_{t_{\cal A} \circ \rho } . \]
The proof is by induction on $M$. 
Since it is very similar to the corresponding part of the proof of \ref{isocalg}, we omit it. 
The only non-trivial point is that in the case for abstraction we need:
\[ \forall a \in A. \, a \models_{\cal A} \phi \; \Longrightarrow \; M, \rho [x \mapsto a ] \models_{\cal A} \psi \]
if and only if
\[ M, \rho [ x \mapsto a_{\phi}] \models_{\cal A} \psi , \]
which is proved similarly to (3) in $(ii) \; \Rightarrow \; (iii)$.

\noindent $(iv) \; \Longrightarrow \; (v)$.

Assuming $(iv)$, $\cal A$ is isomorphic (modulo bisimulation) to a substructure of $D$. 
Since formulas in {\sf HF} are (equivalent to) universal ($\Pi^{0}_{1}$) 
sentences, this yields ${\Im}(D) \subseteq {\Im}({\cal A})$. 
Since ${\cal K}(D) \subseteq {\sf Im}\: t_{\cal A}$, to prove the converse it is sufficient to show, for $H \in {\sf HF}$:

\[ D, \rho \nvDash H \;\; \Longrightarrow \;\; \exists \rho_{0} : {\sf Var} \rightarrow {\cal K}(D). \, D, \rho \nvDash H . \]
Let $H \equiv P \Rightarrow F$, where $P \equiv \bigwedge_{i \in I}M_{i} \Converges \wedge \bigwedge_{j \in J}N_{j} \Diverges$. There are four cases, corresponding to the form of $F$.

Case 1: $F \equiv M \sqsubseteq N$. 
$D, \rho \nvDash P \Rightarrow F$ implies $D, \rho \models P$ and 
$D, \rho \nvDash M \sqsubseteq N$. Since $D$ is algebraic, 
$D, \rho \nvDash M \sqsubseteq N$ implies that for some $b \in {\cal K}(D)$, 
$b \sqsubseteq \lsem M \rsem^{D}_{\rho}$ and  $b \not\sqsubseteq \lsem N \rsem^{D}_{\rho}$. 
Since the expression  $\lsem M \rsem^{D}_{\rho}$ is continuous in $\rho$, 
$b \sqsubseteq \lsem M \rsem^{D}_{\rho}$ implies that for some 
$\rho_{1} : {\sf Var} \rightarrow {\cal K}(D)$, $\rho_{1} \sqsubseteq \rho$ 
and $b \sqsubseteq \lsem M \rsem^{D}_{\rho_{1}}$. 
For all $\rho'$ with $\rho_{1} \sqsubseteq \rho' \sqsubseteq \rho$, 
$\lsem N \rsem^{D}_{\rho'} \sqsubseteq \lsem N \rsem^{D}_{\rho}$, and hence 
$b \not\sqsubseteq \lsem N \rsem^{D}_{\rho'}$. 
Again, since $D$ is algebraic,
\[ D, \rho \models M_{i} \Converges \;\; \Longrightarrow \;\; \exists \rho_{i} : {\sf Var} \rightarrow {\cal K}(D) . \, \rho_{i} \sqsubseteq \rho \: \& \: D, \rho_{i} \models M_{i} \Converges . \]
Now let $\rho_{0} \equiv \bigsqcup_{i \in I}{\rho_{i} \sqcup \rho_{1}}$. 
This is well-defined since $D$ is a lattice. 
Moreover, $\rho_{0} \sqsubseteq \rho$, and $\rho_{0} : {\sf Var} \rightarrow {\cal K}(D)$. 
Since $\rho_{0} \sqsupseteq \rho_{i} \; (i \in I)$, $D, \rho_{0} \models M_{i} \Converges$; 
while since $\rho_{0} \sqsubseteq \rho$, $D, \rho_{0} \models N_{j} \Diverges \; (j \in J)$. 
Since $\rho_{1} \sqsubseteq \rho_{0} \sqsubseteq \rho$, 
$b \sqsubseteq \lsem M \rsem^{D}_{\rho_{0}}$ and  
$b \not\sqsubseteq \lsem N \rsem^{D}_{\rho_{0}}$, and so 
$D, \rho_{0} \nvDash M \sqsubseteq N$. 
Thus  $D, \rho_{0} \nvDash P \Rightarrow F$, as required.

The remaining cases are proved similarly.

\noindent $(v) \; \Longrightarrow \; (i)$ ($\cal A$ sensible).

Consider the formula
\[ H \equiv x \BOmega ({\bf K} \BOmega ) \Converges \wedge x ( {\bf K} \BOmega ) \BOmega \Converges \; \Rightarrow \; x \BOmega \BOmega \Converges . \]
It is easy to see that ${\cal A} \models H$ iff {\sf P} is not definable in $\cal A$. 
Since {\sf P} is definable in $D$, the result follows. \qed

We now turn to the question of when the bisimulation preorder on an lts can be characterised by means of a contextual equivalence, as in \cite{Bar,Plo77,Mil77}.

\begin{definition}
{\rm Let $\cal A$ be an lts, $X, Y \subseteq A$. Then {\em $X$ separates $Y$} if:}
\[ \begin{array}{l}
\forall M, N \in {\BLambda}^{0}(Y). \, {\cal A} \nvDash M \sqsubseteq N \; \Longrightarrow \\ 
\;\; \exists P_{1}, \ldots , P_{k} \in {\BLambda}^{0}(X) . \,
{\cal A} \models M P_{1} \ldots P_{k} \Converges \: \& \: {\cal A} \models N P_{1} \ldots P_{k} \Diverges . 
\end{array} \]
\end{definition}

In particular, if $X$ separates $A$ we say that it is a {\em separating set}. For example, $A$ is always a separating set.

\begin{proposition}
\label{sepl}
Let $\cal A$ be an approximable lts, and suppose $X$ separates $Y$. Then
\[ \forall  M, N \in {\BLambda}^{0}(Y). \, {\cal A} \models M \sqsubseteq N \;\; \Longleftrightarrow \] 
\[ \forall C[\cdot ] \in {\BLambda}^{0}(X). \,
{\cal A} \models C[M] \Converges \; \Rightarrow \; {\cal A} \models C[N] \Converges . \] 
\end{proposition}

\proof\ Suppose ${\cal A} \nvDash M \sqsubseteq N$. 
Then since $X$ separates $Y$, for some $P_{1}, \ldots , P_{k} \in {\BLambda}^{0}(X)$, 
${\cal A} \models M P_{1} \ldots P_{k} \Converges$ and 
${\cal A} \models N P_{1} \ldots P_{k} \Diverges$. 
Let $C[\cdot ] \equiv [\cdot ]P_{1} \cdots P_{k}$. 
For the converse, suppose ${\cal A} \models M \sqsubseteq N$ and ${\cal A} \models C{M} \Converges$. 
Since ${\cal A}$ is approximable and ${\cal A} \models C[M] = \lambda x . C[x] M$, 
for some $\phi$ $\lambda x . C[x] \models_{\cal A} (\phi \rightarrow \lambda )_{\bot}$ and $M \models_{\cal A} \phi$. 
Since ${\cal A} \models M \sqsubseteq N$, by the Characterisation Theorem $N \models_{\cal A} \phi$, and so ${\cal A} \models C[N] \Converges$. \qed

As a first application of this Proposition, we have:

\begin{proposition}
Let $\cal A$ be a sensible, approximable lts in which {\sf C} and {\sf P} are definable. Then $\{ {\sf C}, {\sf P} \}$ is a separating set.
\end{proposition}

\proof\ By the Full Abstraction Theorem, for each $\phi \in {\cal L}$ there is $M_{\phi} \in {\BLambda}^{0}(\{{\sf C}, {\sf P} \})$ such that
\[ M_{\phi} \models_{\cal A} \psi \; \Longleftrightarrow \; {\cal L} \vdash \phi \leq \psi . \]
Now
\[ \begin{array}{ll}
\bullet & {\cal A} \nvDash M \sqsubseteq N  \\ 
\Longrightarrow & \exists \phi . \, M \models_{\cal A} \phi \: \& \: N \nvDash \phi , \;\; 
\mbox{since {\cal A} is approximable} \\
\Longrightarrow & \exists \phi_{1} , \ldots , \phi_{k} . \, M \models_{\cal A} (\phi_{1} 
\rightarrow \cdots (\phi_{k} \rightarrow \lambda )_{\bot} \cdots )_{\bot}  \\ 
& \;\; \& \: N \nvDash_{\cal A} (\phi_{1} \rightarrow \cdots (\phi_{k} \rightarrow \lambda )_{\bot} \cdots )_{\bot}  \\
\Longrightarrow & M M_{\phi_{1}} \ldots M_{\phi_{k}} \Converges \: \& \: N M_{\phi_{1}} \ldots M_{\phi_{k}} \Diverges . \;\;\; \qed
\end{array} \]

The hypothesis of approximability has played a major part in out work. We now give a useful sufficient condition.

\begin{definition}
{\rm Let $\cal A$ be an lts, $X \subseteq A$. Then $\cal A$ is $X$-{\em sensible} if}
\[ \forall M \in {\BLambda}^{0}(X). \, {\cal A} \models M \Converges \; \Rightarrow \; D \models M \Converges . \]
\end{definition}

Here $\lsem M \rsem^{D}$ is the denotation in $D$ obtained by mapping each $a \in X$ to $t_{\cal A}(a)$. Note that if we extend our endogenous program logic to terms in ${\BLambda}^{0}(X)$, with axioms
\[ a, \Gamma \vdash \phi \;\; ( \phi \in {\cal L}(a)) , \]
then the Soundness and Completeness Theorems for $D$ still hold, by a straightforward extension of the arguments used above.

\begin{proposition}
\label{approxl}
Let $\cal A$ be an $X$-sensible lts. Then $\cal A$ is $X$-approximable, i.e.
\[ \forall M, N_{1} , \ldots , N_{k} \in {\BLambda}^{0}(X) . \, {\cal A} \models M N_{1} \ldots N_{k} \Converges \; \Rightarrow \; \exists \phi_{1} , \ldots , \phi_{k} . \]
\[ M \models_{\cal A} (\phi_{1} \rightarrow \cdots (\phi_{k} \rightarrow \lambda )_{\bot} \cdots )_{\bot} \: \& \: N_{i} \models_{\cal A} \phi_{i}, \; 1 \leq i \leq k . \]
\end{proposition}

\proof 
\[ \begin{array}{ll}
\bullet & {\cal A} \models M N_{1} \ldots N_{k} \Converges \\
\Rightarrow & D \models M  N_{1} \ldots N_{k} \Converges \\
\Rightarrow &  \exists \phi_{1} , \ldots , \phi_{k} . \,  M \models_{\cal D} (\phi_{1} \rightarrow \cdots (\phi_{k} \rightarrow \lambda )_{\bot} \cdots )_{\bot} \\
& \;\; \& \: N_{i} \models_{\cal D} \phi_{i}, \; 1 \leq i \leq k , \; \mbox{since {D} is approximable} \\
\Rightarrow & \exists \phi_{1} , \ldots , \phi_{k} . \,  M \vdash (\phi_{1} \rightarrow \cdots (\phi_{k} \rightarrow \lambda )_{\bot} \cdots )_{\bot} \\
& \;\; \& \: N_{i} \vdash \phi_{i}, \; 1 \leq i \leq k , \; \mbox{by extended Completenss} \\
\Rightarrow & \exists \phi_{1} , \ldots , \phi_{k} . \,  M \models_{\cal A} (\phi_{1} \rightarrow \cdots (\phi_{k} \rightarrow \lambda )_{\bot} \cdots )_{\bot} \\
& \;\; \& \: N_{i} \models_{\cal A} \phi_{i}, \; 1 \leq i \leq k , \; \mbox{by extended Soundness} . \;\;\; \qed
\end{array} \]

In particular, if $X$ generates $\cal A$ and $\cal A$ is $X$-sensible, then $\cal A$ is approximable. We now turn to a number of applications of these ideas to syntactically presented lts, i.e. ``programming languages''.

Firstly, we consider the lts $\ell = ({\BLambda}^{0}, eval)$ defined in section 3 (and studied previously in section 2). 
Since $\ell$ is $\varnothing$-sensible by \ref{ltsprop}, and it is generated by $\varnothing$, it is approximable by \ref{approxl}. 
Since $\varnothing$ is a separating set for ${\BLambda}^{0}$, we can apply \ref{sepl} to obtain Theorem \ref{cont}.

Next, we consider extensions of $\ell$.

\begin{definition}
{\rm (i) $\ell_{\sf C}$ is the extension of $\ell$ defined by
\[ \ell_{\sf C} = (\BLambda (\{{\sf C} \}), \_ \Converges \_ ) \]
where $\Converges$ is the extension of the relation defined in \ref{convdef} with the following rules:
\[ \bullet \; {\sf C} \Converges {\sf C} \;\;\;\; \bullet \; \frac{M \Converges}{{\sf C} M \Converges {\bf I}} \]
(ii) $\ell_{\sf P}$ is the extension $(\BLambda (\{{\sf C} \}), \_ \Converges \_ )$ of $\ell$ with the rules}
\[ \bullet \; {\sf P} \Converges {\sf P} \;\;\;\; \bullet \; {\sf P} M \Converges {\sf P} M \;\;\;\; \bullet \; \frac{M \Converges}{{\sf P} M N \Converges {\bf I}} \;\;\;\; \bullet \; \frac{N \Converges}{{\sf P} M N \Converges {\bf I}} \]
\end{definition}

It is easy to see that the relation $\_ \Converges \_$ as defined in both $\ell_{\sf C}$ and $\ell_{\sf P}$ is a partial function. 
Moreover, with these definitions the {\sf C} and {\sf P} combinators have the 
properties required by \ref{Cdef}; while {\sf C} is definable in $\ell_{\sf P}$, by
\[ {\sf C} M \equiv {\sf P} M M . \]

Since $\ell_{\sf C}$ is generated by $\{ {\sf C } \}$, and $\ell_{\sf P}$ by  $\{ {\sf P } \}$, these are separating sets. 
Thus to apply Theorem \ref{sepl}, we need only check that  $\ell_{\sf C}$ is {\sf C}-sensible, and  $\ell_{\sf P}$ {\sf P}-sensible. 

To do this for $\ell_{\sf C}$, we proceed as follows. Define
\[ c \equiv \{ (\lambda \rightarrow ( \phi \rightarrow \phi )_{\bot} )_{\bot} \: | \: \phi \in {\cal L} \}^{\dag} \in {\sf Filt} \: {\cal L}. \]
Then it is easy to see that $c \subseteq t_{\cal A}({\sf C})$, and by monotonicity and the Soundness Theorem,
\[ \lsem M[c / {\sf C} ] \rsem^{D} \subseteq \lsem M \rsem^{D} \]
for $M \in \BLambda^{0}(\{ {\sf C} \} )$. Thus
\[ (\star) \;\; D \models M [ c / {\sf C} ] \Converges \; \Longrightarrow \; D \models M \Converges . \]
Now we prove
\[ \begin{array}{cl}
(\star \star ) & \forall M, N \in {\BLambda}^{0}(\{ {\sf C} \} ) . \\ 
& M \Converges N \;\; \Longrightarrow \;\;
\lsem M [ c / {\sf C} ] \rsem^{D} = \lsem N [ c / {\sf C} ] \rsem^{D} \: \& \: D \models N [ c / {\sf C} ] \Converges , 
\end{array} \]
which by $(\star )$ yields   $\ell_{\sf C} \models M \Converges \; \Rightarrow \; D \models M \Converges$, as required. $(\star \star )$ is proved by a straightforward induction on the length of the proof that $M \Converges N$.

The argument for $\ell_{\sf P}$ is similar, using
\[ p \equiv \{ (\lambda \rightarrow (\true \rightarrow (\phi \rightarrow \phi )_{\bot} )_{\bot} )_{\bot} \wedge (\true \rightarrow (\lambda \rightarrow (\psi \rightarrow \psi )_{\bot} )_{\bot} )_{\bot} : \phi , \psi \in {\cal L} \}^{\dag} . \]

Altogether, we have shown
\begin{theorem}[Contextual Equivalence]
(i) $\forall M, N \in \BLambda^{0}(\{ {\sf C} \} )$:
\[ \ell_{\sf C} \models M \sqsubseteq N \; \Longleftrightarrow \; \forall C[\cdot ] \in {\BLambda}^{0}(\{ {\sf C} \} ). \, \ell_{\sf C} \models C[M] \Converges \; \Rightarrow \; \ell_{\sf C} \models C[N] \Converges . \]
(ii) $\forall M, N \in \BLambda^{0}(\{ {\sf P} \} )$:
\[ \ell_{\sf P} \models M \sqsubseteq N \; \Longleftrightarrow \; \forall C[\cdot ] \in {\BLambda}^{0}(\{ {\sf P} \} ). \, \ell_{\sf P} \models C[M] \Converges \; \Rightarrow \; \ell_{\sf P} \models C[N] \Converges . \]
\end{theorem}

As a further application of these ideas, we have
\begin{proposition}[Soundness of D]
If $\cal A$ is $X$-sensible, and $X$ separates $X$ in $\cal A$, then:
\[ {\Im}^{0}(D, X) \subseteq {\Im}^{0}({\cal A}, X) . \]
\end{proposition}

\proof 
\[ \begin{array}{ll}
\bullet & D \models M \sqsubseteq N \\
\Longrightarrow & \forall C[ \cdot ] \in {\BLambda}^{0}(X). \, D \models C[M] \sqsubseteq C[N] \\
\Longrightarrow & D \models C[M] \Converges \; \Rightarrow \; D \models C[N] \Converges \\
\Longrightarrow & {\cal A} \models C[M] \Converges \; \Rightarrow \; {\cal A} \models C[N] \Converges \\
\Longrightarrow & {\cal A} \models M \sqsubseteq N .
\end{array} \]
The argument for formulae of other forms is similar. \qed

As an immediate corollary of this Proposition,
\begin{proposition}
The denotational semantics of each of our languages is {\em sound} with respect to
the operational semantics:
\[ \begin{array}{rl}
(i) & {\Im}^{0}(D) \subseteq {\Im}^{0}(\ell ) \\
(ii) &  {\Im}^{0}(D, \{ {\sf C} \} ) \subseteq {\Im}^{0}(\ell_{\sf C}, \{ {\sf C} \} ) \\
(iii) & {\Im}^{0}(D, \{ {\sf P} \} ) \subseteq {\Im}^{0}(\ell_{\sf P}, \{ {\sf P} \} ) . 
\end{array} \]
\end{proposition}

We now turn to the question of full abstraction for these languages. Since, as we have seen, $\ell_{\sf P}$ is {\sf P}-sensible, and hence sensible and approximable, and {\sf C} and {\sf P} are definable, we can apply the Full Abstraction Theorem to obtain
\begin{proposition}
D is fully abstract for $\ell_{\sf P}$.
\end{proposition}

We now use the sequential nature of $\ell$ and $\ell_{\sf C}$ to obtain negative full abstraction results for these languages. This will require a few preliminary notions.

\begin{definition}
{\rm The {\em one-step reduction} relation $>$ over terms in $\BLambda$ is the least satisfying the following axioms and rules:
\[ 
\bullet \;\; (\lambda x . M)N > M[N/x] \;\;\;\;
\bullet \;\; \frac{M > M'}{MN > M' N} 
\]
This is then extended to $\BLambda ( \{ {\sf C} \} )$ with the additional rules
\[ \bullet \;\; {\sf C} ( \lambda x . M ) > {\bf I} 
\;\;\;\; \bullet \;\; {\sf CC} > {\bf I} 
\;\;\;\; \bullet \;\; \frac{M > M'}{{\sf C} M > {\sf C} M'} \]
We then define
\[ \begin{array}{crcl}

\bullet & \gg & \equiv & \mbox{the reflexive, transitive closure of $>$} \\
\bullet & M \diverges & \equiv & \exists \{ M_{n} \} . \, M = M_{0} \: \& \: \forall n . \, M_{n} > M_{n + 1} \\
\bullet & M {\not>} & \equiv & M \not\in {\sf dom} {>} \\
\bullet & M {\converges} & \equiv & M \gg N \: \& \: N \not>.
\end{array} \] }
\end{definition}

It is clear that $>$ is a partial function. Note that these relations are being defined over {\em all} terms, not just closed ones. For closed terms, these new notions are related to the evaluation predicate $\_ \Converges \_$ as follows:

\begin{proposition}
\label{one1}
For $M, N \in \BLambda^{0} \; ( \BLambda^{0} ( \{ {\sf C} \} )$:
\[ \begin{array}{rrcl}
(i) & M \Converges N & \Longleftrightarrow & M \converges N \\
(ii) & M \Diverges & \Longrightarrow & M \diverges .
\end{array} \] 
\end{proposition}

We omit the straightforward proof. The following proposition is basic; it says that ``reduction commutes with substitution''.

\begin{proposition}
\label{one2}
$ M \gg N \; \Rightarrow \; M [ P / x ] \gg N [ P / x ]$ .
\end{proposition}

\proof\ Clearly, it is sufficient to show:
\[ M > N \; \Rightarrow \; M [ P / x ] > N [ P / x ] . \]
This is proved by induction on $M$, and cases on why $M > N$. We give one case for illustration:
\[ M \equiv (\lambda y . M_{1})M_{2} > N \equiv M_{1} [ M_{2} / y] . \]
We assume $x \not= y$; the other sub-case is simpler.
\begin{Eqarray}
M [ P / x ] & = & (\lambda y . M_{1} [ P / x ] ) M_{2} [ P / x ] & \\
& > &  M_{1} [ P / x ] [ M_{2} [ P / x ] / y ] & \\
& = & M_{1} [ M_{2} / y] [ P / x ] &  \mbox{by \cite[2.1.16]{Bar}} \\
& = & N [ P / x ] .  & \qed 
\end{Eqarray}

Now we come to the basic sequentiality property of $\ell$ from which various non-definability results can be deduced.

\begin{proposition} 
\label{thrcase}
For $M \in \BLambda$, exactly one of the following holds:
\[ \begin{array}{rl}
(i) & M \diverges \\
(ii) & M \gg \lambda x . N \\
(iii) & M \gg x N_{1} \ldots N_{k} \; (k \geq 0 ) .
\end{array} \]
\end{proposition}

\proof\ Since $>$ is a partial function, the computation sequence beginning with $M$ is uniquely determined. Either it is infinite, yielding $(i)$; or it terminates in a term $N$ with $ N \not>$, which must be in one of the forms $(ii)$ or $(iii)$. \qed

As a consequence of this proposition, we obtain

\begin{theorem}
\label{Cundef}
{\sf C} is not definable in $\ell$. Moreover, $D$ is not fully abstract for $\ell$.
\end{theorem}

\proof\ We shall show that $\ell$ satisfies
\[ (\star ) \;\; x = {\bf I} \;\; \mbox{or} \;\; [x \BOmega \Converges \;\; \Longleftrightarrow \;\; x ({\bf K} \BOmega ) \Converges ] . \]
Indeed, consider any term $M \in {\BLambda}^{0}$. 
Either $M \Diverges$, in which case $M \BOmega \Diverges $ and $M ( {\bf K} \BOmega ) \Diverges $, or $M \Converges$. 
In the latter case, by $(\Converges \eta )$ we have $\lambda \ell \models M = \lambda x . M x$. 
Thus without loss of generality we may take $M$ to be of the form $\lambda x . M'$, with $FV(M) \subseteq \{ x \}$. 
Now applying the three previous propositions to $M'$, we see that in case $(i)$ of \ref{thrcase}, $(\lambda x . M' ) \BOmega \Diverges$ and 
$(\lambda x . M' ) ({\bf K} \BOmega ) \Diverges$; 
in case $(ii)$, $(\lambda x . M' ) \BOmega \Converges$ and $(\lambda x . M' ) ({\bf K} \BOmega ) \Converges$; 
finally in case $(iii)$, if $k = 0$, $\lambda x . M' = {\bf I}$; 
while if $k > 0$, $(\lambda x . M' ) \BOmega \Diverges$ and $(\lambda x . M' ) ({\bf K} \BOmega ) \Diverges$. 
Since ${\sf C} \not= {\bf I}$, ${\sf C} \BOmega \Diverges$ and ${\sf C} ( {\bf K} \BOmega ) \Converges$, this shows that {\sf C} is not definable. 
Moreover, $(\star )$ implies
\[ (\star \star ) \;\;  x \BOmega \Diverges \: \& \: x ( {\bf K} \BOmega ) \Converges \; \Rightarrow \; x = {\bf I} \]
which is not satisfied by $D$, since {\sf C} is definable in $D$, and taking $x = {\sf C}$ refutes $(\star \star )$; hence $D$ is not fully abstract for $\ell$. \qed

Note that since {\sf C} is not definable in $\ell$, we could not apply the Full Abstraction Theorem. 
By contrast, to show that $D$ is not fully abstract for $\ell_{\sf C}$, it suffices to show that {\sf P} is not definable. 
For this purpose, we prove a result analogous to \ref{thrcase}.

\begin{proposition} 
\label{frcase}
For $M \in \BLambda ( \{ {\sf C} \} )$, exactly one of the following conditions holds:
\[ \begin{array}{rl}
(i) & M \diverges \\
(ii) & M \gg \lambda x . N \\
(iii) & M \gg {\sf C} \\
(iv) & M \gg \underbrace{{\sf C}({\sf C} \ldots ({\sf C}}_{n} x N_{1} \ldots N_{k} ) \ldots ) P_{1} \ldots P_{m} \;\; (n, k, m \geq 0)
\end{array} \]
\end{proposition}

\proof\ Similar to \ref{thrcase}. \qed

\begin{theorem}
\label{Pundef}
{\sf P} is not definable in $\ell_{\sf C}$; hence $D$ is not fully abstract for $\ell_{\sf C}$.
\end{theorem}

\proof\ We show that $\ell_{\sf C}$ satisfies
\[ x ({\bf K} \BOmega ) \BOmega \Converges \: \& \: x \BOmega ( {\bf K} \BOmega ) \Converges \; \Rightarrow \; x \BOmega \BOmega \Converges , \]
and hence, as in the proof of the Full Abstraction Theorem, {\sf P} 
is not definable in $\ell_{\sf C}$. 
As in the proof of \ref{Cundef}, without loss of generality we consider closed terms of the form $\lambda y_{1} . \lambda y_{2} . M$. 
Assume $(\lambda y_{1} . \lambda y_{2} . M) ({\bf K} \BOmega ) \BOmega \Converges$ 
and  $(\lambda y_{1} . \lambda y_{2} . M) \BOmega ( {\bf K} \BOmega ) \Converges$. 
Applying \ref{frcase}, we see that case $(i)$ is impossible; 
cases $(ii)$ and $(iii)$ imply that $(\lambda y_{1} . \lambda y_{2} . M) \BOmega \BOmega  \Converges$; 
while in case $(iv)$, if $x = y_{1}$, then  $(\lambda y_{1} . \lambda y_{2} . M) \BOmega ( {\bf K} \BOmega ) \Diverges$, {\em contra hypothesis}; 
and if $x = y_{2}$, $(\lambda y_{1} . \lambda y_{2} . M) ({\bf K} \BOmega ) \BOmega \Diverges$, 
also {\em contra hypothesis}. 
Thus case $(iv)$ is impossible, and the proof is complete. \qed

For our final non-definability result, we shall consider a different style of extension of $\ell$, to incorporate {\em ground data}. 
We shall consider the simplest possible such extension, where a single atom is added. This corresponds to the domain equation
\[ D_{\star} = {\bf 1} + [D_{\star} \rightarrow D_{\star}] \]

(where $+$ is separated sum), which is indeed an extension of our original domain, in the sense that $D$ is a retract of $D_{\star}$. 
$D_{\star}$ is still a Scott domain (indeed, a coherent algebraic cpo), but it is no longer a lattice; 
we have introduced {\em inconsistency} via the sum.

This extension is reflected on the syntactic level by two constants, $\star$ and {\sf C}. We define
\[ \ell_{\star} = (\BLambda^{0} ( \{ \star , {\sf C} \} ), \_ \Converges \_ ) \]
with $\_ \Converges \_$ extending the definition for $\ell$ as follows:
\[ \bullet \;\; \star \Converges \star \]
\[ \bullet \;\; {\sf C} \Converges {\sf C} \]
\[ \bullet \;\; \frac{M \Converges \lambda x . N}{{\sf C} M \Converges {\sf T}} \;\; ({\sf T} \equiv \lambda x . \lambda y . x ) \]
\[ \bullet \;\; \frac{M \Converges {\sf C}}{{\sf C} M \Converges {\sf T}} \]
\[ \bullet \;\; \frac{M \Converges \star}{{\sf C} M \Converges {\sf F}} \;\; ({\sf F} \equiv \lambda x . \lambda y . y ) \]

We see that the {\sf C} combinator introduced here is a natural generalisation 
(not strictly an extension) of the {\sf C} defined previously in the pure case. 
Of course, {\sf C} corresponds to {\em case selection}, which in the unary case --- lifting being unary separated sum --- is just convergence testing.

A theory can be developed for $\ell_{\star}$ which runs parallel to what we have done for the pure lazy $\lambda$-calculus. 
Some of the technical details are more complicated because of the presence of inconsistency, but the ideas and results are essentially the same. 
Our reasons for mentioning this extension are twofold:

\begin{enumerate}

\item To show how the ideas we have developed can be put in a broader context. 
In particular, with the extension to $\ell_{\star}$  the reader should be able to see, at least in outline, 
how our work can be applied to systems such as Martin-L\"{o}f's Type Theory 
under its Domain Interpretation \cite{Cha83}, and (the analogues of) our 
results in this section can be used to settle most of the questions and conjectures raised in \cite{Cha83}.

\item To prove an interesting result which clarifies a point about which there seems to be some confusion in the literature; 
namely, {\em what is parallel or}? 
\end{enumerate}

The {\it locus classicus} for parallel or in the setting of typed $\lambda$-calculus is \cite{Plo77}. 
But what of untyped $\lambda$-calculus? 
In \cite[p.\  375]{Bar}, we find the following definition: 
\[ F M N = \left\{ \begin{array}{ll}
{\bf I} & \mbox{if {\mit M} or {\mit N} is solvable,} \\
{\rm unsolvable} & \mbox{otherwise}
\end{array}
\right. \]
which (modulo the difference between the standard and lazy theories) corresponds to our parallel convergence combinator {\sf P}. 
The point we wish to make is this: in the pure $\lambda$-calculus, where (in domain terms) there are no inconsistent data values 
(since everything is a function), i.e. we have a lattice, 
parallel convergence does indeed play the role of parallel or, as the Full Abstraction Theorem shows. 
However, when we introduce ground data, and hence inconsistency, a distinction reappears between parallel convergence and parallel or, 
and it is definitely {\em wrong} to conflate them. 
To substantiate this claim, we shall prove the following result: even if parallel convergence is added to $\ell_{\star}$, parallel or is still not definable. 
This result is also of interest from the point of view of the fine structure of definability; 
it shows that parallelism is not all or nothing even in the simple, deterministic setting of $\ell_{\star}$.

\begin{definition}
{\rm $\ell_{\star {\sf P}}$ is the extension of $\ell_{\star}$ with a constant {\sf P} and the rules}
\[ \bullet \;\; {\sf P} \Converges {\sf P} \;\;\;\;
\bullet \;\; {\sf P} M \Converges {\sf P} M \;\;\;\;
\bullet \;\; \frac{M \Converges}{{\sf P} M N \Converges {\bf I}} \;\;\;\; 
\bullet \;\; \frac{N \Converges}{{\sf P} M N \Converges {\bf I}} \]
\end{definition}

\begin{definition}
{\rm Let $\ell'$ be an extension of $\ell_{\star}$. 
We say that {\em parallel or is definable in $\ell'$} if for some term $M$
\[ \begin{array}{rl}
(i) & M ({\bf K} \BOmega ) \BOmega ,  M \BOmega ({\bf K} \BOmega ) \;\; 
\mbox{converge to abstractions} \\
(ii) & M \star \star \Converges \star .
\end{array} \] }
\end{definition}

\begin{theorem}
Parallel or is not definable in $\ell_{\star {\sf P}}$.
\end{theorem}

\proof\ We proceed along similar lines to our previous non-definability results. Firstly, we extend our definition of $>$ as follows:
\[ \bullet \;\; {\sf constructor}(M) \equiv M \; \mbox{is an abstraction, {\sf P}, {\sf C} or $\star$} \]
\[ \bullet \;\; {\sf constructor}(M) \: \& \: M \not= \star \; \Rightarrow \; {\sf C} M > {\sf T} \]
\[ \bullet \;\; {\sf C} \star > {\sf F} \]
\[ \bullet \;\; \frac{M > M'}{{\sf C} M > {\sf C} M'} \]
\[ \bullet \;\; {\sf constructor}(M) \; \mbox{or} \; {\sf constructor}(N) \; \Rightarrow \; {\sf P} M N > {\bf I} \]
\[ \bullet \;\; \frac{M > M' \;\; N > N'}{{\sf P} M N > {\sf P} M' N'} \]

With these extensions, $>$ is still a partial function, and \ref{one1}, \ref{one2} still hold. For each $M \in \BLambda ( \{ \star , {\sf C} , {\sf P} \} )$, one of the following two disjoint conditions must hold:
\[ \begin{array}{cl}
\bullet & M \diverges \\
\bullet & M \gg N \: \& \: N \not> .
\end{array} \]

We now define $\cal T$ to be the set of all terms $M$ in $\BLambda ( \{ \star , {\sf C} , {\sf P} , \bot \} )$, where $\bot$ is a new constant, such that:
\[ \begin{array}{cl}
\bullet & FV(M) \subseteq \{ y_{1}, y_{2} \} \\
\bullet & M \; \mbox{contains no {$>$}-redex.}
\end{array} \]
Note that $\cal T$ is closed under sub-terms.

\subsection*{Lemma A}
For all $M \in {\cal T}$:
\[ \begin{array}{c}
M[ {\bf K} \BOmega / y_{1} , \BOmega / y_{2} ] \converges a \; \& \; 
M[ \BOmega / y_{1} , {\bf K} \BOmega / y_{2} ] \converges b \; \& \; 
M[ \star / y_{1} , \star / y_{2} ] \converges c \\
\Rightarrow \; a = b = c = \star \; \mbox{or} \;\star \not\in \{ a, b, c \} .
\end{array} \] 

\proof\ By induction on $M$. Since terms in $\cal T$ contain no $>$-redexes, $M$ must have one of the following forms:
\[ \begin{array}{rl}
(i) & x N_{1} \ldots N_{k} \;\; ( x \in \{ y_{1}, y_{2} \} , k \geq 0 ) \\
(ii) & \star N_{1} \ldots N_{k} \;\; (k \geq 0 ) \\
(iii) & \lambda x . N \\
(iv) & {\sf C} \;\; (v) \; {\sf P} \;\; (vi) \; {\sf P} N \\
(vii) & {\sf C} N N_{1} \ldots N_{k} \;\; (k \geq 0 ) \\
(viii) & {\sf P} M_{1} M_{2} N_{1} \ldots N_{k} \;\; (k \geq 0 ) \\
(ix) & \bot N_{1} \ldots N_{k} \;\; (k \geq 0 )
\end{array} \]

Most of these cases can be disposed of directly; we deal with the two which use the induction hypothesis.

$(vii)$. Firstly, we can apply the induction hypothesis to $N$ to conclude that $N[c_{1} / y_{1}, c_{2} / y_{2}]$ 
converges to the same result (i.e. either an abstraction or $\star$) 
for all three argument combinations $c_{1}, c_{2}$; 
we can then apply the induction hypothesis to either $N_{1} N_{3} \ldots N_{k}$ or $N_{2} N_{3} \ldots N_{k}$.

$(viii)$. Under the hypothesis of the Lemma, we must have 
\[ ({\sf P} M_{1} M_{2})[c_{1} / y_{1}, c_{2} / y_{2}] \Converges {\bf I} \]
for all three argument combinations $c_{1}, c_{2}$; hence we can apply the induction hypothesis to $N_{1} \ldots N_{k}$. \qed

\subsection*{Lemma B}
Let $M \in \BLambda\ ( \{ \star , {\sf C} , {\sf P} \} )$, with $FV(M) \subseteq \{ y_{1}, y_{2} \}$. 
Then for some $M' \in {\cal T}$, for all $P, Q \in {\BLambda}^{0} ( \{ \star , {\sf C} , {\sf P} \} )$:
\[ M [ P / y_{1} , Q / y_{2} ] \converges {\star} \;\; \Longleftrightarrow \;\;  M' [ P / y_{1} , Q / y_{2} ] \converges {\star} . \]

\proof\ Given $M$, we obtain $M'$ as follows; working in an inside-out fashion, we replace each sub-term $N$ by:
\[ \left\{ \begin{array}{ll}
N' & \mbox{if $N \converges N'$} \\
\bot & \mbox{if $N \diverges$.} \;\;\; \qed
\end{array} \right. \]

Now suppose that we are given a putative term in ${\BLambda}^{0} ( \{ \star , {\sf C} , {\sf P} \} )$ defining parallel or. 
As in the proof of \ref{Pundef}, we may take this term to have the form $\lambda y_{1} . \lambda y_{2} . M$. 
Applying Lemma B, we can obtain $M' \in {\cal T}$ from $M$; but then applying Lemma A, we see that $\lambda y_{1} . \lambda y_{2} . M'$ 
cannot define parallel or. 
Applying Lemma B again, we conclude that $\lambda y_{1} . \lambda y_{2} . M$ cannot define parallel or either. \qed

\section{Variations}

Throughout this Chapter, we have focussed on the lazy $\lambda$-calculus. 
We round off our treatment by briefly considering the varieties of function space.

\subsection*{1. The Scott function space}
$[D \rightarrow E]$, the standard function space of all continuous functions from $D$ to $E$, 
which we treated in Chapters 3 and 4. 
In terms of our domain logic $\cal L$, we can obtain this construction by adding the axiom
\[ (1) \;\; \true \leq ( \true \rightarrow \true ) . \]
Note that with (1), $\cal L$ collapses to a single equivalence class (corresponding to the trivial one-point solution of $D = [ D \rightarrow D]$). 
For this reason, Coppo {\it et al.} have to introduce atoms in their work on Extended Applicative Type Structures \cite{CDHL84}.

\subsection*{2. The strict function space}
$[D \rightarrow_{\bot} E]$, all {\em strict} continuous functions. 
This satisfies (1), and also
\[ (2) \;\; ( \true \rightarrow_{\bot} \phi ) \leq \false \;\; (\phi \converges ) . \]

\subsection*{3. The lazy function space}
$[D \rightarrow E]_{\bot}$, which satisfies neither (1) nor (2). 
This has of course been our object of study in this Chapter.

\subsection*{4. The Landin-Plotkin function space}
$[D \rightarrow_{\bot} E]_{\bot}$, the lifted strict function space. 
This satisfies (2) but not (1). 
The reason for our nomenclature is that this construction in the category of 
domains and strict continuous functions corresponds to Plotkin's 
$[D \rightharpoonup E]$ construction in his (equivalent) category of 
predomains and partial functions \cite{Plo85}.
Moreover, this may be regarded as the formalisation of Landin's applicative-order $\lambda$-calculus, 
with abstraction used to protect expressions from evaluation, as illustrated extensively in \cite{Lan64,Lan65,Bur75}.

The intriguing point about these four constructions is that (1) and (2) are {\em mathematically} natural, 
yielding cartesian closure and monoidal closure in e.g. {\bf CPO} and ${\bf CPO}_{\bot}$ respectively (the latter being analogous to partial functions over sets); 
while (3) and (4) are {\em computationally} natural, as argued extensively for (3) in this Chapter, 
and as demonstrated convincingly for (4) by Plotkin in his work on predomains \cite{Plo85}. 
Much current work is aimed at providing good categorical descriptions of generalisations of (4) \cite{Ros86,RR87,Mog86,Mog87a,Mog87b}; it remains to be seen if a similar programme can be carried out for (3).

\chapter{Further Directions}
Our development of the research programme adumbrated in Chapter~1 has been fairly extensive, but certainly not complete.
There are many possibilities for extension and generalisation of our results.
In this Chapter, we shall try to pick out some of the most promising topics for future research.
\begin{enumerate}
\item A first, very basic extension would be to rework the material of Chapters~3 and~4 for {\bf SFP} rather than {\bf SDom}.
In terms of the meta-language, the extension would be to incorporate the Plotkin powerdomain and the associated term constructions.
Our treatment of the Plotkin powerdomain in a specific instance in Chapter~5 should convey the general flavour of what is involved.
The extension to {\bf SFP} is conceptually straightforward; we remain within the sphere of coherent spaces.
However, there are some technical intricacies which arise with the meta-predicates, 
to do with the fact that the identification of primes is more subtle in the 
{\bf SFP} case; this should be clear from our work on normal forms in Chapter~5 section~4.
These intricacies are negotiable, and indeed I claim that all our work in this 
thesis {\em does} carry over (a detailed account, taking Chapters~3 and~4 of  
the present thesis as its starting point, is being worked out by a student of 
Glynn Winskel's \cite{Zha87}).
\item All our work in this thesis has been based on Domain Theory, simply because this is the best established and most successful foundation for denotational semantics, and a wealth of applications are ready to hand.
However, our programme is really much more general than this.
{\em Any} category of topological spaces in which a denotational metalanguage can be interpreted, and for which a suitable Stone duality exists, could serve as the setting for the same kind of exercise as we carried out in Chapter~4.
As one example of this: the main alternatives to domains in denotational 
semantics over the past few years have been {\em compact ultrametric spaces} 
\cite{Niv81,deBZ82,Mat85}.
These spaces in their metric topologies are Stone spaces, and indeed the category of compact ultrametric spaces and continuous maps is {\em equivalent} to the category of second-countable Stone spaces \cite{Abr85?}.
A restricted denotational metalanguage comprising product, (disjoint) sum and powerdomain (the Vietoris construction \cite{Joh85,Smy83}, which in this context is induced by the Hausdorff metric \cite{Niv81,deBZ82,Mat85}), can be interpreted in {\bf Stone}, together with the corresponding sub-language of terms (with {\em guarded} recursion, leading to {\em contracting} maps, and hence unique fixpoints \cite{Niv81,deBZ82,Mat85}).
Under the classical Stone duality as expounded in Chapter~1, the corresponding logical structures are Boolean algebras, and a {\em classical} logic can be presented for this metalanguage in entirely analogous fashion to that of Chapter~4.
Since the meta-language is rich enough to express a domain equation for 
synchronisation trees, a case study along the same lines as that of Chapter~5 can be carried through.
Moreover, there is a satisfying relationship between the Stone space of 
synchronisation trees (which is the metric topology on the ultrametric 
space constructed in \cite{deBZ82}), and the corresponding domain studied in 
Chapter~5; namely, the former is the {\em subspace of maximal elements} of the latter. 
This is in fact an instance of a general relationship, as set out in \cite{Abr85?}.
The important point here is that our programme is just as applicable to the metric-space  approach to denotational semantics as to the domain-theoretic approach.
\item A further kind of generalisation would be to structures other than topological spaces.
Many Stone-type dualities in such alternative contexts are known; e.g. Stone-Gelfand-Naimark duality for $C^{\star}$-algebras, Pontrjagin duality for topological groups, Gabriel-Ulmer duality for locally finitely presented categories, etc. \cite{Joh82}.
Particularly promising for Computer Science applications are the measure-theoretic dualities studied by Kozen \cite{Koz83} as a basis for the semantics and logic of probabilistic programs.
A very interesting feature of these dualities is that whereas the purely 
topological dualities have the Sierpinski space $\Oh$ as their 
``schizophrenic object'' (see \cite[Chapter 6]{Joh82}), i.e. the fundamental relationship $P \models \phi$ takes values in $\{ 0, 1 \}$, the measure-theoretic dualities take their ``characters'' in the reals; satisfaction of a measurable function by a measure is expressed by {\em integration} \cite{Koz83}.
The richer mathematical structure of these dualities should deepen our understanding of the framework.
Furthermore, there are intriguing connections with Lawvere's concept of ``generalised logics'' \cite{Law73}.
\item The logics of compact-open sets considered in this thesis have been very weak in expressive power, and are clearly inadequate as a specification formalism.
For example, we cannot specify such properties of a stream computation as ``emits an infinite sequence of ones''.
Thus we need a language, with an accompanying semantic framework, which permits us to go beyond compact-open sets.
A first step would be to allow the expression of more general open sets, e.g. by means of  a least fixed point operator on formulae $\mu p . \phi$, permitting the finite description of infinite disjunctions $\bigvee_{i \in \omega} \phi^{i}(\false )$.
This would have the advantage of not requiring any major extension of our semantics, but would still not be sufficiently expressive for specification purposes, as the above example shows.
What is needed is the ability to express infinite {\em conjunctions}, e.g. by {\em greatest} fixpoints $\nu p . \phi$, corresponding to $\bigwedge_{i \in \omega} \phi^{i}(\true )$.
Such an extension of our logic would necessarily take us beyond open sets.
An important topic for further investigation is whether such an extension can be smoothly engineered and given a good conceptual foundation.

Another reason for extending the logic is the tempting proximity of locale theory to topos theory.
Could this be the basis of the junction between topos theory and Computer Science which many researchers have looked for but none has yet convincingly demonstrated?
We must leave this point unresolved.
If there {\it is} a natural extension of our work to the level of topos theory, we have not (yet) succeeded in finding it.
\item Another variation is to change the {\em morphisms} under consideration.
Stone dualities relating to the various powerdomain constructions
(i.e. dualities for {\em multi-functions} rather than functions) are
interesting for a number of reasons: they generalise
{\em predicate transformers} in the sense of Dijkstra \cite{Dij76,Smy83};
dualities for the Vietoris construction provide a natural setting
for intuitionistic modal logic, with interesting differences to
the approach recently taken by Plotkin and Stirling;
while there are some remarkable {\em self-dualities} arising from the
Smyth powerdomain \cite{Vic87c}.
These turn out, quite unexpectedly, to provide a model for
Girard's classical linear logic \cite{Gir87}; more speculatively,
they also suggest the
possibility of a homogeneous logical framework in which programs and
properties are interchangeable.
This may turn out to provide the basis for a unified and systematic
treatment of a number of existing  {\it ad hoc} formalisms \cite{GS86,Win85}.
\item Turning now to the first of our case studies, a number of interesting further developments suggest themselves.
Firstly, from the results of Chapter~5, we can define a fully abstract denotational semantics for SCCS in our denotational metalanguage, and faithfully interpret Hennessy-Milner logic into our domain logic.
Thus we should {\em automatically} get a compositional proof theory for HML.
It would be particularly worthwhile to demonstrate this in detail, as the construction of compositional proof systems for HML by Stirling \cite{Sti87} and Winskel \cite{Win85} is one of the most impressive examples to date of the exercise of {\it ad hoc} ingenuity in the design of program logics.

Other useful extensions of our work would be to equivalences other then bisimulation (hard); and to countable non-determinism, using Plotkin's powerdomain for countable non-determinism \cite{Plo82}.
An interesting point about this construction is that  we lack a good representation for it, and a logical description might help.
\item Our development of the lazy $\lambda$-calculus represents no more than a beginning.
An extensive study is being undertaken by Luke Ong; anyone interested in pursuing the subject further is strongly recommended to read his forthcoming thesis 
(Imperial College, University of London; expected 1988).
\item Some more general points concerning the two case studies.
Firstly, the operational models we study---labelled transition systems in Chapter~5 and lambda transition systems in Chapter~6---are {\em almost} derived in a systematic way from our domain equations.
Namely, a labelled transition system is a map
\[ {\rm Proc} \longrightarrow \wp (({\sf Act} \times {\rm Proc} ) \cup \{ \bot \} ) \]
i.e. a coalgebra of the functor (on {\bf Set})
\[ X \mapsto \wp (({\sf Act} \times X ) \cup \{ \bot \} ) . \]
Similarly, an applicative transition system is a coalgebra of the {\bf Set}-functor
\[ X \mapsto (X \rightarrow X) \cup \{ \bot \} . \]
Since ${\sf Act} \times {\cal D} \cup \{ \bot \}$ can be put in natural 
bijection with $\sum_{a \in {\sf Act}} {\cal D}$, and $( {\cal D} \rightarrow {\cal D} ) \cup \{ \bot \}$ with $( {\cal D} \rightarrow {\cal D} )_{\bot}$, we see that our domain equations give rise to essentially the {\em same} functors, but over domains rather than sets.
Moreover, because of the limit-colimit coincidence in Domain theory \cite{SP82}, we can take the {\em initial solution} of a domain equation (with respect to embeddings) as the {\em final coalgebra} (with respect to  projections).
Thus our results can in some sense be seen as concerning the interpretation and ``best approximation'' of {\bf Set}-based structures in topological ones.
Clearly some general theory is called for here.
\item Finally, one of our aims in Chapters~5 and~6 was to place the study of functional languages and concurrency on as similar a footing as possible.
Much remains to be done here, although we hope to have made a useful first step.
\end{enumerate}

\addcontentsline{toc}{chapter}{References}
\bibliography{biblio}
\end{document}